%% file: PPNP-june2010.tex
\documentclass[twoside,12pt]{article}
\usepackage{epsfig}
\usepackage{graphicx}

\newcommand{\be}{\begin{equation}}
\newcommand{\ee}{\end{equation}}
\newcommand{\bea}{\begin{eqnarray}}
\newcommand{\eea}{\end{eqnarray}}

\begin{document}

\title{ \vspace{1cm} Charmonium and bottomonium in heavy-ion
collisions}
\author{R.~Rapp,$^1$ D.~Blaschke$^{2,3}$ and P.~Crochet$^{4}$\\
$^1$Cyclotron Institute and Department of Physics \& Astronomy\\
Texas A{\&}M University, College Station, Texas 77843-3366, USA\\
Email: rapp@comp.tamu.edu\\
$^2$Institute of Theoretical Physics, University of Wroc{\l}aw\\
Max Born pl. 9, 50-204 Wroc{\l}aw, Poland\\
Email: blaschke@ift.uni.wroc.pl\\
$^3$Bogoliubov Laboratory of Theoretical Physics\\
JINR Dubna, 141980 Dubna, Russia\\
$^4$Clermont Universit\'e, Universit\'e Blaise Pascal \\
CNRS/IN2P3, LPC, BP 10448, F-63000 Clermont-Ferrand, France\\
Email: crochet@clermont.in2p3.fr}
\date{\today}
\maketitle
\begin{abstract}
We review the present status in the theoretical and phenomenological  
understanding of charmonium and bottomonium production in heavy-ion
collisions. We start by recapitulating the basic notion of ``anomalous 
quarkonium suppression'' in heavy-ion collisions and its recent 
amendments involving regeneration reactions. We then survey in some 
detail concepts and ingredients needed for a comprehensive approach
to utilize heavy quarkonia as a probe of hot and dense matter. The
theoretical discussion encompasses recent lattice QCD computations of 
quarkonium properties in the Quark-Gluon Plasma, their interpretations 
using effective potential models, inelastic rate calculations and 
insights from analyses of electromagnetic plasmas. We illustrate the 
powerful techniques of thermodynamic Green functions ($T$-matrices) 
to provide a general framework for implementing
microscopic properties of heavy quarkonia into a kinetic theory of 
suppression and regeneration reactions. The theoretical concepts are 
tested in applications to heavy-ion reactions at SPS, RHIC and LHC. 
We outline perspectives for future experiments on charmonium and
bottomonium production in heavy-ion collisions over a large range in
energy (FAIR, RHIC-II and LHC). These are expected to provide key insights
into hadronic matter under extreme conditions using quarkonium observables.
\end{abstract}
\setcounter{page}{0}
\newpage
\tableofcontents
\newpage

\section{Introduction}
\input{intro}


\section{Equilibrium Properties}
\label{sec_equil}
As emphasized in the Introduction, the basic quantity encoding the 
equilibrium properties of a quarkonium state of given quantum number 
$\alpha$, is its spectral function, $\rho_\alpha$. Roughly speaking, 
$\rho_\alpha(\omega)$ characterizes the spectral distribution
of strength as a function of energy, $\omega$. Bound or resonance states
manifest themselves as peaks with well defined mass, $m_\alpha$, and
spectral width, $\Gamma_\alpha$ (in principle, more than one state is 
possible for a given $\alpha$, e.g., in vacuum $J/\psi$, $\psi'$ etc.
for $J^{PC}=1^{--}$, recall Tab.~\ref{tab_psi-vac}).  
Note, however, that in the context of charmonium production in
heavy-ion collisions, it is unlikely that spectral modifications
can be directly detected in the dilepton invariant-mass spectrum of
(vector-meson) quarkonium decays\footnote{E.g., the $J/\psi$ lifetime 
in free space is 
$\tau_{J/\psi}^{\rm vac}=1/\Gamma_{J/\psi}^{\rm tot,vac}
\simeq 2000$\,fm/$c$, compared to a typical fireball lifetime of
$\sim$10\,fm/$c$. This means that the ratio of $J/\psi$ decays inside
to outside the medium is about 1/200. If the average in-medium $J/\psi$
width is $\sim$200\,MeV, a detection of this effect would require a mass
resolution of about 1\,MeV, corresponding to an unrealistic 0.03\% at
the $J/\psi$ mass.}. Thus, relevant connections of the in-medium 
spectral properties of quarkonia to URHIC phenomenology are:
(a) their masses, $m_\Psi$, which determine the equilibrium abundances,
(b) their inelastic widths, $\Gamma_\Psi^{\rm inel}$, which determine 
formation and destruction rates (or chemical equilibration times),
(c) their dissolution temperatures, $T_\Psi^{\rm diss}$, which determine 
the absence of formation processes for $T>T_\Psi^{\rm diss}$, and,
(d) their elastic widths, $\Gamma_\Psi^{\rm el}$, which affect 
momentum spectra (and determine the kinetic equilibration times).

First-principle information on quarkonium properties can be 
obtained from numerical lattice-discre\-tized computations in QCD at 
finite temperature. We start our discussion from this perspective by 
reviewing the current status of this approach in Sec.~\ref{ssec_lqcd}. 
As we will see, lattice-QCD (lQCD) simulations do not 
directly provide the physical spectral function, nor do they readily 
allow for insights into the mechanisms underlying ``observable" medium 
effects. Effective models thus play an important role in the  
interpretation of lQCD results, as well as in furnishing quantitative 
input for applications in heavy-ion reactions. A natural starting point 
is the extension of the potential model, which works well for quarkonium 
spectroscopy in vacuum, to finite temperatures; this is discussed in 
Sec.~\ref{ssec_pot}, with emphasis on recent developments incorporating 
information and constraints from lQCD. Potential models are typically
used to assess how medium effects in the 2-body interaction affect the 
bound-state spectrum (i.e., the location of the peaks in the spectral 
function). The thermodynamic $T$-matrix approach enables to assess 
both bound and continuum states. Of equal importance for phenomenological 
applications is the determination of inelastic reaction rates, both in the 
QGP and hadron gas (HG), which is elaborated in Secs.~\ref{ssec_diss-qgp} 
and \ref{ssec_diss-hg}, respectively. 
The problem of in-medium quarkonia shares several features with the
well-studied fermionic two-body problem in electromagnetic plasmas.
Techniques developed for the latter may thus provide useful insights
which we attempt to exhibit in Sec.~\ref{ssec_plasma}.
Regarding notation, we adopt 3 variants of
the single heavy-quark (HQ) mass with the following
meaning: $m_Q^0$: bare mass, $m_Q$: constant effective mass,
$m_Q^*$: temperature-dependent effective mass.

\subsection{Correlation and Spectral Functions from Lattice QCD}
\label{ssec_lqcd}
The properties of the $Q\bar Q$ interaction in a given hadronic 
(color-singlet) channel with quantum numbers $\alpha$ are encoded in 
the two-point correlation function of the pertinent current, $j_\alpha$. 
In coordinate space the correlation function (or correlator) is defined 
by the amplitude of creating the current at the origin, propagating it 
to point $(\tau,\vec r)$ and absorbing it,
\begin{equation}
G_\alpha (\tau,\vec r) =
\langle\langle j_\alpha(\tau,\vec r) j_\alpha^\dagger(0,\vec 0)
 \rangle\rangle \ .
\label{G_x}
\end{equation}
In a heat bath of temperature $T$ the amplitude corresponds to a 
thermal average over the partition function of the system.
For meson correlators, the QCD currents are given by quark bilinears,
$j_\alpha = \bar Q \Gamma_\alpha Q$, where $\Gamma_\alpha$ specifies
the spin-flavor channel. In momentum space the imaginary part of the 
(retarded) correlation function, $G_\alpha^R (\omega,\vec p)$, is 
commonly referred to as the spectral function,
\begin{equation}
\rho_\alpha(\omega,p) = -2~{\rm Im} G_\alpha^R(\omega,p) \ ,
\label{rho_alpha}
\end{equation}
where $\omega$ and $p$ denote the energy and 3-momentum modulus, 
respectively (in the literature, the spectral function is also denoted 
by $\sigma_\alpha=\rho_\alpha/2\pi$; to avoid redundancy in notation with 
the string tension and cross sections, we will not use this notation in
the present article). In the timelike regime, $\omega^2-p^2>0$, $\rho_\alpha$ 
characterizes the physical excitation spectrum in the channel $\alpha$,
which, in principle, can be measured in experiment.

In field theory, the implementation of temperature into the partition 
function involves the transformation of the vacuum transition rate in 
real time to imaginary time. This amounts to replacing time evolutions 
with thermal averages. The temporal direction is thereby restricted to 
the interval $[0,\beta]$ with the upper limit being identified with the
inverse temperature of the system, $\beta =1/T$.
The evaluation of thermal expectation values of
correlation functions, Eq.~(\ref{G_x}), is routinely performed in
lQCD. At large $r$, an exponential decay of the spatial correlator
renders the lowest mass state dominant,
$G_\alpha(r)\propto{\rm e}^{-m_\alpha^{scr} r}$, which can be used
to extract hadronic screening masses, $m_\alpha^{scr}$, at finite
temperature~\cite{DeTar:1987xb,Born:1991zz}.
For temporal correlators, usually projected onto a fixed 3-momentum,
$G(\tau,p)$, the limited extent of the $\tau$ interval renders
an extraction of physical ground-state masses much more difficult.
For a better comparison of quenched ($N_f$=0) and full ($N_f$=2,3) 
lattice QCD, it is customary to discuss temperature dependencies in units 
of $T_c$ in the respective simulation. This accounts at least qualitatively 
for the difference in the absolute values of critical temperature 
by comparing the results at roughly equal parton densities (which is 
often the relevant quantity in the discussion of medium effects). For
example, for $T_c^{\rm quench}$$\simeq1.5T_c^{\rm full}$, the factor of 
$\sim$3 larger gluon density in quenched QCD is roughly compensated by 
the extra quarks in $N_f$=3 QCD. In addition, the quantitative values 
for the (pseudo-) critical temperatures computed in unquenched lQCD, for 
a given number of quark flavors, are currently beset with a systematic 
uncertainty of ca.~$\pm$10-15\% corresponding to 
$T_c\simeq$~180$\pm$30\,MeV~\cite{Karsch:2007dt} (the 
uncertainty is smaller in quenched QCD with $T_c\simeq$~270\,MeV). 

As mentioned above, the information on the excitation spectrum in a 
given hadronic channel is encoded in the spectral function. The 
temporal correlator is related to the spectral function via
\begin{equation}
G_\alpha(\tau,p;T)=
\int\limits_0^\infty \frac{d\omega}{2\pi} \ \rho_\alpha(\omega,p;T) \
K(\omega,\tau;T)
\label{G-tau}
\end{equation}
with the finite-$T$ kernel
\begin{equation}
K(\omega,\tau;T)= \frac{\cosh[(\omega(\tau-1/2T)]}{\sinh[\omega/2T]} \ .
\end{equation}
The extraction of the spectral function thus requires an inverse 
integral transform, which, on a finite number of lattice points, is, 
in principle, not possible in a unique way.
In Ref.~\cite{Asakawa:2000tr} it has been suggested to
employ the so-called Maximum Entropy Method (MEM) to perform a
probabilistic reconstruction of the most likely spectral function.
Subsequently, this method has been widely used to extract hadronic
spectral functions from lQCD
correlators~\cite{Asakawa:2003re,Datta:2003ww,Morrin:2005zq,
Jakovac:2006sf,Aarts:2007pk}. 
\begin{figure}[!tb]
\hspace{-0.2cm}
\begin{minipage}{0.5\linewidth}
\includegraphics[width=1.0\textwidth]{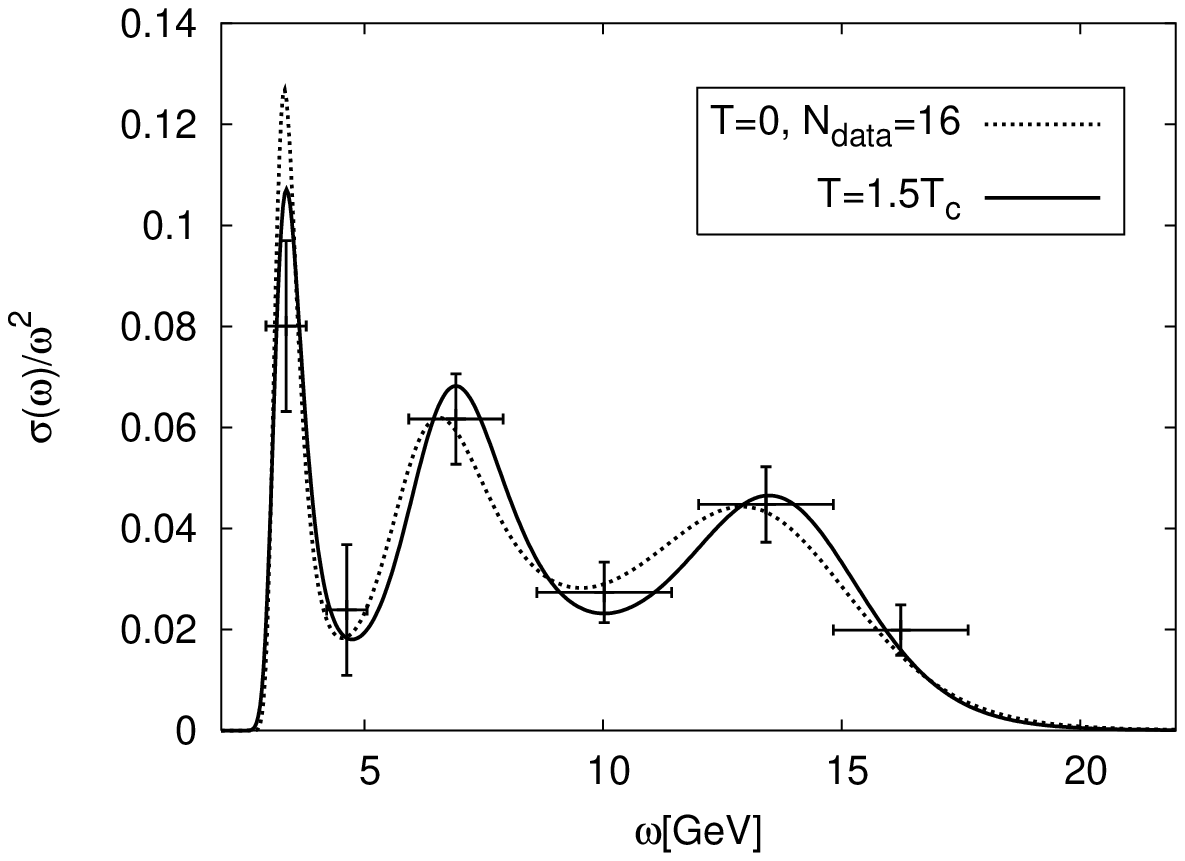}
\end{minipage}
\hspace{-0.4cm}
\begin{minipage}{0.5\linewidth}
\includegraphics[width=1.0\textwidth]{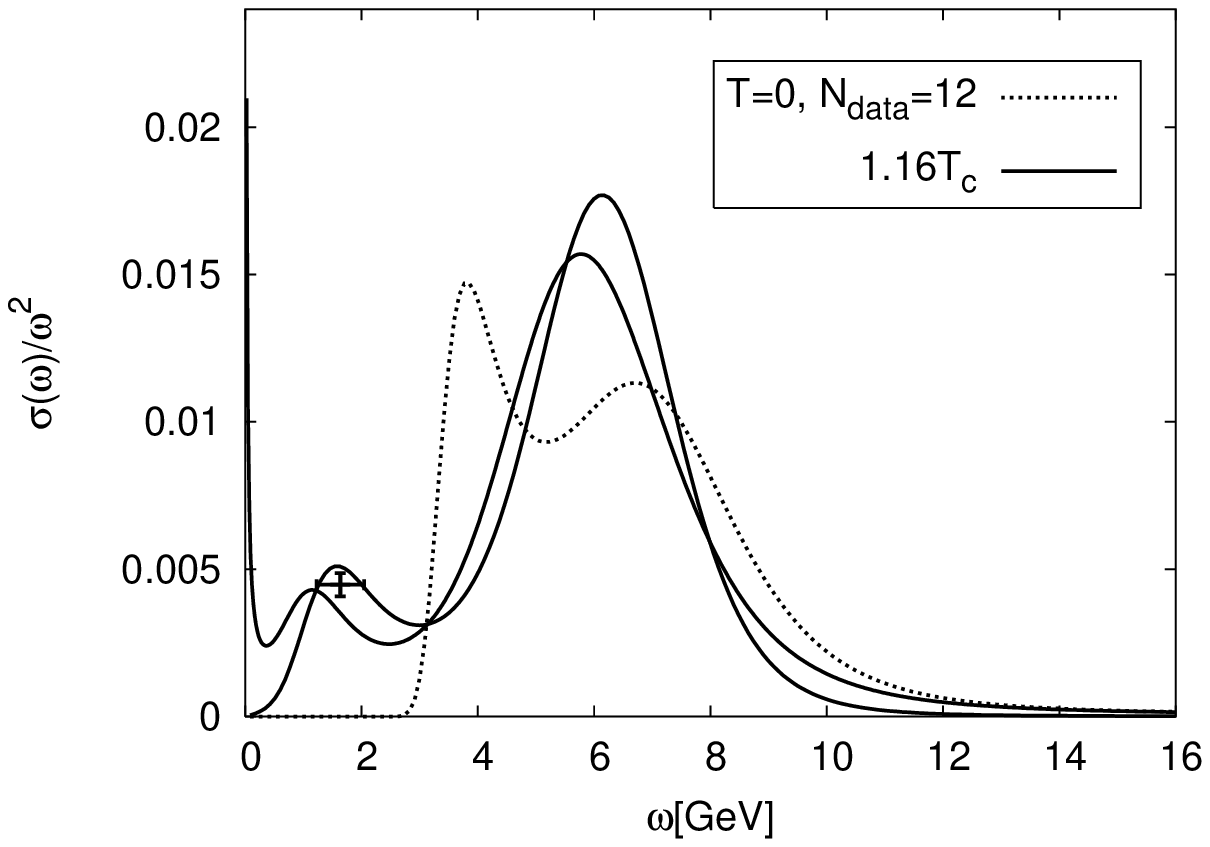}
\end{minipage}
\vspace{0.6cm}

\begin{minipage}{0.5\linewidth}
\hspace{0.6cm}
\includegraphics[width=.88\textwidth]{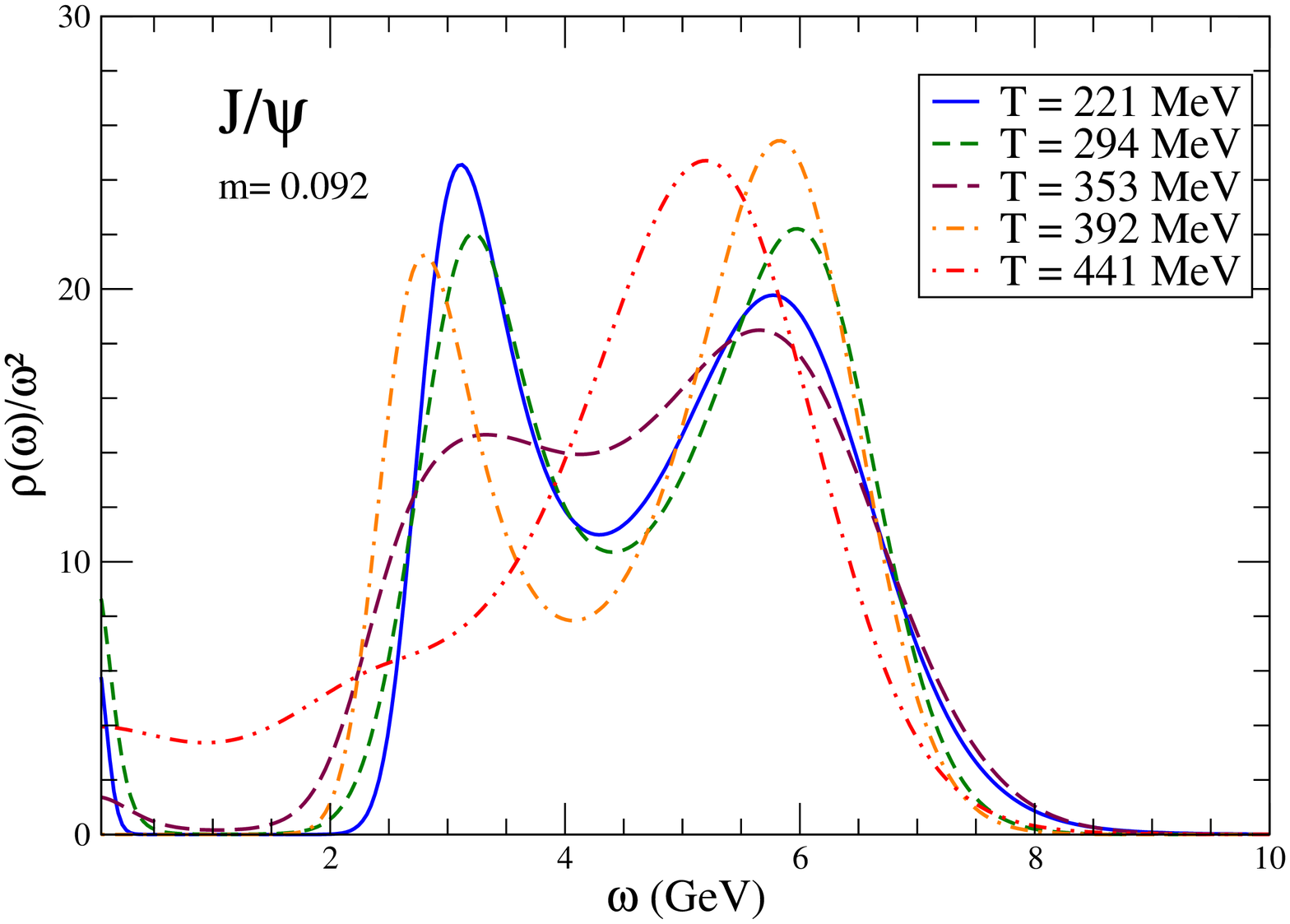}
\end{minipage}
\hspace{0.4cm}
\begin{minipage}{0.5\linewidth}
\includegraphics[width=.88\textwidth]{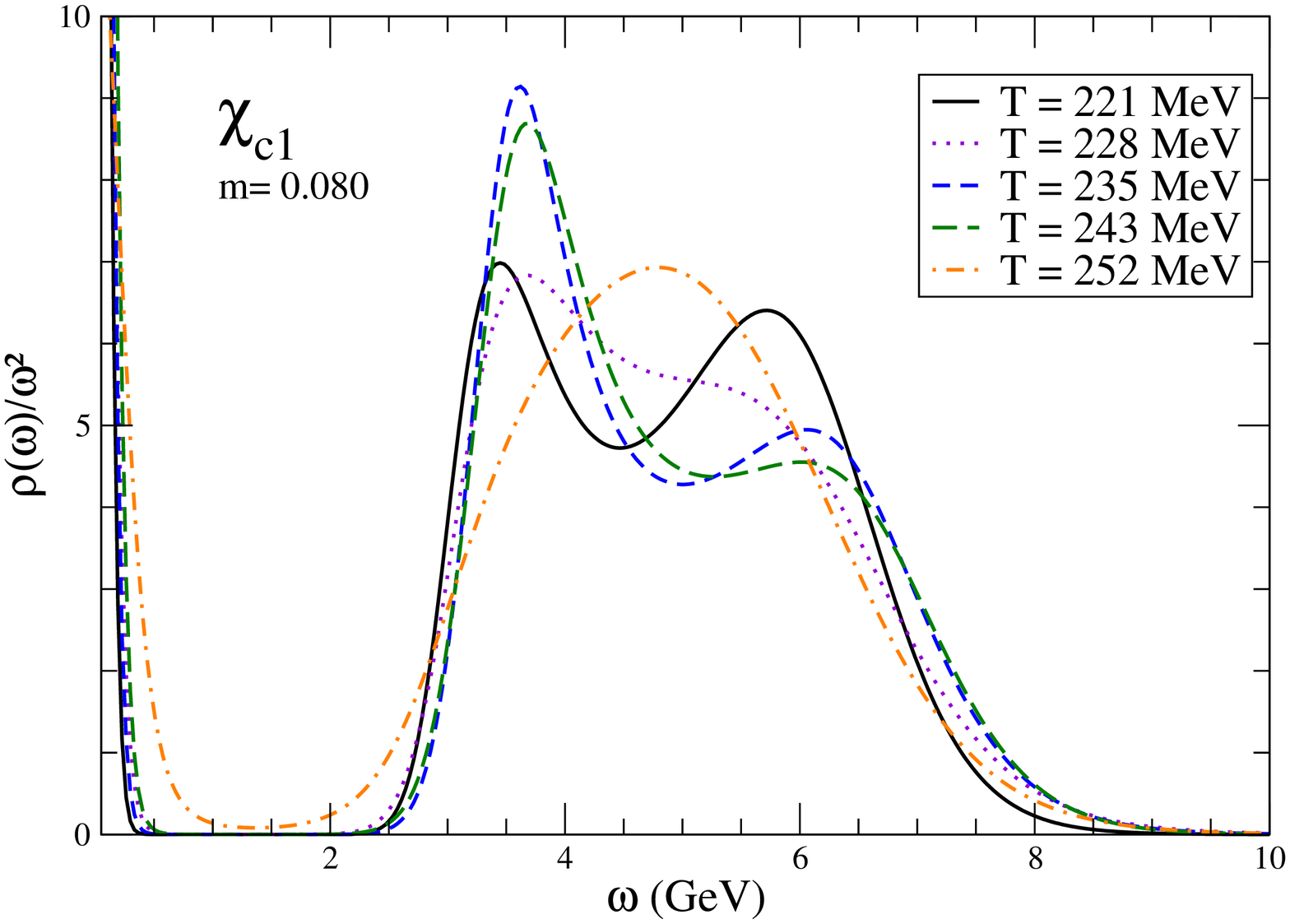}
\end{minipage}
\caption{Charmonium spectral functions as a function of energy 
($\omega$) and at vanishing 3-momentum ($p$=0) as evaluated
from lQCD correlation functions using the maximum entropy method (MEM).
Upper panels: $\eta_c$ (left) and $\chi_c$ (right) in quenched QCD
(with $T_c$=295\,MeV),
compared to their vacuum spectral functions (dotted
lines)~\cite{Jakovac:2006sf}; the $S$-wave ($\eta_c$) signal at
1.5\,$T_c$ is essentially unmodified compared to the vacuum, while the
$P$-wave ($\chi_c$) signal at 1.16~$T_c$ is largely distorted (the 
``data'' points in the upper left panel, as well as the two solid lines
in the right panel, indicate the uncertainty in the MEM reconstruction
procedure).
Lower panels: $J/\psi$ (left) and $\chi_c$ (right) computed in full QCD
for 2 light flavors~\cite{Aarts:2007pk} (the $y$ scale is in arbitrary units); 
the critical temperature in these simulations is $T_c\simeq210$\,MeV (slightly
larger than typical values in the range of 160-190\,MeV); a $J/\psi$ signal 
around $\omega$$\simeq$3\,GeV survives up to temperatures of at least
392\,MeV$\simeq$1.9\,$T_c$ while the $\chi_c$ peak is essentially smeared
out at $T$=252\,MeV$\simeq$1.2\,$T_c$ (the quark masses, $m$, quoted in 
the lower panels are unrenormalized input values; after renormalization,
they give rise to physical zero-temperature charmonium masses
within 10-15\,\% of the experimental values).}
\label{fig_SF-lqcd}
\end{figure}
A few examples in various
charmonium channels are collected in Fig.~\ref{fig_SF-lqcd}.
The general picture emerging from these calculations is that the ground
state $S$-wave charmonia, i.e., the $\eta_c$ ($J^P$=$0^-$) and $J/\psi$
($J^P$=$1^-$), are surviving in the QGP up to temperatures of
$\sim$2\,$T_c$, while the $P$-wave states, $\chi_c$ ($J^P$=$0^+,1^+,2^+$),
dissolve at temperatures slightly above the critical one,
$\sim$1.2\,$T_c$ or so. 
The first excited $S$-wave state ($\psi'$), cannot be resolved numerically,
but presumably dissolves earlier than the stronger bound
$\chi_c$ states, possibly even below $T_c$. The $\psi'$ should therefore
be quite sensitive to modifications in the hadronic phase of a
heavy-ion collision. Ground-state bottomonia seem to survive to
even higher temperatures, 2.5-3~$T_c$ or
more~\cite{Jakovac:2006sf,Petrov:2005ej}.

\begin{figure}[!tb]
\begin{minipage}{0.5\linewidth}
\includegraphics[width=0.95\textwidth]{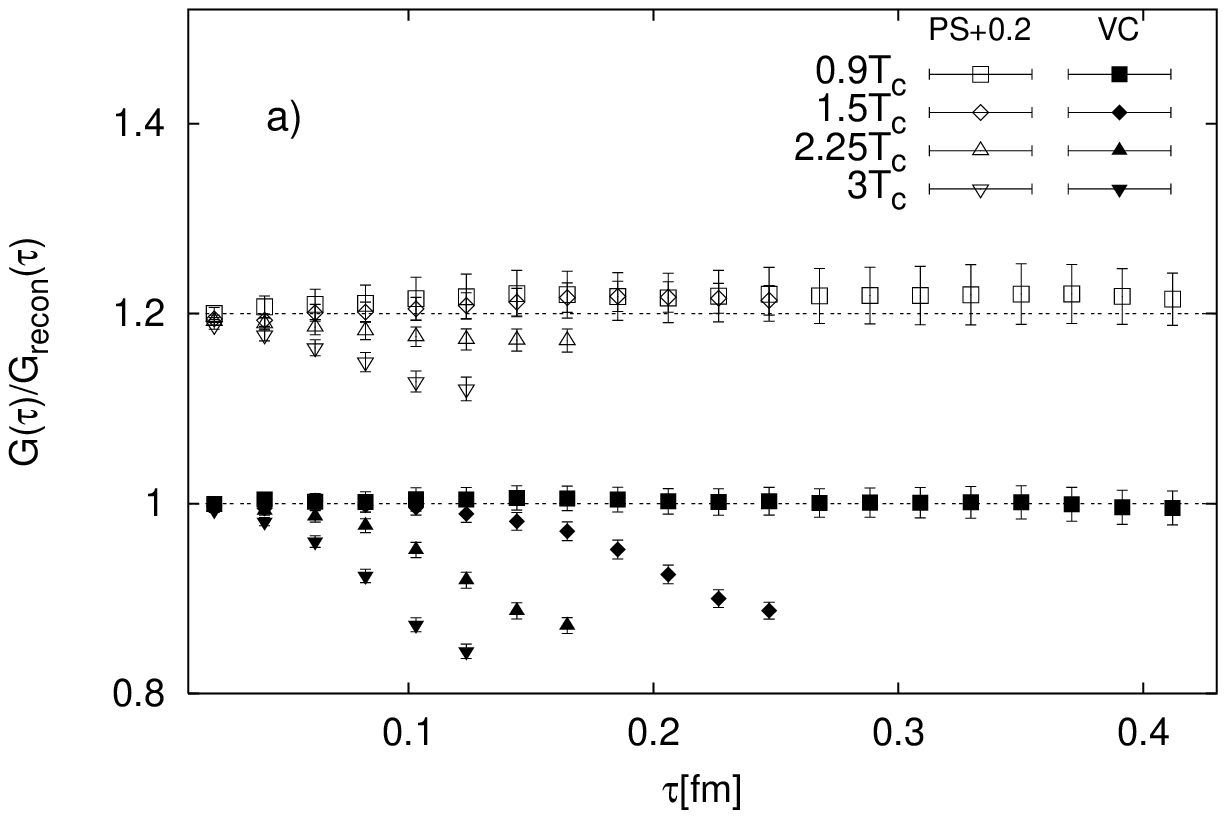}
\end{minipage}
\begin{minipage}{0.5\linewidth}
\includegraphics[width=0.95\textwidth]{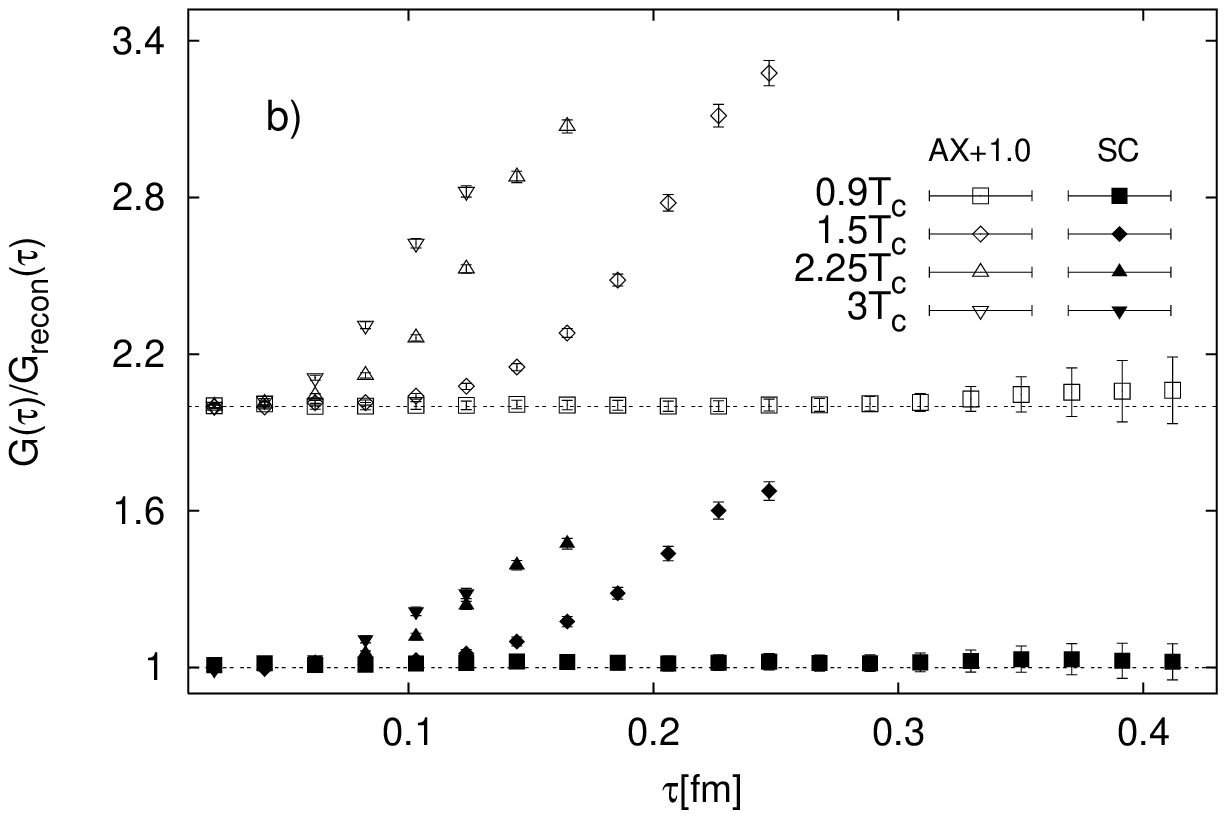}
\end{minipage}
\vspace{0.6cm}

\hspace{0.5cm}
\begin{minipage}{0.5\linewidth}
\includegraphics[width=0.88\textwidth]{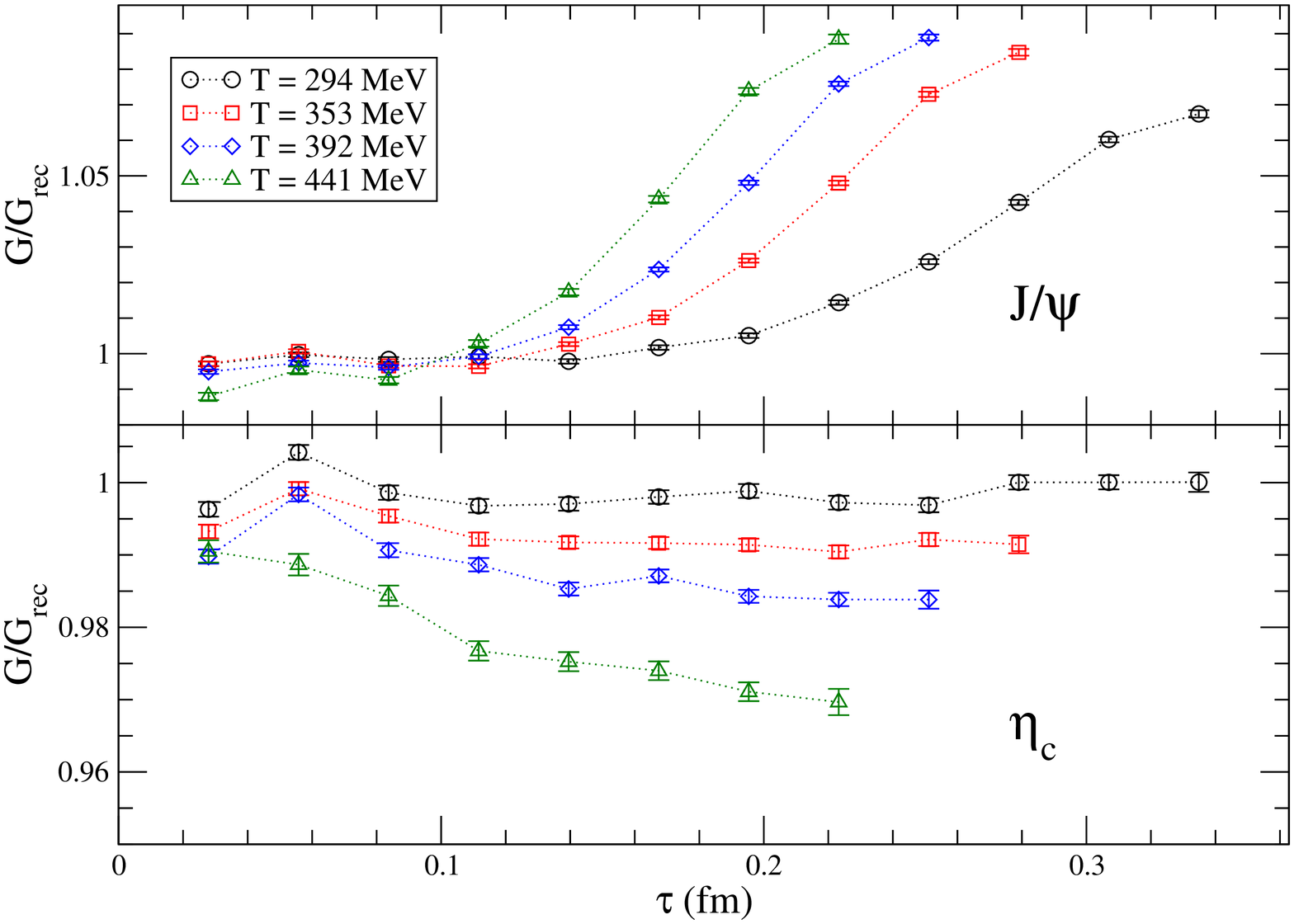}
\end{minipage}
\begin{minipage}{0.5\linewidth}
\includegraphics[width=0.88\textwidth]{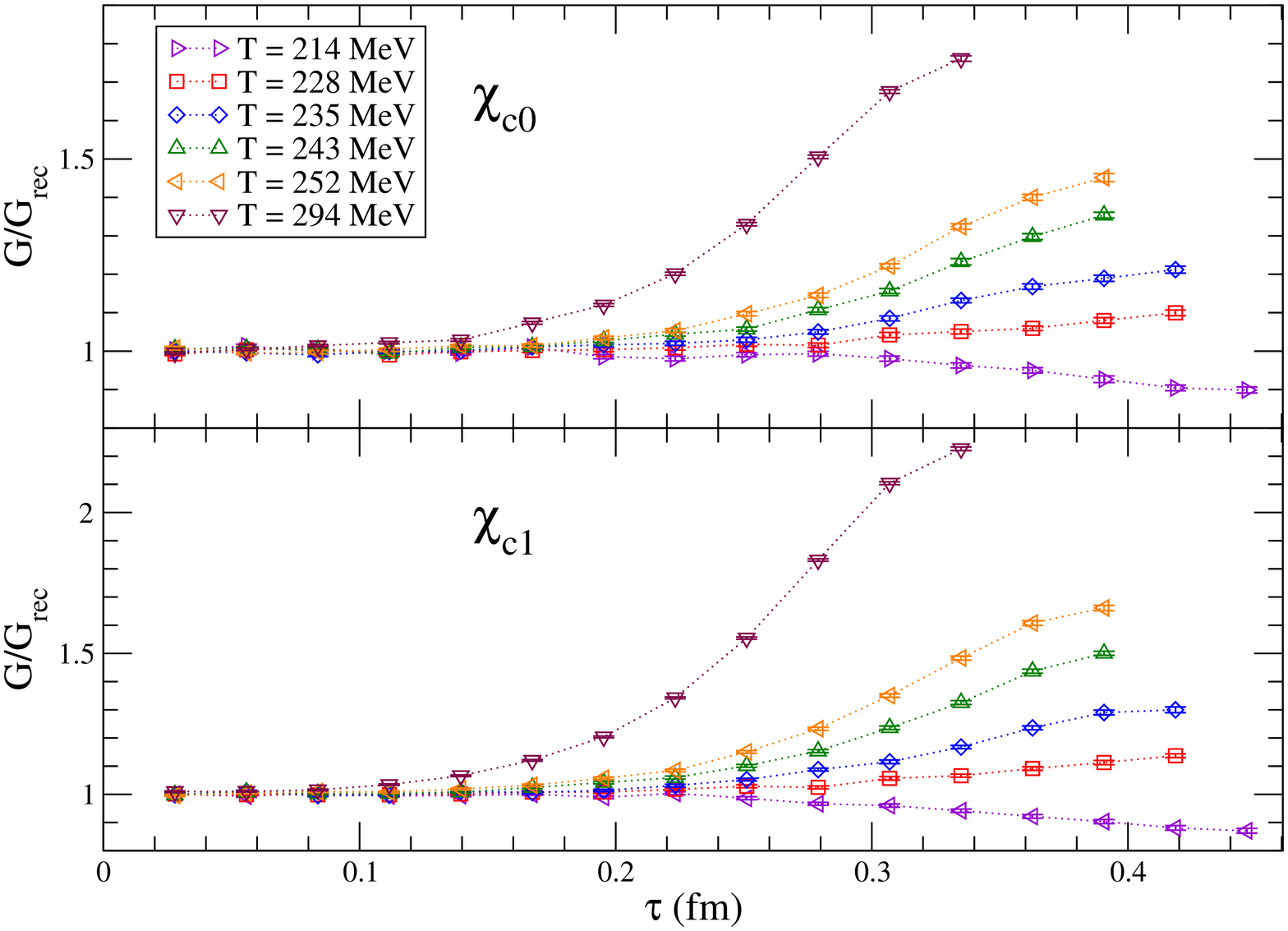}
\end{minipage}
\caption{Temporal charmonium correlators at zero 3-momentum computed for 
$VC$=$J/\psi$,  $PS$=$\eta_c$ (left panels) and  $SC$=$\chi_{c0}$,
$AX$=$\chi_{c1}$ (right panels) in quenched (upper panels)~\cite{Datta:2003ww}
and $N_f$=2 (lower panels)~\cite{Aarts:2007pk} lattice QCD.
The critical temperature in these simulations is $T_c\simeq270$\,MeV
and 210\,MeV, respectively. Note the comparatively small variations
in the $S$-waves (left panels), and the rather large increase in the
$P$-waves (right panels) indicative for additional low-energy strength
relative to the reconstructed correlator. For clarity, the $PS$ and
$AX$ correlators in the upper panels have been offset by the indicated
constant.} 
\label{fig_G-lqcd}
\end{figure}
It is instructive to examine the underlying temporal correlators,
examples of which are displayed in Fig.~\ref{fig_G-lqcd} (at vanishing
3-momentum, $p$=0). To facilitate their interpretation, they have
been normalized to a so-called reconstructed correlator,
which is evaluated with the kernel at temperature $T$,
\begin{equation}
G_{\rm rec}^\alpha(\tau;T) = \int\limits_0^\infty \frac{d\omega}{2\pi} 
\ \rho_\alpha(\omega;T^*)
\ K(\omega,\tau;T) \ ,
\end{equation}
but employs a spectral function at a low temperature, $T^*$, where no
significant medium effects are expected. The correlator ratio,
\begin{equation}
R_G^\alpha(\tau,T)=G_\alpha(\tau;T) / G_{\rm rec}^\alpha(\tau;T) \ ,
\label{RG}
\end{equation}
is then an indicator of medium effects in $G_\alpha(\tau;T)$ through 
deviations from 1 (note that a normalization to $G_\alpha(\tau;T^*)$ 
is not meaningful due to different $\tau$ ranges for $T^*$ and $T$).
The stability of the temporal correlator ratios, i.e., their small
deviations from one, in the vector and pseudoscalar channels has been 
associated with the survival of the
ground-state $S$-wave charmonia. The variations for the $\chi_c$ (which
are $P$ waves) are much larger and set in much closer to $T_c$ (e.g., up 
to 40-50\% at $T$=1.2~$T_c$). The opposite trend of the $J/\psi$ correlators
in the upper left and lower left panel could be due to the different 
vector components considered ($\sum_{\mu=0-3} G_V^{\mu\mu}$ in the upper 
panel vs. $\sum_{i=1-3} G_V^{ii}$ in the lower panel).

\begin{figure}[!tb]
\includegraphics[width=.35\textwidth]{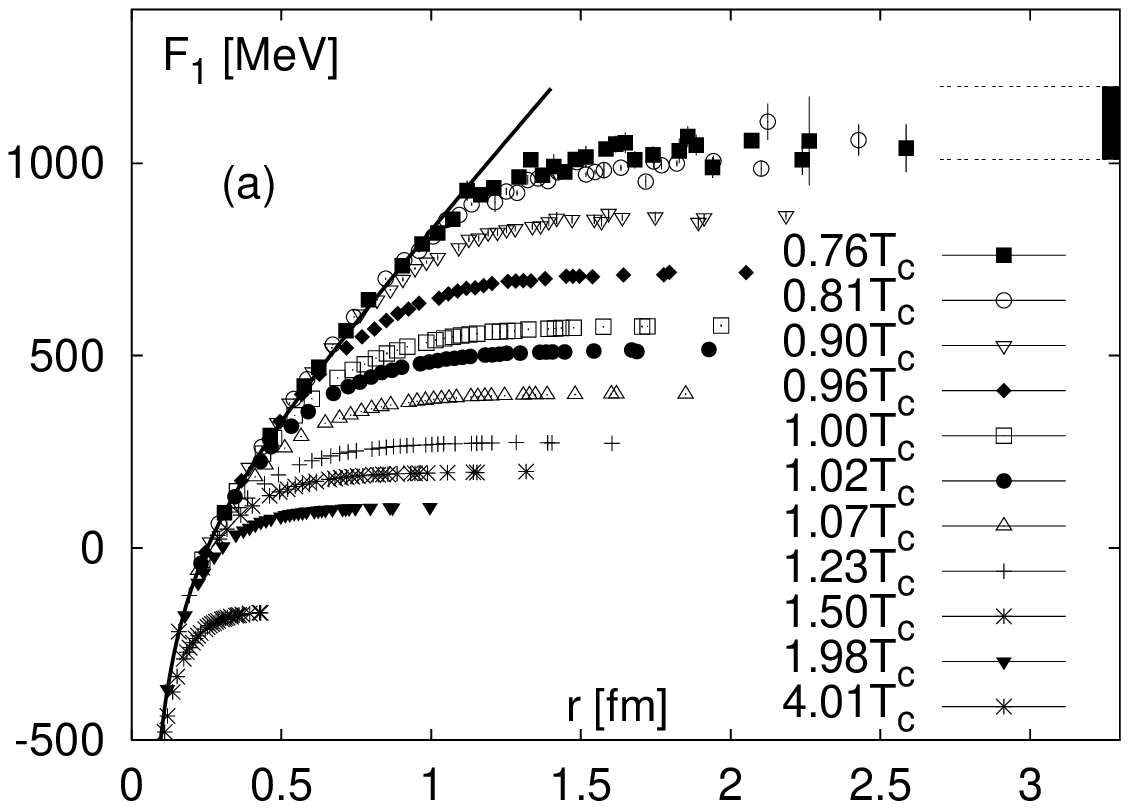}
\hspace{-0.7cm}
\includegraphics[width=.35\textwidth]{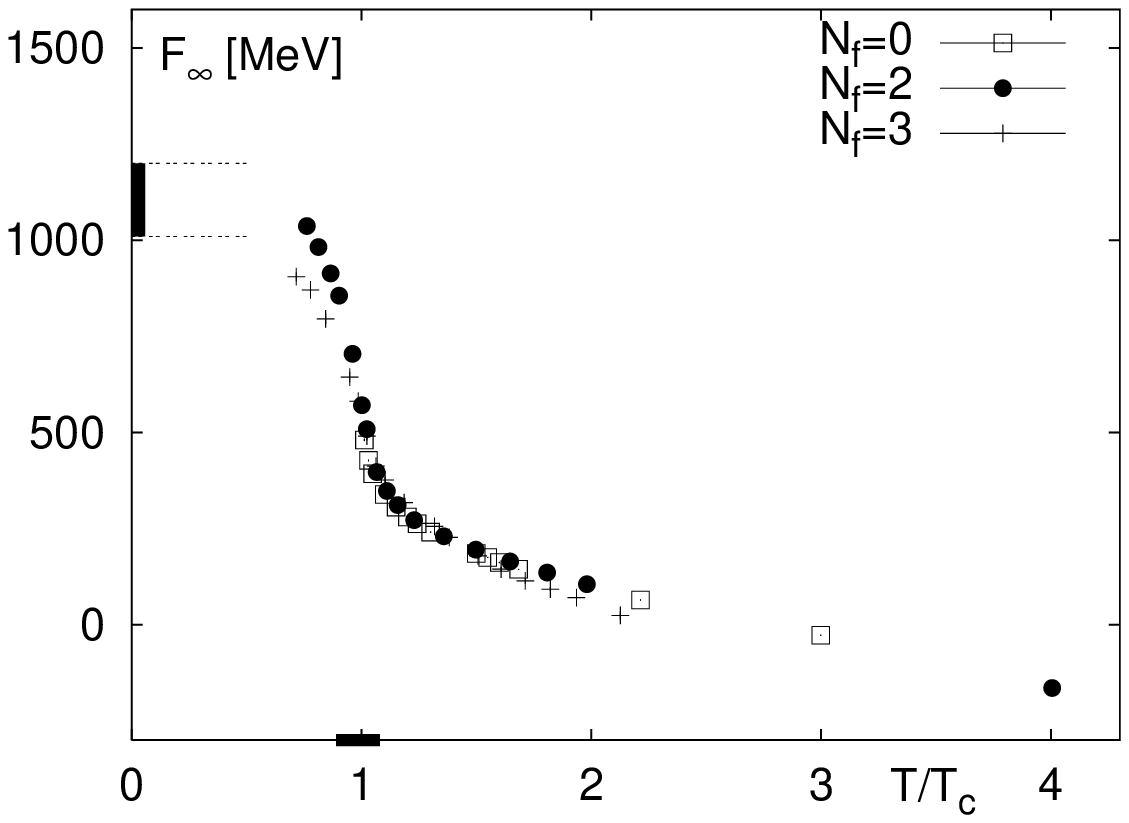}
\hspace{-0.7cm}
\includegraphics[width=.35\textwidth]{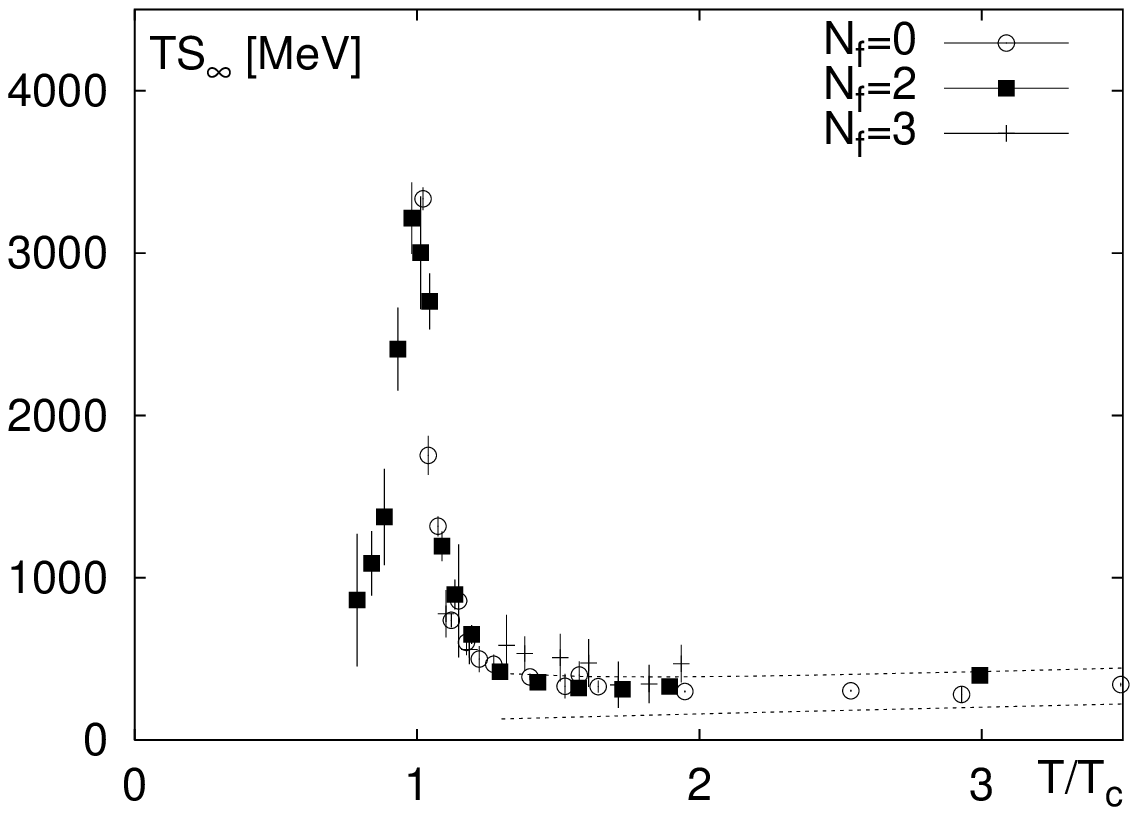}
\caption{Left panel: free energy of a color-singlet heavy
quark-antiquark pair as a function of its size for various temperatures
as computed in $N_f$=2 lattice QCD~\cite{Kaczmarek:2005ui}; the vertical
bar in the upper left part indicates the estimated string-breaking
energy, $V_{Q\bar Q}^{(1)}(r_{\rm sb})\simeq 1.1$\,GeV, in vacuum.
Middle panel: temperature dependence of the asymptotic value of the 
color-averaged free energy, $F_{Q\bar Q}^\infty(T)$, in quenched
($N_f$=0)~\cite{Kaczmarek:2002mc}, $N_f$=2~\cite{Kaczmarek:2005ui} and
$N_f$=3~\cite{Petreczky:2004pz} lQCD; the pertinent critical
temperatures are $T_c$=270, 200 and 193\,MeV, respectively.
Right panel: entropy contribution to the color-averaged free energy
at asymptotically large quark-antiquark separation corresponding
to the free energies of the middle panel. The band enclosed by the
dotted lines for $T\ge1.3T_c$ indicates a perturbative estimate of
$TS_{Q\bar Q}^\infty(T)$.}
\label{fig_F-lat}
\end{figure}

Another quantity that can be computed with good accuracy in lQCD, and
which encodes information about charmonium properties at zero and finite 
$T$, is the free energy, $F_{Q\bar Q}(r)$, of a static pair of heavy 
quark and antiquark (more precisely, lQCD obtains $F_{Q\bar Q}(r)$ as 
the difference between free energies for a thermal system with and 
without the $Q\bar Q$ pair).
In gluo-dynamics the large distance limit of the free energy,
$F_{Q\bar Q}^\infty \equiv F_{Q\bar Q}(r\to\infty)$, can be related to
the expectation value of the Polyakov loop~\cite{McLerran:1980pk},
\begin{equation}
{\rm e}^{-F_{Q\bar Q}^\infty/T} = |\langle L \rangle |^2  \  ;
\end{equation}
$L$ characterizes a heavy-quark (HQ) source whose expectation value is 
zero (finite) in the (de-) confined phase, and thus serves as an order
parameter of deconfinement. At zero temperature, the free energy
computed in the color-singlet channel (which we denote by a superscript
``(1)") closely follows the form expected for a phenomenological 
Cornell-type potential~\cite{Bali:2000gf,Brambilla:2004wf,Kaczmarek:2005ui}, 
\begin{equation}
F_{Q\bar Q}^{(1)}(r;T=0) = -\frac{4}{3} \frac{\alpha_s}{r} + \sigma r \ ,
\label{Fqq-vac}
\end{equation}
recall Eq.~(\ref{Vqq-vac})\footnote{Strictly speaking, the projection of 
the color-averaged free energy as computed in lattice QCD, onto color-singlet 
and -octet channels is not gauge invariant~\cite{Philipsen:2008qx}. However,
they can be constrained at sufficiently short distances where temperature 
effects are not operative, or guided by Casimir scaling relations 
inferred from perturbative QCD.}. 
This finding, in connection with the development of HQ effective theories, 
has confirmed a posteriori the suitability of potential models to
quantitatively compute charmonium and bottomonium spectroscopy in free
space based on first principles~\cite{Brambilla:2004wf}. In the presence
of light quarks, string breaking occurs, being characterized by
$F_{Q\bar Q}^{(1)}(r;T=0)$ leveling off at a finite value. This may be 
interpreted as an effective quark (or $D$-meson) mass according to 
$m_c = m_c^0 + F_{Q\bar Q}^{(1),\infty}/2$ (with a bare quark mass of
$m_c^0\simeq$~1.2-1.3\,GeV).
Numerically, with a typical string-breaking separation of 
$r_{\rm sb}\simeq 1.2$\,fm, one finds 
$F_{Q\bar Q}^{(1)}(r_{\rm sb};0)\simeq 1.1$\,GeV, which roughly
recovers the empirical $D$-meson mass.
At finite temperature, lQCD computations for $F_{Q\bar Q}^{(1)}(r;T)$
find the expected color-Debye screening, which gradually penetrates
to smaller distances as the temperature increases, cf.~left panel
of Fig.~\ref{fig_F-lat}.
The implications for HQ bound states (or spectral functions) are, 
however, quite subtle. First, at finite $T$ the free energy
receives an extra contribution due to an entropy term,
\begin{equation}
F_{Q\bar Q}(r;T)= U_{Q\bar Q}(r;T) - T S_{Q\bar Q}(r;T)  \ . 
\label{Fqq-med}
\end{equation}
It is currently not clear whether the internal energy,
$U_{Q\bar Q}(r;T)$, or the free one, $F_{Q\bar Q}(r;T)$, is a more
appropriate quantity to be identified with a finite-temperature HQ
potential. Second, the temperature introduces an extra scale into the
problem which needs to be properly implemented into the construction of
a HQ effective theory~\cite{Escobedo:2008sy,Brambilla:2008cx}. Third, 
the presence of the entropy term, which
does not vanish at large separation $r$, renders the identification of
an effective HQ mass more problematic, especially close to $T_c$, where
$TS_{Q\bar Q}^{\infty}$ becomes very large, see right panel of
Fig.~\ref{fig_F-lat}.
Nevertheless, the application of potential models
at finite temperature has provided valuable insights into heavy
quarkonium properties at finite $T$. In particular, the synthesis of
independent information on correlators and free
energies from lQCD has enabled interesting insights, as
we discuss in the following Section.

\subsection{Potential Models in the QGP}
\label{ssec_pot}
\begin{figure}[!tb]
\begin{center}
\includegraphics[width=.45\textwidth]{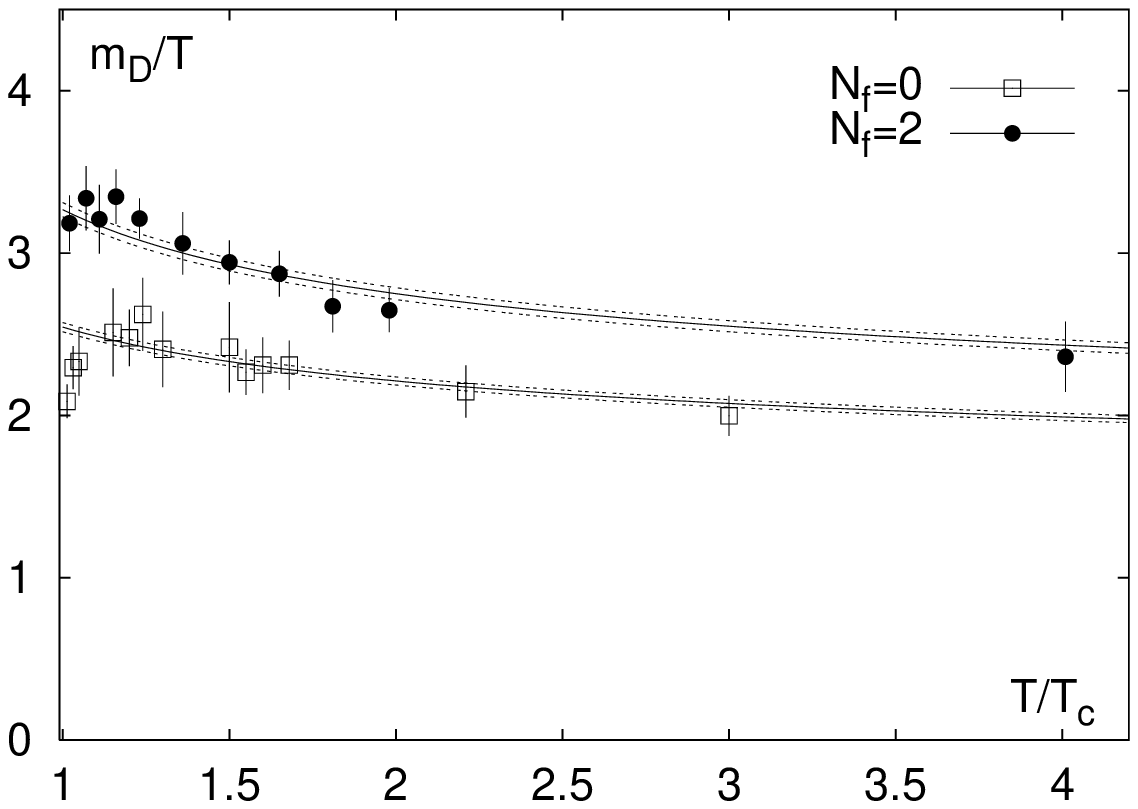}
\end{center}
\caption{Nonperturbative extraction of the Debye mass ($m_D$$\equiv$$\mu_D$) 
based on exponential
fits of a screened Coulombic term to the in-medium color-singlet HQ free
energy in $N_f$=2 (solid circles)~\cite{Kaczmarek:2005ui} and quenched 
($N_f$=0, open squares)~\cite{Kaczmarek:2004gv} lQCD. The bands are
analytic fits using the perturbative form of the Debye mass,
$\mu_D/T=A (1+N_f/6)^{1/2} g_{\rm 2-loop}$ with a 2-loop finite-$T$
running coupling and a multiplicative constant $A$$\simeq$1.5 to
account for nonperturbative effects.
Absolute units are obtained by using the pertinent critical
temperatures of $T_c\simeq200(270)$\,MeV for $N_f$=2(0).}
\label{fig_muD}
\end{figure}

Early applications~\cite{Karsch:1987pv} of in-medium heavy-quark 
potentials have employed a phenomenological ansatz to implement 
color-screening effects in a deconfined plasma into the Cornell 
potential, Eq.~(\ref{Vqq-vac}), via 
\begin{equation}
V_{Q\bar Q}(r;T)=\frac{\sigma}{\mu_D(T)} \left( 1-{\rm e}^{-\mu_D(T)r}\right)
-\frac{4\alpha_s}{3r} \ {\rm e}^{-\mu_D(T)r}  \ . 
\label{Vqq-med}
\end{equation}
The entire temperature dependence is encoded in the Debye mass, 
$\mu_D(T)$ (the vacuum potential is recovered for $\mu_D\to 0$).
Defining the Hamiltonian as
\begin{equation}
\hat{H}(r;T)= 2m_Q - \frac{\vec\nabla^2}{m_Q} + V_{Q\bar Q}(r;T) \ ,
\label{H_QQ}
\end{equation}
the Schr\"odinger equation for the bound state problem,
\begin{equation}
\hat{H} \ \phi_{nl}(r;\mu_D) = \omega_{nl}(\mu_D) \ \phi_{nl}(r;\mu_D) \ ,
\end{equation}
has been solved for the eigen-energies, $\omega_{nl}$, as a function of the
Debye mass. A state $(n,l)$ was considered to be dissolved at a ``critical"
Debye mass, $\mu_D^{\rm diss}$, if the ``dissociation" energy,
\begin{equation}
\omega_{nl}^{\rm diss} \equiv 2m_Q+ \frac{\sigma}{\mu_D} - \omega_{nl} \ ,
\end{equation}
reaches zero. Note that the term $\frac{\sigma}{\mu_D}$ corresponds to
the large-separation limit of the in-medium potential, Eq.~(\ref{Vqq-med}), 
cf.~also the discussion in the text following
Eq.~(\ref{Fqq-vac}) in the previous Section. The critical Debye masses for 
dissolution were found to be $\mu_D^{\rm diss}(n=1)\simeq0.7(1.6)$\,GeV 
for ground state ($n$=1) charmonium (bottomonium), and
$\mu_D^{\rm diss}(n=2)\simeq0.35(0.6)$\,GeV for the first excited states, 
$\psi'$ and $\chi_c$ ($\Upsilon'$ and $\chi_b$)~\cite{Karsch:1987pv}. 
Recent lQCD computations of
the in-medium Debye mass based on the Coulombic term in the color-singlet
free energy are displayed in Fig.~\ref{fig_muD}~\cite{Kaczmarek:2005ui}.
Comparing these to the values for $\mu_D^{\rm diss}$ extracted from the 
screened Cornell potential, Eq.~(\ref{Vqq-med}), suggests dissociation
temperatures of all charmonia at or slightly above $T_c$.

The next step became possible with the availability of quantitative
lQCD results for heavy-quark free energies. In Ref.~\cite{Digal:2001ue},
the color-singlet free energy, $F_{Q\bar Q}^{(1)}$ was used in a
Schr\"odinger equation, and $J/\psi$ dissolution occurred at about
1.1~$T_c$, while $\chi_c$ ($\psi'$) were found to melt (well) below
$T_c$. The recent lQCD spectral function results, indicative for
$J/\psi$ states surviving well above $T_c$, appear to challenge this
conclusion (recall Fig.~\ref{fig_SF-lqcd}).
In Ref.~\cite{Shuryak:2004tx} it has been suggested to employ the
internal, rather than the free, energy as an interquark potential.
It was argued that $U_{Q\bar Q}$ is the appropriate quantity in the 
limit where the typical bound state (formation or life) time is much 
smaller than the typical time for heat exchange with the thermal
environment, $\tau_{\rm bound} \ll \tau_{\rm heat}$. In this case, the
entropy term is not active and should be removed from the free energy,
leading to the internal energy. While, according to Eq.~(\ref{Fqq-med}), 
one has $U_{Q\bar Q} > F_{Q\bar Q}$, the actually used
potentials become deeper upon subtraction of the asymptotic values of
the internal energy (recall that the force is given by the derivative 
of the potential). As noted in connection with 
Fig.~\ref{fig_F-lat}, the subtraction procedure is problematic in the 
immediate vicinity of $T_c$. This problem is currently the weakest link in
converting lQCD results to effective heavy-quark potentials. From a
pragmatic point of view, the use of $U_{Q\bar Q}$ in a Schr\"odinger
equation~\cite{Shuryak:2004tx,Wong:2004zr,Alberico:2005xw} improved
the qualitative agreement of the potential approach with the
lQCD spectral functions in that $J/\psi$ bound states are supported up
to temperatures of $\sim$2-2.5~$T_c$, while the $\chi_c$ dissolves at
around 1.1-1.2~$T_c$. In Ref.~\cite{Mannarelli:2005pz}, the internal
energies have been implemented into a $T$-matrix approach for
$Q$-$\bar Q$ (and $q$-$\bar q $) interactions, which has the important
advantage over the Schr\"odinger equation that it provides a unified
description of bound and scattering states (i.e., above the nominal
$Q$-$\bar Q$ threshold). For $S$-wave charmonia, it was found that
the lowest bound state moves into the continuum at $\sim$2~$T_c$,
after which it rapidly dissolves. However, it was also noted that
the extractions of the internal energy from different lQCD calculations
are not unique, thus adding to the uncertainty in the potential
definition. For example, in perturbative 
approaches~\cite{Laine:2006ns,Beraudo:2007ky} the free energy has 
been identified with the heavy-quark potential. In the vicinity
of a phase transition and/or in the presence of a confining term, this 
notion is less obvious. In Ref.~\cite{Satz:2008zc} it has been pointed 
out that, in the static limit, the thermal expectation value of the 
interaction part of the $\bar QQ$ Hamiltonian is precisely the internal
energy of the $\bar QQ$ pair and thus the appropriate quantity to serve
as in-medium two-body potential. Another alternative has been suggested 
in Ref.~\cite{Wong:2004zr} in terms of a linear combination of $F$ and 
$U$, based on the idea of subtracting the internal energy of the induced
gluon cloud. Clearly, the question of extracting a two-body potential 
from the lQCD free energy deserves further study. 

Model comparisons to lattice QCD ``data" can be made more quantitative
by computing the pertinent spectral functions, $\rho_\alpha(\omega)$,
within a given approach (see, e.g., Ref.~\cite{Rapp:2002pn} for an
early application in the light quark sector): the integral in
Eq.~(\ref{G-tau}) can then be easily carried out and
compared to lQCD results for Euclidean correlation functions.
In the following paragraphs we discuss such line of work within
potential models. One idea is that the different binding properties 
resulting from assuming either $F$ or $U$ as potential can be 
distinguished by such comparisons.

The first applications of a finite-$T$ potential model for charmonia
to lQCD correlators has been conducted in Ref.~\cite{Mocsy:2005qw}.
For the bound-state part a Schr\"odinger equation was solved using a
screened Cornell potential, Eq.~(\ref{Vqq-med}), as well as lQCD
internal energies, with fixed heavy-quark masses. The continuum
was modeled as non-interacting with a temperature dependent threshold 
$s_0=s_0(T)$, leading to a spectral function
\begin{equation}
\rho_\alpha(\omega;T)
=4\pi \sum\limits_i M_i F_i^2 \delta(\omega^2-M_i^2)
+\frac{3}{4\pi} \omega^2 \Theta(\omega-s_0) f_\alpha(\omega,s_0) \ 
\label{sig-schem}
\end{equation}
(the kinematic coefficients $f_\alpha$ depend on the meson channel 
$\alpha$, $F_i$ are wave-function overlaps at $r$=0, and the sum over
$i$ includes all bound states for a fixed $\alpha$).
While some trends of the lattice correlator ratios were reproduced, a
comprehensive agreement could not be achieved; similar results have
been obtained in Ref.~\cite{Alberico:2006vw}.

In Ref.~\cite{Cabrera:2006wh} the $T$-matrix approach introduced in
Ref.~\cite{Mannarelli:2005pz} has been further developed for a systematic 
analysis of heavy-quarkonium correlator ratios. Starting from the
covariant Bethe-Salpeter equation for two-body scattering, one can
apply standard reduction schemes to obtain a 3-dimensional
Lippmann-Schwinger equation. After a partial-wave expansion, the latter
becomes a 1-dimensional integral equation for the $Q$-$\bar Q$ $T$-matrix 
which for vanishing total 3-momentum ($p$=0) takes the form
\begin{equation}
T_l(\omega;q',q) = V_l(q',q) + \frac{2}{\pi} \int_0^{\infty} dk \,
k^2 \, V_l(q',k)\, G_{\bar{Q}{Q}}(\omega;k) \, T_l(\omega;k,q) \,
\hat f^{Q\bar{Q}}(\omega_k) \ ;
\label{LS}
\end{equation}
$G_{Q\bar Q}(\omega,k)$ denotes the intermediate heavy
quark-antiquark propagator,  $\hat f^{Q\bar{Q}}$=1$-$$f^Q$$-$$f^{\bar Q}$
with $f^{Q,\bar Q}$ HQ Fermi distributions, and $q$, $q'$ and $k$ are 
the (off-shell) 3-momenta of the quarks in the initial, final and 
intermediate state, respectively. 
\begin{figure}[!tb]
\begin{center}
\includegraphics[width=0.8\textwidth]{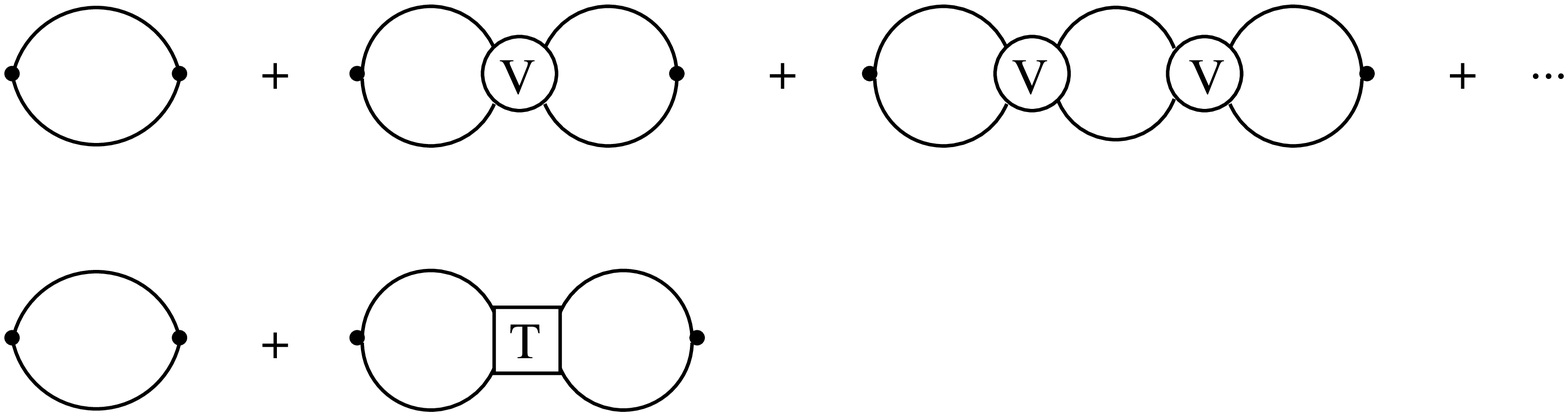}
\end{center}
\caption{Diagrammatic representation of the $Q$-$\bar{Q}$ correlation
function as obtained from resumming the HQ potential (upper panel)
resulting in a folding with the $T$-matrix (lower panel) corresponding 
to Eq.~(\ref{GTG}). The solid dots represent operators projecting into 
different mesonic channels $\alpha$ (giving rise to the coefficients
$a_l$ in Eqs.(\ref{G0}) and (\ref{delG})) and the 2 connecting lines 
represent the non-interacting $\bar QQ$ propagator $G_{\bar QQ}$; figure
taken from Ref.~\cite{Cabrera:2006wh}.}
\label{fig_gtg}
\end{figure}
The in-medium $T$-matrix equation 
can be derived from a finite-temperature Green's function 
approach~\cite{Ebeling:1986} and constitutes a consistent many-body 
framework to compute
in-medium 1- and 2-particle correlations (it has been widely applied,
e.g., in nuclear many-body theory~\cite{Machleidt:1989tm}, as well
as in the analysis of electromagnetic plasmas, as will be discussed
in Sec.~\ref{ssec_plasma}).
It is particularly suited for the problem at
hand since the potentials can be directly identified with the ones
extracted from the lattice (after Fourier transformation and
partial-wave expansion with angular-momentum quantum number $l$).
It treats bound and scattering states on the
same footing, which is mandatory for situations where
bound states gradually dissolve into a continuum. At the same time,
rescattering effects in the continuum, including possible resonance
formation, are accounted for by a full resummation of the potential,
without the need for matching procedures or $K$ factors. The 2-particle
propagator allows for the implementation of medium effects
on the heavy quarks via (complex) single-particle self-energies encoding
mass changes (real part) and finite widths (imaginary part). Charmonia
widths are important for phenomenological applications in heavy-ion
reactions and thus should be accounted for in correlator analyses.
\begin{figure}[!tb]
\begin{minipage}{0.5\linewidth}
\includegraphics[width=0.95\textwidth]{ImGccS-tmat.eps}
\end{minipage}
\begin{minipage}{0.5\linewidth}
\includegraphics[width=0.95\textwidth]{ImGccP-tmat.eps}
\end{minipage}
\begin{minipage}{0.5\linewidth}
\vspace{0.3cm}
\includegraphics[width=0.95\textwidth]{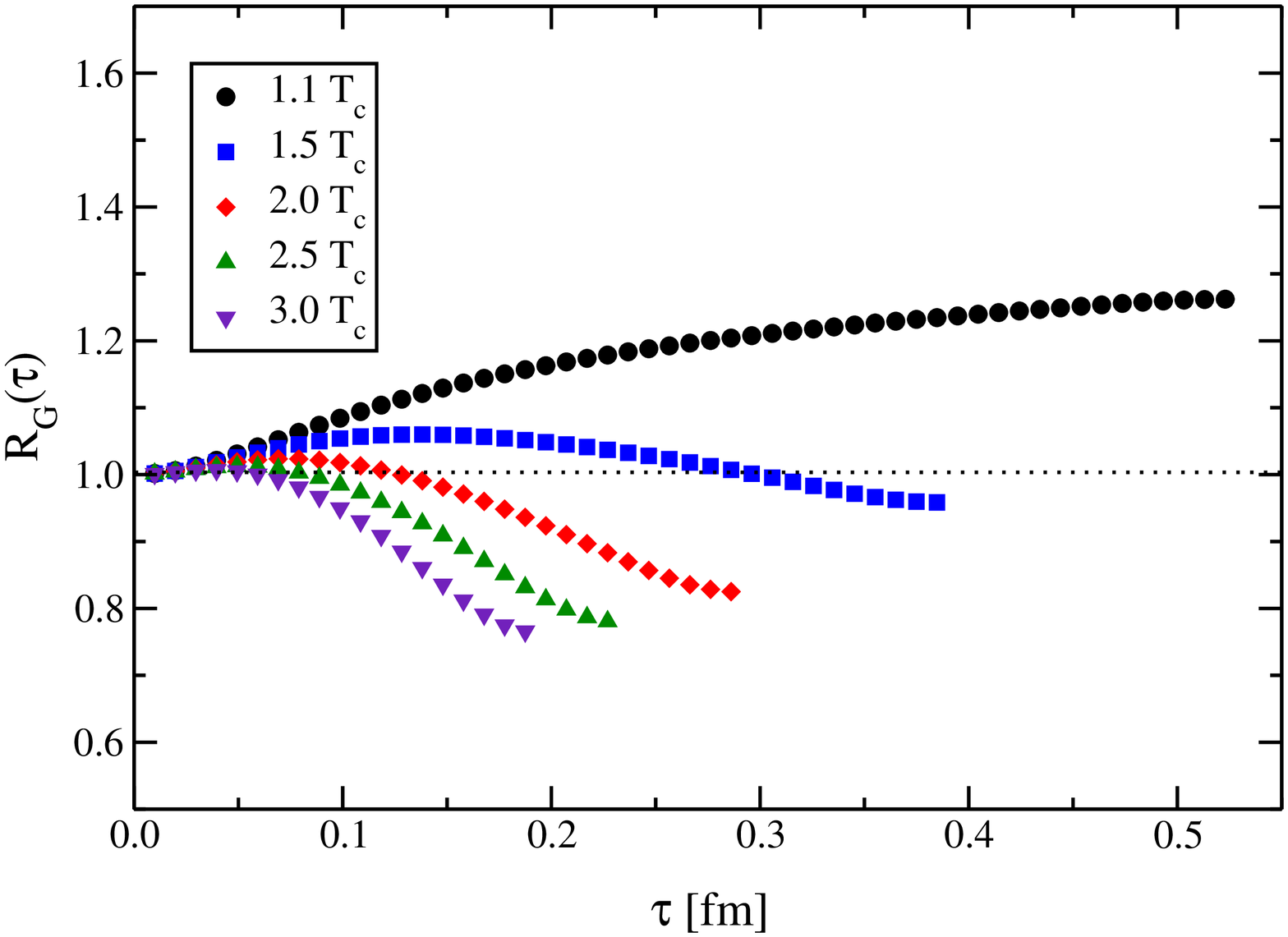}
\end{minipage}
\begin{minipage}{0.5\linewidth}
\vspace{0.3cm}
\includegraphics[width=0.95\textwidth]{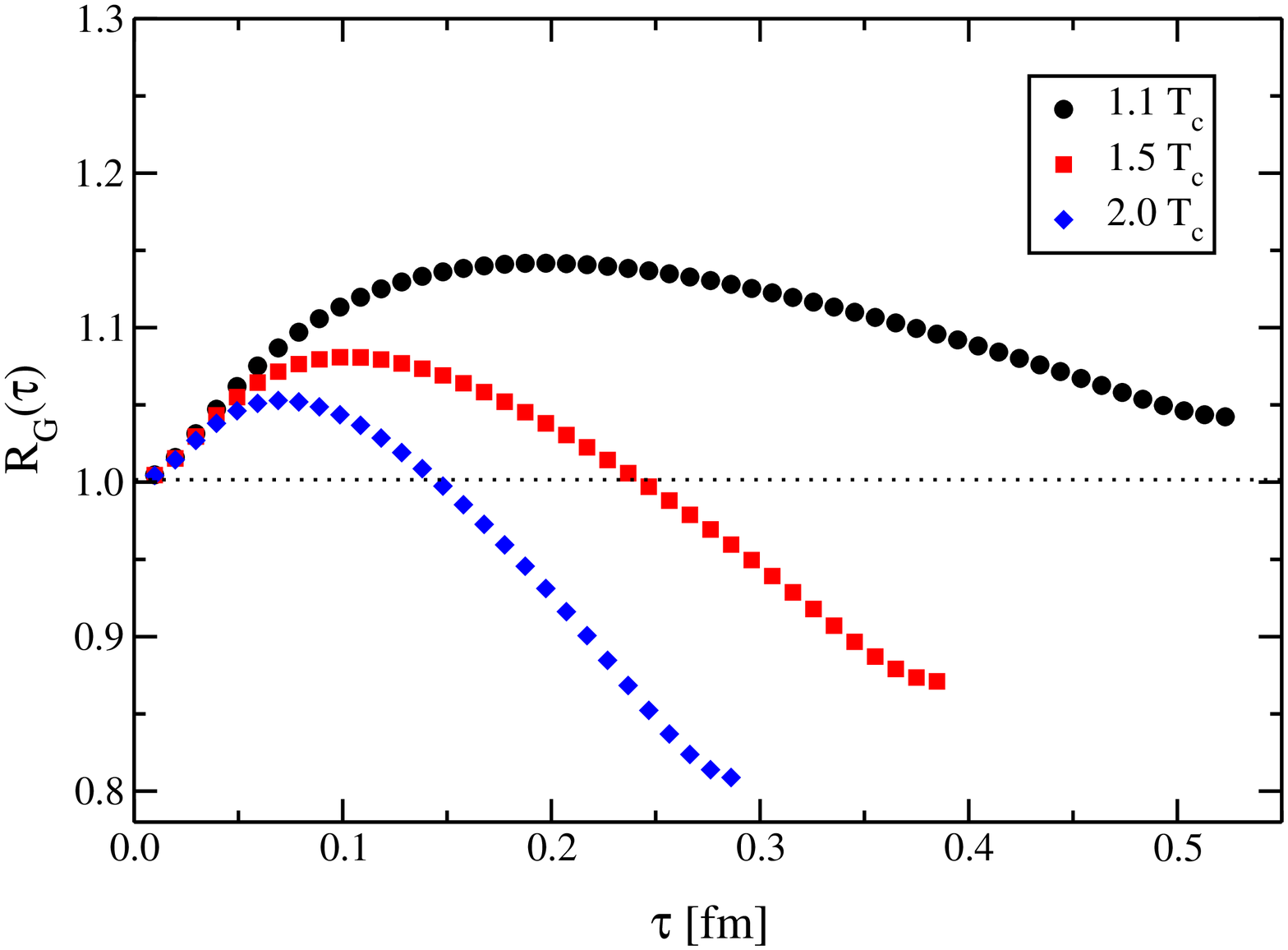}
\end{minipage}
\caption{Charmonium spectral functions (upper panels) and pertinent 
Euclidean-time ($\tau$) correlator ratios (lower panels), Eq.~(\ref{RG}), 
within the $T$-matrix approach of Ref.~\cite{Cabrera:2006wh} using
the internal energy extracted from $N_f$=3 lattice 
QCD~\cite{Petreczky:2004pz} as input potential. Note that the energy range 
in the upper panels is smaller than in Figs.~\ref{fig_SF-lqcd} and 
\ref{fig_MP08}, to better resolve the bound-state and threshold regime.
The correlator ratios are normalized using the calculated vacuum 
spectral function.
The left panels represent $S$-wave states ($J/\psi$,  $\eta_c$) while
the right ones correspond to $P$-waves ($\chi_{c0}$, $\chi_{c1}$). The
charm-quark mass has been fixed at $m_c$=1.7\,GeV and a small width
of 20\,MeV has been introduced in the $Q$ and $\bar Q$ propagator for
better display of the bound-state regime.}
\label{fig_G-tmat}
\end{figure}
The $T$-matrix is directly related to the correlation
function in momentum space by a double folding over external
vertices, schematically given by
\begin{equation}
G =   G^0 + G^0 \ T \ G^0 \equiv  G^0 + \Delta G \ ,
\label{GTG}
\end{equation}
and illustrated in Fig.~\ref{fig_gtg}.
Performing a partial wave expansion and keeping the leading terms
in $1/m_Q$ the explicit form of the correlator becomes
\begin{eqnarray}
G^0_l(\omega)&=& \frac{2N_c}{\pi} \int k^2 dk
 \, a_l(k,k) \,  G_{\bar{Q}Q}(\omega;k) \, \hat f^{Q\bar{Q}}(\omega_k)
\label{G0}
\\
\Delta G_l(\omega) &=& \frac{N_c}{\pi^3} \int k^2 dk
\, G_{\bar{Q}Q}(\omega;k) \, \hat f^{Q\bar{Q}}(\omega_k)
\int k'^2 dk'\, G_{\bar{Q}Q}(\omega;k') \, \hat f^{Q\bar{Q}}(\omega_{k'})
  \, a_l(k,k') \, T_l(\omega,k,k') \ ,
\label{delG}
\end{eqnarray}
with coefficients $a_{0,1}(k,k')$=2, 2$kk'/m_Q^2$ in the
$S$- and $P$-wave channels, respectively.
The imaginary part of the correlator is just the spectral function
(up to a constant factor). To leading order in $1/m_Q$, HQ spin symmetry
implies degeneracy of different spin states within a partial wave
(i.e., for $S$-waves $\eta_c$ and $J/\psi$, as well as for the
$P$-wave $\chi_c$ states). Results of the $T$-matrix approach for
spectral and correlation functions are shown in Fig.~\ref{fig_G-tmat}
for a constant effective $c$-quark mass ($m_c$=1.7\,GeV) and a small 
width ($\sim$20\,MeV) in the quark propagators (for better resolution
we here focus on the energy regime around the 
open-charm threshold, $\omega_{thr}=2m_c$).
For the potential the (subtracted) $N_f$=3 internal
energy of Ref.~\cite{Petreczky:2004pz} has been employed. The spectral
functions displayed in the upper panels of Fig.~\ref{fig_G-tmat} 
indicate that the $J/\psi$ (or $\eta_c$) dissolves around $\sim$3\,$T_c$ 
($\sim$2.5\,$T_c$ when using the internal energy of 
Ref.~\cite{Kaczmarek:2005gi}), while the $\chi_c$ states melt at lower 
temperatures, $\sim$1.3\,$T_c$.
These values should be considered as upper limits, since
larger widths will lead to a melting at smaller temperatures.
The continuum part of the spectral functions in both $S$- and
$P$-waves exhibits a large enhancement over the non-interacting case;
this effect cannot be neglected in quantitative applications to
correlation functions.
\begin{figure}[!tb]
\begin{center}
\includegraphics[width=.45\textwidth]{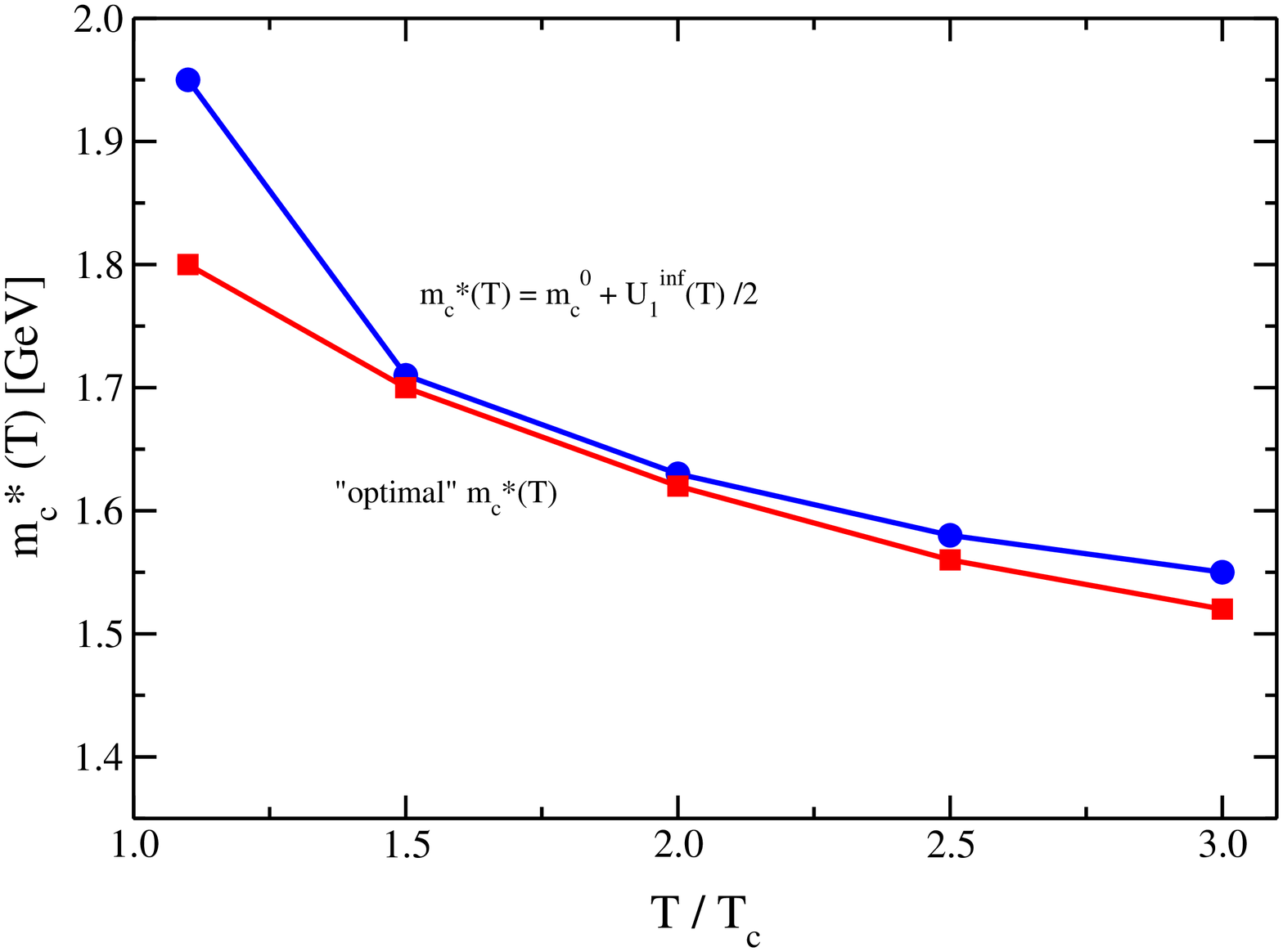}
\hspace{1cm}
\includegraphics[width=.40\textwidth]{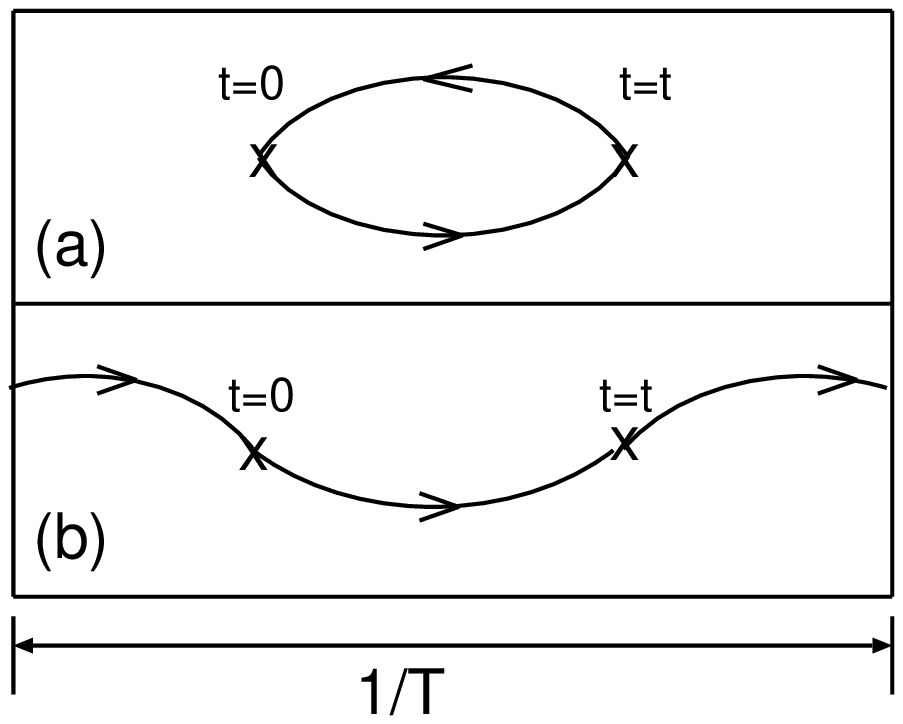}
\end{center}
\caption{Two effects at the single HQ level relevant to quarkonium 
correlators.  Left panel: in-medium charm-quark mass following from
the asymptotic value of the internal energy (upper
curve)~\cite{Petreczky:2004pz,Cabrera:2006wh} and as inferred from a 
``fit" to the $\eta_c$ correlator~\cite{Rapp:2008fv} as shown in 
the upper right of Fig.~\ref{fig_G-AB-CR}.
Right panel: contributions to temporal correlation functions computed in
finite-temperature lattice QCD~\cite{Umeda:2007hy}. In addition to the
standard mesonic point-to-point correlator (upper panel), the periodic
boundary conditions in $\tau$ direction allow for a quark scattering 
diagram (lower panel).
}
\label{fig_mc-zero}
\end{figure}
The pertinent temporal correlator ratios, $R_G^\alpha$ (recall 
Eq.(\ref{RG})), are plotted in the lower panels of Fig.~\ref{fig_G-tmat}.
In the $S$-wave they are qualitatively similar to the ones in lQCD, 
while in the $P$-waves they are quite different in both magnitude and 
temperature dependence, recall Fig.~\ref{fig_G-lqcd}. In particular, the 
decreasing trend at large $\tau$ indicates that the model spectral 
functions are missing low-energy strength with increasing temperature. In 
the $\eta_c$ channel, this problem is largely overcome when the in-medium 
charm-quark mass, as following from the asymptotic value of the internal 
energy, is included, i.e., 
\begin{equation}
m_c^*(T)=m_c^0+U^{(1),\infty}_{Q\bar Q}(T)/2 \ , 
\label{mcstar}
\end{equation}
displayed in the left panel of Fig.~\ref{fig_mc-zero}. This stabilizes the 
$T$-dependence of the $\eta_c$ correlator ratio~\cite{Cabrera:2006wh}, 
but it is not enough to quantitatively improve on the agreement with the 
lQCD results for $J/\psi$ and $\chi_c$ . 

\begin{figure}[!tb]
\hspace{0.0cm}
\begin{minipage}{0.5\linewidth}
\includegraphics[width=0.99\textwidth]{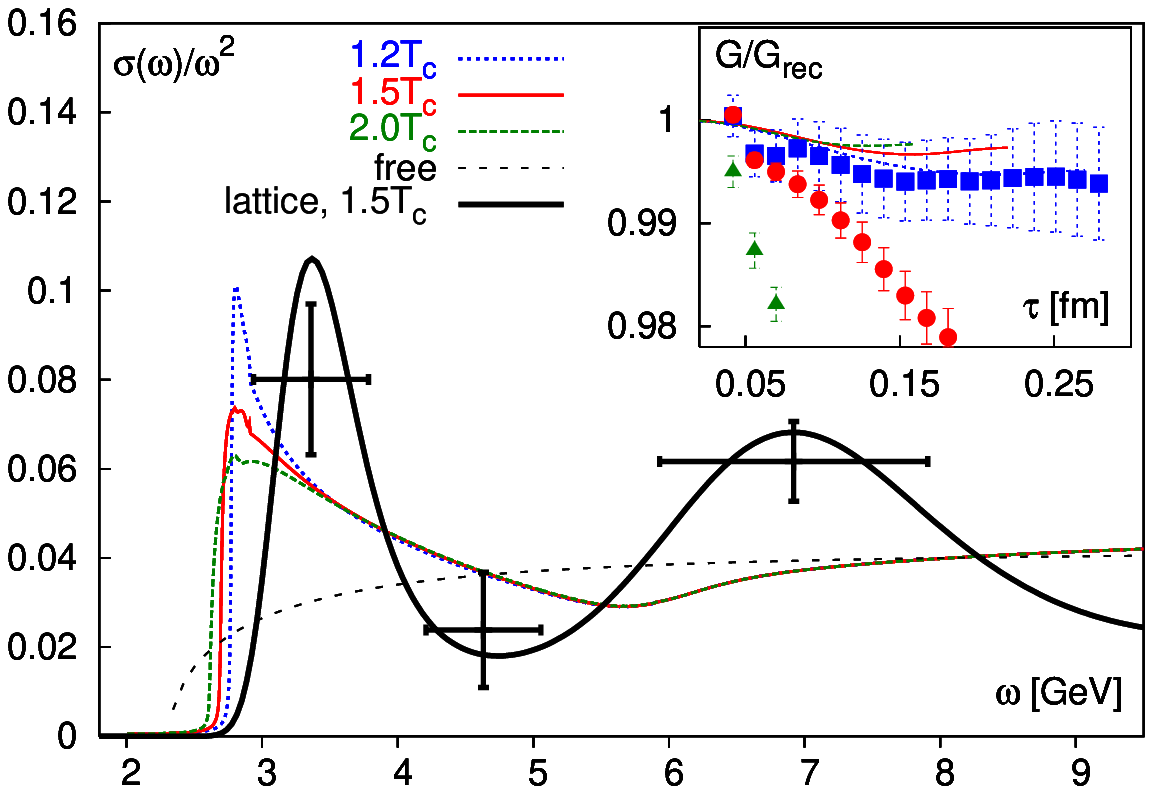}
\end{minipage}
\begin{minipage}{0.5\linewidth}
\includegraphics[width=0.99\textwidth]{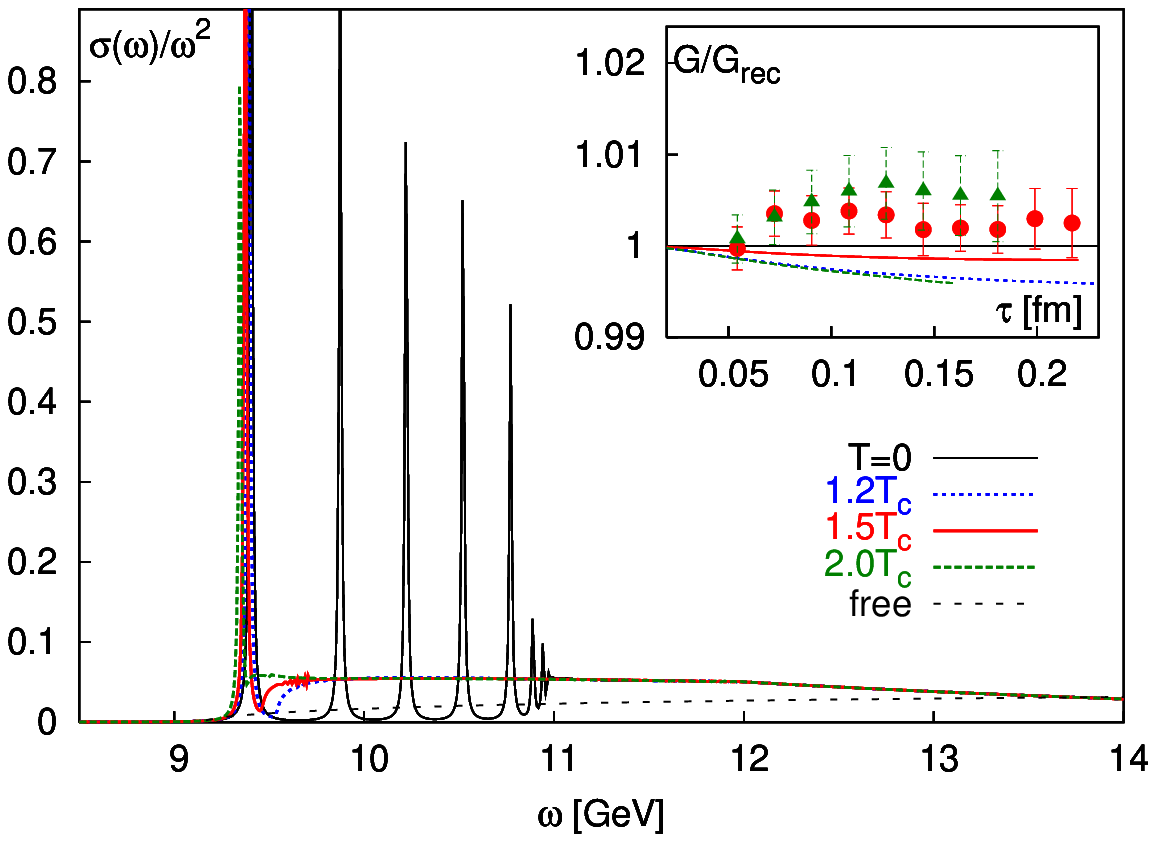}
\end{minipage}
\begin{minipage}{0.5\linewidth}
\includegraphics[width=0.99\textwidth]{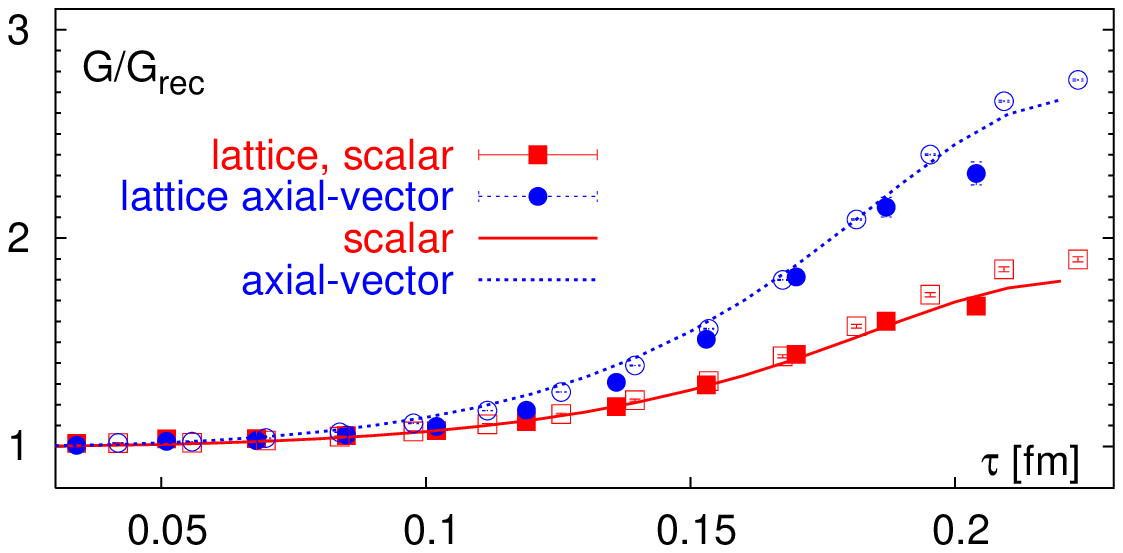}
\end{minipage}
\begin{minipage}{0.5\linewidth}
\includegraphics[width=0.99\textwidth]{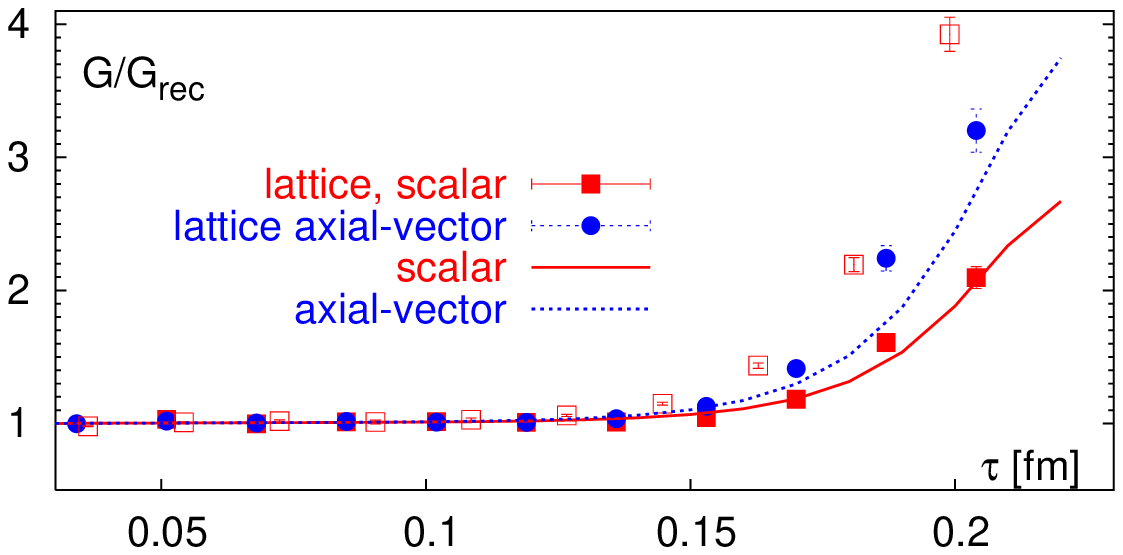}
\end{minipage}
\caption{Quarkonium spectral functions and correlator ratios in a Gluon 
Plasma (quenched QCD) within a nonrelativistic Green's function approach 
employing a screened Cornell potential~\cite{Mocsy:2007jz}, supplemented 
with a perturbative QCD continuum including a $K$ factor at high energies.
Upper panels: spectral functions and correlator ratios (insets) for 
$S$-wave charmonium ($\eta_c$, left) and bottomonium ($\eta_b$, right) 
at various temperatures (the data points in the inset are lQCD results 
with the same color code as for the calculations represented by the 
lines); the lattice results are for quenched QCD~\cite{Jakovac:2006sf},
with the data points 
in the spectral functions indicating the statistical uncertainty (vertical 
error bar) of the MEM over the energy interval indicated by the horizontal 
bar (same as in upper left panel of Fig.~\ref{fig_SF-lqcd}).
Lower panels: Euclidean correlator ratios
for $P$-wave charmonia (left) and bottomonia (right) at 1.5~$T_c$;
closed and open symbols are quenched lQCD results using 
isotropic~\cite{Datta:2003ww,Datta:2006ua} and 
anisotropic~\cite{Jakovac:2006sf} lattices.}
\label{fig_MP08}
\end{figure}

\begin{figure}[!tb]
\begin{minipage}{0.5\linewidth}
\includegraphics[width=0.91\textwidth]{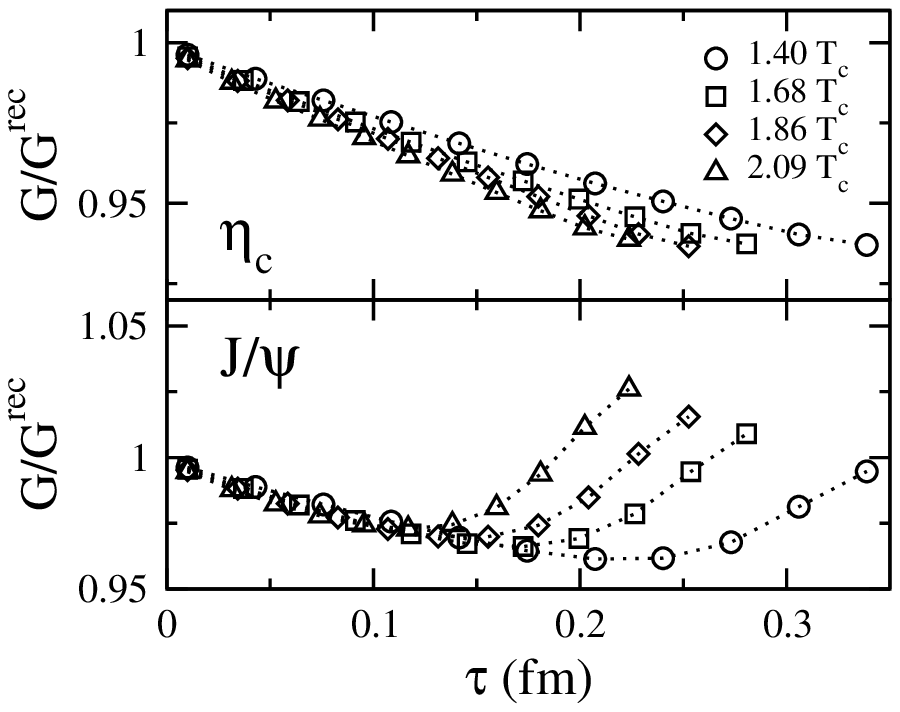}
\end{minipage}
\begin{minipage}{0.5\linewidth}
\includegraphics[width=0.93\textwidth]{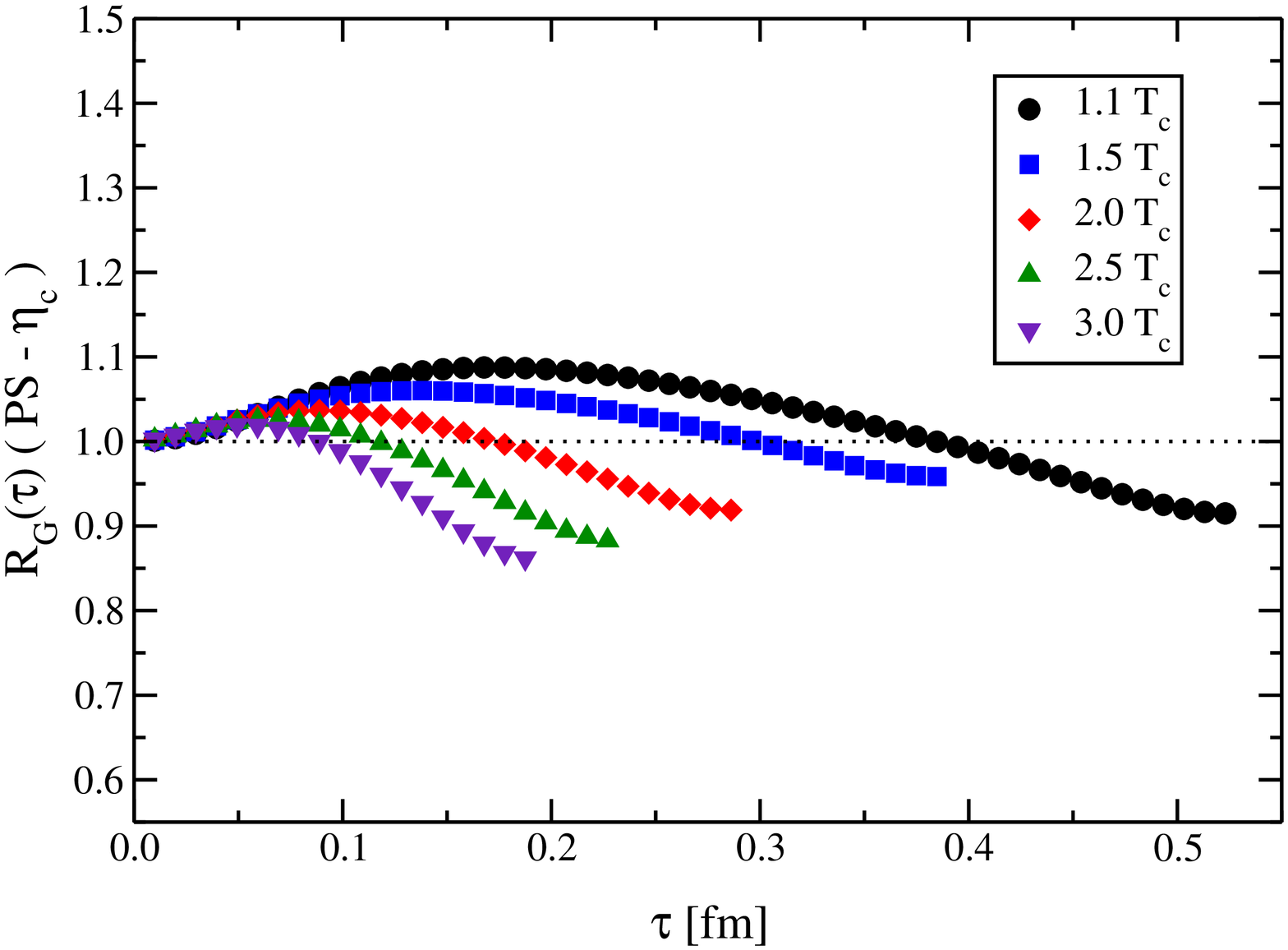}
\end{minipage}
\begin{minipage}{0.5\linewidth}
\includegraphics[width=0.9\textwidth]{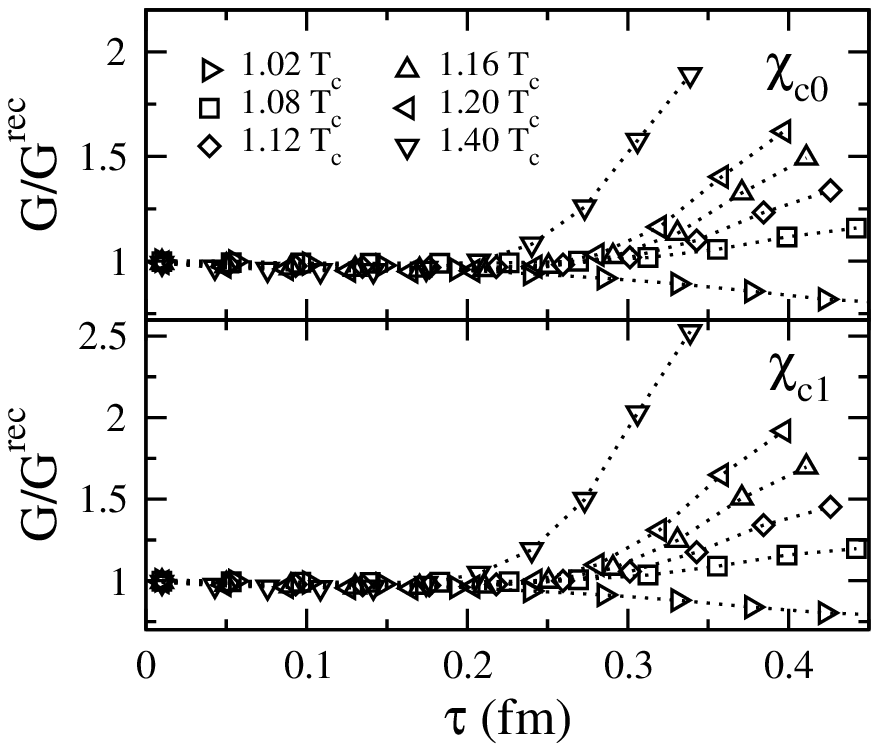}
\end{minipage}
\begin{minipage}{0.5\linewidth}
\includegraphics[width=0.93\textwidth]{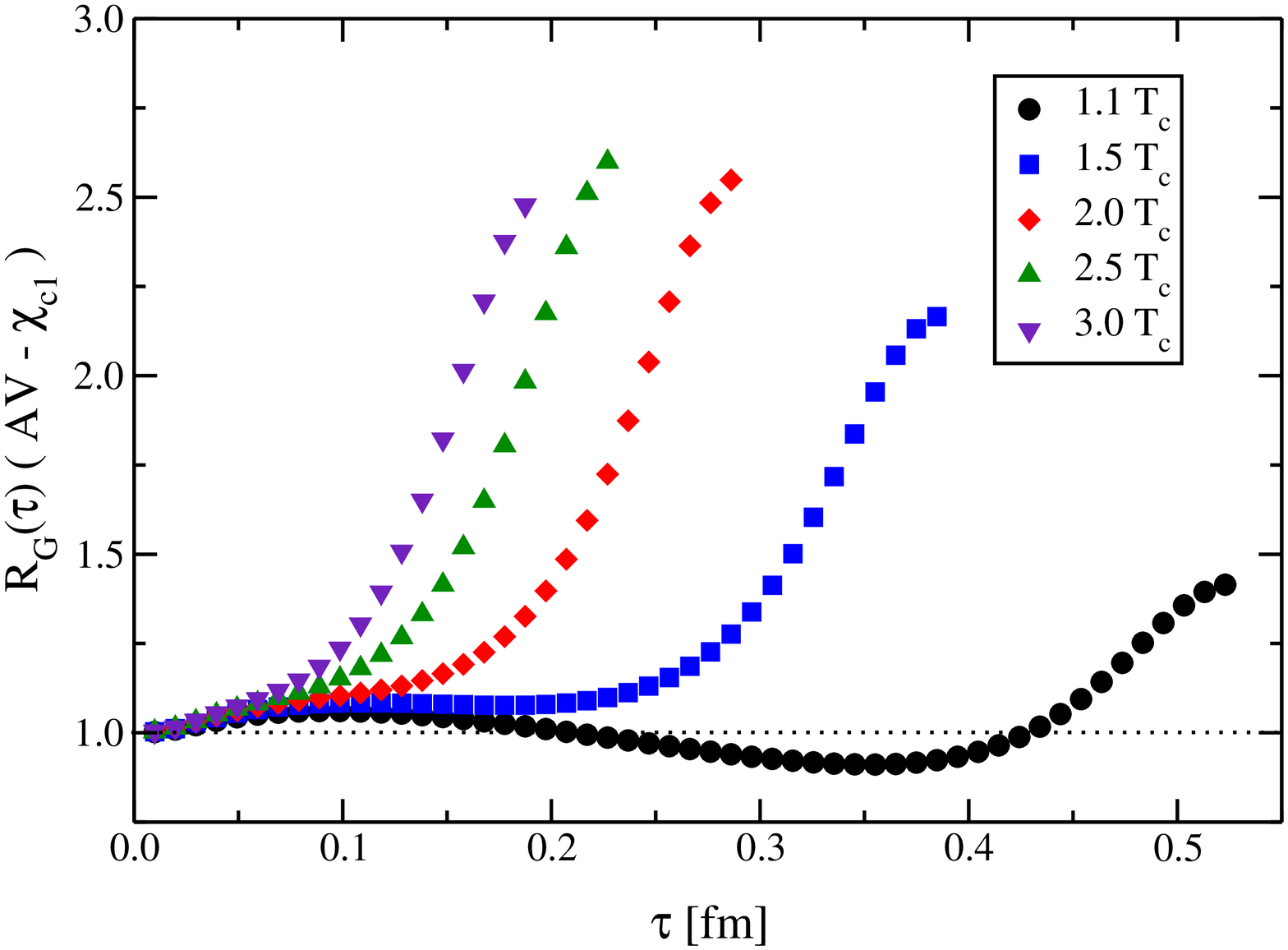}
\end{minipage}
\caption{Charmonium correlator ratios, $R_G=G/G_{\rm rec}$ as in 
Eq.~(\ref{RG}), including zero-mode contributions.
Left panels: $S$- and $P$-waves (upper and lower left, respectively) in
the Schr\"odinger-equation approach of Ref.~\cite{Alberico:2007rg} where
the HQ potential is based on a combination~\cite{Wong:2004zr} of free and 
internal energy extracted from unquenched lattice QCD.
Right panels: $\eta_c$ and $\chi_{c1}$ (upper and lower right, respectively)
in the $T$-matrix approach of Refs.~\cite{Cabrera:2006wh,Rapp:2008fv}
(employing $N_f$=3 internal energies from lQCD) 
including an in-medium charm-quark mass according
to the left panel of Fig.~\ref{fig_mc-zero}.}
\label{fig_G-AB-CR}
\end{figure}

An essential part of this puzzle has been identified in 
Ref.~\cite{Umeda:2007hy}, in terms of an energy-independent (but 
$T$-dependent) contribution to the correlators which arises due to
the periodic boundary conditions in the (finite) $\tau$ direction at
finite temperature (see right panel of Fig.~\ref{fig_mc-zero}). 
This term may be interpreted as a scattering contribution of a single 
heavy quark in the medium which generates a low-energy peak in the 
quarkonium spectral functions that can be related to a contribution
from a heavy-quark susceptibility, $\chi_\alpha$, to the lattice QCD 
correlators~\cite{Petreczky:2005nh}.
These contributions have been readily implemented into potential model
analyses and are found to largely resolve the discrepancies with
the lQCD correlator ratios~\cite{Rapp:2008fv,Mocsy:2007jz,Alberico:2007rg}.
While the pseudoscalar ($\eta_c$) correlator appears to be free of the 
zero-mode contribution, the $P$-wave correlators in the QGP are enhanced 
appreciably, in particular at large $\tau$ (the $J/\psi$ is mildly 
affected). This is due to the additional very-low energy strength in 
the spectral functions which, in quasiparticle approximation, 
is generated by extra terms $\sim\chi_\alpha \omega \delta(\omega)$.
When implementing these into the correlation functions and evaluating 
the susceptibilities, $\chi_\alpha$, for a free gas, the description 
of the lQCD $P$-wave correlator ratios (Fig.~\ref{fig_G-lqcd}) is much
improved~\cite{Rapp:2008fv,Mocsy:2007jz,Alberico:2007rg},
see lower panels in Figs.~\ref{fig_MP08} and \ref{fig_G-AB-CR}.
However, the agreement of the potential approaches with lQCD correlator
ratios still allows for a significant redundancy in the underlying 
spectral functions, leading to rather different conclusions on the 
``melting" temperatures of the various charmonia, as we will now discuss.

In Ref.~\cite{Mocsy:2007jz}, a screened Cornell potential has been
employed within a nonrelativistic Green's function approach to describe 
the low-energy part of the spectral function. The latter has been 
combined with a high-energy part which has been approximated with a 
perturbative continuum multiplied by a $K$ factor (simulating radiative 
corrections) to match the large-$\tau$ behavior of the lQCD data. 
Above $T_c$ the $S$-wave charmonium and
bottomonium spectral functions exhibit a moderate variation with $T$
close to threshold, cf.~upper panels in Fig.~\ref{fig_MP08}. The
$\eta_c$ state melts at a temperature as low as $T\simeq 1.3T_c$,
but the corresponding $S$-wave correlator changes little with
temperature, roughly consistent with the small variations found in 
the lQCD correlator ratios (cf.~the insets in the upper left panels).
It was therefore concluded that the lQCD correlators
are consistent with a melting of all charmonium states at temperatures
below 1.3~$T_c$.
Essential to this argument is a reduction of the in-medium
charm-anticharm quark threshold to $2m_c^*\simeq2.7$\,GeV, together with a
nonperturbative threshold enhancement in the spectral function, as first
emphasized in the $T$-matrix approach of Ref.~\cite{Cabrera:2006wh}.

In Ref.~\cite{Alberico:2007rg} the in-medium HQ potential was taken as a 
linear combination of $F_{Q\bar Q}^{(1)}$ and $U_{Q\bar Q}^{(1)}$ (as 
advocated in Ref.~\cite{Wong:2004zr}), in connection with a constant HQ 
mass and a noninteracting perturbative continuum, corresponding to a 
spectral function as in Eq.~(\ref{sig-schem}). Here, the (semi-) 
quantitative description of the lQCD correlator ratios (see left panels 
of Fig.~\ref{fig_G-AB-CR}) goes along with a $J/\psi$ melting at
$\sim$1.6~$T_c$. 

The right panels of Fig.~\ref{fig_G-AB-CR} depict $S$-
and $P$-wave charmonium correlator ratios within the $T$-matrix
approach based on the spectral functions in the upper panels of
Fig.~\ref{fig_G-tmat}, where the HQ potential is taken as the
(subtracted) internal energy. The in-medium charm-quark mass has been
calculated including the asymptotic value of the internal energy,
$m_c^*=m_c^0+U_{Q\bar{Q}}^{(1),\infty}(T)/2$ (except close to $T_c$ 
where the large
entropy contribution, $TS_{Q\bar{Q}}^{(1),\infty}$, is problematic,
cf.~Fig.~\ref{fig_mc-zero}).
As in Refs.~\cite{Mocsy:2007jz,Alberico:2007rg}, the lQCD correlator
ratios can be approximately reproduced, only that in this approach the
dissolution temperature of the $J/\psi$ is around $\sim$2.5$T_c$.

In a slightly different line of work, potential models have recently 
been developed utilizing perturbative techniques, starting with a suitable 
identification of the static finite-$T$ HQ 
potential~\cite{Escobedo:2008sy,Brambilla:2008cx,Laine:2006ns,Beraudo:2007ky}. 
It has been found that, even at leading order in $\alpha_s$, the 
potential develops an imaginary part related to Landau damping in 
$t$-channel gluon exchange (cf.~Ref.~\cite{Laine:2007qy} for an
analysis using classical lQCD). When applied within a Schr\"odinger 
equation in the bottomonium sector, bound-state solutions 
emerge which gradually melt with temperature~\cite{Laine:2007gj}; 
pertinent spectral functions exhibit a large threshold enhancement, not 
unlike the ones displayed in Fig.~\ref{fig_G-tmat}~\cite{Cabrera:2006wh}.
In Ref.~\cite{Burnier:2007qm} it has been shown that, in weak coupling, 
quarkonium resonances survive up to temperatures {\em parametrically} 
given by $T\sim g^2 m_Q$, but have melted at $T\sim g m_Q$ (the melting 
occurs at $T\sim g^{4/3}m_Q$~\cite{Escobedo:2008sy}).

Let us briefly reflect on the main points of Secs.~\ref{ssec_lqcd} and 
\ref{ssec_pot}. The main strength (and best accuracy) of lQCD lies in 
the computation of euclidean correlators, which, however, are difficult 
to transform unambiguously into spectral functions containing the 
information on the physical excitation spectrum (such as the existence,
masses and widths of the quarkonium states). Potential models,
on the other hand, are less general in principle (e.g., based on a
static approximation), but have the advantage of enabling direct 
calculations of spectral functions and providing insights into
the underlying mechanisms (e.g., color screening and quark-mass
dependencies). Moreover, the spectral functions can be readily used
to calculate pertinent euclidean correlators. 
It is thus the synergy of lQCD and effective models which provides
the best prospects to better understand the properties of heavy
quarkonia in the QGP.  
More concretely, lQCD computations of Euclidean correlator ratios 
and associated spectral functions (Figs.~\ref{fig_SF-lqcd}, 
\ref{fig_G-lqcd}) exhibit a weak temperature dependence for $S$-wave 
states ($\eta_c$, $J/\psi$), while $P$-wave ($\chi_c$) correlators 
increase substantially. 
Potential models employing lQCD-motivated input potentials (or 
screened Cornell potentials) can essentially reproduce the lQCD results
(Figs.~\ref{fig_MP08}, \ref{fig_G-AB-CR}),
but the interpretation is not conclusive: the required low-energy 
strength in the $S$-wave spectral functions can be provided by either 
bound states surviving up to $\sim$2$T_c$, or by a
rather strong reduction in the open-charm threshold in connection with
a nonperturbative rescattering enhancement. The increase in the $P$-wave
correlators can be accounted for by a low-energy transport peak which
is not directly related to bound-state physics (but may be sensitive
to the charm-quark mass~\cite{Petreczky:2008px}).
A quantitative determination of the in-medium HQ mass will therefore
be important to make further progress.
Another ingredient that has not received much attention in the context
of correlators is the (inelastic) width of the charmonium spectral
function, which is a very relevant quantity for heavy-ion phenomenology.
This is the topic of the following Section.

\subsection{Quarkonium Dissociation Reactions in the QGP}
\label{ssec_diss-qgp}
\begin{figure}[!tb]
\begin{center}
\includegraphics[width=.33\textwidth]{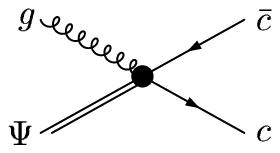}
\hspace{2cm}
\includegraphics[width=.33\textwidth]{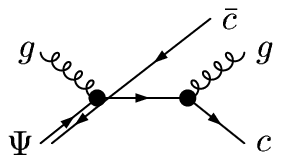}
\end{center}
\caption{Charmonium dissociation reactions via parton impact;
left panel: gluo-dissociation~\cite{Bhanot:1979vb,Peskin:1979va}
representing the leading-order QCD process;
right panel: quasifree dissociation~\cite{Grandchamp:2001pf}
representing a next-to-leading order process.}
\label{fig_diss-dia}
\end{figure}

The survival of a charmonium state in the QGP does not imply that 
it is protected from suppression, since it can be dissociated by 
inelastic collisions with surrounding partons\footnote{In the 
following, we refer to $\Psi=\eta_c,J/\psi, \psi',\chi_c, ...$ as a 
generic charmonium state, and $Y=\eta_b,\Upsilon, \Upsilon',\chi_b, 
...$ as a bottomonium state. To streamline the discussion, we 
concentrate on charmonia. Most of the arguments also apply to 
bottomonia although regeneration effects are expected to be less 
important than for charmonia.}.
By detailed balance the inverse of these reactions leads to charmonium 
{\em formation}, and therefore the pertinent inelastic width,
$\Gamma_\Psi=(\tau_\Psi)^{-1}$, is directly related to the relaxation
time, $\tau_\Psi$, of the charmonium abundance toward its thermal
equilibrium number, $N_\Psi^{\rm eq}$. This is immediately borne out
of the pertinent kinetic rate equation for the time evolution of
the charmonium number $N_\Psi(\tau)$, which for a spatially homogeneous
system takes the simple form
\begin{equation}
\frac{dN_\Psi}{d\tau} = -\Gamma_\Psi ( N_\Psi - N_\Psi^{\rm eq}) \
\label{rate-eq}
\end{equation}
(its solutions in the context of heavy-ion collisions will be discussed 
in Sec.~\ref{ssec_transport} below). As usual, such a rate equation is
applicable if a well-defined quasiparticle state $\Psi$ exists and
if the deviation from thermal equilibrium is not too large so that
the relaxation time approximation is valid.  
In general, the dissociation of a composite quasi-particle in a
medium can be classified into the following processes:
\begin{itemize}
\item[(1)] direct decays, $\Psi\to c +\bar c$ (or $\Psi\to c +\bar c + g$);
\item[(2)] inelastic collisions with partons, most notably
\begin{itemize}
\item[(2i)\,] $\Psi +g \to c + \bar c$ (cf.~left panel in
Fig.~\ref{fig_diss-dia})
\item[(2ii)]
$\Psi +i \to c + \bar c + i$, with $i=g,q,\bar q$ (cf.~right panel
in Fig.~\ref{fig_diss-dia}).
\end{itemize}
\end{itemize}
Note that the distinction between a bound and a resonance state from
the point of view of the rate equation (\ref{rate-eq})
is immaterial; it matters, however, for quantitative assessments of the
dominant contributions to the inelastic width. E.g., if a $\Psi$ state
moves into the continuum, i.e., above the $c\bar c$ threshold, at a
certain temperature, the opening of the direct-decay channel usually 
renders it the dominant process. In this sense the dissolution of a 
state due to screening corresponds to the $\Gamma_\Psi\to \infty$ limit 
of direct decays. The potential-model
calculations discussed in the previous section indeed suggest that
once a charmonium state becomes unbound, resonance states can no longer
be supported. On the other hand, for binding energies of the order of
100\,MeV, inelastic widths of a similar magnitude could induce a
``premature" melting of the state (the dissociation temperatures 
quoted above are therefore to be considered as upper limits).

\begin{figure}[!tb]
\begin{center}
\includegraphics[width=.45\textwidth]{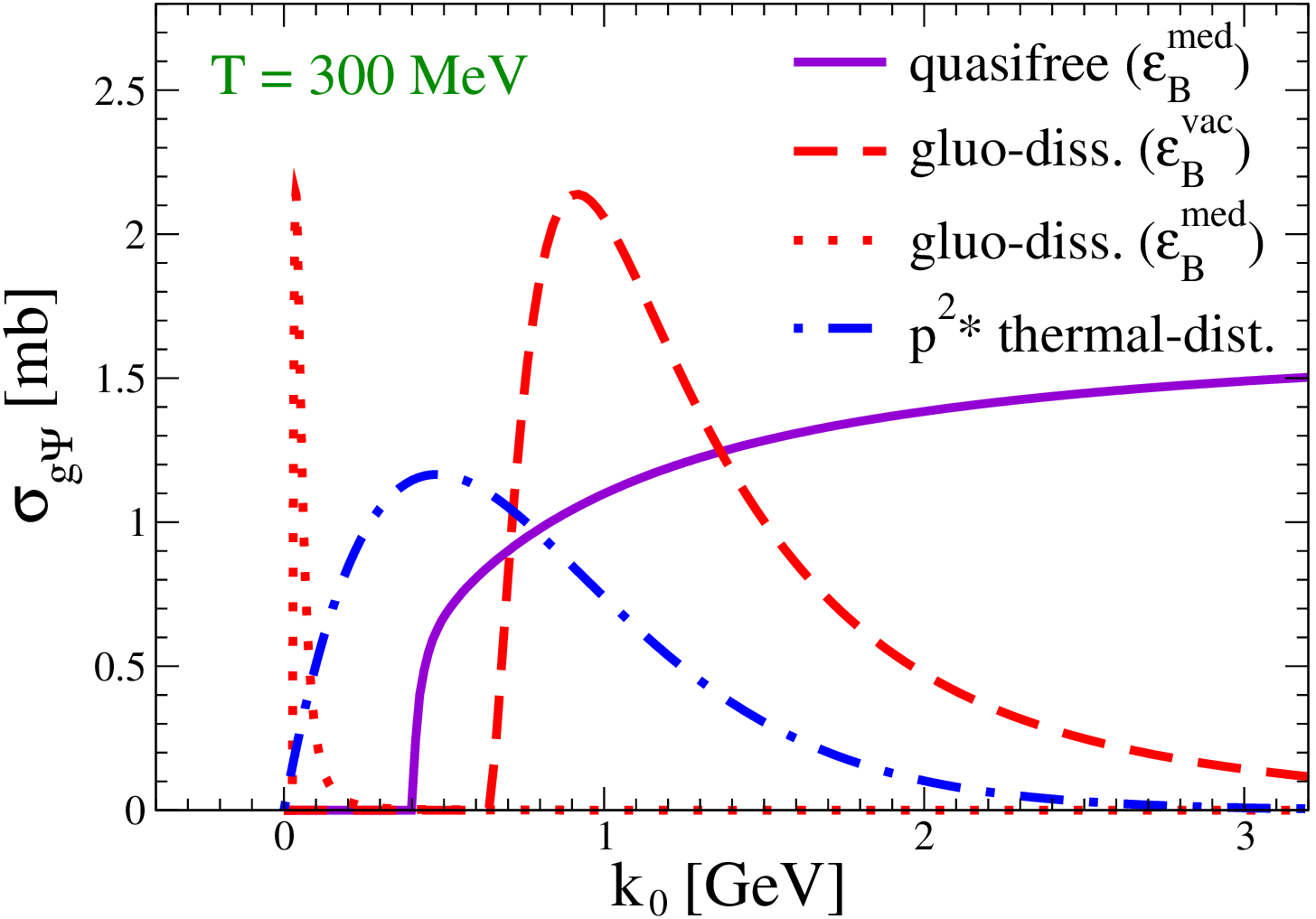}
\hspace{1cm}
\includegraphics[width=.42\textwidth]{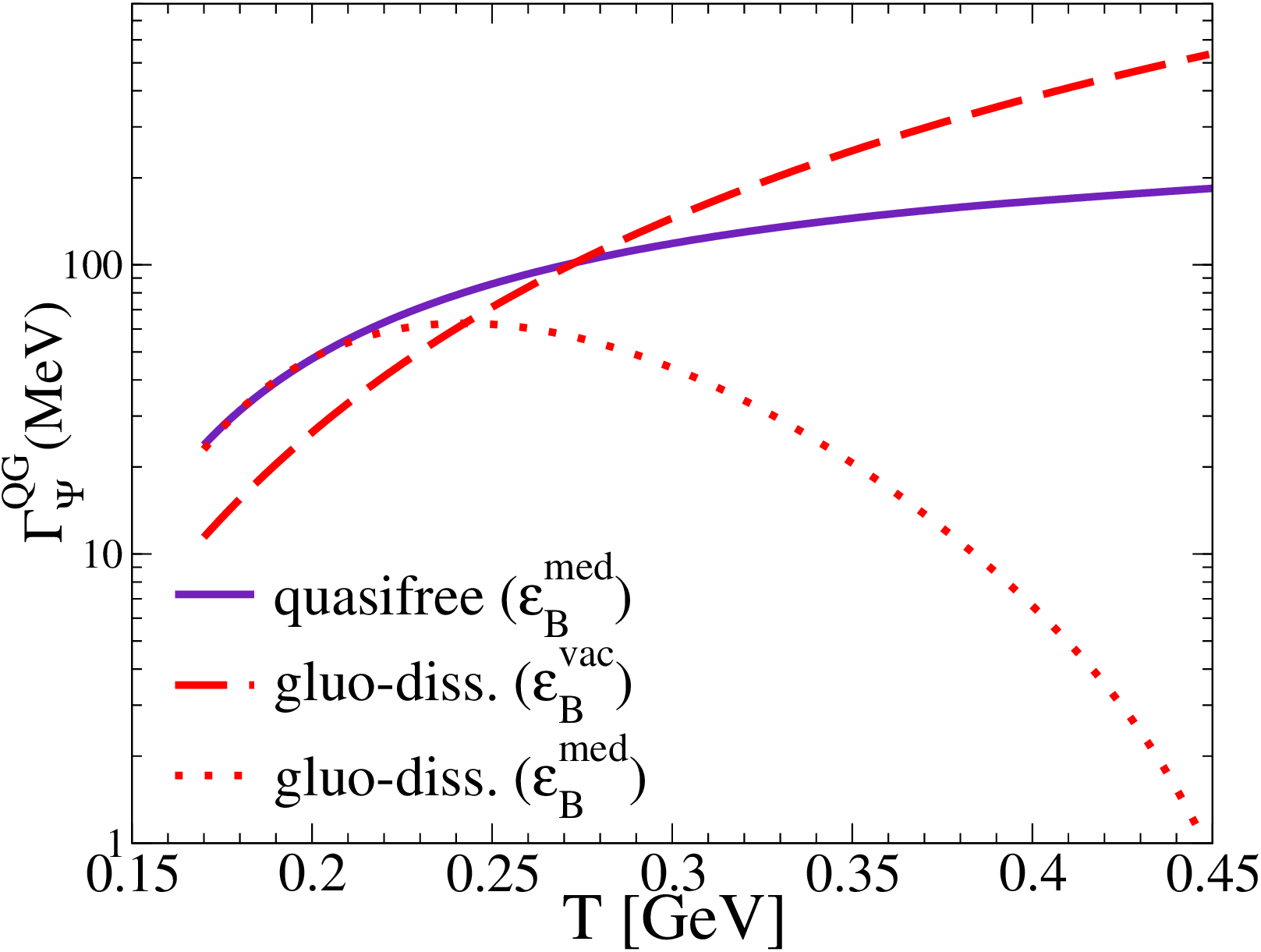}
\end{center}
\caption{Total cross sections (left panel) and pertinent dissociation
rates (right panel) for $J/\psi$ dissociation
reactions in the QGP: gluo-dissociation, $g+J/\psi\to c+ \bar c$,
according to Eq.~(\ref{gdiss}) with vacuum ($\varepsilon_B$=640\,MeV,
dashed line) and in-medium ($\varepsilon_B$$\simeq$30\,MeV, dotted line)
binding energies (the pertinent phase distribution, $p^2 f^B(p;T)$,
for massless gluons is illustrated by the dash-dotted line);
parton-induced quasifree dissociation, $i+J/\psi\to c +\bar c +i$,
using an in-medium binding energy, $\mu_D=gT$ and thermal parton masses
$m_i \propto gT$~\cite{Grandchamp:2001pf,Grandchamp:2002wp}.}
\label{fig_diss-rates}
\end{figure}

A standard approach to compute inelastic reaction rates in a thermal
medium is to calculate the cross section for a given dissociation
reaction and fold it over the thermal distribution of the (on-shell)
medium particles,
\begin{equation}
\Gamma_\Psi = \int \frac{d^3k}{(2\pi)^3} \ f^{q,g}(\omega_k,T)  \
v_{\rm rel} \ \sigma^{\rm diss}_\Psi(s) \
\label{rate}
\end{equation}
($v_{\rm rel}$: relative velocity of the incoming particles in the
center of mass of the collision; $s$=$(p+k)^2$: center-of-mass energy
squared of the parton-$\Psi$ collision).
The cross section for gluo-dissociation, $g+J/\psi\to c+\bar c$, has
been evaluated in Refs.~\cite{Bhanot:1979vb,Peskin:1979va,Kharzeev:1994pz}, 
using hydrogen-like wave functions. The latter are expected to be a good
approximation as long as the Coulombic part of the Cornell potential
dominates the $Q$-$\bar Q$ bound-state properties. As noted in
Refs.~\cite{Bhanot:1979vb,Peskin:1979va} this approximation is not
really (marginally) satisfied for the $J/\psi$
($\Upsilon$) with vacuum binding energies of $\sim$0.6(1.0)\,GeV.
Within these limitations, the gluo-dissociation cross section for 
massless gluons of energy $k_0$ incident on a ground-state quarkonium 
reads
\begin{eqnarray}
\sigma_{g\Psi,Y}(k_0) &=& \frac{2\pi}{3} \left(\frac{32}{3}\right)^2
\left(\frac{m_Q}{\varepsilon_B}\right)^{1/2} \frac{1}{m_Q^2}
\frac{(k_0/\varepsilon_B-1)^{3/2}}{(k_0/\varepsilon_B)^5}
\nonumber\\
 &=&  2\pi \left(\frac{16}{3}\right)^{2}  \alpha_s \ a_0^2 \
\frac{(k_0/\varepsilon_B-1)^{3/2}}{(k_0/\varepsilon_B)^5}
\ ,
\label{gdiss}
\end{eqnarray}
where $\varepsilon_B$ denotes the quarkonium binding energy and $m_Q$ 
the effective heavy-quark mass (as used, e.g., in fits to vacuum 
quarkonium spectra). The second line has been
obtained by using the Coulomb-potential expressions for
binding energy, $\varepsilon_B=(\frac{3}{4}\alpha_s)^2 m_Q$, and
Bohr radius, $a_0=4/(3\alpha_sm_Q)$. It illustrates the leading-order
(LO) character (${\cal O}(\alpha_s)$) of the cross section,
being proportional to the ``geometric size"
of the bound state.  The cross section has a threshold at
$k_0^{\rm thr}=\varepsilon_B$ and acquires its maximum at
$k_0^{\rm max}=\frac{10}{7}\varepsilon_B$, cf.~left panel of
Fig.~\ref{fig_diss-rates}.
For QGP temperatures of 200-300\,MeV, typical for SPS and RHIC energies,
the average thermal gluon energy of 3$T$ amounts to 600-900\,MeV, providing
good overlap with the vacuum gluo-dissociation cross section and resulting
in dissociation rates of $\Gamma_{J/\psi}\simeq30-200$\,MeV, see right panel
of Fig.~\ref{fig_diss-rates}. However, if the binding energy decreases with
temperature, the available phase space of gluo-dissociation shrinks
substantially. E.g., in the screened potential model of
Ref.~\cite{Karsch:1987pv}, corresponding to Eq.~(\ref{Vqq-med}), a 
moderate screening mass of $\mu_D\simeq gT$ with $\alpha_s$$\simeq$0.25
($g$$\simeq$1.75), yields a $J/\psi$ binding energy of only about
200(30)\,MeV at $T$=180(300)\,MeV, reaching zero at around 360\,MeV. 
Under these conditions the gluo-dissociation rate decreases with increasing 
temperature for $T\ge250$\,MeV, cf.~dotted line in the right panel of 
Fig.~\ref{fig_diss-rates}. This signals that other dissociation processes
become important (see also Refs.~\cite{Wong:2004zr,Arleo:2004ge,Blaschke:2004dv}).
For the same reason, the gluo-dissociation process cannot provide a realistic 
description for the dissociation of excited charmonia whose binding
energies are small already in the vacuum.

An early nonperturbative approach to compute heavy quarkonium break-up
from quark impact (corresponding to the category (2ii) of the above
classification) has been suggested in Ref.~\cite{Ropke:1988zz} in
analogy to electron-induced atom ionization;
the pertinent ``thermal activation" cross section,
\begin{equation}
\label{sigma-diss}
\sigma(T) =\pi r_{Q\bar Q}^2 \ {\rm e}^{-k_0^{\rm diss}/T}  \ ,
\end{equation}
is characterized by a temperature dependent dissociation energy,
$k_0^{\rm diss}\equiv\varepsilon_B$, while the coefficient has been
estimated by the ``geometric" transverse size of the bound state,
$\pi r_{Q\bar Q}^2$ (cf.~Ref.~\cite{Kharzeev:1995ju} for an alternative
derivation of a similar result, which was shown to apply to sufficiently
large dissociation energies, $k_0^{\rm diss}\gg T$). To obtain more
quantitative results, microscopic approaches are required.

\begin{figure}[!tb]
\begin{minipage}{0.55\linewidth}
\includegraphics[width=.95\textwidth]{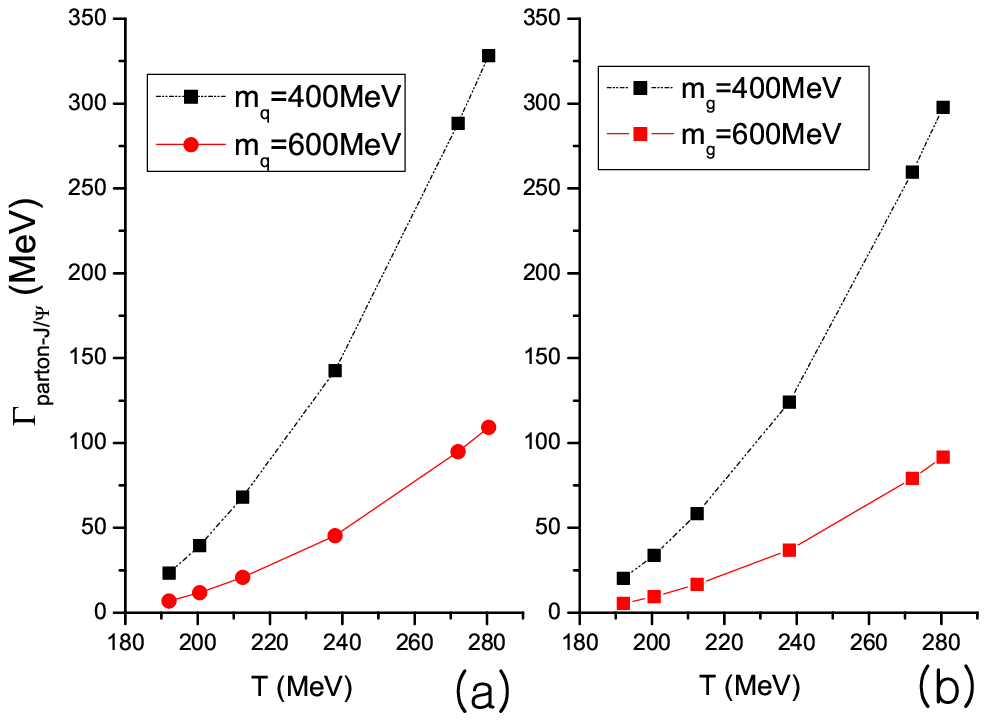}
\end{minipage}
\begin{minipage}{0.45\linewidth}
\vspace{2.5cm}
\includegraphics[width=0.85\textwidth]{gam-psi-p.eps}
\end{minipage}
\begin{minipage}{0.5\linewidth}
\begin{center}
\includegraphics[width=.55\textwidth]{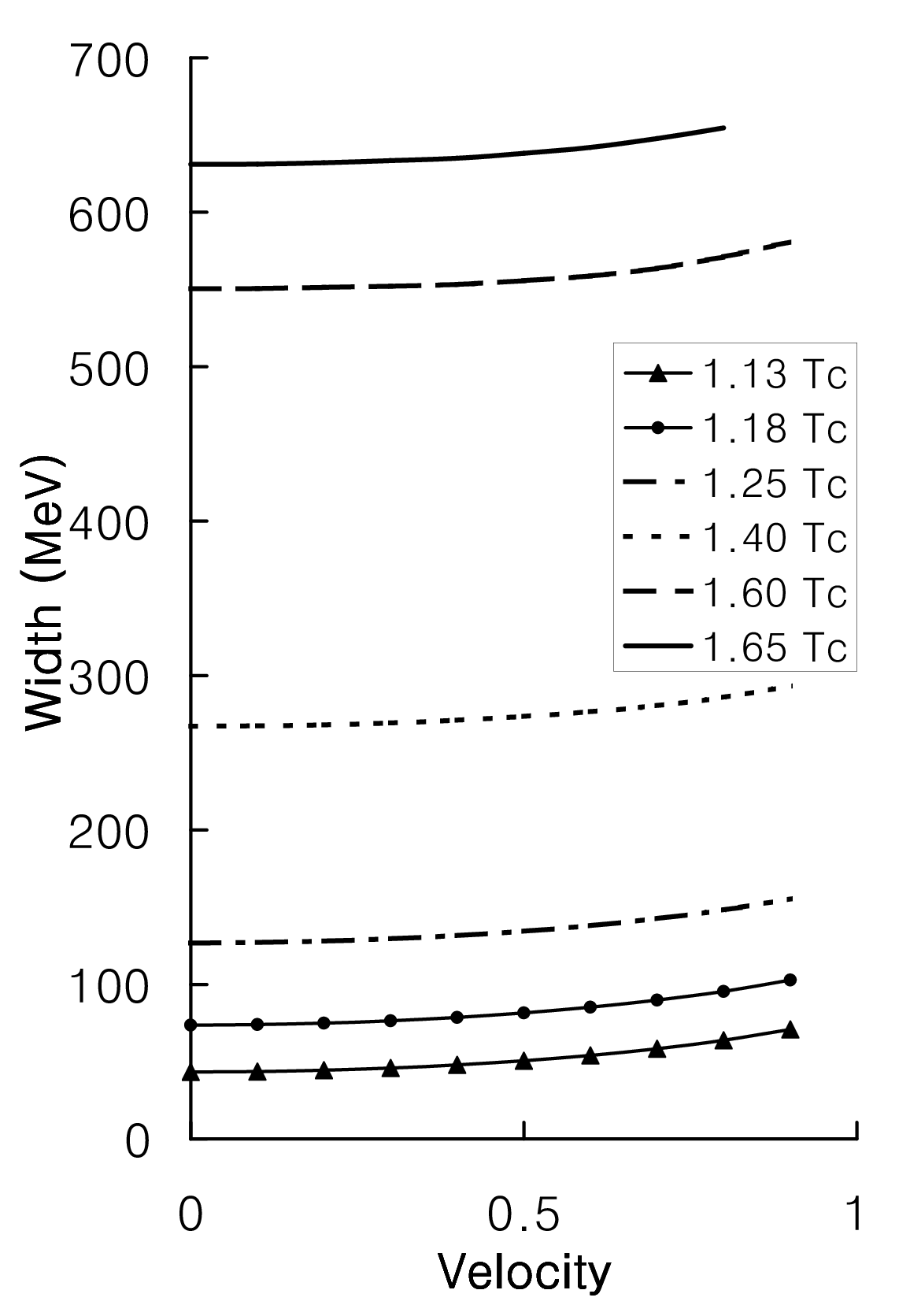}
\end{center}
\end{minipage}
\hspace{0.85cm}
\begin{minipage}{0.45\linewidth}
\vspace{-2.2cm}
\includegraphics[width=0.85\textwidth]{gam-chi-p.eps}
\end{minipage}
\caption{Parton-induced $J/\psi$ dissociation rates in the QGP.
Left panels: NLO results
using in-medium binding energies ranging from $\sim$40$\to$0\,MeV 
for $T$=190$\to$280\,MeV (upper left: 
$T$ dependence of $\Gamma_{J/\psi}$($p$=0) for (a) quark- 
plus antiquark- and (b) gluon-induced reactions for constant 
thermal parton masses~\cite{Park:2007zza};
lower left: 
$\Gamma_{J/\psi}^{\rm tot}$=$\Gamma_{J/\psi q+\bar q}$+$\Gamma_{J/\psi g}$,
as a function of the $J/\psi$ 3-velocity, $v=p/\omega_p$, 
relative to the heat bath for different temperatures with 
constant thermal parton masses of 600\,MeV~\cite{Song:2007gm} and
$T_c$=170\,MeV).
Right panels: comparison of the 3-momentum dependence of $J/\psi$
(upper right) and $\chi_c$ (lower right) dissociation rates at 3
different temperatures using either gluo-dissociation with vacuum
binding energies~\cite{Bhanot:1979vb,Peskin:1979va} (dashed lines)
or quasifree destruction with in-medium binding
energies~\cite{Grandchamp:2001pf,Zhao:2007hh} (solid lines).}
\label{fig_diss-rates-2}
\end{figure}
In Refs.~\cite{Grandchamp:2001pf,Grandchamp:2002wp} the dissociation
reactions of weakly bound charmonia have been treated within
a quasifree approximation for parton impact on either one of the heavy
quarks within the bound state, $i+\Psi\to c +\bar c +i$. The basic
process is (quasi-) elastic $i+c\to i+c$
scattering~\cite{Combridge:1978kx}, including the proper dissociation
kinematics (i.e., finite $\varepsilon_B$ and 4-momentum conservation),
as well as thermal parton and Debye masses (the latter are essential to
render the dominant $t$-channel gluon-exchange contributions finite).
The key quantities controlling the rate are $\alpha_s$, $\mu_D$ and 
$\varepsilon_B$, while the dependence on $m_c$ is weak.  
Formally, these reactions are of next-to-leading order (NLO) compared 
to LO gluo-dissociation, but the available phase space in the cross 
section is substantially increased (see solid line in the left panel 
of Fig.~\ref{fig_diss-rates}), leading to significantly larger 
dissociation rates at temperatures where the charmonium binding is 
weak (see solid line in the right panel of Fig.~\ref{fig_diss-rates}).
In fact, the infrared singularity in the $t$-channel exchange of a 
(Debye-screened) gluon for inelastic quarkonium scattering off
thermal partons renders this process of order $g^2$. 

A complete NLO calculation has recently been performed in
Ref.~\cite{Park:2007zza} confirming the prevalence of
$i+J/\psi\to c +\bar c +i$ over $g+J/\psi\to c +\bar c$ processes
for weak binding. More quantitatively, for $\alpha_s$=0.5, 
$T$=250\,MeV and thermal parton (Debye) masses $m_{q,g}=400$\,MeV, 
the total dissociation width of a weakly bound $J/\psi$ has been 
found to be $\Gamma_{J/\psi}\simeq 350$\,MeV. This is in reasonable 
agreement with Refs.~\cite{Grandchamp:2001pf,Grandchamp:2002wp}
where, using $\alpha_s$$\simeq$0.25, 
$\Gamma_{J/\psi}$($T$=250\,MeV)$\simeq$80\,MeV, since the NLO rate 
is proportional to $\alpha_s^2$ (modulo screening-mass corrections 
in $t$-channel gluon exchange diagrams for forward scattering; e.g., 
at $T$=250\,MeV, the Debye mass in 
Refs.~\cite{Grandchamp:2001pf,Grandchamp:2002wp} is $\mu_D$=440\,MeV).
For a constant Debye mass, the NLO results of Refs.~\cite{Park:2007zza}
exhibit a rather strong increase of $\Gamma_{J/\psi}$  with
temperature (upper left panel in Fig.~\ref{fig_diss-rates-2}),
which is, however, tamed if a perturbative temperature dependence,
$\mu_D\propto g(T) T$, were included (since the rates decrease with
increasing $\mu_D$). The dependence of the NLO rate
on the 3-momentum, $p$, of the $J/\psi$ is rather flat, see lower
left panel in Fig.~\ref{fig_diss-rates-2}~\cite{Song:2007gm}. This
is quite different from the LO gluo-dissociation process whose rate
drops substantially with $p$ (even more so for weak binding for which
the cross section is concentrated at small gluon energies), see
dashed lines in the right panels of Fig.~\ref{fig_diss-rates-2}.
The $p$ dependence following from the quasifree dissociation
rate~\cite{Zhao:2007hh} (solid lines in the right panels of
Fig.~\ref{fig_diss-rates-2}) is quite similar to the full NLO
calculation (recall that $\alpha_s$ differs by a factor of $\sim$2
in these two calculations).

\begin{figure}[!tb]
\begin{minipage}{0.45\linewidth}
\includegraphics[width=0.95\textwidth]{gam-ups-qfree-mass.eps}
\end{minipage}
\begin{minipage}{0.5\linewidth}
\includegraphics[width=.95\textwidth]{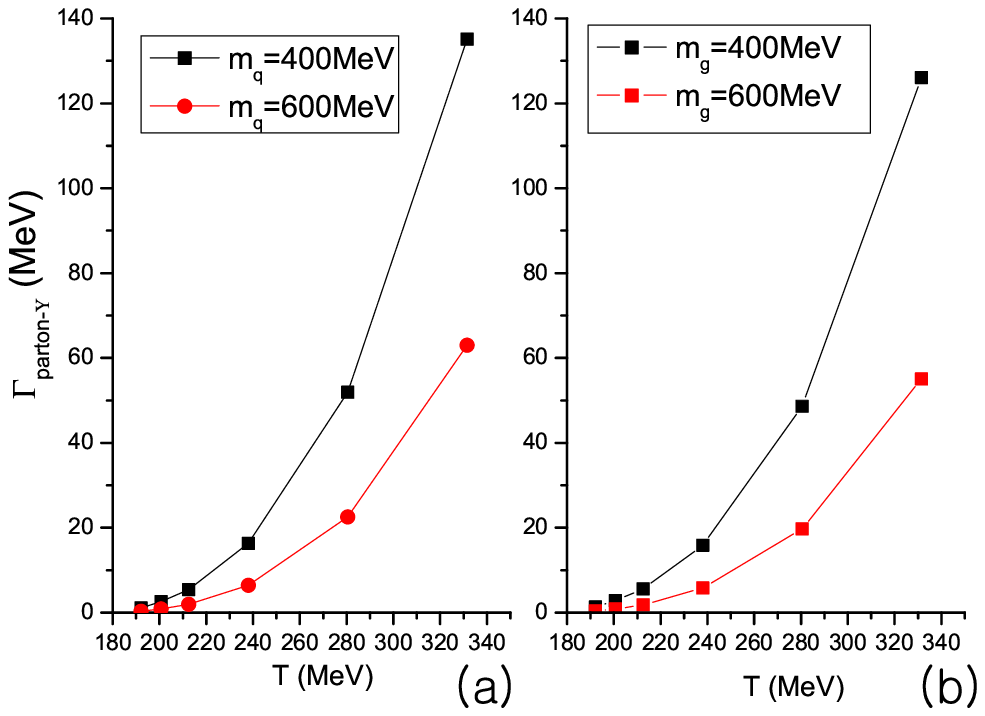}
\end{minipage}
\caption{Bottomonium dissociation rates in the QGP. Left panel:
quasifree calculation~\cite{Grandchamp:2005yw},
$\Upsilon + i \to b + \bar b + i$ ($i$=$q$, $\bar q$, $g$), using
$\alpha_s$$\simeq$0.25 with vacuum (lower curves) and in-medium
(upper curves) binding energies for $\Upsilon$, $\chi_b$ and $\Upsilon'$
(the in-medium binding energies are based on Ref.~\cite{Karsch:1987pv},
with $\varepsilon_B^\Upsilon$$\simeq$550(200)\,MeV at $T$=180(350)\,MeV).
Right panels: NLO calculations for quark- and gluon-induced break-up
of the ground-state $\Upsilon$~\cite{Park:2007zza} (sub-panels (a) and (b),
respectively) including $\Upsilon + g \to b + \bar b$ and
$\Upsilon + i \to b + \bar b + i$ with in-medium binding energies,
$\alpha_s$=0.5 and two different values for thermal parton masses.
}
\label{fig_diss-rates-ups}
\end{figure}
Inelastic collision rates of the ground-state $\Upsilon$ are compiled
in Fig.~\ref{fig_diss-rates-ups}. Analogous to charmonia, in-medium 
reduced binding energies render the quasifree process dominant over
gluo-dissociation, see Ref.~\cite{Grandchamp:2005yw} for a detailed 
discussion\footnote{For the vacuum binding
energy, the gluo-dissociation rate of $\Upsilon$ is comparable to the
quasifree one.}; this has also been
confirmed in a complete NLO calculation~\cite{Park:2007zza}.
Since the vacuum binding energy of the $\Upsilon$ is quite large
($\varepsilon_\Upsilon^0$$\simeq$1.1\,GeV) compared to typical temperatures
at RHIC ($T$$\simeq$0.3\,GeV), its dissociation rates are rather sensitive
to color-screening (which has important consequences for
$\Upsilon$ suppression in heavy-ion collisions), e.g., $\sim$7\,MeV
for $\varepsilon_\Upsilon^0$$\simeq$1.1\,GeV vs. $\sim$80\,MeV for
$\varepsilon_\Upsilon^{med}$$\simeq$0.25\,GeV at $T$=0.3\,GeV.
For comparable values of the parameters (strong coupling constant,
binding energy as well as Debye and thermal masses), the widths within
the quasifree calculation~\cite{Grandchamp:2005yw} 
($\varepsilon_B(T)$-curves in the left panel of
Fig.~\ref{fig_diss-rates-ups}) tend to be larger than within the NLO 
analysis~\cite{Park:2007zza} (right panel of 
Fig.~\ref{fig_diss-rates-ups}) when accounting for the factor 2
difference in $\alpha_s$. This discrepancy could be due a reduced
accuracy of the quasifree approximation for more tightly bound
systems~\cite{Laine:2006ns,Beraudo:2007ky}. As discussed in the latter 
paper, in perturbation theory, the imaginary part of the finite-temperature 
heavy-fermion potential referred to in the previous section is closely 
related to quasifree dissociation. In particular, it has been shown that 
in a QED plasma the infinite-distance limit of the imaginary part, 
Im\,$V_{\bar QQ}(r\to\infty) = -g^2 T/4\pi$, precisely corresponds to 
(twice) the damping factor of a single heavy fermion in the static 
limit, $\Gamma_Q = -g^2T/8\pi$ (in pQCD, an additional Casimir factor,
$C_F$=4/3, appears). For small distances (i.e., for tightly bound states), 
interferences occur which suppress this quarkonium width from the 
quasifree limit. 

Recent calculations have investigated the effects of anisotropic 
momentum distributions in the QGP on the quarkonium 
width~\cite{Dumitru:2007hy,Burnier:2009yu,Philipsen:2009wg}. 
The emerging consensus seems to be that anisotropies do not have a 
large impact on the quarkonium dissociation temperatures and widths 
as long as the thermal parton density (or entropy density) is kept 
constant. This conclusion may depend on the microscopic mechanism 
underlying the dissociation. For example, if the dissociation rate 
increases (decreases) with 3-momentum, harder momentum spectra of the 
medium partons may lead to an increase (decrease) in width over the 
isotropic thermal case.

Finally, let us address the impact of finite-width effects
on the charmonium correlator ratios discussed in the previous Section.
In the $T$-matrix approach of Ref.~\cite{Cabrera:2006wh}, a $J/\psi$
width has been implemented via an elastic charm-quark width, 
$\Gamma_c$, in the two-particle propagator, $G_{Q\bar Q}$, figuring 
into the Lippmann-Schwinger Eq.~(\ref{LS}) (see also 
Ref.~\cite{Riek:2010fk} for a more recent calculation). As 
discussed above, this generates a charmonium width in the spirit 
of the quasifree dissociation mechanism where the ``elastic" 
scattering of thermal
partons essentially occurs with one of the charm quarks within
the bound state. Using $\Gamma_c=\Gamma_{\bar c}=50$\,MeV, the
$J/\psi$ spectral function acquires a width of about 100\,MeV, and
the corresponding correlator ratio with this width is compared
to the calculations shown in Fig.~\ref{fig_G-tmat} (corresponding
to a total $J/\psi$ width of $\sim$40\,MeV), using the correlator
at 1.5~$T_c$ as the reconstructed one.
\begin{figure}[!t]
\begin{center}
\vspace{0.4cm}
\includegraphics[width=.40\textwidth]{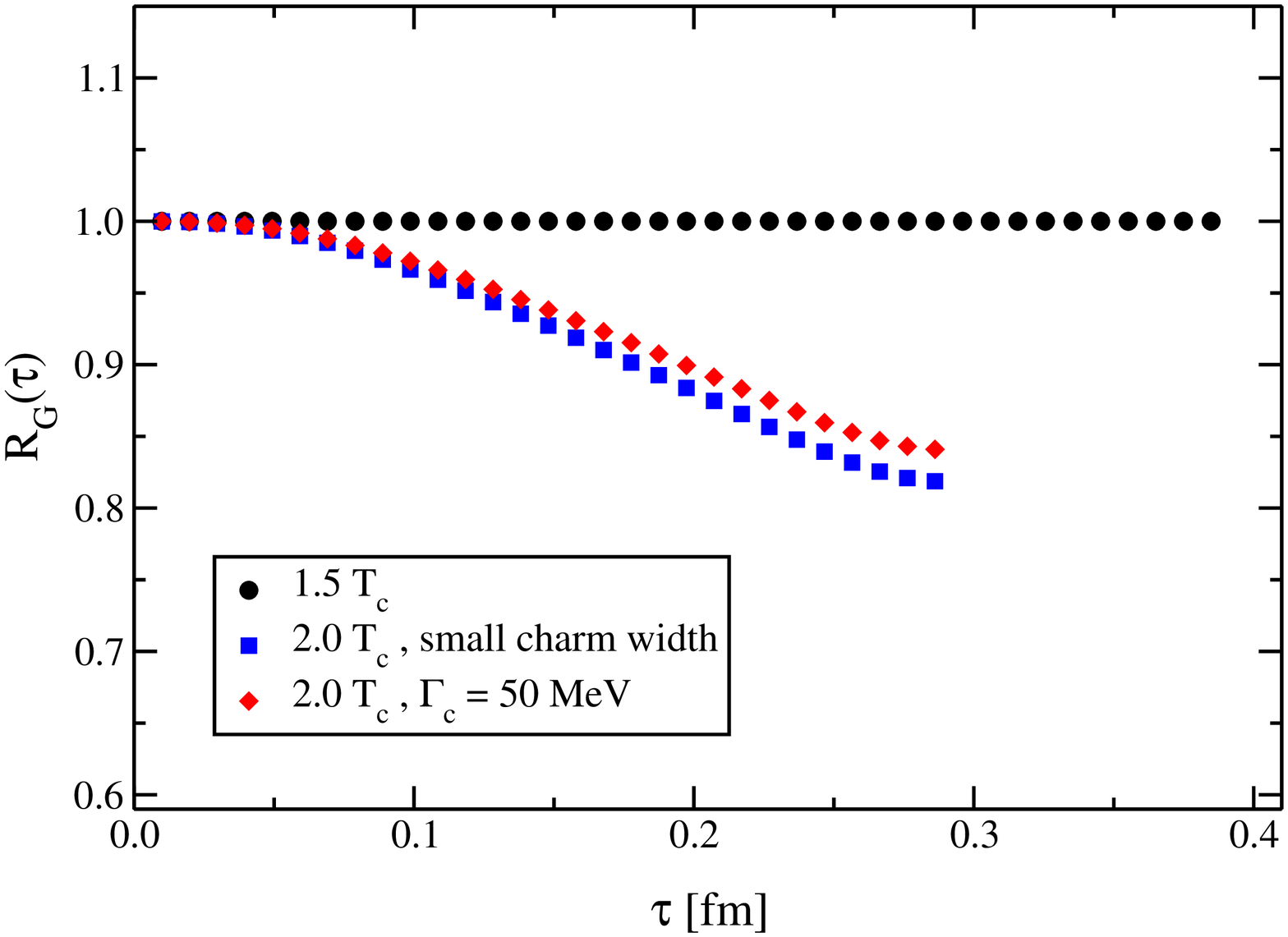}
\end{center}
\vspace{-0.4cm}
\caption{$S$-wave charmonium correlator ratios obtained within the
$T$-matrix approach~\cite{Cabrera:2006wh} using 2 different
values for the in-medium charm-quark width corresponding to
total $\eta_c$ (or $J/\psi$) widths of 40\,MeV (lower squares) and
100\,MeV (diamonds); the reconstructed correlator used for normalization
is the one at 1.5\,$T_c$, and the charm-quark mass has been held fixed
at $m_c$=1.7\,GeV (no zero-mode contributions are included).}
\label{fig_Rcc-width}
\end{figure}
One finds that the moderate increase of the $J/\psi$ width increases
the large-$\tau$ regime of the correlator ratio by a few percent,
due to the slight increase of strength in the low-energy part of the
spectral function caused by the broadening. More systematic
investigations of width effects, especially close to the ``melting"
temperature, remain to be carried out, e.g., by implementing microscopic
dissociation width calculations into the $T$-matrix approach. Generally, 
finite-width effects tend to accelerate the dissociation of the bound 
state, i.e., lower its dissociation temperature~\cite{Riek:2010fk}.

\subsection{Quarkonium Dissociation in Hadronic Matter}
\label{ssec_diss-hg}
\input{diss-hg}

\subsection{Insights from Plasma Physics}
\label{ssec_plasma}
\input{plasma}

\section{Quarkonia in Heavy-Ion Collisions}
\label{sec_hics}
The utilization of heavy quarkonia as a probe of the medium created in 
heavy-ion reactions has to rely on the analysis of their final number and 
momentum spectra.  Since the equilibration of quarkonia is generally not 
warranted, the interface between their in-medium properties and observables 
is usually based on transport approaches (Sec.~\ref{ssec_transport}). If 
the bulk medium is in local thermal equilibrium, the in-medium properties 
(as elaborated in the previous Section) enter in the form of relaxation 
times (dissociation rates) and asymptotic limits for the yields (equilibrium 
abundances). With realistic initial conditions, usually inferred from 
$p$-$A$ collisions (Sec.~\ref{ssec_nuc-abs}), phenomenological studies over 
a wide range of energies can be performed (Sec.~\ref{ssec_obs}). 

\subsection{Charmonium Transport}
\label{ssec_transport}
The evolution of the abundance and spectrum of a quarkonium state, $\Psi$, 
in a hot and dense fireball can be tracked via its phase space distribution, 
$f_\Psi$, satisfying a transport equation of the form 
\begin{equation}
p^\mu \partial_\mu f_\Psi(\vec r,t;\vec p) =
- \omega_p \ \Gamma_\Psi(\vec r,t;\vec p) \ f_\Psi(\vec r,t;\vec p) +
\omega_p \ \beta_\Psi(\vec r,t;\vec p)  \ .
\label{transport}
\end{equation}
Here, $\Gamma_\Psi$ denotes the dissociation width discussed in
Secs.~\ref{ssec_diss-qgp} and \ref{ssec_diss-hg}, $\beta_\Psi$ the gain
term responsible for regeneration (discussed in more detail below), and
$p_0$=$\omega_p$=$(m_\Psi^2+{\vec p}^2)^{1/2}$ the on-shell 
energy of state $\Psi$ implying that a well-defined quasiparticle state 
exists (mean-field terms have been neglected which is justified if
mass corrections are small). The right-hand-side of this equation
is the collision term which accounts for both dissociation and
formation reactions (possibly including transitions between different
charmonium states). Detailed balance (and the correct equilibrium limit)
requires that both terms include the same reaction mechanisms.

Let us first discuss the situation where the gain term in the transport
equation can neglected; this can be realized in scenarios (or
evolution phases) where either no charmonium states can be formed (are
not supported in the medium) and/or the actual charmonium number is much 
larger than the
equilibrium abundance in connection with relatively small reaction
rates. Under these circumstances, the evolution of the charmonium
number is given by a sequence of suppression stages for which the
solution of the rate equation may be schematically written in terms
of a (3-momentum dependent) survival probability (or suppression
factor)
\begin{eqnarray}
S_\Psi &=& S_{\rm HG} \ S_{\rm QGP}  \ S_{\rm nuc}
\nonumber\\
& \simeq & \exp\left(-\int\limits_{T_c}^{T_{fo}}\Gamma_{\rm HG}(T)
\frac{dT}{\dot{T}}\right) \
\exp\left(-\int\limits_{T_{0}}^{T_c}\Gamma_{\rm QGP}(T)
\frac{dT}{\dot{T}}\right) \
\exp\left(-n_N \sigma_{\rm abs} L\right) \ .
\label{supp-fac}
\end{eqnarray}
In this approximation the key ingredients are the inelastic reaction
rates in hadron gas (HG) and QGP phases, as well as primordial nuclear
absorption (which will be discussed in more detail in the following
Section, \ref{ssec_nuc-abs}). If the latter can be accurately determined
from $p$-$A$ data, and if the hadronic suppression is small, $S_\Psi$ 
would provide direct information on the (temperature-dependent)
inelastic reaction rate in the QGP (recall that the complete
melting of a charmonium state (e.g. due to color screening) may be
considered as the limit $\Gamma_\Psi\to \infty$). As discussed in
the introduction, the suppression approach has been widely applied
in the interpretation of SPS data on charmonium production in light-
and heavy-ion reactions, encompassing a wide range of models
for both the charmonium dissociation mechanisms (hadronic
comovers~\cite{Gavin:1988hs,Vogt:1988fj,Capella:2000zp,Cassing:1996zb,
Spieles:1999pm}, screening-based models~\cite{Matsui:1986dk,
Blaizot:1996nq,Chaudhuri:2003nj}, as well as thermal dissociation-rate 
calculations) and the evolution of the medium (transport, fireball
and hydrodynamic simulations). An updated account of the
state-of-affairs in this enterprise will be given in
Sec.~\ref{ssec_obs} below.
The suppression factor, Eq.~(\ref{supp-fac}), encodes no notion of
the charmonium masses, of their equilibrium number, nor of any coupling
to the open-charm content in the system. These aspects, however,
become essential upon inclusion of regeneration processes, i.e., the
gain term in Eq.~(\ref{transport}), as we will now discuss.

Early evaluations of secondary charmonium production in heavy-ion 
collisions were focusing on hadronic regeneration reactions in the
hadronic~\cite{Ko:1998fs} and mixed phases~\cite{BraunMunzinger:2000dv}  
under LHC conditions. In 
Refs.~\cite{BraunMunzinger:2000px,Gorenstein:2000ck,Gazdzicki:1999rk}
(see, e.g., Ref.~\cite{Andronic:2006ky} for a recent update) the
statistical hadronization model has been invoked assuming that charmonia 
thermally equilibrate (and diffuse into a homogeneous distribution) upon
completion of the hadronization phase transition
(on the hadronic side), without further modifications in their
abundances in the subsequent hadronic phase. In terms of the rate
equation (\ref{transport}) this implies that the thermal relaxation
time (or inelastic width) of charmonia changes rather rapidly across
the phase transition, which is not inconsistent with the microscopic
calculations discussed in the previous two Sections (at least for the
$J/\psi$). It also implies that the open-charm states forming the
charmonia have kinetically equilibrated. This assumption is less certain,
with current estimates of the thermal relaxation time of charm quarks
of around $\tau_c^{\rm eq}\simeq$~5-7\,fm/$c$~\cite{vanHees:2007me}. However, 
it is presently an open question by how much the $J/\psi$ regeneration yield
is affected by deviations of charm-quark distributions from thermal
equilibrium. E.g., when comparing the limiting cases of either primordial
spectra (i.e., no reinteractions) or a thermal source (+flow), a factor 
of $\sim$3 difference in the $J/\psi$ yield in central Au-Au at RHIC has 
been found in the coalescence model of Ref.~\cite{Greco:2003vf}, while 
the variations in the transport model of Ref.~\cite{Yan:2006ve} are smaller.
In Ref.~\cite{Gazdzicki:1999rk} it was additionally assumed that charm
equilibrates chemically with the surrounding medium,
so that the charmonium density is determined
solely by its mass and and the ambient temperature. With a hadronization
temperature of $T_c=175$\,MeV, a surprisingly good description of the
$J/\psi$-over-pion ratio at the SPS was found.
In Ref.~\cite{BraunMunzinger:2000px} the assumption of chemical
equilibration of charm-quarks was relaxed based on theoretical
expectations~\cite{Levai:1994dx,Rafelski:1996hf,BraunMunzinger:2000dv}  that
$c\bar c$ pair production is restricted to primordial $N$-$N$ collisions
and frozen thereafter (available experimental data at
RHIC~\cite{Adler:2004ta} and SPS~\cite{Shahoyan:2007zz} support this
assertion). Under these conditions, charmonium production at the
hadronization transition corresponds to a coalescence of available
charm and anticharm quarks. In thermal equilibrium, this leads to
a charmonium density
\begin{equation}
n_\Psi^{\rm eq}(m_\Psi; T,\gamma_c) = d_\Psi \gamma_c^2
\int \frac{d^3p}{(2\pi)^3} f_\Psi(\omega_p;T) \ ,
\label{npsi}
\end{equation}
where $\gamma_{\bar c}$=$\gamma_c$ denote a (anti-) charm-quark fugacity 
(particle densities and numbers are denoted by $n$ and $N$, 
respectively). In {\em relative} chemical equilibrium, the fugacity is 
determined by matching the number of charm-quark carriers to the total 
number, $N_{c\bar c}$, of $c\bar c$ pairs in the system,
\begin{equation}
N_{c\bar c}=\frac{1}{2} N_{op}\frac{I_1(N_{op})}{I_0(N_{op})}+
V_{FB} \sum\limits_{\eta_c,J/\psi,\dots} n_\Psi^{\rm eq}(T)
 \ ,
\label{Ncc}
\end{equation}
where $N_{op}$ denotes the number of all open-charm states in a fireball
of volume $V_{FB}$. In a QGP, where the charm-quark number is carried by
deconfined $c$-quarks, one
has $N_{op}$=$V_{FB}\gamma_c 2 n_c^{\rm eq}(m_c^*,T)$, whereas in the
HG, $N_{op}$=$V_{FB}\gamma_c \sum_\alpha n_\alpha^{\rm eq}(T,\mu_B)$
with $\alpha$ running over all known charmed hadrons ($D$, $\bar D$,
$\Lambda_c$, $\dots$; note that for a finite baryon chemical potential, 
$\mu_B>0$, 
$N_{\Lambda_c}\ne N_{\bar\Lambda_c}$ implying $N_D \ne N_{\bar D}$,
etc.). The ratio of modified Bessel functions, $I_1/I_0$, in
Eq.~(\ref{Ncc}) approaches one for large $N_{op}$, but goes to
$0.5 N_{op}$ for $N_{op}\ll 1$. In the latter limit, $I_1/I_0$ 
can be interpreted as the probability of producing an extra 
(anti-) charm state, i.e., it enforces exact charm
conservation in the canonical ensemble ($c$ and $\bar c$ can only
be produced together)~\cite{Gorenstein:2000ck}.
Note that the charmonium equilibrium limit does not depend on
nuclear absorption or pre-equilibrium effects. The QGP evolution prior
to hadronization is, however, imprinted on the collective properties
of the statistically produced charmonia, i.e., their $p_T$ spectra
and elliptic flow~\cite{Greco:2003vf,Yan:2006ve,Bugaev:2001sj}.
If $c$-$\bar c$ coalescence occurs off equilibrium,
the pertinent charmonium spectra reflect on the
collective properties of the charm quarks, which, in turn, may
serve as an additional means to identify a coalescence component
in the spectra~\cite{Greco:2003vf,Thews:2005vj}.
In Refs.~\cite{Grandchamp:2001pf,Grandchamp:2002wp}, a 2-component
model has been proposed where statistical production at hadronization
is combined with primordially produced charmonia subject to suppression
in the QGP, followed by hadronic suppression of both components.

Solutions of the transport equation (\ref{transport}) for $J/\psi$
production in heavy-ion collisions under inclusion
of the gain have first been evaluated in Ref.~\cite{Thews:2000rj};
neglecting spatial dependencies in the temperature profile
as well as in the $J/\psi$ and open-charm densities,
a simplified rate equation of the form
\begin{equation}
\frac{dN_{J/\psi}}{d\tau}  = \lambda_F N_c n_{\bar c}
- \lambda_D N_{J/\psi} n_g \ ,
\label{rate-g}
\end{equation}
has been employed, very similar to Eq.~(\ref{rate-eq}).
The reactivity $\lambda_D = \langle v_{\rm rel} \sigma_D \rangle$ is
given in terms of the $J/\psi$ dissociation cross section,
for which the gluo-dissociation process, Eq.~(\ref{gdiss}), has been
used (and detailed balance for the formation reactivity, $\lambda_F$).
As emphasized in Ref.~\cite{Thews:2000rj} (as well as in the
statistical models with charm
conservation~\cite{BraunMunzinger:2000px,Gorenstein:2000ck}), the
$J/\psi$ formation rate depends quadratically on the number,
$N_{\bar cc}$, of (anti-)
charm quarks in the system, rendering an accurate knowledge of the
charm-production cross section an essential ingredient for reliable
predictions of charmonium regeneration. In
the equilibrium limit, the quadratic dependence is encoded in the
squared fugacity factor in the charmonium densities, Eq.~(\ref{npsi}).
\begin{figure}[!tb]
\begin{center}
\includegraphics[width=.50\textwidth]{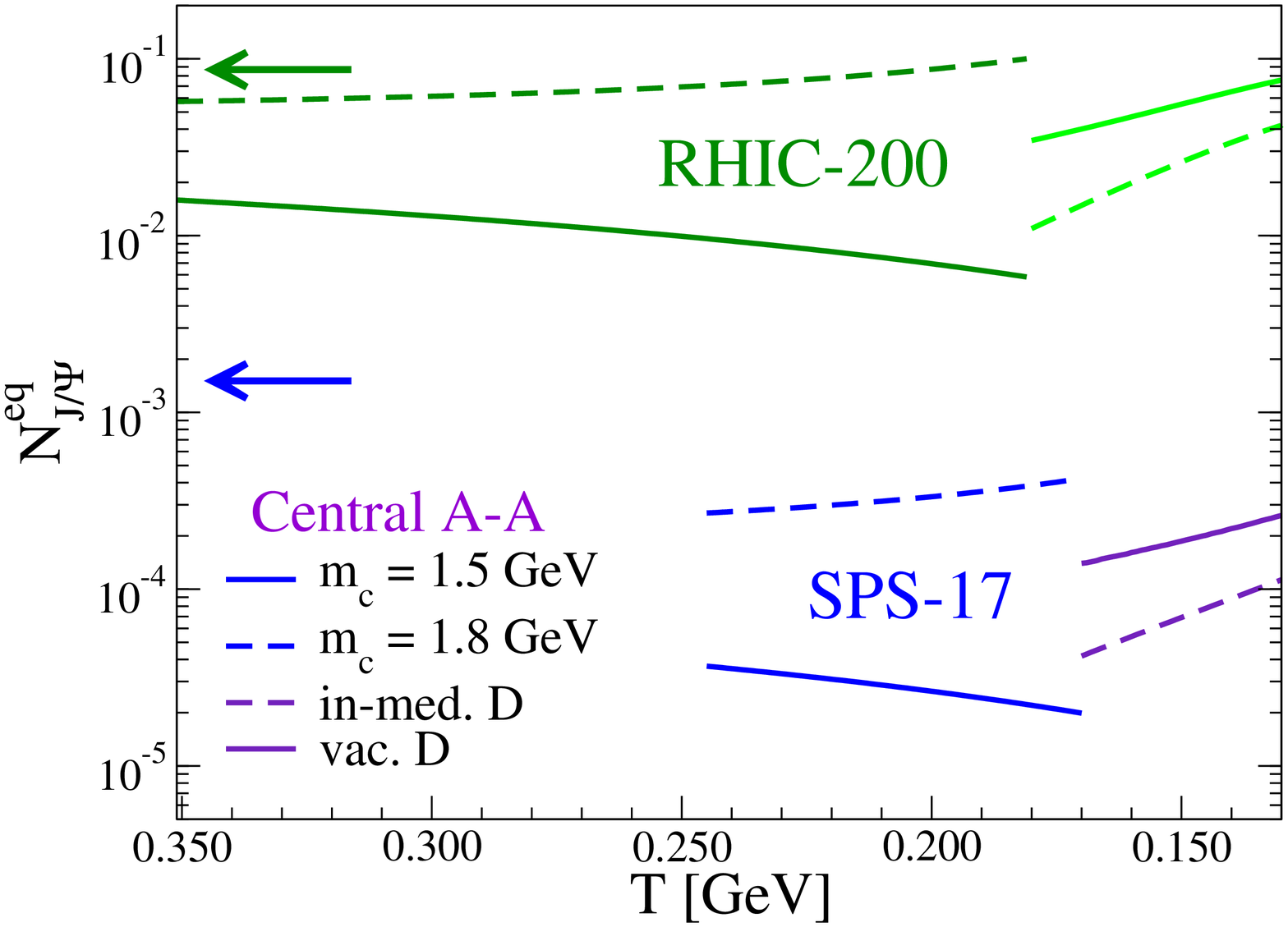}
\end{center}
\caption{Equilibrium abundance of $J/\psi$ mesons as a function of
temperature in an isotropic fireball representative for central Pb-Pb
and Au-Au collisions at SPS and RHIC,
respectively~\cite{Grandchamp:2003uw}. While the variation
in temperature is comparatively moderate, a large sensitivity of the
equilibrium numbers with respect to the charm-quark
(charm-hadron) masses in the QGP (HG) is found. The absolute
numbers are based on open-charm cross sections in $p$-$p$ collisions 
of $\sigma_{pp}^{c\bar c}=570~\mu$b and 5.5~$\mu$b at RHIC and SPS, 
respectively (these values are roughly consistent with PHENIX
measurements~\cite{Adler:2004ta,Adare:2006hc} and compilations for
fixed target energies~\cite{Frixione:1997ma,Lourenco:2006vw}), 
extrapolated to central $A$-$A$ using binary-collision
scaling, and for a rapidity interval of 3.6 units around midrapidity
(corresponding to 2 thermal fireballs). The arrows to the left
indicate the $J/\psi$ abundance from primordial (hard) production.}
\label{fig_npsi-eq}
\end{figure}
As elaborated in Refs.~\cite{Grandchamp:2003uw,Rapp:2003vj,Andronic:2007zu}, 
the equilibrium limit of the $J/\psi$ abundance is sensitive to
the masses of both open- and hidden-charm states. E.g., for fixed charmonium
masses, a reduced charm-quark mass in the QGP (or reduced charm-hadron
masses in the HG) implies a reduced $\gamma_c$ for fixed $N_{c\bar c}$,
which in turn leads to a smaller $N_\Psi^{\rm eq}$; 
in other words,
reduced (increased) open-charm masses make it thermodynamically more
(less) favorable to allocate a given number of $c\bar c$ pairs into
open-charm states rather than into charmonia, see
Fig.~\ref{fig_npsi-eq}.
This interplay has been quantitatively worked out in
Ref.~\cite{Grandchamp:2003uw} based on the rate equation (\ref{rate-eq})
with in-medium charm-quark and -hadron masses in QGP and HG phases,
respectively. In particular, $m_c^*(T_c)\simeq$~1.6-1.7\,GeV on the QGP
side has been chosen to provide a continuous transition of $\gamma_c$
into the hadronic phase at the (pseudo-) critical temperature 
(this range of $c$-quark masses is somewhat below the value
of $m_c^*(T_c)\simeq1.8$\,GeV as inferred from the lattice correlator
analysis using the internal energy as the heavy-quark potential,
recall Fig.~\ref{fig_mc-zero}). As discussed in Sec.~\ref{ssec_pot},
the screening of lattice-based potentials
leads to reduced binding energies, while the in-medium change in
the $J/\psi$ mass is expected to be rather small. The latter is 
rendered approximately constant by a decrease in the in-medium 
charm-quark mass $m_c^*$, as seems to be required in potential-model 
analyses of lattice QCD correlators (recall 
also the paragraph prior to Eq.(\ref{E-W})). 
In Ref.~\cite{Zhao:2007hh}, the rate-equation approach
of Ref.~\cite{Grandchamp:2003uw}, which was restricted to inclusive
charmonium production (i.e., for $p$=0), has been extended to finite
3-momentum. The suppression part has been treated explicitly
(including spatial dependencies, such as leakage effects) in an
expanding fireball background, whereas the momentum dependence of
the regeneration part was approximated with a blast-wave model
(within the same fireball). The latter implies the underlying
charm-quark spectra to be thermalized, at least in the relevant 
regime of low momentum~\cite{vanHees:2007me,Greco:2007sz}.

A more advanced treatment of the gain term in the rate
equation~(\ref{transport}) within a
hydrodynamically evolving background medium was carried out in
Ref.~\cite{Yan:2006ve} (see also Ref.~\cite{Gossiaux:2004qw}).
Assuming the prevalence of the 2$\leftrightarrow$2 dissociation
rate from gluo-dissociation, $g+\Psi \leftrightarrow c +\bar c$,
the gain term takes the form
\begin{eqnarray}
\beta_\Psi(\vec p_T;\vec r_t,\tau) = \frac{1}{2\omega_p} \int
\frac{d^3k}{(2\pi)^3 2\omega_k} \frac{d^3p_c}{(2\pi)^3 2\omega_{p_c}}
\frac{d^3p_{\bar c}}{(2\pi)^3 2\omega_{p_{\bar c}}}
f^c(\vec p_c;\vec r_t,\tau) \, f^{\bar c}(\vec p_{\bar c};\vec r_t,\tau)
\,  W_{c\bar c}^{g\Psi}(s) \, \Theta(T(\vec r_t,\tau)-T_c)
\nonumber\\
\times  (2\pi)^4 \, \delta^{(4)}(p+k-p_c-p_{\bar c}) \ ,
\label{gain}
\end{eqnarray}
where the charmonium phase space has been restricted to the transverse
plane at central rapidity ($p_z$=0=$y$). With
$W_{g\Psi}^{c\bar c}(s)=\sigma_{g\Psi}^{\rm diss} v_{\rm rel}
(2\omega_k 2 \omega_p)$ representing the $\Psi$ dissociation 
probability, the formation probability, $W^{g\Psi}_{c\bar c}(s)$, is
inferred from detailed balance. $\omega_k$, $\omega_p$, $\omega_{p_c}$
and $\omega_{p_{\bar c}}$ denote the on-shell energies of gluon, 
charmonium, charm and anticharm quark, respectively, with charm(onium) 
masses of $m_c$=1.87\,GeV and $m_{J/\psi}$=3.1\,GeV (corresponding
to a ``vacuum" binding energy of $\varepsilon_B^{J/\psi}$=0.64\,GeV).
The $\Theta(T-T_c)$ function in Eq.~(\ref{gain}) restricts the
inelastic processes to the QGP phase, with $T_c$=165\,MeV in
Ref.~\cite{Yan:2006ve}.
Note that an explicit treatment of NLO dissociation processes
in the regeneration term requires the evaluation of a 3-body
initial state.
The initial conditions for the charmonium distribution function,
$f_\Psi(\vec p;\vec r,\tau_0)$, for its evolution in the thermally
evolving medium are typically taken from a Glauber collision profile
for production in primordial $N$-$N$ collisions, plus the effects
of nuclear absorption. In the ``pre-equilibrium" phase, i.e., for
times earlier than the thermalization time, $\tau_0$, of the fireball,
the gain term is switched off. While the production of a $c\bar c$ pair
is expected to occur on a short time scale,
$\tau_{c\bar c}\sim1/2m_c^0\approx0.1$\,fm/$c$, the development of the
charmonium wave function in its rest frame requires a time duration
on the order of its inverse binding energy,
$\tau_\Psi\sim 1/\varepsilon_B\approx0.3$\,fm/$c$.
For a slowly moving $J/\psi$, this time is comparable or smaller to
commonly assumed thermalization times of $\tau_0=$~0.3-1\,fm/$c$. However,
for $J/\psi$'s at large transverse momentum, the formation time in the
fireball frame is time-dilated by the (transverse) Lorentz factor
$\gamma_{T}=m_{T}/m$, and therefore may well reach into the QGP phase, 
presumably reducing the suppression 
effects~\cite{Spieles:1999pm,Karsch:1987uk,Blaizot:1987ha,Chu:1987sy,Karsch:1990wi,Cugnon:1995fu}
(in a naive picture, the $c\bar c$ wave package is of smaller size than
a fully formed bound state, thus reducing dissociation cross sections
with thermal partons). The interplay of charmonium-formation and
fireball-thermalization time scales will be further pronounced for
in-medium reduced binding energies, as well as for excited charmonia
(which even in vacuum have rather small binding energies).

An alternative approach to thermal descriptions of the bulk medium
are microscopic transport models where the evolution of essentially
all particle distributions is treated by numerical simulations of
the Boltzmann equation. On the one hand, transport models have the
important advantage that no explicit assumption on equilibration is
required, which in particular eliminates the introduction of
thermalization and freezeout times (or conditions)\footnote{Much like
for charmonia, primordial particle production in hadronic transport
models account for a formation time, $\tau_F^h\simeq 1/\Lambda_{\rm QCD}
\simeq$~0.5-1\,fm/$c$, in the rest frame of the hadron.}.
On the other hand, the treatment of phase transitions is more
problematic than in thermal models, where pertinent aspects are
readily encoded in the equation of state (EoS). The main input for
charmonium interactions are  2$\leftrightarrow$2 dissociation reactions
(plus detailed balance), i.e., gluo-dissociation ($g+\Psi\to c+\bar c$)
in the partonic phase and meson dissociation ($M+\Psi\to D+\bar D$ with
$M$=$\pi$, $K$, $\rho$, $\dots$) in the
hadronic phase (sometimes augmented with elastic cross sections).
Especially hadronic cross sections at low energy, as relevant for
the typical reactions in the hadronic phase (corresponding to temperatures
$T<200$\,MeV), are currently beset with rather large theoretical
uncertainties (and, in the case of meson dissociation, not accessible
experimentally), recall Sec.~\ref{ssec_diss-hg}. This problem applies, 
of course, to both transport and thermal descriptions
of the expanding medium. Systematic comparisons of charmonium
observables within complementary descriptions of the bulk medium are
thus valuable to extract robust information on charmonium
properties in heavy-ion collisions.
Recent calculations of charmonium production within transport models,
including regeneration reactions, can be found in
Ref.~\cite{Zhang:2002ug,Linnyk:2006ti,Linnyk:2007zx}, and will be
reiterated in the context of heavy-ion data in Sec.~\ref{ssec_obs}.
E.g., in Ref.~\cite{Zhang:2002ug}, it was found that an increase in the
masses of open-charm states ($c$-quarks and $D$-mesons in the partonic
and hadronic phase, respectively) entails an enhanced $J/\psi$
production, which is consistent with the systematics of the thermal
model as illustrated in Fig.~\ref{fig_npsi-eq}. In
Ref.~\cite{Bratkovskaya:2003ux}, the uncertainty in meson-induced
dissociation reactions was circumvented by introducing a single
matrix element for all dissociation reactions, augmented by
spin-isospin factors and phase space factors (subsequently used
in Refs.~\cite{Linnyk:2006ti,Linnyk:2007zx}).
In a similar spirit, the hadronic comover model~\cite{Capella:2000zp}
has recently been extended to include regeneration
reactions~\cite{Capella:2007jv}.

\subsection{Nuclear Absorption and Initial-State Effects}
\label{ssec_nuc-abs}
\input{nuc-abs}

\subsection{Applications at SPS, RHIC and LHC}
\label{ssec_obs}
In this section we attempt to convert the theoretical and
phenomenological developments described above (charmonium equilibrium
properties as well as implementations into transport
approaches) into interpretations of available data from
ultrarelativistic heavy-ion collisions (URHICs) at SPS and RHIC.
As a rough guideline, let us start by recalling some of the basic
relations which have been proposed to associate different observables
with key properties of charmonia in medium.
\begin{itemize}
\item[(A)] Centrality dependence of $J/\psi$ production, which is
 typically
  normalized to the expected yield in the absence of any medium effects,
  $R_{AA}(N_{\rm part})$ -- based on an increasing matter density
  with centrality (i.e., decreasing impact parameter),
  this is the classic observable to search for an anomalous suppression
  of charmonia possibly linked with the onset of QGP formation.
\item[(B)] Transverse-momentum ($p_T$) spectra, normalized to the
  expected spectra without medium effects (i.e., in $p$-$p$ collisions),
  $R_{AA}(p_T)$ -- here, the original objective is the extraction of the
  $p_T$ dependence of the charmonium dissociation rates. In addition, 
  $p_T$-spectra from regeneration are expected to be softer than the 
  primordial power-law spectra, which provides a handle to disentangle 
  secondary and (suppressed) primordial production. The discrimination 
  power may be further augmented in the elliptic flow, $v_2(p_T)$.
\item[(C)] Excitation function, $R_{AA}(\sqrt{s})$ 
  -- the suppression effect on primordial
  charmonia is expected to increase with increasing collision energy,
  providing complementary information on the temperature and density
  dependence of charmonium disintegration (in addition to centrality
  dependencies at fixed energy). However, the
  possibly largest virtue of an excitation function is the stronger 
  increase in charm-quark production relative to light particles,
  leading to a large variation in the charm-quark densities in the fireball
  which (quadratically) enhances regeneration mechanisms with increasing
  energy. The excitation function thus promises to be a prime observable 
  to study the interplay of suppression and regeneration mechanisms.
\item[(D)] Rapidity dependence, $R_{AA}(y)$ -- 
  the idea behind this observable is
  reminiscent of (C) in that light- and charm-quark production are
  characterized by different rapidity distributions (narrower for heavy
  quarks); regeneration, being proportional to $N_{c\bar c}^2$ 
  (grand-canonical limit), should thus lead to a narrowing
  of the charmonium distributions compared to $p$-$p$ collisions.
\item[(E)] Excited charmonia ($\chi_c$, $\psi'$) -- different binding
energies, dissociation rates and dissolution temperatures are hoped
to provide a systematic suppression/production pattern that can serve
as a ``thermometer" upon varying control parameters such as centrality,
system size and collision energy (all of the $J/\psi$ observables
listed under (A)-(D) apply here).
\item[(F)] Bottomonia -- due to stronger binding energies, bottomonium
suppression could provide a more sensitive measure of color screening
via an associated increase in dissociation rates. This notion is 
corroborated by the small open-bottom cross sections which presumably
suppresses bottomonium regeneration.
\end{itemize}
We recall that feed-down contributions
via the decays of excited charmonia into lower lying ones need to be
considered. For the $J/\psi$ about 30\% of its inclusive production 
yield in $p$-$p$ collisions arises from decays of $\chi_{c}$ 
states~\cite{Abt:2002vq}, and $\sim$10\% from 
$\psi'$~\cite{Abt:2006va,Leitch:2008vw}. Such fractions cannot
be neglected in quantitative calculations of $J/\psi$ observables in
heavy-ion reactions, rendering a simultaneous and consistent treatment of
medium effects on $\chi_c$ and $\psi'$ production mandatory.
At high $p_T$$>$5\,GeV,
relevant for collider energies, the inclusive $J/\psi$ yield
is expected to receive additional feed-down contributions from $B$-meson
decays, on the order of 20-40\%~\cite{Acosta:2004yw,Leitch:2008vw}.

In the remainder of this Section we discuss the 6 classes of observables
in the order given above, focusing on the recent status and
developments, and with preference for approaches which have been applied
at both SPS and RHIC energies.

\subsubsection{Centrality Dependence}
We start our discussion with the centrality dependence of
$J/\psi$ production at SPS energies, specifically with implementations
of the $J/\psi$ transport equation into a thermally expanding background
medium.

\begin{figure}[!tb]
\begin{minipage}{0.5\linewidth}
\includegraphics[width=1.0\textwidth]{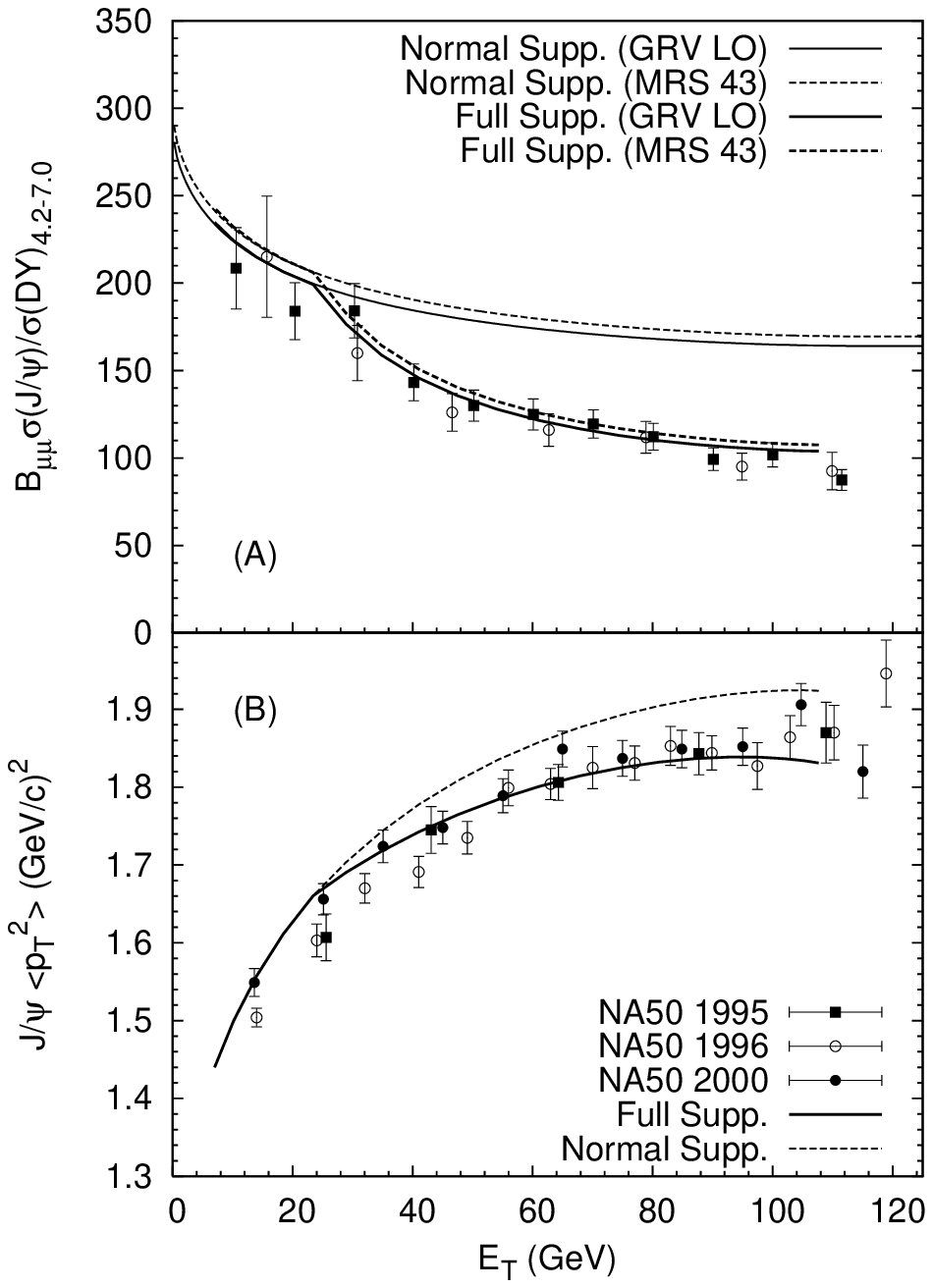}
\end{minipage}
\hspace{0.2cm}
\begin{minipage}{0.5\linewidth}
\vspace{1.2cm}
\includegraphics[width=.88\textwidth]{RAA-j-sps-zhao.eps}
\includegraphics[width=.88\textwidth]{pt2-j-sps-zhao.eps}
\end{minipage}
\caption{$J/\psi$ production in Pb-Pb collisions at SPS employing the
rate equation~(\ref{transport}) within a hydrodynamic evolution with 
gluo-dissociation in the QGP (left panels, gain term 
neglected)~\cite{Zhu:2004nw}, and within an expanding thermal fireball
with quasifree dissociation in the QGP and meson-induced break-up in the
hadronic phase (right panels)~\cite{Grandchamp:2003uw,Zhao:2007hh}.
Upper panels: nuclear modification factor as a function of centrality
quantified by the measured transverse energy, $E_T$, of produced
particles. Lower panels: centrality dependence of the average
transverse momentum squared of the $J/\psi$ spectra.}
\label{fig_raa-na50}
\end{figure}

In Ref.~\cite{Zhu:2004nw} the gluo-dissociation rates for $J/\psi$ and 
$\chi_c$ (using vacuum binding energies) have been folded over a 2+1
hydrodynamic evolution of Pb(158\,AGeV)-Pb collisions, employing an EoS
consisting of a 2+1-flavor QGP and a hadron resonance gas
(connected via a mixed phase at $T_c\simeq165$\,MeV).
Hadronic suppression of the $J/\psi$ and
regeneration effects (expected to be small at SPS energies
due to small open charm production) have been neglected.
The primordial nuclear absorption cross section has been fixed at
$\sigma_{\rm abs}=4.3$\,mb.
The overall magnitude of the suppression in the NA50 data has been
reproduced by introducing a minimal initial entropy density of
$s_0^{\rm min}\simeq 32$\,fm$^{-3}$, below which the matter is assumed to
decouple instantaneously (i.e., not to become part of the thermally
evolving medium); with a thermalization time of $\tau_0=0.8$\,fm/$c$ (the 
moment when the hydrodynamic evolution is assumed to start), the
centrality dependence of the $J/\psi$ yield
(upper left panel of Fig.~\ref{fig_raa-na50}), as well as its average
momentum squared (lower left panel), are well reproduced.

Within a similar spirit, but adopting a slightly different approach,
solutions of the rate equation (\ref{rate-eq}) have been evaluated
in Refs.~\cite{Grandchamp:2003uw,Zhao:2007hh}. The thermal evolution
of the medium has been treated in a more simplistic isotropic fireball
expansion, parametrized to resemble hydrodynamic models, with an EoS 
including QGP, mixed and hadronic phases and $T_c=170(180)$\,MeV at SPS
(RHIC)\footnote{A smaller value of $T_c$ at SPS energies is expected due 
to its decrease with $\mu_B$, while $T_c$=180\,MeV at $\mu_B$=0 is within 
the current range of lattice results for full QCD~\cite{Karsch:2007dt}.}, 
similar to Ref.~\cite{Zhu:2004nw}. The inelastic charmonium
reactions in the QGP include an in-medium reduction of the binding
energies, incorporated using the quasifree dissociation
process~\cite{Grandchamp:2001pf} (rather than gluo-dissociation,
recall Fig.~\ref{fig_diss-dia}) corresponding to the rates displayed
in the right panel of Fig.~\ref{fig_diss-rates}.
The dissociation rates in the hadronic phase have
been estimated based on $SU(4)$ effective theory, leading to
rather small effects on the $J/\psi$ (somewhat more
significant for $\chi_c$ and $\psi'$). Fixing $\sigma_{\rm abs}=4.4$\,mb
and $\tau_0=1$\,fm/$c$ (translating into an average initial temperature 
of $\bar T_0\simeq 210$\,MeV in
central Pb-Pb), the coupling constant in the quasifree process has
been adjusted to $\alpha_s$$\simeq$0.25 to fit the suppression in
central Pb-Pb. Also here the centrality dependence of inclusive $J/\psi$ 
production and its mean $\langle p_T^2\rangle$ are well described. The 
contribution from regeneration turns out to be small (with a total 
open-charm cross section of $\sigma_{c\bar c}^{tot}$=5.5\,$\mu$b 
distributed over 2 thermal fireballs, i.e., $\sim$3.5 units in 
rapidity); it is further reduced by a 
schematic implementation of incomplete charm-quark thermalization via
a relaxation-time factor ${\cal R}(\tau)$ (with a smaller relaxation
time, $\tau_c^{\rm eq}$, leading to a larger regeneration yield), see below.

Hadronic comover-interaction~\cite{Capella:2000zp} and
transport~\cite{Spieles:1999kp} models have also been successful
in describing the $J/\psi$ suppression pattern observed by NA50. With
a typical comover dissociation cross section of 0.65-1\,mb, the required
comover energy densities in central Pb-Pb are rather large, (well)
above the typical value of 1\,GeV/fm$^3$ associated with the phase
transition.  The comover interaction is thus to be understood as
at least partially of partonic origin, consistent with the rather small 
value of the dissociation cross section. In the
hadronic transport calculations of Ref.~\cite{Spieles:1999kp}, the
hadronic comover interactions are significantly larger (a few
millibarns for baryon-induced dissociation, and 2/3 of that for
mesons), but ``pre-hadronic" degrees of freedom still play a role.

\begin{figure}[!t]
\begin{minipage}{0.33\linewidth}
\includegraphics[width=.9\textwidth]{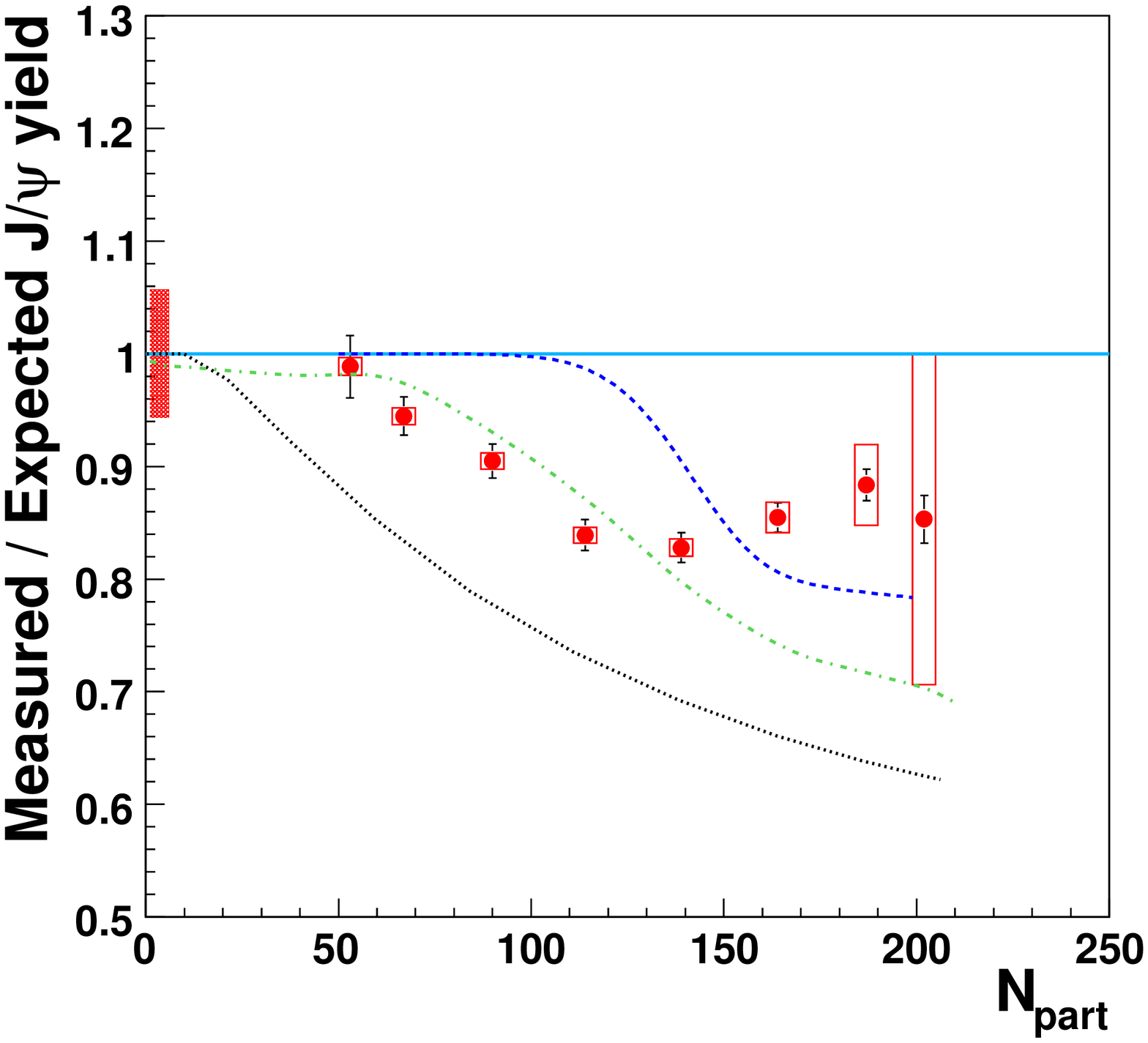}
\end{minipage}
\begin{minipage}{0.74\linewidth}
\includegraphics[width=.9\textwidth]{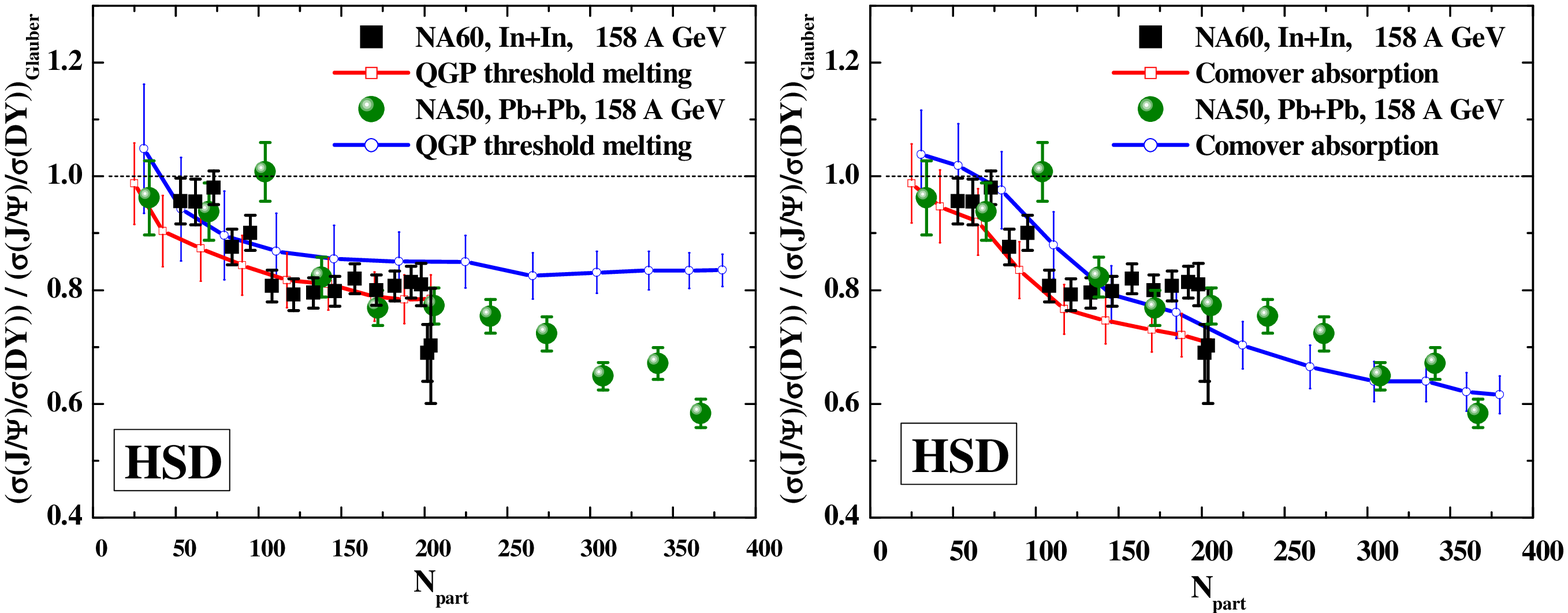}
\end{minipage}
\caption{Left panel: NA60 data for $J/\psi$ production in
In(158\,AGeV)-In collisions at SPS energies~\cite{Arnaldi:2007zz}
compared to theoretical predictions based on the QGP
threshold-suppression scenario~\cite{Digal:2003sg} (upper dashed
line), QGP and hadronic suppression in a thermal
fireball~\cite{Rapp:2005rr} (middle dashed-dotted line) and the
comover model~\cite{Capella:2005cn} (lower dotted line);
all models are in fair agreement with NA50 data in Pb-Pb collisions.
Middle and right panels: postdictions for Pb-Pb~\cite{Alessandro:2004ap}
and In-In~\cite{Arnaldi:2006it} data computed in the HSD transport
model~\cite{Linnyk:2006ti} using either a QGP threshold-melting
scenario (middle panel) or comover model (right panel); note that the 
NA60 data are the preliminary ones from Ref.~\cite{Arnaldi:2006it}. }
\label{fig_raa-na60}
\end{figure}
An often discussed issue in the interpretation of the NA50 data is the
(non-) presence of a threshold behavior in semi-central Pb-Pb collisions,
around $E_T\simeq$~30-40\,GeV. To address this question experimentally,
the successor experiment NA60 has measured $J/\psi$ production for
an intermediate-size system, i.e., In-In collisions ($A_{In}=115$). 
The pertinent
data are compared theoretical {\em predictions} in the left panel of
Fig.~\ref{fig_raa-na60}, all 3 of which are in reasonable agreement with
the Pb-Pb data. While the comover model~\cite{Capella:2005cn} and the
threshold-melting scenario~\cite{Digal:2003sg} do not reproduce the
data, the thermal rate equation approach in QGP + hadronic
phase~\cite{Grandchamp:2003uw,Rapp:2005rr} shows
a fair agreement (except for central collisions where the data suggest
a reduced suppression). The suppression pattern at SPS energies
has been revisited within the Hadron String Dynamics (HSD) transport
model in Ref.~\cite{Linnyk:2006ti}, where charmonium suppression was
studied within the QGP threshold-melting scenario (middle panel of
Fig.~\ref{fig_raa-na60}) and a comover-interaction model (right panel
of Fig.~\ref{fig_raa-na60}). The former is characterized by critical
energy densities above which the corresponding state does not
form\footnote{It may be thought of as a sudden transition of
the quarkonium width from zero to infinity.}
while the latter has been implemented via a single matrix element for
all charmonium states (including phase space and spin effects).
The threshold melting leads to a rather pronounced drop in $J/\psi$ 
production at small centralities ($N_{part}$$\le$75) together with a 
leveling off at large 
$N_{part}$, whereas the comover scenario produces a more gradual 
suppression pattern which reproduces the trend in the data 
somewhat better~\cite{Linnyk:2006ti}. Note that the level of agreement 
of the HSD+comover scenario with NA60 data is quite comparable to the
thermal rate-equation prediction~\cite{Grandchamp:2003uw,Rapp:2005rr}
in the left panel of Fig.~\ref{fig_raa-na60} (the latter is sometimes 
quoted as not reproducing the NA60 
data~\cite{Linnyk:2006ti,Arnaldi:2007zz,Lourenco:2005zz}).

\begin{figure}[!t]
\begin{minipage}{0.61\linewidth}
\includegraphics[width=.98\textwidth]{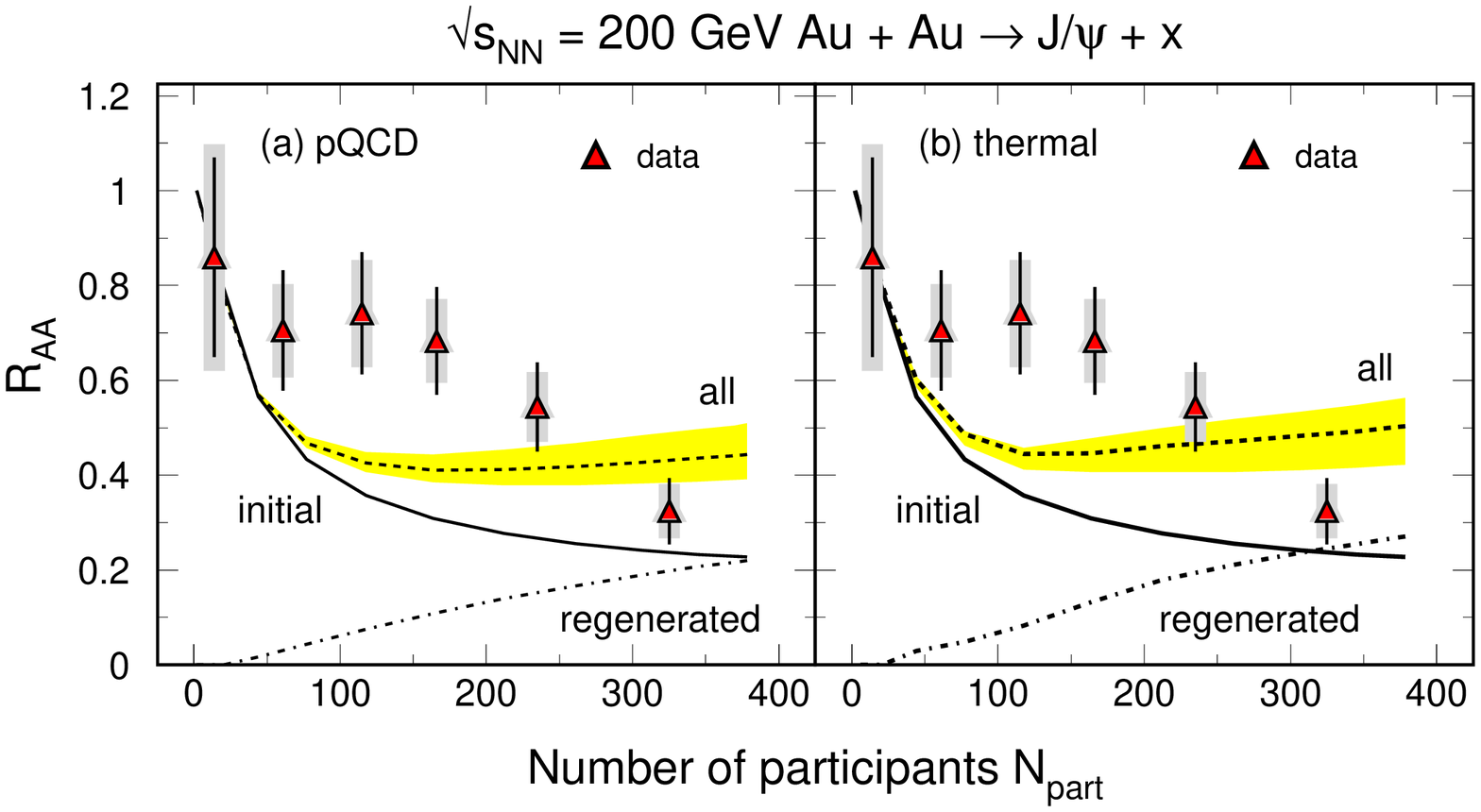}
\end{minipage}
\begin{minipage}{0.61\linewidth}
\includegraphics[width=.6\textwidth]{RAA-j-rhic-zhao.eps}
\end{minipage}
\begin{minipage}{0.50\linewidth}
\vspace{-0.5cm}
\includegraphics[width=1.2\textwidth]{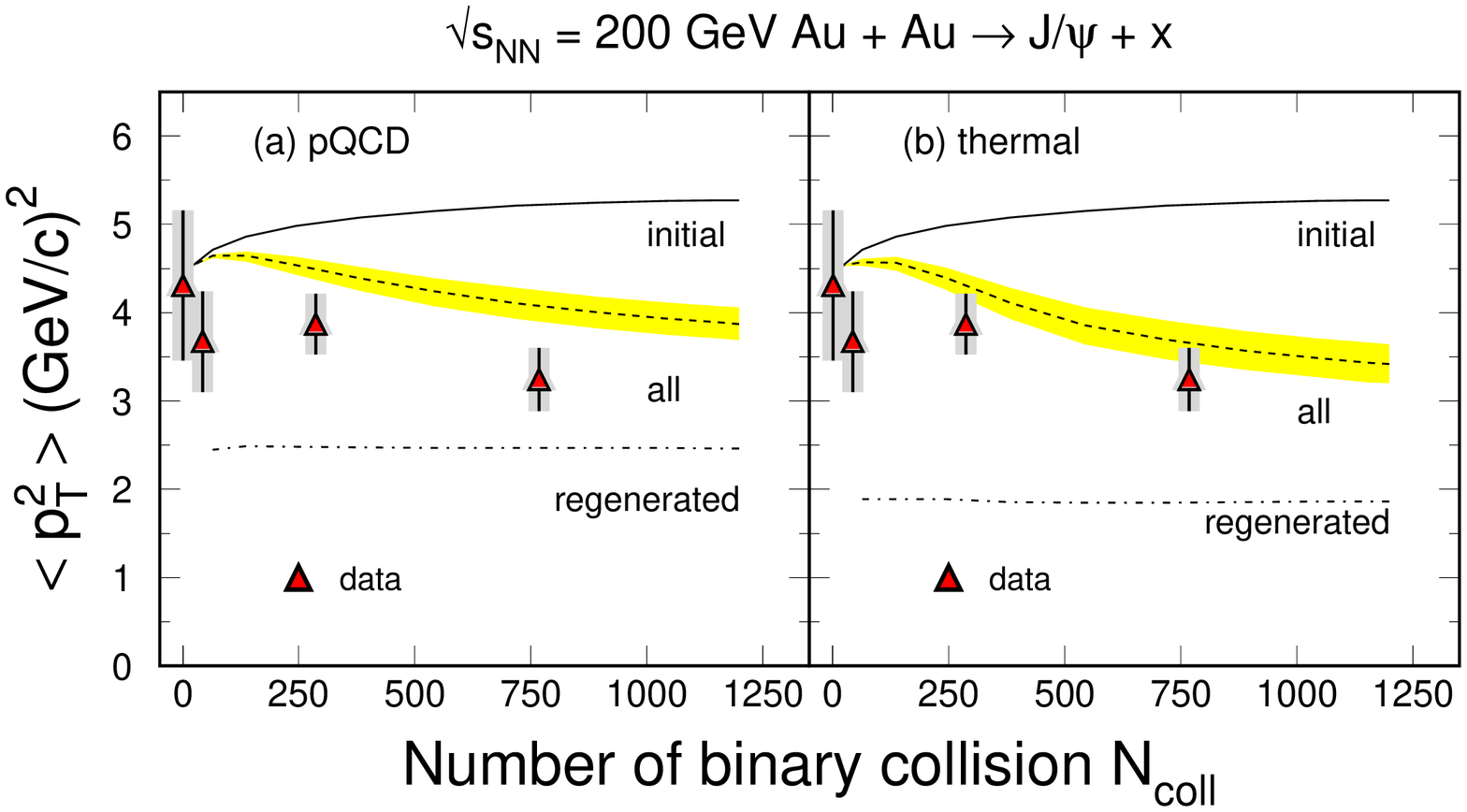}
\end{minipage}
\hspace{1.5cm}
\begin{minipage}{0.5\linewidth}
\vspace{-0.5cm}
\includegraphics[width=.78\textwidth]{pt2-j-rhic-zhao.eps}
\end{minipage}
\caption{$J/\psi$ production in Au-Au collisions at RHIC computed
within kinetic rate equations in a thermally evolving background
medium, corresponding to the approaches displayed in 
Fig.~\ref{fig_raa-na50} (left panels: gluo-dissociation
with vacuum binding energies in a hydrodynamic
simulation~\cite{Yan:2006ve,Zhu:2004nw}, right panels: quasifree
dissociation with in-medium binding energies in a thermal fireball
expansion~\cite{Grandchamp:2003uw,Zhao:2007hh}). The upper panels
show the nuclear modification factor, $R_{AA}^{J/\psi}(N_{part})$
vs. centrality for inclusive $J/\psi$ yields, while the lower
panels display the centrality dependence of the $J/\psi$'s
average momentum squared (vs. $N_{coll}$ in the left panels and vs. 
$N_{part}$ in the right panel). The left panels contain preliminary
data~\cite{PereiraDaCosta:2005xz}, the right panels final 
ones~\cite{Adare:2006ns}.}
\label{fig_raa-phenix}
\end{figure}
Let us now turn to RHIC energies, focusing on the approaches discussed
above in the SPS context. The hydrodynamic model with gluo-dissociation
has been extended to include regeneration reactions,
$c+\bar c \to g + \Psi$, in Ref.~\cite{Yan:2006ve}. The charm-quark
distributions figuring into the gain term, Eq.~(\ref{gain}), have
been approximated by two limiting scenarios: (a) primordial spectra
in $p$-$p$ given by perturbative QCD (pQCD),
or (b) fully thermalized spectra including
the collective flow of the hydrodynamic background. The corresponding
difference in the inclusive $J/\psi$ yields turns out to be rather
small, cf.~upper left panels in Fig.~\ref{fig_raa-phenix}.
The underlying open-charm production cross section has been adopted
from the PHENIX measurement in Ref.~\cite{Adler:2004ta}, 
$\sigma_{pp}^{c\bar c}$=622$\pm$57\,$\mu$b\footnote{An update of the 
PHENIX open-charm cross section amounts to 
$\sigma_{pp}^{c\bar c}$=567$\pm$57$\,\mu$b\cite{Adare:2006hc}; for
both measurements the additional systematic uncertainty is about 
150-200\,$\mu$b; STAR data~\cite{Zhong:2007iq} give significantly 
larger values, $\sigma_{pp}^{c\bar c}$=1.3$\pm$0.1$\pm$0.2\,mb, 
which is a currently unresolved discrepancy.}.
The experimental uncertainty translates into an uncertainty in
the regeneration yield reflected by the colored bands in the left
panels of Fig.~\ref{fig_raa-phenix}
(the cross section has been scaled by the number of binary $N$-$N$
collision at given impact parameter).  The overall 
magnitude of preliminary PHENIX $J/\psi$ data~\cite{PereiraDaCosta:2005xz} 
is roughly reproduced (the agreement is somewhat better with updated PHENIX 
data~\cite{Adare:2006ns} displayed in the upper right panel). The 
contribution from regeneration is moderate but significant, reaching 
ca.~50\% of the total yield in central Au-Au.

The right panels of Fig.~\ref{fig_raa-phenix} shown results of the thermal 
rate-equation approach~\cite{Grandchamp:2003uw}, extended to include the 
finite 3-momentum dependence of the inelastic charmonium reaction
rates~\cite{Zhao:2007hh}. Here, the impact of incomplete charm-quark
thermalization has been schematically modeled by a ``thermal relaxation 
time" factor, ${\cal R}(\tau) = 1-{\rm e}^{-\tau/\tau_c^{\rm eq}}$,
multiplying the charmonium equilibrium abundance, $N_\Psi^{\rm eq}$,
which figures into the gain term of the rate equation (\ref{rate-eq}). This 
implementation leads to a stronger dependence of the regeneration yield
on charm-quark equilibration than in Ref.~\cite{Yan:2006ve};
the original predictions of Ref.~\cite{Grandchamp:2003uw} were
done with $\tau_c^{\rm eq}=3$\,fm/$c$ and
$\sigma_{pp}^{c\bar c}=570\,\mu$b~\cite{Grandchamp:2002wp} 
(consistent with the PHENIX cross section), which somewhat overestimate 
the $J/\psi$ yield in central Au-Au as compared to the most recent 
PHENIX measurement~\cite{Adare:2006ns} (cf.~dash-double-dotted line in 
the upper right panel of Fig.~\ref{fig_raa-phenix}). 
Using $\tau_c^{\rm eq}$=5-7\,fm/$c$~\cite{Zhao:2007hh} improves the 
agreement, implying a decomposition into primordial 
and regenerated component rather comparable to the one shown in the 
upper left panels~\cite{Yan:2006ve}. A potential problem for both 
calculations are the (two) data points at $N_{part}=$~100-200, 
and has been addressed in threshold-melting scenarios~\cite{Gunji:2007uy}. 
HSD calculations at RHIC based on suppression-only scenarios (comover
or threshold melting)~\cite{Linnyk:2007zx} underpredict existing
PHENIX data, thus supporting the presence of a regeneration component.

The $J/\psi$ regeneration yield is
rather sensitive to the open-charm content of the system. First
and foremost, this pertains to the number of charm-anticharm pairs,
$N_{c\bar c}$, in the system. In the grand-canonical limit
($N_{op}\gg1$) the charmonium equilibrium number depends
quadratically on $N_{c\bar c}$. But even at SPS energies, where
charm-quark production is small and the canonical limit applies
(implying that $N_\Psi^{\rm eq}$ depends linearly on $N_{c\bar c}$),
it has been argued that the (semi-) central NA50 $J/\psi$ data can be 
accounted for by statistical production alone~\cite{Andronic:2006ky} 
using a factor of $\sim$2 increase of the pQCD open-charm cross section 
(amounting to $d\sigma_{c\bar c}/dy=5.7\,\mu$b; a recent data 
compilation of open-charm data~\cite{Lourenco:2006vw} results in a value 
of  $\sigma_{c\bar c}^{tot}(E_{lab}=158{\rm AGeV})\simeq$~3.6-5.2\,$\mu$b 
for the total cross section).
It furthermore matters over what range in rapidity the statistical model 
is applied. In the canonical limit, it follows from Eqs.~(\ref{npsi}) 
and (\ref{Ncc}) that $N_\Psi\propto 1/V_{FB}(N_{part})$, which 
reproduces the observed centrality dependence for semi-/central 
collisions.  The width of the considered rapidity window also affects 
the total yield due to different $y$-distributions of $N_{ch}$ and 
$N_{c\bar{c}}$. Finally, recalling
Fig.~\ref{fig_npsi-eq}, even for fixed $N_{c\bar c}$, the
charm-quark fugacity, and thus the equilibrium charmoniun numbers,
are sensitive to the underlying open-charm spectrum (e.g., due to medium 
effects in charm hadrons and/or quarks).

Finally, we comment on two recent developments.
In Ref.~\cite{Young:2008he} the regeneration processes at RHIC have
been studied by performing Langevin simulations for both $c$ and 
$\bar c$ quarks in a hydrodynamically evolving medium and treating 
their mutual interaction after production via a potential taken as 
the heavy-quark internal energy provided by lQCD. The combimation of a 
small spatial $c$-quark diffusion coefficient with a strong $c$-$\bar c$
attraction (deep potential) keeps the original $c\bar c$ pair, whose
primordial production essentially occurs at the same space-time point,
close together and results in a sizable probability to form a 
charmonium bound state in the subsequent QGP and mixed-phase evolution.
This ``diagonal" recombination of $c$ and $\bar c$ from the same
primordial $c\bar c$ pair is not unlike the ``canonical limit" of
the statistical model, augmented by a ``correlation volume" to localize
it in space. The correlation volume is of course replaced by a dynamical
calculation using a microscopic diffusion process. The centrality
dependence of the PHENIX data for the inclusive $J/\psi$ yield can be 
approximately reproduced, and the $p_t$ spectra exhibit a significant   
softening caused by the softening of the (partial) thermalization of
the $c$-quark spectra.
\\
In Ref.~\cite{Zhao:2010} the thermal rate-equation approach (underlying
the right panels in Figs.~\ref{fig_raa-na50} and \ref{fig_raa-phenix})
has been employed to study
the consequences of weak- and strong-binding scenarios for charmonia.
Pertinent charmonium spectral functions have been constructed for
these two scenarios based on the $T$-matrix calculations of 
Ref.~\cite{Riek:2010fk} for either the free ($F$) or internal energy 
($U$) as potential, respectively. In quasiparticle approximation, 
the main quantities characterizing the spectral functions and figuring 
into the rate equation are the temperature-dependent charmonium binding 
energies, charm-quark mass and inelastic width (the former two define
the charmonium masses). The resulting spectral functions were
constrained to yield correlator ratios close to one (in line with lQCD).
In addition, updated nuclear absorption cross sections have been
implemented into the initial conditions, i.e., increasing from 
$\sigma_{\rm abs}\simeq 4.5$\,mb to $\sim$7.5\,mb at SPS energy 
in line with recent $p$-$A$ data by NA60 at 158\,GeV projectile 
energy~\cite{Arnaldi:2010ky}, and from 1.5\,mb to $\sim$4\,mb at RHIC 
energy (which also accounts for shadowing)~\cite{Adare:2006kf}. 
Some of the findings are as follows: the strong-binding scenario (with 
a $J/\psi$ dissociation temperature of $T_{\rm diss}\simeq1.8T_c$)
entails a very small ``anomalous suppression" of primordial $J/\psi$'s
at SPS; most of the observed anomalous suppression (which is 
small due to the large $\sigma_{\rm abs}\simeq7.5$\,mb) is due
to $\chi_c$ and $\psi'$ suppression, and the regeneration contribution
is small (not unlike the right panels in Fig.~\ref{fig_raa-na50}).
In central Au-Au at RHIC, the partition of (suppressed) primordial 
and regenerated components is comparable (not unlike the right
panels in Fig.~\ref{fig_raa-phenix}). On the other hand,
in the weak-binding scenario (with a $J/\psi$ dissociation 
temperature of $T_{\rm diss}\simeq1.1T_c$), the regeneration
component exceeds the suppressed primordial one already for central
Pb-Pb at SPS, and dominates the yield for central Au-Au at RHIC.
While both scenarios can be compatible with the centrality dependence 
of the total yields, i.e., $R_{AA}(N_{\rm part})$, the different $p_T$ 
dependencies of the two components offer a more promising observable for 
discriminating weak- and strong-binding scenarios. 

\subsubsection{Transverse-Momentum Spectra and Elliptic Flow}
Charmonium $p_T$-spectra are believed to shed
more light on the suppression mechanisms (in particular,
its momentum dependence, recall Fig.~\ref{fig_diss-rates-2}), as
well as to disentangle (suppressed) primordial and regenerated 
production components.
Since latter is expected to follow an exponential shape
(plus effects from collective flow), vs. the power-law spectra
of primordial production, it should primarily contribute at low 
momentum (where the open-charm phase space density is 
largest). The $p_T$-dependence of suppression may thus be more
directly studied toward higher momenta (unless regeneration is absent 
(or small) altogether, as is expected at lower collision energies). 
At high momenta, additional effects may lead to a reduction in 
charmonium suppression, e.g., escape from the fireball (``leakage 
effect")~\cite{Ropke:1988zz,Karsch:1987uk,Blaizot:1987ha,Zhuang:2003fu},
delayed formation time due to Lorentz time 
dilation~\cite{Karsch:1987uk,Gavin:1990gm,Blaizot:1988ec} (coupled with 
an inhibited formation in the QGP zone not unlike a leakage effect, or 
using reduced dissociation cross sections during the build-up of the 
charmonium wave package~\cite{Cugnon:1995fu,Gerland:1998bz}), or the 
Cronin effect (a broadening of hadron $p_T$ spectra in $p$-$A$ relative 
to $p$-$p$ collisions)~\cite{Blaizot:1988hh,Hufner:1988wz,Gavin:1988tw}.

In lower panels of Fig.~\ref{fig_raa-na50} the centrality dependence
of the $J/\psi$'s average $p_T^2$ in the hydro+gluo-dissociation (left)
and fireball+quasifree dissociation (right) calculations is compared
to NA50 data. Most of the observed increase of $\langle p_T^2\rangle$
with centrality is accounted for by the Cronin effect. In both calculations 
the latter is implemented into the initial condition for the $p_T$ spectra 
(at the start of the thermal evolution, $\tau=\tau_0$) using an 
impact-parameter dependent average nuclear pathlength, 
$\langle l \rangle(b)$, of the parton prior to fusing into charmonium, 
$\langle p_T^2\rangle(b;\tau_0)=\langle p_T^2\rangle_{pp} +
a_{gN} \cdot \langle l \rangle(b)$ with $a_{gN}$=0.76\,GeV$^2$/fm as
extracted from $p$-$A$ data~\cite{Abreu:2000xe}. The reduction of the
final $\langle p_T^2\rangle$ toward central collisions in the lower right
panel of Fig.~\ref{fig_raa-na50}~\cite{Zhao:2007hh} is mostly caused
by the 3-momentum dependence of the quasifree dissociation rate
(cf.~right panels in Fig.~\ref{fig_diss-rates-2}), leading to
larger suppression for higher momentum charmonia (the smallness of
the regeneration yield, evaluated in blast-wave approximation at $T_c$,
renders its impact on $\langle p_T^2\rangle$ practically negligible
at SPS). A similar reduction is observed in the lower left panel of
Fig.~\ref{fig_raa-na50}~\cite{Zhu:2004nw}, despite the decrease
of the underlying dissociation rate with 3-momentum.
In Ref.~\cite{Gorenstein:2001ti}, the $J/\psi$ and $\psi'$ spectra in
central Pb(158\,AGeV)-Pb collisions have been analyzed using a blast-wave
description alone; the extracted temperature ($T\simeq170$\,MeV) and collective
expansion velocity ($\bar v \simeq 0.2c$) are consistent with the
charmonia forming at the hadronization transition (quite similar to the
regeneration component in the lower left panel of Fig.~\ref{fig_raa-na50}).
However, at high $p_T>3$\,GeV, and toward more peripheral collisions,
this description most likely needs to be supplemented by non-thermal
(primordial) contributions.

\begin{figure}[!tb]
\begin{minipage}{0.5\linewidth}
\includegraphics[width=.47\textwidth]{raa-pt-60-92.eps}
\includegraphics[width=.47\textwidth]{raa-pt-40-60.eps} \\
\includegraphics[width=.47\textwidth]{raa-pt-20-40.eps}
\includegraphics[width=.47\textwidth]{raa-pt-0-20.eps}
\end{minipage}
\begin{minipage}{0.5\linewidth}
\includegraphics[width=.6\textwidth,angle=90]{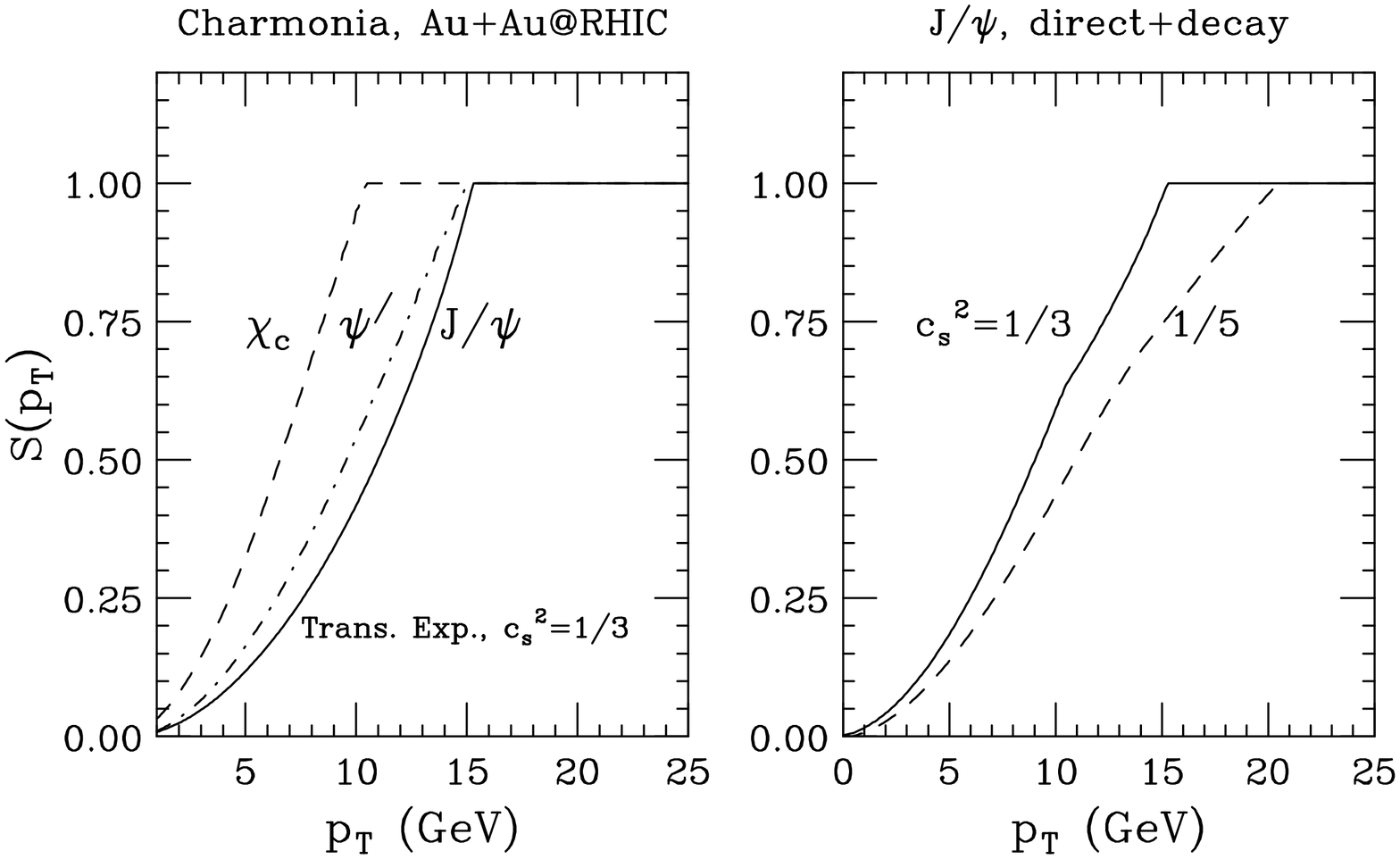}
\end{minipage}
\caption{$J/\psi$ nuclear modification factor as a function of
transverse momentum in Au-Au($\sqrt{s_{NN}}$=200\,GeV) collisions.
Left panels: PHENIX data~\cite{Adare:2006ns} compared to
calculations in a fireball+quasifree dissociation model (including
regeneration)~\cite{Zhao:2007hh} in various centrality classes
(dotted lines: primordial production including Cronin effect and nuclear
suppression, dashed line: primordial production after QGP and hadronic
suppression, dash-double-dotted line: primordial suppressed production
without leakage effect, dash-dotted line: regeneration yield in
blast-wave approximation, full line: total). Right panels: suppression
of primordial charmonium states (second from right) and inclusive $J/\psi$
(far right; including feed-down as well as the sensitivity to the speed of
sound employed in the hydrodynamical expansion)
within a hydrodynamic model employing a QGP threshold-melting
scenario supplemented by formation time effects~\cite{Pal:2000zm}.
}
\label{fig_pt-rhic}
\end{figure}
At RHIC energies, the regeneration component in both the
hydro+gluo-dissociation and the fireball+quasifree dissociation calculations
predicts a softening of  $\langle p_T^2\rangle$ with
centrality, relative to the Cronin-enhanced initial production, see
the lower panels of Fig.~\ref{fig_raa-phenix}.
Current PHENIX data support this scenario, but one should note
that the Cronin effect is not yet well constrained from available
$d$-$A$ data. The analysis of $J/\psi$ $p_T$-spectra in terms of
a blast-wave description does not yet allow for a quantitative
identification of the temperature and transverse flow associated with
the kinetic decoupling of charmonia. However, the current PHENIX
data suggest that a decoupling at the phase boundary results in spectra
which are too soft~\cite{Andronic:2006ky}, especially for noncentral
collisions. This is illustrated more
explicitly in Fig.~\ref{fig_pt-rhic}, where the $J/\psi$ $R_{AA}(p_T)$
is displayed in 4 centrality bins and compared to the
fireball+quasifree dissociation approach~\cite{Zhao:2007hh}.
The rather soft regeneration spectra (corresponding to decoupling at
$T_c$ with a flow velocity increasing for more central collisions),
together with the much harder primordial spectra (including suppression),
result in a rather flat $R_{AA}(p_T)$ (suggesting that both primordial 
and regeneration components are relevant) 
Recent measurements of high-$p_T$
$J/\psi$'s in Cu-Cu($\sqrt{s_{NN}}$=200\,GeV) indicate that
$R_{AA}$($p_T$$>$5\,GeV) is compatible with one~\cite{Tang:2008uy}.
At first sight, and in view
of the strong suppression of other measured hadrons thus far
($R_{AA}^{\pi,\eta}\simeq0.25$), this appears to be a surprise.
However, estimates of bottom feed-down ($B\to J/\psi+ X$) and 
formation-time effects (via a reduced absorption cross section at early
times), indicate that this observation may be understood in the
framework of the rate-equation approach~\cite{Zhao:2008vu}. Formation
time effects coupled with threshold suppression
scenarios~\cite{Karsch:1987uk,Blaizot:1987ha} have indeed predicted
$R_{AA}^{\Psi}\to 1$ at high $p_T$ $\sim$20 years ago, see, e.g., right 
panel of Fig.~\ref{fig_pt-rhic}~\cite{Pal:2000zm}. A rather unique 
signature of this effect is a stronger suppression of $J/\psi$ relative 
to $\chi_c$, which is opposite to both conventional suppression and
statistical hadronization scenarios.

\begin{figure}[!tb]
\begin{minipage}{0.5\linewidth}
\includegraphics[width=.9\textwidth]{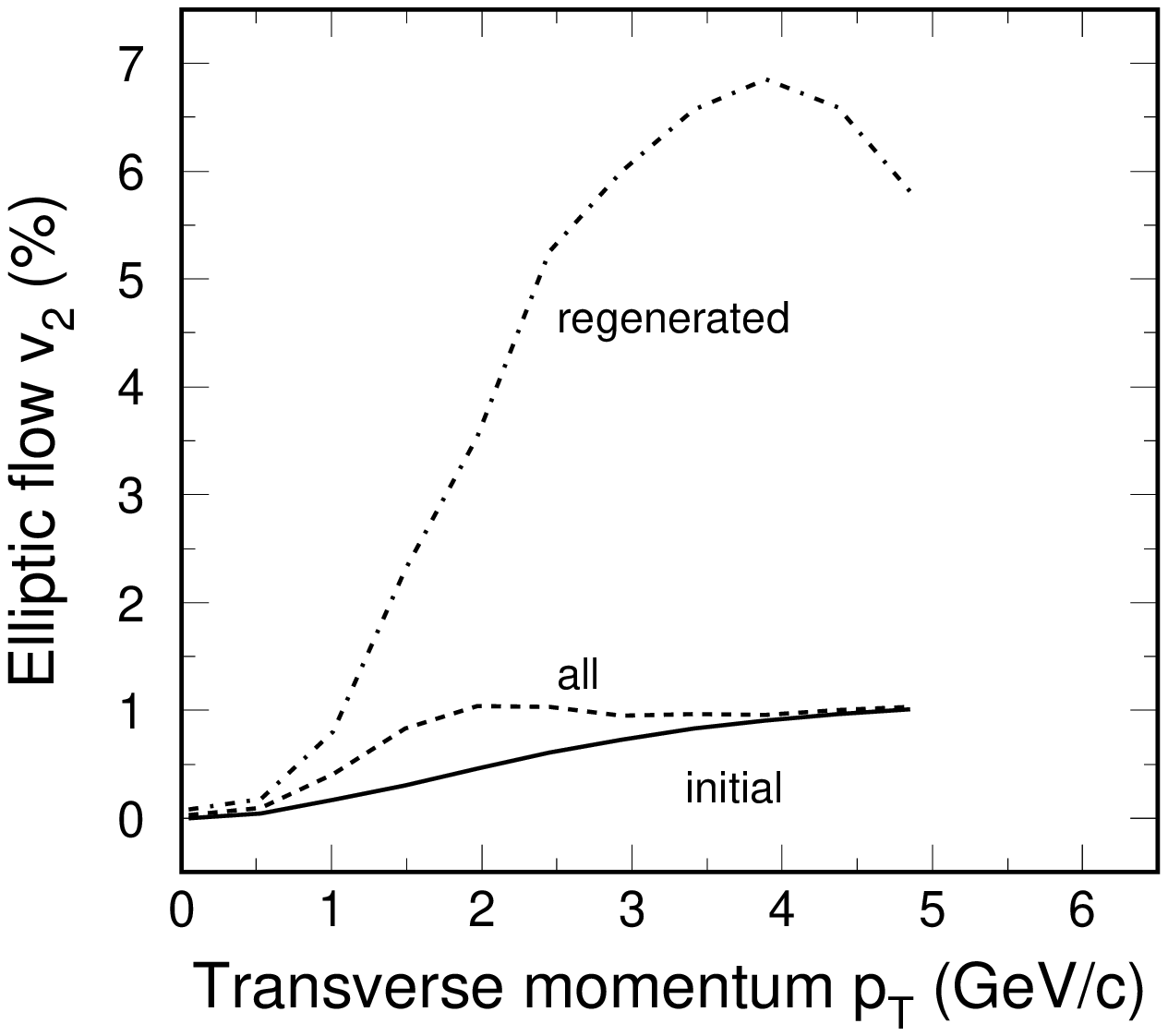}
\end{minipage}
\begin{minipage}{0.5\linewidth}
\vspace{-0.6cm}
\includegraphics[width=.9\textwidth]{v2-j-rhic-comp.eps}
\end{minipage}
\vspace{-0.9cm}
\caption{Theoretical predictions for elliptic flow of $J/\psi$
mesons in 20-40\% central Au-Au($\sqrt{s_{NN}}$=200\,GeV) collisions
(corresponding to an average impact parameter of $b\simeq7.8$\,fm).
Left panel: transport calculations in the hydro+gluo-dissociation
approach~\cite{Yan:2006ve}; right panel: weighted average of
coalescence model and suppression calculations using the
$p_T$-dependent weights computed in the fireball+quasifree
dissociation approach~\cite{Zhao:2007hh,Zhao:2008vu}.}
\label{fig_v2-rhic}
\end{figure}
A promising observable to discriminate suppressed primordial
production from regeneration is believed to be the azimuthal asymmetry
of $J/\psi$ production in the transverse plane, quantified by
the second Fourier coefficient, $v_2(p_T)$, in the azimuthal angle,
$\phi$, relative to the direction of the impact parameter vector
(commonly identified with the positive $x$-axis),
\begin{equation}
\frac{dN_\Psi}{d^2p_T} = \frac{dN_\Psi}{2\pi p_T dp_T}
      \left( 1 + 2 v_2(p_T) \cos (2\phi) + \dots \right) \ .
\end{equation}
Direct and regenerated charmonia are expected to exhibit a large 
difference in $v_2$. For the former, the azimuthal asymmetry is solely 
due to different absorption path lengths of the $J/\psi$ traversing
the almond-shaped overlap zone in a noncentral nuclear collision:
a shorter path length along the ``short" ($x$-) axis of the almond
implies less suppression, while there is larger suppression along the
``long" ($y$-) axis. Calculations typically predict a corresponding $v_2$
of up to 1-3~\%~\cite{Yan:2006ve,Wang:2002ck}, see Fig.~\ref{fig_v2-rhic}.
On the contrary, if charmonia are regenerated from $c$-$\bar c$ 
coalescence, their $v_2$ is largely determined by the underlying 
charm-quark elliptic flow, $v_2^{c,\bar c}$, and expected to 
approximately ``scale" as $v_2^\Psi(p_T)\simeq 2v_2^c(p_T/2)$ (assuming 
$v_2^c$=$v_2^{\bar c}$)~\cite{Lin:2003jy,Greco:2003vf,Ravagli:2007xx}\footnote{More 
precisely, the $Q$-value of the reaction should be small, i.e., the meson mass 
should be close to the quark-antiquark threshold~\cite{Ravagli:2007xx}; this 
should be a good approximation for loosely bound $J/\psi$'s.}. 
Such an approximate 
scaling has been observed at RHIC for light and strange mesons and baryons, 
with the inferred light-quark $v_2$ reaching up to $\sim$7-8\%. If the 
charm-quark $v_2$ reaches similar values (as suggested by theoretical
predictions~\cite{vanHees:2005wb} for semileptonic electron-decay
spectra~\cite{Adare:2006nq}), the elliptic flow of regenerated $J/\psi$'s
may reach up to $\sim$15\%~\cite{Lin:2003jy,Greco:2003vf,Ravagli:2007xx} 
(see right panel of Fig.~\ref{fig_v2-rhic}), about an order of magnitude 
larger than in suppressed primordial production.
In a two-component picture, the combined $v_2$ is given by
the weighted sum, $v_2^{\rm tot}(p_T)=f^{\rm prim}(p_T)\ v_2^{\rm prim}(p_T) +
f^{\rm reg}(p_T)\ v_2^{\rm reg}(p_T)$, where 
$f^{\rm prim}(p_T)+f^{\rm reg}(p_T)$=1 describe the fraction of primordial 
and regenerated $J/\psi$'s at each value of $p_T$.
The left panel of Fig.~\ref{fig_v2-rhic} displays the results of the
hydro+gluo-dissociation transport model (employing thermal $c$-quark
distributions)~\cite{Yan:2006ve}.
While the $v_2$ of the regenerated component is large,
it carries a small weight (decreasing toward higher $p_T$), leading to
a total $v_2$ of around 1\%. This will be very difficult to detect
experimentally. In the right panel of Fig.~\ref{fig_v2-rhic}, the total 
$J/\psi$ $v_2$ (represented by the band) has been estimated by combining 
blast-wave results for the regenerated 
component~\cite{Greco:2003vf,Ravagli:2007xx}
with the suppression calculations of Refs.~\cite{Yan:2006ve,Wang:2002ck},
using the weighting following from the fireball+quasifree dissociation
model~\cite{Zhao:2007hh}. The maximum value for the total $v_2$ of
$\sim$3$\pm$1\% is somewhat larger than in Ref.~\cite{Yan:2006ve}, mostly
due to the larger $v_2$ in the coalescence component. Most of the
difference in the $v_2$ of the regenerated component in
Refs.~\cite{Yan:2006ve} and \cite{Zhao:2008vu} is presumably due to
the fact that the former accounts for continuous regeneration
throughout the QGP while the latter approximates the production at
$T_c$. Another difference could be due to the underlying formation
reaction, which is $c+\bar c \to J/\psi +g$ in Ref.~\cite{Yan:2006ve}
compared to $c+\bar c + p \to J/\psi + p$ ($p$=$q$, $\bar q$ or $g$)
in Ref.~\cite{Zhao:2007hh}.
The main point, however, is the overall smallness of the $J/\psi$
elliptic flow in both approaches. Note that for semi-central
collisions ($N_{part}\simeq150$), for which the $v_2$ is evaluated,
both calculations underestimate the inclusive yield (cf.~upper panels
of Fig.~\ref{fig_raa-phenix}), leaving room for extra regeneration and
thus larger elliptic flow.

\subsubsection{Rapidity Distributions}
The large mass of charm and bottom quarks implies that their rapidity 
($y$) distributions in hadronic (and nuclear) collisions are narrower 
than those of light and strange quarks/hadrons, even at the LHC. 
Since the equilibrium number of charmonia is given by 
$N_\Psi^{\rm eq}\propto N_{c\bar c}^2/V_{FB}$ 
(grand-canonical limit), and the volume of the fireball 
at a given temperature (e.g., $T_c$) is determined by the light 
particles, the $y$-distribution of the equilibrium number of charmonia
will be narrower than that from primordial production, typically
given by $N_\Psi^{\rm prim}\propto N_{c\bar c}$. This is to be contrasted 
with the suppression of primordial production, which is expected to become 
{\em weaker} with decreasing light-particle density, thus generating a 
{\em broader} $\Psi$ rapidity distribution than the 
primordial one (more suppression at mid- relative to forward rapidity).

\begin{figure}[!tb]
\begin{minipage}{0.49\linewidth}
\vspace{0.6cm}
\includegraphics[width=1.01\textwidth]{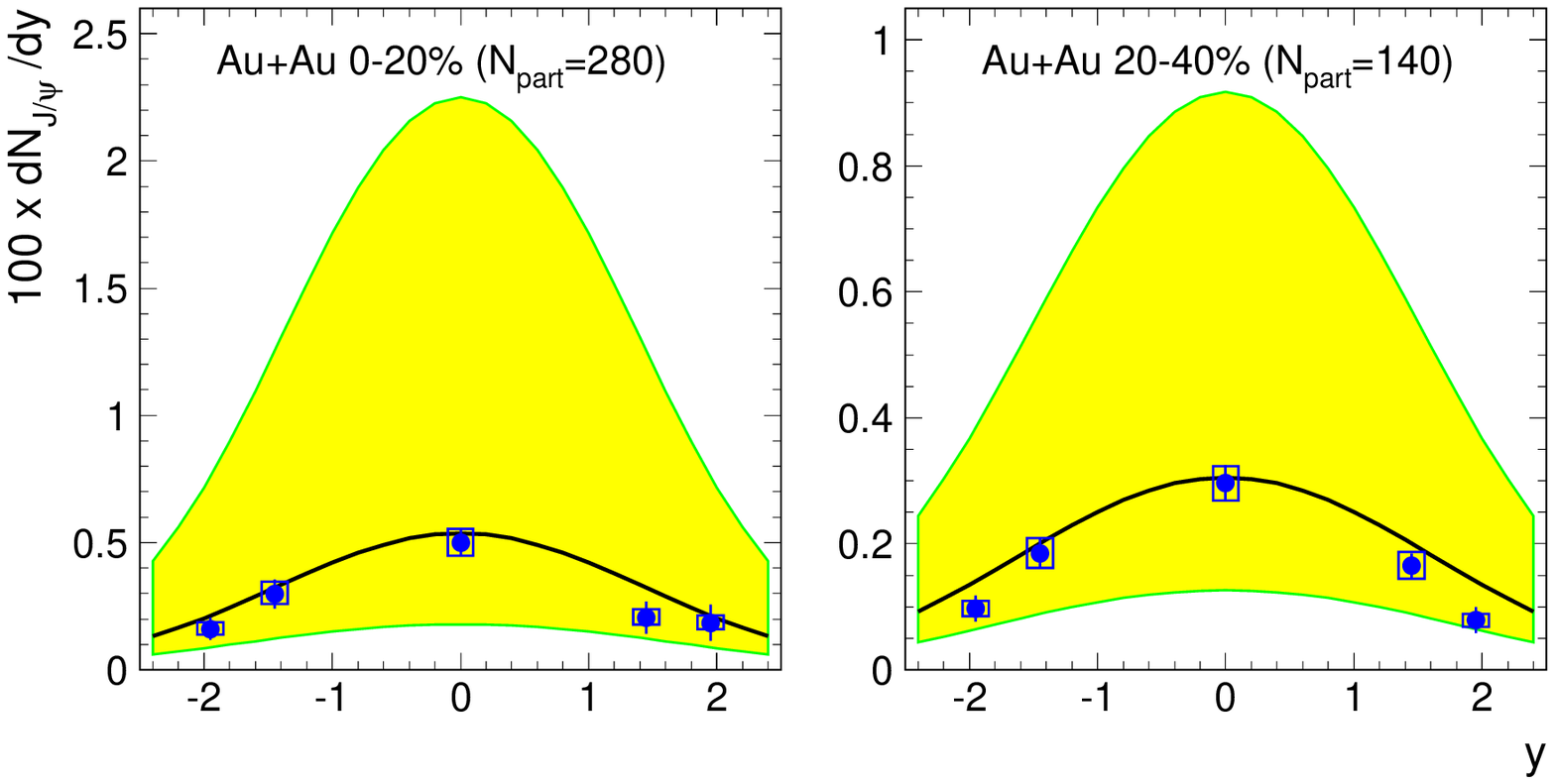}
\end{minipage}
\begin{minipage}{0.25\linewidth}
\includegraphics[width=1.03\textwidth]{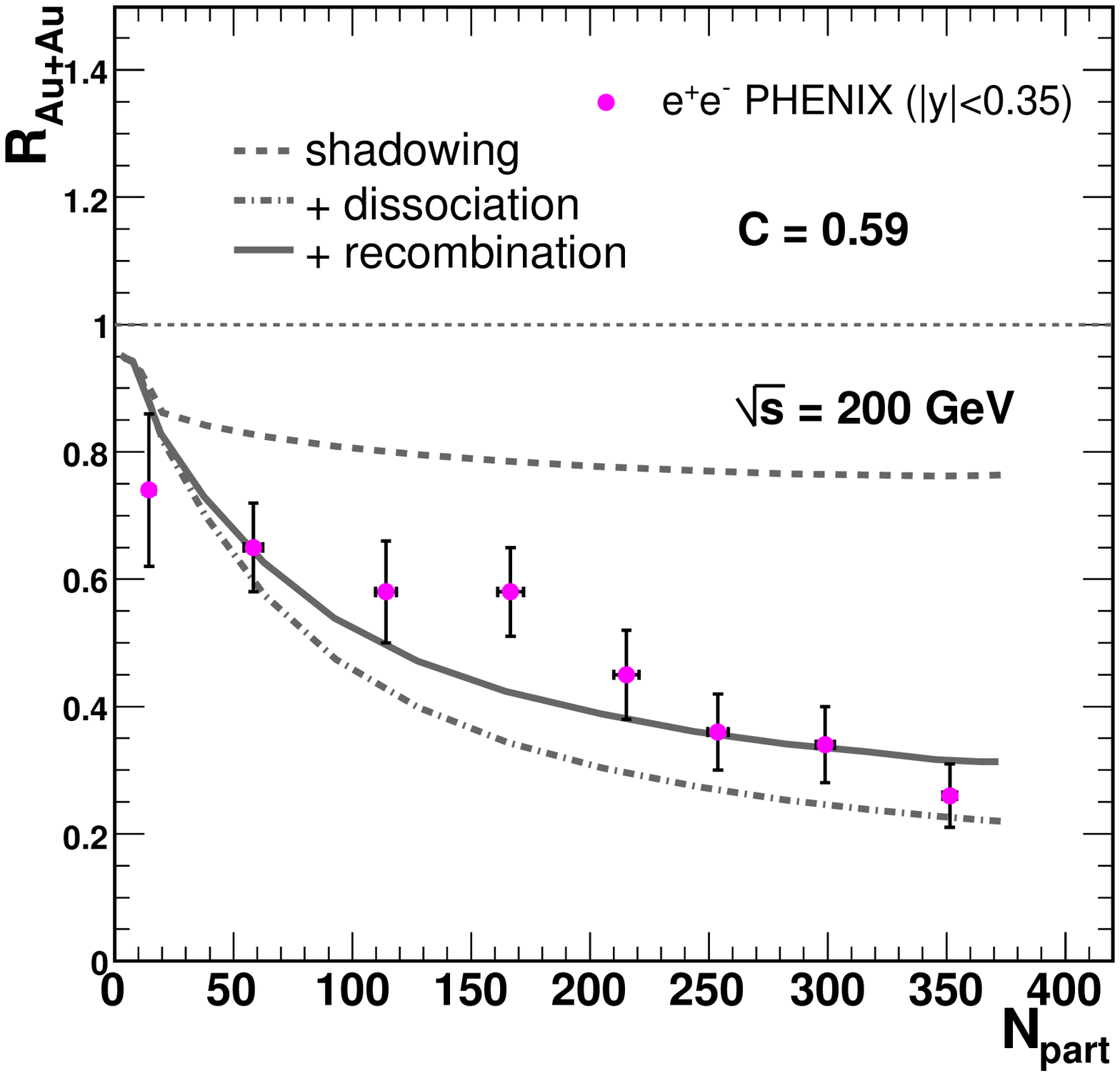}
\end{minipage}
\begin{minipage}{0.25\linewidth}
\hspace{-0.5cm}
\includegraphics[width=1.03\textwidth]{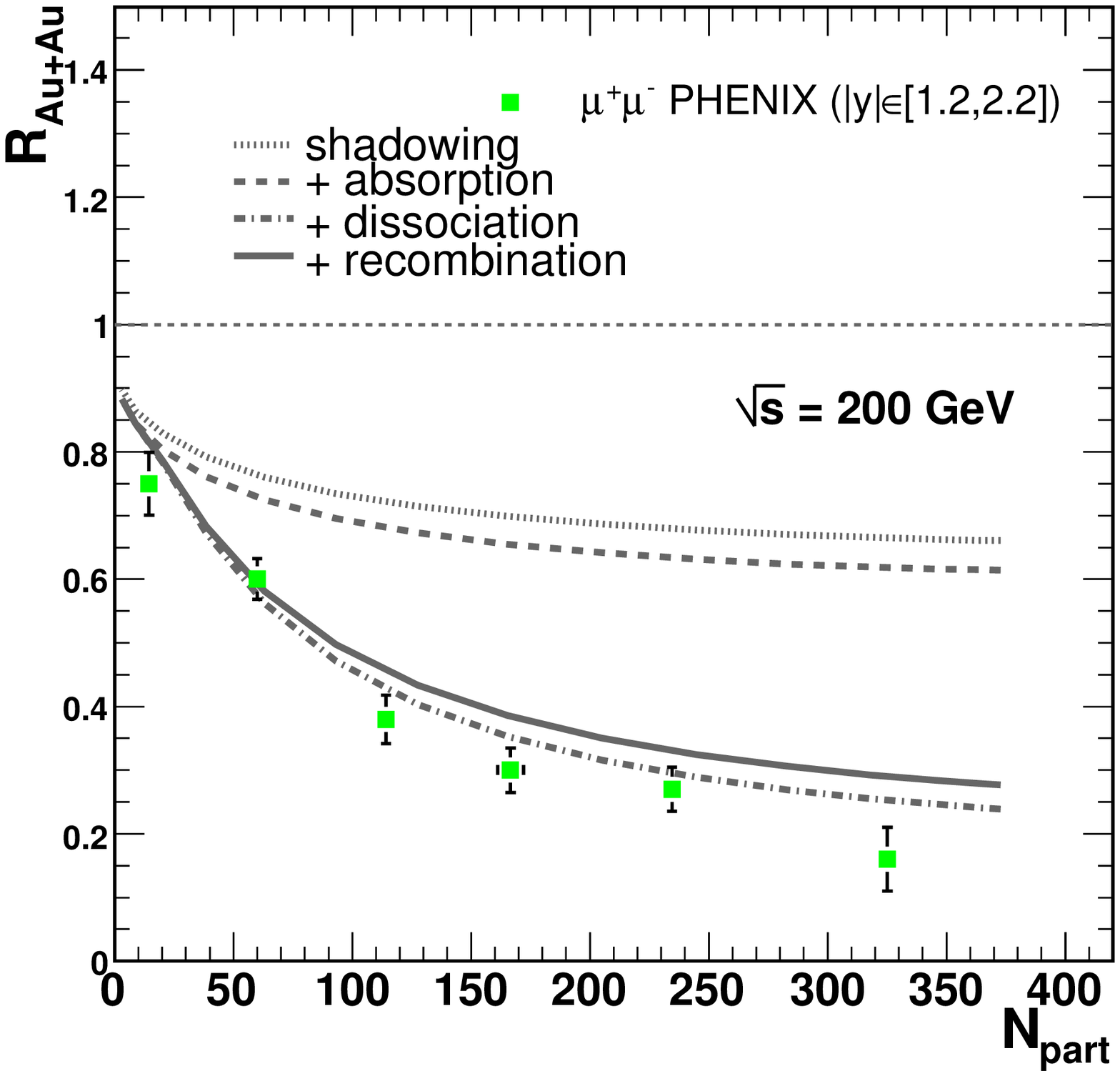}
\end{minipage}
\caption{Rapidity dependence of $J/\psi$ production in
Au-Au($\sqrt{s_{NN}}$=200\,GeV) collisions at RHIC. Left panels:
calculations for the rapidity density in central (first left) and
semi-central (second left) Au-Au within the statistical
hadronization model~\cite{Andronic:2006ky}, compared to PHENIX
data~\cite{Adare:2006ns}. The central lines (and shaded error bands)
correspond to (the uncertainty in) pQCD charm cross sections,
$\sigma_{pp}^{c\bar c}$=256$^{+400}_{-146}$~$\mu$b~\cite{Cacciari:2007xx}.
Right panels: calculations for $R_{AA}(N_{part})$ within the
comover-interaction model (including regeneration) at mid- (second
from right) and forward (far right) rapidity~\cite{Capella:2007jv},
compared to PHENIX dielectron and dimuon data~\cite{Adare:2006ns}.}
\label{fig_y-rhic}
\end{figure}
The rapidity distributions of the PHENIX $J/\psi$ data in Au-Au
exhibit a narrowing relative to $p$-$p$ collisions which is quite
consistent with regeneration only, as demonstrated by calculations
in the statistical hadronization model~\cite{Andronic:2006ky},
cf.~the two left panels in Fig.~\ref{fig_y-rhic}. Since the
$J/\psi$ abundance in the statistical model is directly proportional
to the (squared) open-charm number in the system, the open-charm
cross section is the key input to compute the $J/\psi$
rapidity distribution. 
Alternatively, the rapidity dependence has been evaluated in the
comover-interaction model in Ref.~\cite{Capella:2007jv} (two right
panels in Fig.~\ref{fig_y-rhic}). Including shadowing and comover 
dissociation on the primordial component, the total
suppression of the latter is about equal at mid- and forward
rapidity (a larger comover density around $y$=0 is essentially
compensated by larger shadowing at $y$=1.2-2.2). Only upon inclusion
of a gain term does the $J/\psi$ production at mid-rapidity
slightly exceed the one at forward $y$. Very similar results are obtained
in the thermal rate-equation approach~\cite{Zhao:2008pp}. Recent
d-Au data indicate that CNM effects may account
for a large part of the observed rapidity dependence~\cite{Frawley:2009}.

\subsubsection{Excitation Function}
\begin{figure}[!tb]
\begin{minipage}{0.5\linewidth}
\includegraphics[width=.9\textwidth]{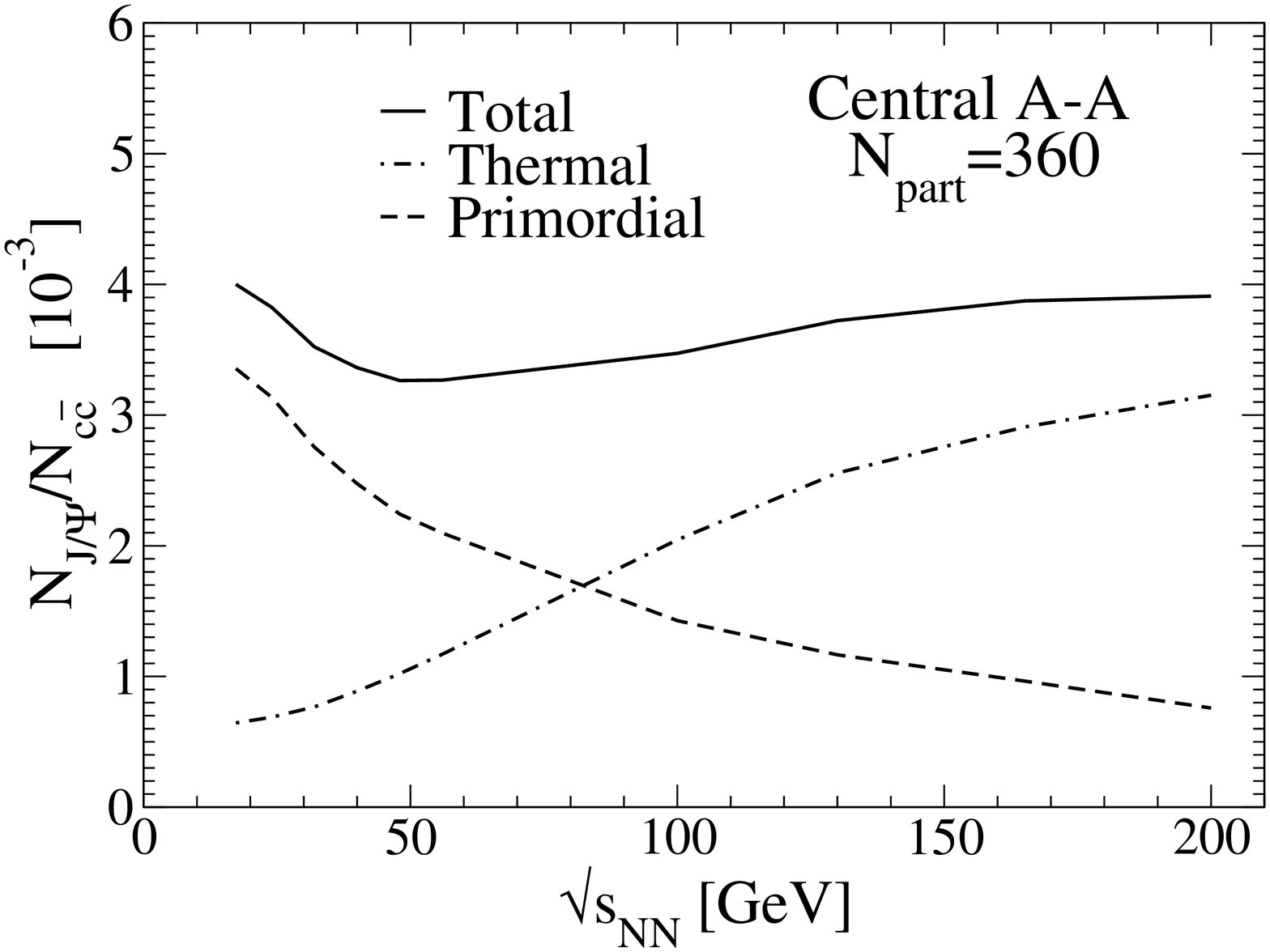}
\end{minipage}
\begin{minipage}{0.5\linewidth}
\vspace{-0.6cm}
\includegraphics[width=.9\textwidth]{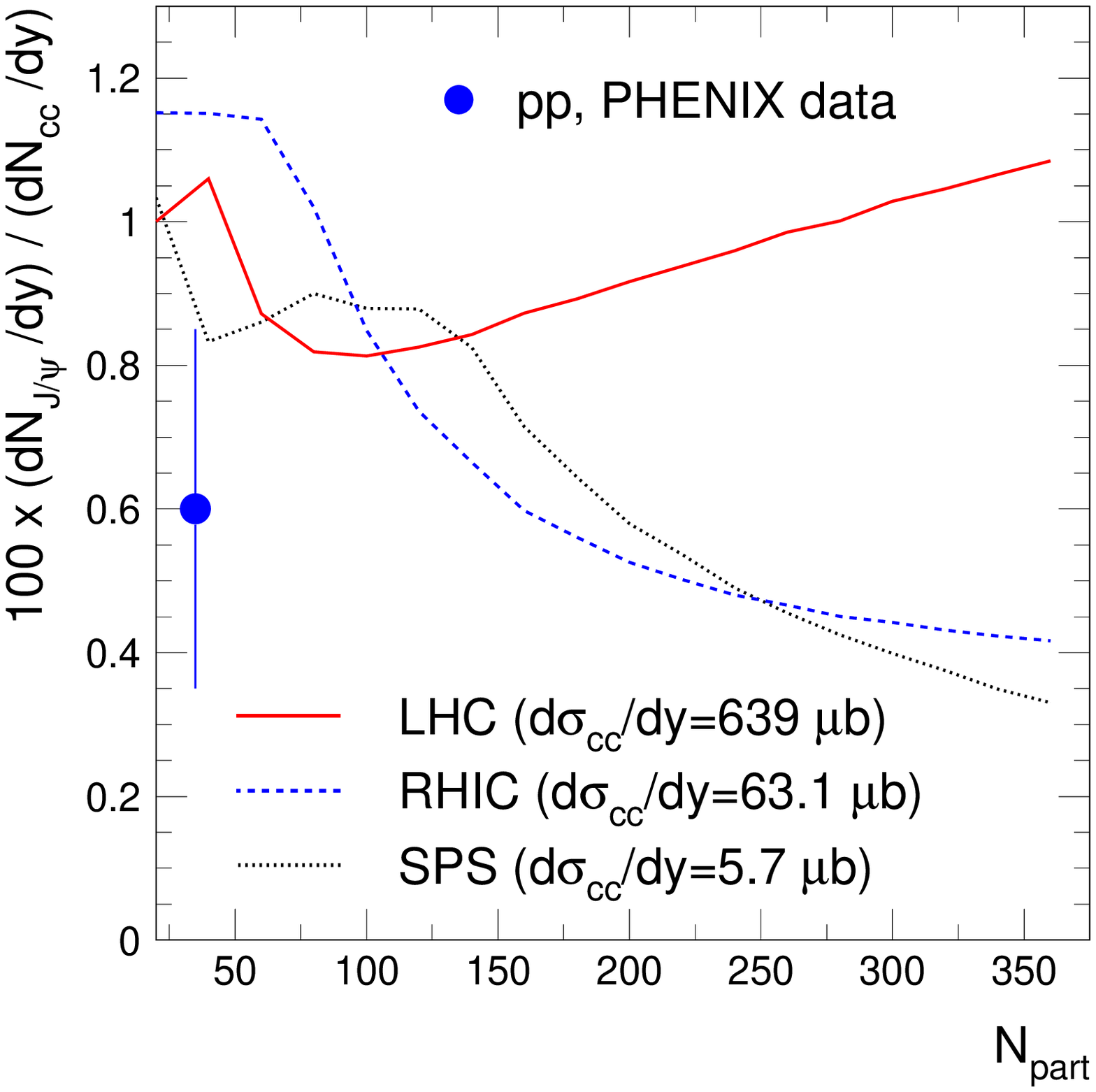}
\end{minipage}
\vspace{-0.5cm}
\caption{Collision-energy dependence of $J/\psi$ production, normalized
to the number of $c\bar c$ pairs, in ultrarelativistic heavy-ion
collisions. Left panel: original predictions of the 2-component
model~\cite{Grandchamp:2001pf,Grandchamp:2002wp} for central $A$-$A$
collisions from SPS to RHIC energies (using 2 thermal fireballs). 
Right panel: predictions within the statistical hadronization model 
for the centrality dependence at SPS, RHIC and LHC (for $\Delta y$=1) 
including corona effects for primordial 
production~\cite{Andronic:2006ky}; the data point is obtained from
PHENIX $p$-$p$ measurements~\cite{Adare:2006hc,Adare:2006kf}.}
\label{fig_excit}
\end{figure}

A promising observable to identify the interplay of (suppressed)
primordial  and secondary charmonium suppression is 
an excitation function. This is due to the much stronger increase
of charm- relative to light-quark production over a large range
in energy. Original predictions of a 2-component 
model~\cite{Grandchamp:2001pf} for central $A$-$A$ collisions
envisaged a transition from a suppression-dominated regime at SPS
to a largely regeneration-dominated one at top RHIC energy,
cf.~left panel of Fig.~\ref{fig_excit}.  While the more recent
theoretical analyses (recall upper panels of Fig.~\ref{fig_raa-phenix})
suggest that the current RHIC data contain
a smaller fraction of regeneration than originally predicted, it
is important to realize the following (experimental) uncertainty
pertaining to secondary production in the commonly used nuclear
modification factor. The latter is generically defined as the yield 
(or spectrum) of hadron $h$ in $A$-$A$ normalized to the
binary-collision scaled yield in $p$-$p$,
$R_{AA}^h = N_{AA}^h / (N_{coll} N_{pp}^h)$.
For (suppressed) primordial $J/\psi$ production, the dependence on the
initial production thus drops out (to the extent that primordial
production follows binary collision scaling). This is {\em not}
the case for the (absolutely normalized) regeneration component, which in
$R_{AA}$ plots is therefore beset with additional uncertainty due to
the input cross section from $p$-$p$ collisions (at RHIC the pertinent
experimental uncertainty currently amounts to
$\sim$20-25\%~\cite{Adare:2006kf}).

The right panel of Fig.~\ref{fig_excit} shows calculations of the
statistical hadronization model (augmented by primordial production 
from a corona region) for the centrality dependence of $J/\psi$ over
open-charm numbers at midrapidity for top SPS, RHIC and LHC 
energies~\cite{Andronic:2006ky}. The open-charm production cross 
section is taken from pQCD (using $N_{coll}$-scaling for $A$-$A$) 
along with assuming a fixed value of 
$N_{J/\psi}/N_{c\bar c}$=1\% in $p$-$p$. The increasing trend 
toward more peripheral collisions at
SPS and RHIC is largely induced by the canonical suppression factor
(for central collisions, the results are in fair agreement with the 
excitation function in the left panel of
Fig.~\ref{fig_excit}). At LHC, however, the transition to
the grand-canonical ensemble is essentially completed at
$N_{part}$$\simeq$100, so that for larger centralities a distinct
increase of $N_{J/\psi}/N_{c\bar c}$ occurs, signaling the
dominance of the recombination of uncorrelated $c$ and $\bar c$
quarks into $J/\psi$'s.

\subsubsection{Excited Charmonia}
In all phenomenological applications discussed in this section
thus far the number of $J/\psi$'s refers to inclusive production,
i.e., contains the feed-down contributions of excited charmonia
($\sim$30(10)\% from $\chi_c$($\psi'$) in $p$-$p$ reactions).
E.g., a comprehensive interpretation of SPS and RHIC data in 
terms of a schematic ``sequential melting" scenario has
recently been presented in Ref.~\cite{Karsch:2005nk}.
In addition to their role in $J/\psi$ production, excited charmonia
provide complementary information on the produced medium
as their (in-medium) binding energies, dissociation widths and
dissolution temperatures (are expected to) differ significantly
from the $J/\psi$. Consequently, future experimental programs
put a large emphasis on direct measurements of $\chi_c$ and
$\psi'$, as elaborated in Sec.~\ref{sec_exp}.
E.g., ratios of different charmonia can be used to test the
occurrence of relative chemical equilibrium.

\begin{figure}[!tb]
\begin{minipage}{0.32\linewidth}
\includegraphics[width=.94\textwidth]{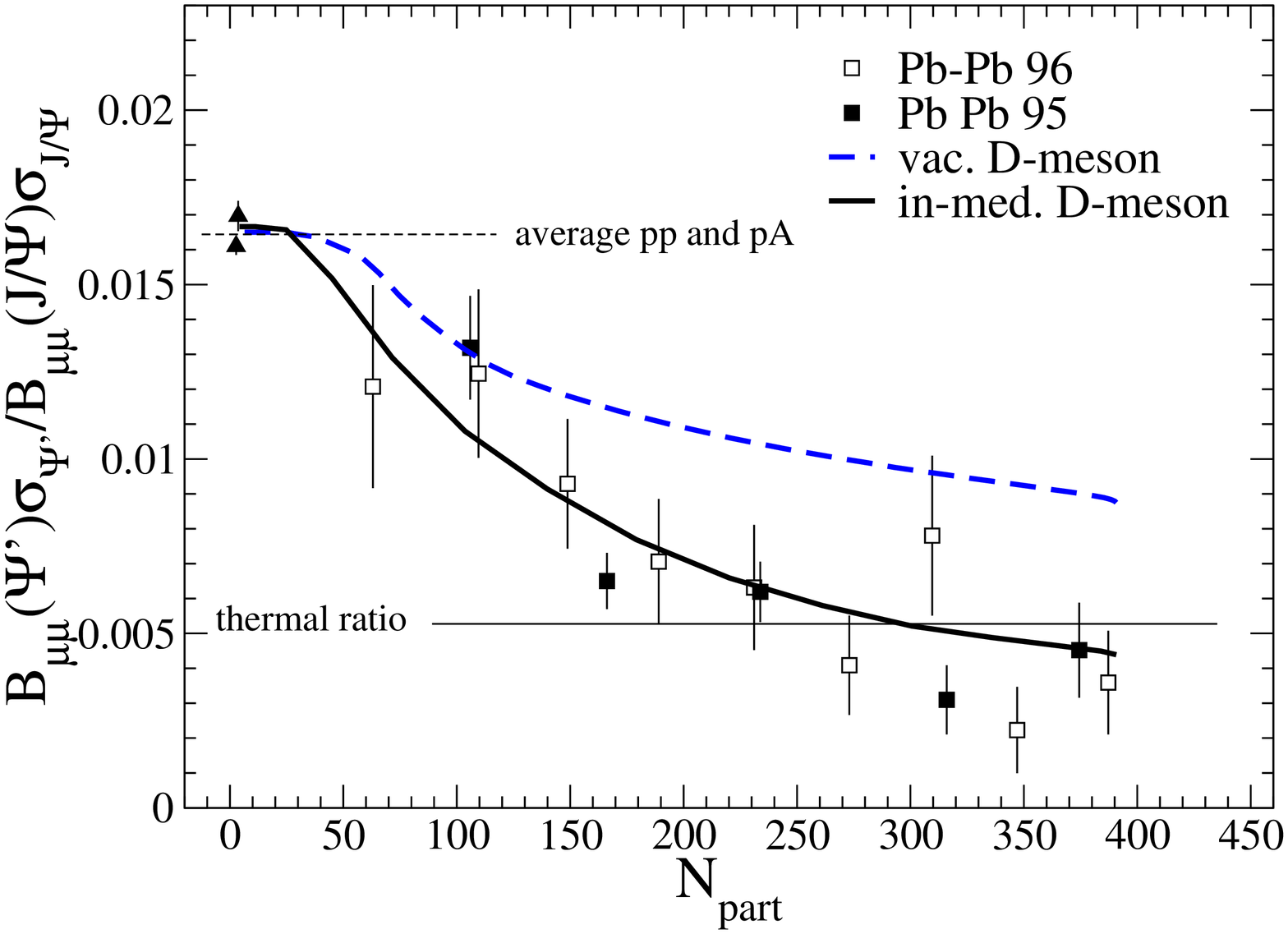}
\end{minipage}
\begin{minipage}{0.32\linewidth}
\includegraphics[width=.94\textwidth]{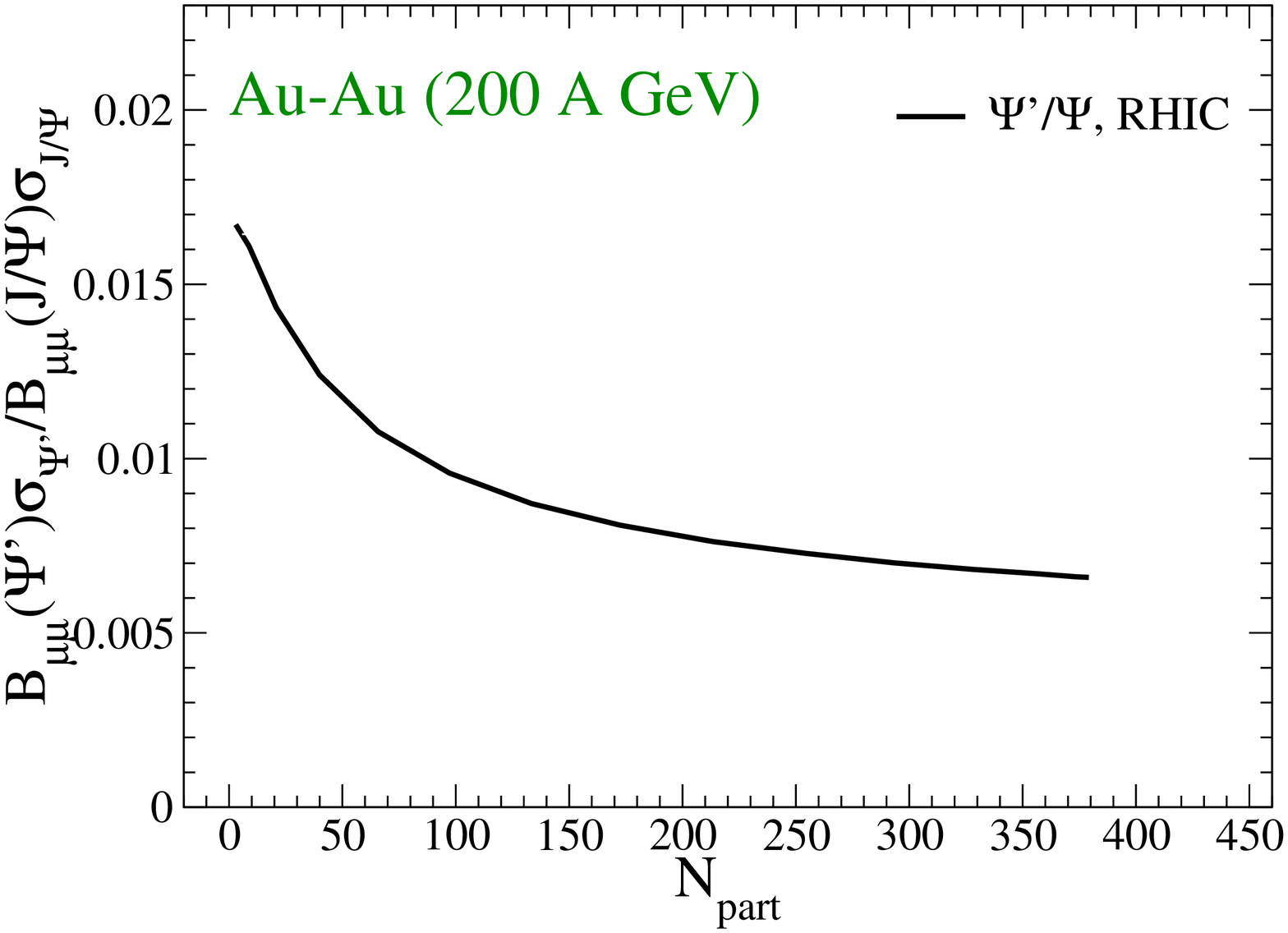}
\end{minipage}
\begin{minipage}{0.32\linewidth}
\includegraphics[width=.99\textwidth]{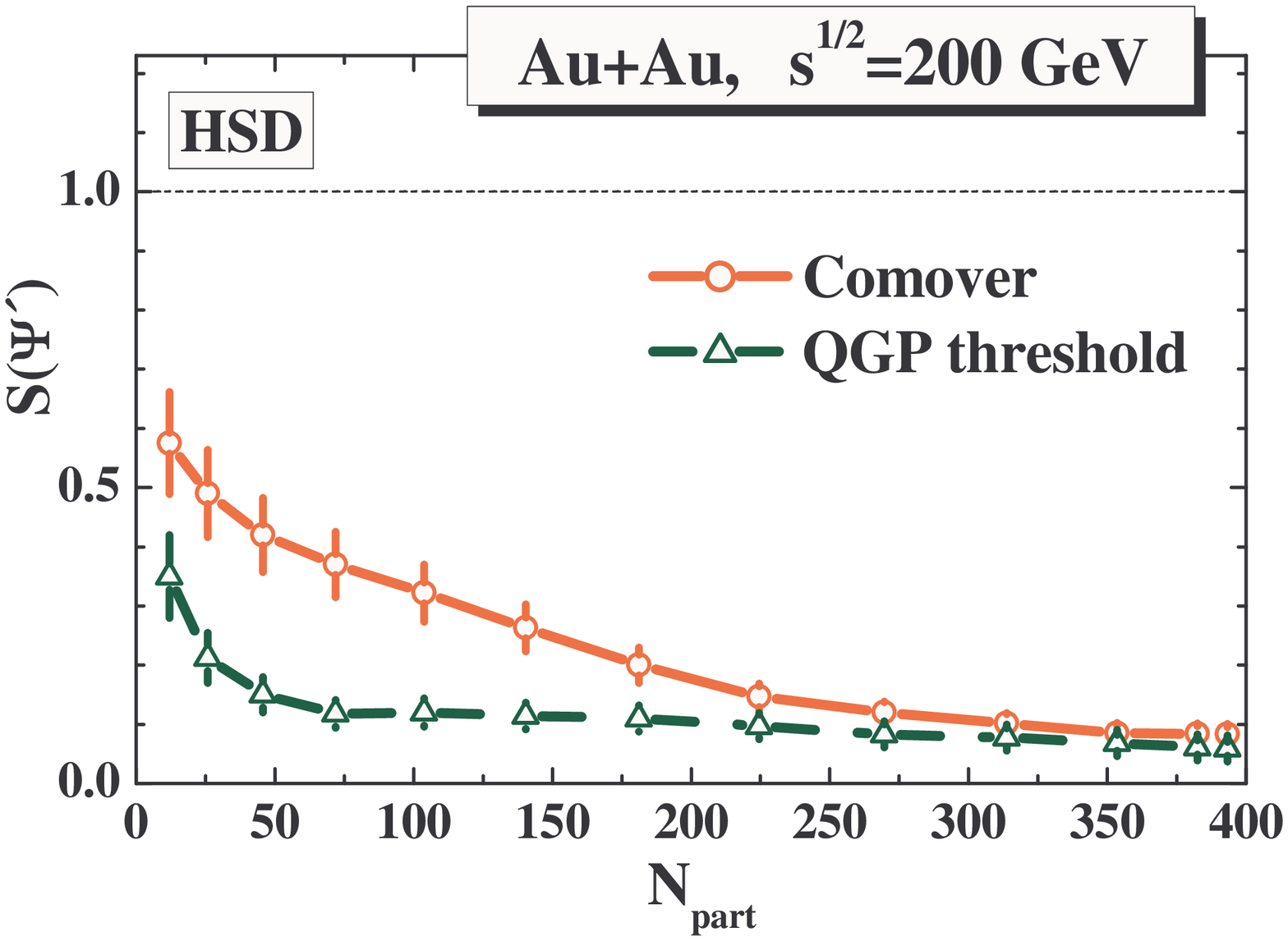}
\end{minipage}
\caption{$\psi'$ production in ultrarelativistic heavy-ion collisions.
Left and middle panel: $\psi'/(J/\psi)$ ratio as computed in the thermal
rate-equation approach with quasifree dissociation in the QGP and 
hadronic dissociation with (solid lines) and without (dashed lines)
in-medium $D$-meson masses at SPS (left; data from 
Ref.~\cite{Abreu:1998vw})~\cite{Grandchamp:2003uw} and
RHIC (middle).
Right panel: $\psi'$ suppression (nuclear modification) factor
in the HSD transport approach in the comover-interaction (circles) and 
QGP threshold-suppression (triangles) scenarios~\cite{Linnyk:2008uf};
upon dividing the $\psi'$ suppression factor by the one for the $J/\psi$ 
(see Fig.~9 in Ref.~\cite{Linnyk:2008uf}), the comover result 
is rather compatible with the rate-equation curve in the middle panel.}
\label{fig_psip}
\end{figure}

NA50 data for the $\psi'/(J/\psi)$ ratio in Pb-Pb collisions at SPS exhibit 
a pronounced decrease with centrality up to a factor of $\sim$3, see
left panel of Fig.~\ref{fig_psip}~\cite{Abreu:1998vw}. It has been 
suggested~\cite{Sorge:1997bg} that a dropping $\sigma$-meson mass (as a 
consequence of (partial) chiral symmetry restoration)  accelerates the 
transition rate for $\psi' \leftrightarrow J/\psi+\sigma$ close to 
$T_c$ and can account for the $\psi'/(J/\psi)$ data. For semi-/central 
Pb-Pb collisions the data can also be explained within the 
statistical hadronization model~\cite{BraunMunzinger:2000px}, although 
here the underlying mechanism is different, i.e., a QGP converting 
into a hadron gas in thermal and (relative) chemical equilibrium (with 
conserved charm- and anticharm-quark number).
In the thermal rate-equation approach of Ref.~\cite{Grandchamp:2003uw},
regeneration at SPS energies is a small effect, and the QGP phase is too
short to provide sufficient $\psi'$ suppression to describe the NA50 data
for the $\psi'/(J/\psi)$ ratio (dashed line in the left panel of
Fig.~\ref{fig_psip}). However, if effects of (partial) chiral restoration
are implemented via an in-medium  reduction of $D$-mesons masses
(associated with their light-quark content), the direct $\psi'\to D\bar D$
channel opens and increases the $\psi'$ dissociation
rate in the hadronic phase close to $T_c$ (in the vacuum, the
$\psi'$ mass is only $\sim$50\,MeV below the $D\bar D$ threshold). This
allows for a good description of the NA50 $\psi'$ data (solid line in 
the left panel of Fig.~\ref{fig_psip}). Alternatively, a broadening of 
the in-medium $D$-meson spectral function can induce a similar effect
as reduced masses, since it also opens phase space for direct 
$\psi'\to D\bar D$ decays. Either mechanism is not mutually 
exclusive to a relative equilibration via 
$J/\psi+\sigma\leftrightarrow \psi'$ (neither is statistical
hadronization, which, however, requires the prevalence of regeneration).
At full RHIC energy, both the thermal rate-equation approach (middle
panel in Fig.~\ref{fig_psip}) and HSD transport calculations
using pre-hadronic and comover interactions with regeneration (right
panel in Fig.~\ref{fig_psip}) predict a $\psi'/(J/\psi)$ ratio which is
very similar to SPS energies. On the other hand, threshold-melting
scenarios predict significantly sharper suppression patterns in the
centrality dependence. The $\chi_c/(J/\psi)$ ratio is expected to drop 
with centrality in both rate-equation and comover approaches, and to 
decrease more sharply for threshold-melting scenarios. 

\subsubsection{Bottomonia}
\label{ssec_bottom}
Bottomonium observables add at least two aspects, relative to charmonia, 
to the study of the medium in heavy-ion collisions at RHIC and LHC:
(i) larger binding energies (parametrically, for Coulombic
wave functions, $\varepsilon_B^{Q\bar Q}\propto \alpha_s^2 m_Q$),
implying larger dissociation temperatures and thus different
dissociation patterns;
(ii) a larger mass of the bottom quark, reducing its primordial
production (e.g., $N_{b\bar b}\approx 0.1$ per central Au-Au at 
RHIC~\cite{Morino:2008nc}),
which suggests that recombination effects are suppressed. The
latter assertion, however, has to be taken with care since primordial
$\Upsilon$ production is also relatively small (typically
$N_{\Upsilon}/N_{b\bar b}\simeq$~0.1\,\%~\cite{Morino:2008nc,Cosentino:2008qn}
compared to 1\,\% for charm/onium),
and bottom-quark fugacities can be rather large.

Early studies of bottomonium production in heavy-ion collisions
have focused on suppression effects~\cite{Ropke:1988bx,Karsch:1990wi}. 
Similarly, in Ref.~\cite{Gunion:1996qc}, the threshold-melting 
scenario, combined with formation-time effects, has been utilized to 
compute the $p_T$ dependence of suppression factors,
$S(p_T)\equiv R_{AA}(p_T)$, at the LHC.
Assuming initial temperatures of $T_0\simeq$~0.8-1.2\,GeV, a large
suppression for all states has been predicted at low $p_T$, which,
however, gradually disappears in the range of $p_T\simeq$~5-30\,GeV
(qualitatively similar to the right panel in Fig.~\ref{fig_pt-rhic}),
depending on the Debye mass and surface effects in the fireball. An
extension of the threshold-melting scenario to RHIC~\cite{Pal:2000zm}
suggests that directly produced $\Upsilon$'s do not undergo any
suppression, while inclusive $\Upsilon$'s (which receive about 40-50\%
feed-down from $\chi_b$ and $\Upsilon'$) are suppressed by up to
45\% at low $p_T$ due to an almost complete melting of the excited
states. As mentioned above, the threshold-melting picture represents
a sharp transition between stable and dissolved quarkonia, i.e.,
a sudden ``jump" of the inelastic width from zero to infinity.
This is a simplification of a more realistic situation where the 
width increases with temperature more gradually, as found in
microscopic calculations discussed in Sec.~\ref{ssec_diss-qgp}
(in the context of Fig.~\ref{fig_diss-rates-ups}). On the other
hand, a rather abrupt dissolution of a quarkonium bound-state as
a function of temperature may be justified by potential-model 
calculations of spectral functions, recall Sec.~\ref{ssec_pot}.
The reason is that, once the binding approaches zero, the $Q\bar Q$
state in the QGP moves into the continuum and the structureless 
potential can no longer support resonance states.
In addition, a rapid increase of the dissociation width is expected
once the direct decay channel opens, $\Upsilon \to b +\bar b$ or
$\Psi\to c +\bar c$, i.e., above the quark-antiquark threshold in 
the QGP.

Quantitative studies of bottomonium suppression and regeneration have
been performed in Ref.~\cite{Grandchamp:2005yw} within the thermal
rate equation (\ref{rate}), solved in an expanding fireball background.
When assuming vacuum binding energies for the $\Upsilon$ throughout the
QGP evolution at RHIC (with an initial temperature of $T_0$=370\,MeV,
dropping to 300\,MeV within the first 0.5\,fm/$c$),
the results of Ref.~\cite{Pal:2000zm} have been confirmed in that
no significant suppression of direct $\Upsilon$'s (i.e. those not 
from feed-down) occurs (nor regeneration). This is understandable due to 
its long lifetime of several
tens of fm/$c$ for temperatures below 300\,MeV (recall left panel in
Fig.~\ref{fig_diss-rates-ups}).
\begin{figure}[!tb]
\begin{minipage}{0.5\linewidth}
\includegraphics[width=.9\textwidth]{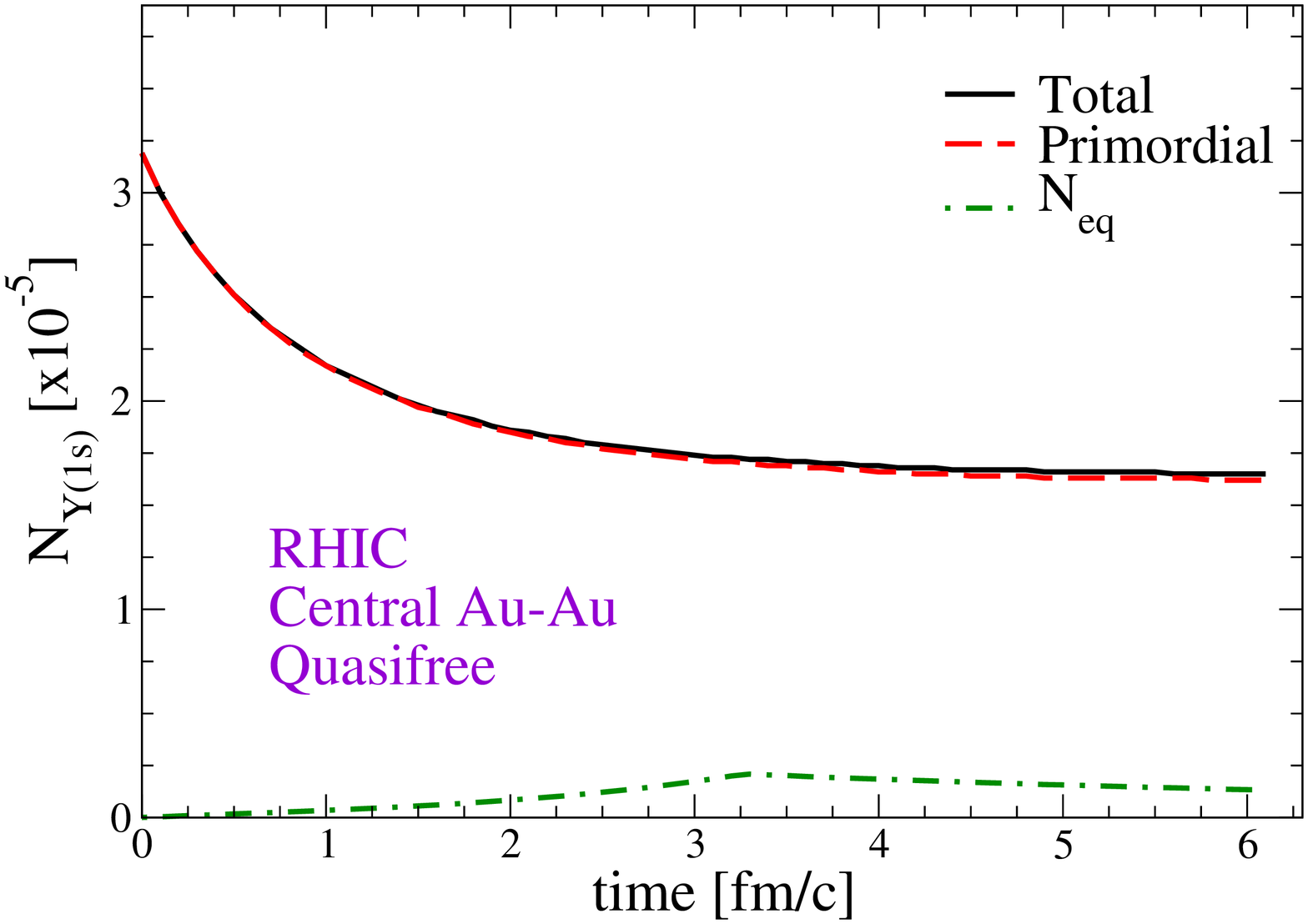}
\end{minipage}
\begin{minipage}{0.5\linewidth}
\includegraphics[width=.9\textwidth]{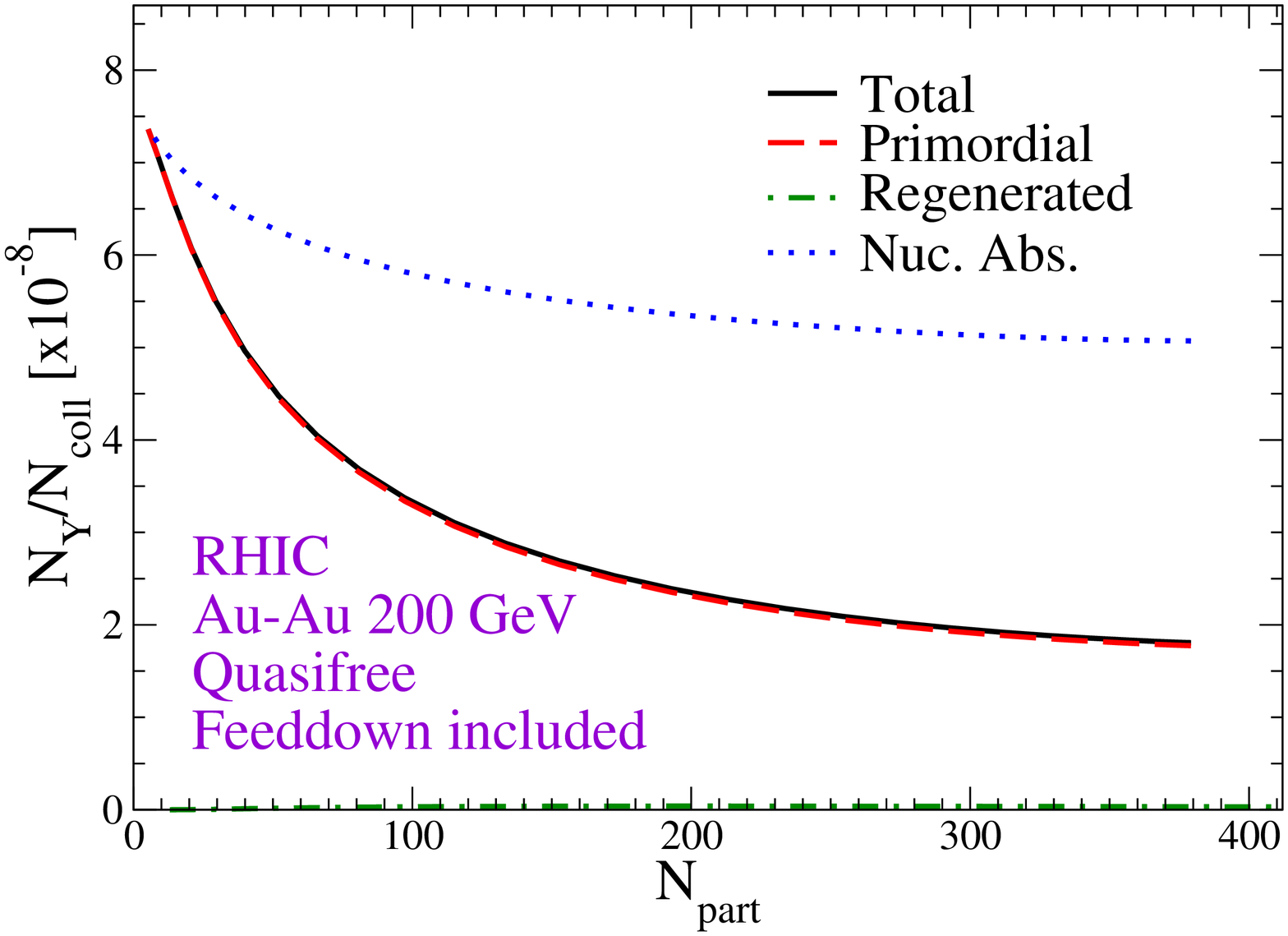}
\end{minipage}
\vspace{-0.5cm}
\caption{Predictions for $\Upsilon$ production at RHIC
within the thermal rate-equation approach using quasifree dissociation
with in-medium binding~\cite{Grandchamp:2005yw} in an expanding fireball.
Left panel: time evolution of $\Upsilon$ abundance in central Au-Au;
right panel: centrality dependence of inclusive $\Upsilon$ yield.}
\label{fig_ups-rhic}
\end{figure}
The situation changes appreciably when in-medium reductions of the
bottomonium binding energies are included (based on the in-medium rates
displayed in the left panel of Fig.~\ref{fig_diss-rates-ups}). In central
Au-Au at RHIC, direct $\Upsilon$'s are suppressed by $\sim$ 40\%, with
most of the suppression occurring in the first 1\,fm/$c$, i.e., for
temperatures above 250\,MeV (cf.~left panel of Fig.~\ref{fig_ups-rhic}). In
the centrality dependence of inclusive $\Upsilon$ production (right panel
in Fig.~\ref{fig_ups-rhic}) the suppression is further amplified, due to 
$\sim$50\% feed-down contributions~\cite{Affolder:1999wm} from excited
$\chi_b$ ($\sim$30\% $\chi_{b1}$, $\sim$10\% $\chi_{b2}$) and $\Upsilon'$ 
($\sim$10\%) states; regeneration contributions are (very) small. 
An important point here is that direct $\Upsilon$ suppression at RHIC is 
a rather direct measure of the Debye screening in the heavy-quark 
potential, especially since regeneration is absent. The binding energies 
figuring into the calculations of Ref.~\cite{Grandchamp:2005yw} are 
somewhat smaller than potential models based on the internal energy 
(but significantly larger than those based on the free energy), as 
discussed in Sec.~\ref{ssec_pot}.
In either case, one could envisage a situation where the production of
$\Upsilon$ is as (or even more) suppressed as for $J/\psi$. This would be
a rather unique QGP signature~\cite{Grandchamp:2005yw}, and provide
indisputable evidence of charmonium regeneration and color-Debye screening.

\begin{figure}[!tb]
\begin{minipage}{0.5\linewidth}
\includegraphics[width=.9\textwidth]{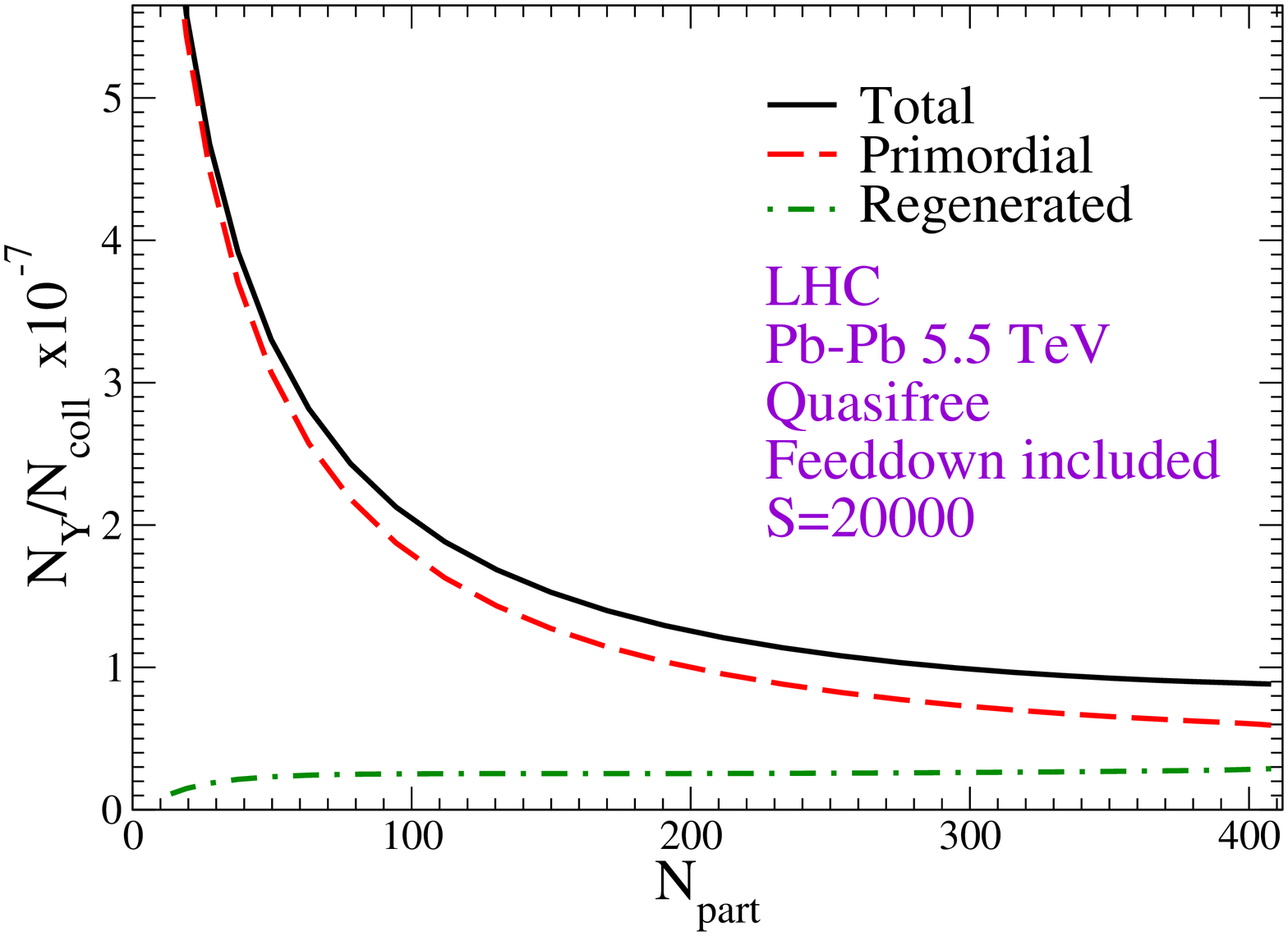}
\end{minipage}
\begin{minipage}{0.5\linewidth}
\vspace{-0.6cm}
\includegraphics[width=.9\textwidth]{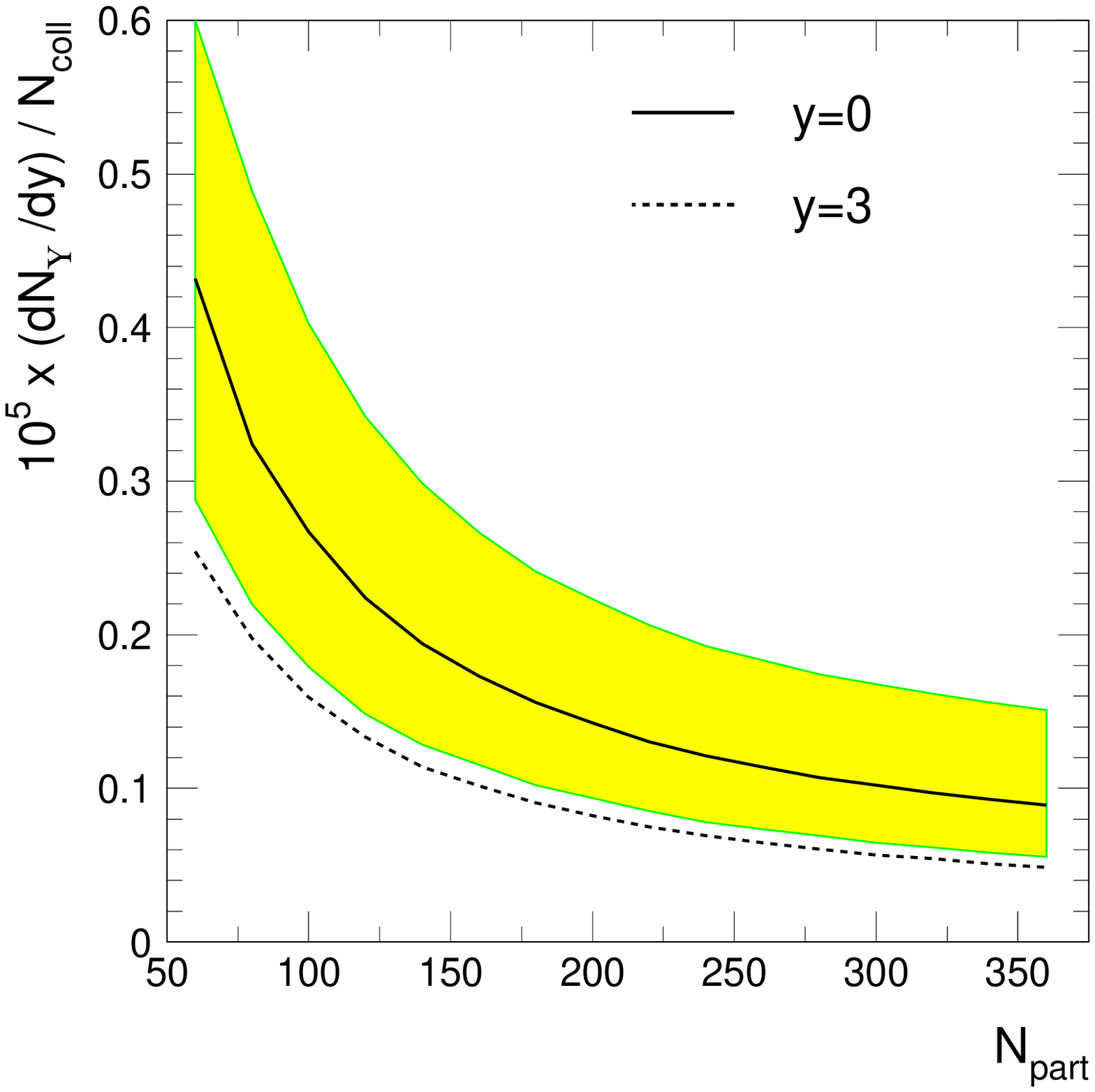}
\end{minipage}
\vspace{-0.5cm}
\caption{Predictions for the centrality dependence of
inclusive $\Upsilon$ production at LHC.
Left panel: thermal rate-equation approach~\cite{Grandchamp:2005yw}
displaying the final yield, $N_\Upsilon$, within a thermal fireball of 
rapidity width $\Delta y=1.8$ (the total fireball entropy of $S=20000$ 
in central Pb-Pb converts into $dN_{ch}/dy\simeq1600$).
Right panel: statistical hadronization model~\cite{Andronic:2006ky}
(based on $dN_{ch}/dy\simeq1800$); the yellow uncertainty band reflects
a variation of the input pQCD $b\bar b$ cross section by a factor of
1.5 up and down.  }
\label{fig_ups-lhc}
\end{figure}
At the LHC, the situation becomes more involved. The rate-equation
approach (left panel in Fig.~\ref{fig_ups-lhc})~\cite{Grandchamp:2005yw}
predicts a suppression of up to a factor of $\sim$10 stronger than the
statistical hadronization model (right panel in 
Fig.~\ref{fig_ups-lhc})~\cite{Andronic:2006ky} in central Pb-Pb 
collisions. Besides moderate differences in input cross sections 
(e.g., a factor of 0.8 shadowing correction in 
Ref.~\cite{Grandchamp:2005yw}), there are several sources of this 
difference:
(a) the equilibrium limit in the rate equation is computed in the QGP
with (in-medium) bottom-quark masses (rather than bottom hadrons),
leading to smaller bottom-quark fugacities; the resulting equilibrium
number is up to  a factor of 3-6 smaller, depending on the (not
very well known) spectrum of bottom hadrons.
(b) the equilibrium limit in the rate equation is reduced by a
schematic thermal relaxation-time factor, ${\cal R}(\tau)<1$, to mimic
incomplete $b$-quark thermalization; at the end of the QGP phase this
factor has reached to about 0.3-0.5 (it is smaller earlier in the 
evolution). However, close
to $T_c$, the inelastic reaction rates for the bottomonia are already
rather small; (c) as mentioned for charmonia at SPS, there is a
question over which range in rapidity the thermal ensemble of
bottom quarks should be defined (e.g., $\Delta y$=1.8 and 1 in
Refs.~\cite{Grandchamp:2005yw} and \cite{Andronic:2006ky}, 
respectively).  More work is needed to clarify these questions.


\section{Future experiments}
\label{sec_exp}
\input{exp} 

\section{Summary}
\label{sec_sum}
\input{sum} 

\input{references}

\end{document}

%% file: intro.tex
The properties of heavy quarkonium states (charmonium and bottomonium)
in a hot and dense QCD\footnote{Quantum Chromodynamics, the theory of the strong 
force and part of the Standard Model of elementary particle physics.} 
medium have been intensely studied for over 20 years 
now, both experimentally and theoretically. 
Current and upcoming heavy-ion collision experiments at the Relativistic 
Heavy-Ion Collider (RHIC, at Brookhaven National Laboratory (BNL) in New 
York), at the Large Hadron Collider (LHC, at the European Organization 
for Nuclear Research (CERN) in Geneva) and at the Facility for Antiproton 
and Ion Research (FAIR, at the Helmholtz Center for Heavy-Ion Research 
(GSI) in Darmstadt) put a large emphasis on heavy quarkonium programs 
in their campaigns. The interest in heavy quarkonia in medium 
is motivated by their unique role in the diagnostics of the highly 
excited medium created in ultrarelativistic heavy-ion collisions (URHICs).  
Early on, $J/\psi$ suppression in URHICs was suggested as a signal of the 
formation of a Quark-Gluon Plasma (QGP)~\cite{Matsui:1986dk}.
This idea was instrumental in triggering a corresponding experimental 
program at the CERN-SPS.
The experimental results have been accompanied and 
pushed forward by a broad spectrum of theoretical work (see, e.g.,  
Refs.~\cite{Gerschel:1998zi,Vogt:1999cu,Satz:2005hx,Kharzeev:2007ej}
for various reviews). 
After many years of analysis and interpretation of the SPS 
data~\cite{Kluberg:2005yh}, with a first round of RHIC results 
completed and with new insights from the theoretical side
(including thermal lattice QCD, effective models and phenomenology),  
it is timely to assess the current state of affairs to help facilitate 
the next stage of developments. 
In the remainder of this introduction we will give an initial view 
of the physics of quarkonia in a hot and dense medium, illustrating some 
of the difficulties in the interpretation of (charmomium) observables in 
URHICs at SPS and RHIC. 

\begin{table}[!b]
\begin{center}
\begin{tabular}{|c|c|c|c|c|}
\hline
State ($nL$) & $J^{PC}$ & $m_\Psi$ [MeV] & $\Gamma_{tot}$ [MeV] & $m_\Psi-2m_D$ [MeV] 
\\
\hline  \hline
$\eta_c$ (1$S$)    & $0^{-+}$ & 2980$\pm$1    & 27$\pm$3    &  -750 \\
\hline
$J/\psi$ (1$S$)    & $1^{--}$ & 3097          & 0.093$\pm$0.002 &  -633 \\
\hline
$\chi_{c0}$ (1$P$) & $0^{++}$ & 3415          & 10.2$\pm$0.7    &  -315\\
\hline
$\chi_{c1}$ (1$P$) & $1^{++}$ & 3511          & 0.89$\pm$0.05   &  -219\\
\hline
$h_{c}$ (1$P$)     & $1^{+-}$ & 3526          &     $<$1        &  -204\\
\hline
$\chi_{c2}$ (1$P$) & $2^{++}$ & 3556          & 2.03$\pm$0.12   &  -174\\
\hline
$\eta_c'$ (2$S$)   & $0^{-+}$ & 3637$\pm$4    & 14$\pm$7        &   -92 \\
\hline
$\psi'$ (2$S$)    & $1^{--}$ & 3686           & 0.32$\pm$0.01   &  -44\\
\hline
$\psi''$ (3$S$)   & $1^{--}$ & 3773$\pm$3     & 27.3$\pm$1   &  +43 \\
\hline
\end{tabular}
\end{center}
\caption{Selected properties of $c\bar c$ mesons (generically 
referred to as $\Psi$) in the vacuum, 
as extracted from the recent review of particle physics~\cite{Amsler:2008zzb}.
The particle name in the first column is supplemented by a nonrelativistic
classification of principal quantum number ($n$) and orbital angular
momentum ($L$), the second column gives the total spin, parity and 
charge-conjugation quantum numbers, the third column the meson's mass (errors 
below 1~MeV are not quoted), the fourth column the total decay width, and the 
last column the mass difference to the open-charm threshold (taken as 
2$m_{D^0}\simeq3730$\,MeV).}
\label{tab_psi-vac}
\end{table}
As a starting point, we collect in Tabs.~\ref{tab_psi-vac} and 
\ref{tab_ups-vac} basic properties of the bound-state spectrum of a 
heavy quark ($Q$=$c$,$b$) and its antiquark ($\bar Q$) in the vacuum 
(note that the lifetime of the top quark is too short to allow 
for developing a $t\bar t$ bound-state spectrum). These spectra can be 
well understood in terms of a potential-model approach, where the underlying 
potential is of the so-called Cornell-type~\cite{Eichten:1979ms}, 
\begin{equation}
V_{Q\bar Q}(r) = -\frac{4}{3} \frac{\alpha_s}{r} + \sigma r \ ,
\label{Vqq-vac}
\end{equation}
consisting of a (color-) Coulomb term dominant at small $Q$-$\bar Q$ 
separation, $r$, and a linearly rising (``confining") term at 
large $r$. The potential 
description is now understood as a low-energy effective theory of Quantum 
Chromodynamics (QCD) utilizing an expansion in the inverse heavy-quark mass 
($1/m_Q$)~\cite{Bali:2000gf,Brambilla:2004wf}. Moreover, the pertinent 
vacuum potential has been computed in lattice QCD~\cite{Kaczmarek:2005ui} 
and found to agree well with the phenomenological Cornell potential
as deduced from applications to quarkonium spectroscopy.
\begin{table}[!t]
\begin{center}
\begin{tabular}{|c|c|c|c|c|}
\hline
State ($nL$) & $J^{PC}$ & $m_\Psi$ [MeV] & $\Gamma_{tot}$ [MeV] & $m_\Psi-2m_B$ [MeV]
\\
\hline  \hline
$\Upsilon$ (1$S$)    & $1^{--}$ & 9460          & 0.054$\pm$0.001 &  -1100 \\
\hline
$\chi_{b0}$ (1$P$) & $0^{++}$ & 9859          & ?    &  -700 \\
\hline
$\chi_{b1}$ (1$P$) & $1^{++}$ & 9893          & ?   &  -665\\
\hline
$\chi_{b2}$ (1$P$) & $2^{++}$ & 9912          & ?   &  -645\\
\hline
$\Upsilon'$ (2$S$)    & $1^{--}$ & 10023           & 0.032$\pm$0.003  &  -535\\
\hline
$\chi_{b0}$ (2$P$) & $0^{++}$ & 10233          &  ?    &  -325 \\
\hline
$\chi_{b1}$ (2$P$) & $1^{++}$ & 10255          &  ?    &  -305 \\
\hline
$\chi_{b2}$ (2$P$) & $2^{++}$ & 10269          & ?   &  -290\\
\hline
$\Upsilon''$ (3$S$)   & $1^{--}$ & 10355       & 0.020$\pm$0.002  &  -205 \\
\hline
$\Upsilon'''$ (4$S$)  & $1^{--}$ & 10579$\pm$1  & 20.5$\pm$2.5  &  +20 \\
\hline
\end{tabular}
\end{center}
\caption{Same as Table~\ref{tab_psi-vac}, but for low-lying $b\bar b$ mesons
(generically referred to as $\Upsilon$) and an open-bottom threshold of 
2$m_{B^\pm}\simeq10558$\,MeV (mass differences have been rounded to steps 
in 5~MeV).}
\label{tab_ups-vac}
\end{table}

The good understanding of the vacuum properties of heavy quarkonia in a 
relatively simple framework is one of the reasons why they are believed 
to be a good probe of medium effects. The latter can be roughly categorized
into screening effects in the two-body potential and dissociation reactions 
with constituents of the heat bath. Since the heavy-quark mass is large
compared to the typical temperatures realized in a heavy-ion reaction, 
a further practical benefit emerges: heavy-quark production is believed
to be largely restricted to the earliest phase of the collision, i.e., in 
primordial ``hard"  (high-momentum transfer) collisions of the incoming 
nucleons. On the one hand, this implies a separation of the (hard) production 
process from the subsequent (soft) medium effects, and, on the other hand, it
provides a baseline to determine the initial abundance prior to the formation
of the medium. For total charm production, this picture is consistent with 
current experimental information~\cite{Adler:2004ta,Shahoyan:2007zz} and also 
supported by theoretical estimates~\cite{Levai:1994dx}.

The above setup defines the basic framework to analyze modifications  
in the production of heavy quarkonia in URHICs, due to final-state 
interactions induced by the surrounding hot/dense medium. 
The first and most widely discussed probe in this context is the $J/\psi$.
Somewhat contrary to the expectations at the time, which considered the 
{\em production} of charmonium as a primordial plasma 
probe~\cite{Shuryak:1978ij,Cleymans:1984zs}, the suggestion by Matsui and 
Satz~\cite{Matsui:1986dk} asserted that with increasing 
centrality in URHICs a suppression of the $J/\psi$ peak in the dimuon 
invariant-mass spectrum should occur. As a suitable reference Drell-Yan 
dileptons at high mass were suggested, as a well-established ``hard" 
process. The underlying mechanism for the dissociation of the charmonium 
bound-state in a dense medium was associated with the ``Debye screening" 
of the binding potential, largely driven by a deconfinement of color 
charges and thus intimately connected to the formation of a QGP.  This 
phenomenon is rather general, very similar to the dissolution of atomic 
bound states in electromagnetic plasmas, or to the Mott transition in 
semiconductors~\cite{Mott:1968,Redmer:1997} where under high pressure 
electrons become delocalized and a conduction band emerges, signaling 
the plasma state.

\begin{figure}[!t] 
\begin{center}
\begin{minipage}[t]{16.5 cm} 
\centering\epsfig{file=intro_fig1.epsi,width=0.8\textwidth}
\end{minipage} 
\caption{First observation of the $J/\psi$ suppression effect in 
O(200\,AGeV)-U collisions in the NA38 experiment at CERN-SPS. 
When comparing the invariant-mass spectrum of muon pairs produced 
in peripheral collisions (characterized by a small transverse energy, 
$E_T<34$\,GeV; left panel) with that in central collisions 
(at high transverse energy, $E_T>85$\,GeV; right panel), a reduction of 
the $J/\psi$ signal relative to the Drell-Yan continuum is apparent 
(figures from Ref.~\cite{Baglin:1990iv}).}
\label{fig_0-U}
\end{center} 
\end{figure} 
\begin{figure}[h] 
\begin{center} 
\begin{minipage}[t]{16.5 cm} 
\centering\epsfig{file=intro_fig2.epsi,width=0.8\textwidth} 
\end{minipage} 
\caption{Early comparisons of nuclear-absorption calculations for 
$J/\psi$ suppression to data in $p$-$A$ and $A$-$B$ collisions
with light-ion projectiles~\cite{Gerschel:1993uh}. 
Recent data and analyses are discussed in the main text.}
\label{suppression3}
\end{center} 
\end{figure} 
Shortly after the prediction of $J/\psi$ suppression, the NA38 experiment
at the CERN-SPS found evidence for this effect in collisions of 200\,AGeV
oxygen (O) projectiles with uranium (U) nuclei~\cite{Baglin:1990iv},
see Fig.~\ref{fig_0-U}.
The interpretation of this observation in terms of QGP formation was 
immediately challenged by more conventional explanations in terms of 
inelastic $J/\psi$ scattering on ``primordial" target and projectile 
nucleons~\cite{Gerschel:1988wn},  
and/or on secondary produced hadronic 
``comovers"~\cite{Gavin:1988hs,Vogt:1988fj}.
The systematic experimental analysis of the nuclear mass-number dependence 
of $J/\psi$ production in proton-nucleus ($p$-$A$) collisions at CERN
(NA3~\cite{Badier:1983dg}, NA38~\cite{Baglin:1990iv,Abreu:1998ee}) and 
FNAL (E772~\cite{Alde:1990wa}), as well as in collision systems with
light-ion projectiles (O-U, S-U, O-Cu) at CERN (NA38~\cite{Baglin:1991vb}), 
indeed suggests that the $J/\psi$ suppression in all these experiments can be 
understood in a unified way by primordial ``nuclear absorption" of the 
charmonium state (with no further reinteractions before being
detected via its dimuon decay). The magnitude of this suppression has 
been characterized by an empirical mean path length, $L$, of the charmonium
traveling through cold nuclear matter at normal density, 
$n_0=0.16$ fm$^{-3}$. The suppression systematics could then be
quantitatively described following the simple absorption law, 
$S_{\rm nucl}=\exp{(-\sigma_{\rm abs}n_0 L)}$ \cite{Gerschel:1993uh},
with the nuclear absorption cross section $\sigma_{\rm abs}$ extracted
from $p$-$A$ collisions\footnote{A more detailed account underlying such 
analysis, based on the Glauber model including realistic nuclear density 
profiles, will be given in Sec.~\ref{ssec_nuc-abs} of the main text.}.
In fact, nuclear absorption systematics could not only account for $p$-$A$ 
but also for $A$-$B$ collision systems, see Fig.~\ref{suppression3}, where 
$L=L_A+L_B$ is a schematic measure for the combined path length through 
projectile and target nuclei.
The QGP signal had seemingly vanished~\cite{Gerschel:1993uh}! 

\begin{figure}[!htb]
\begin{minipage}{0.5\textwidth}
\includegraphics[width=1.0\textwidth]{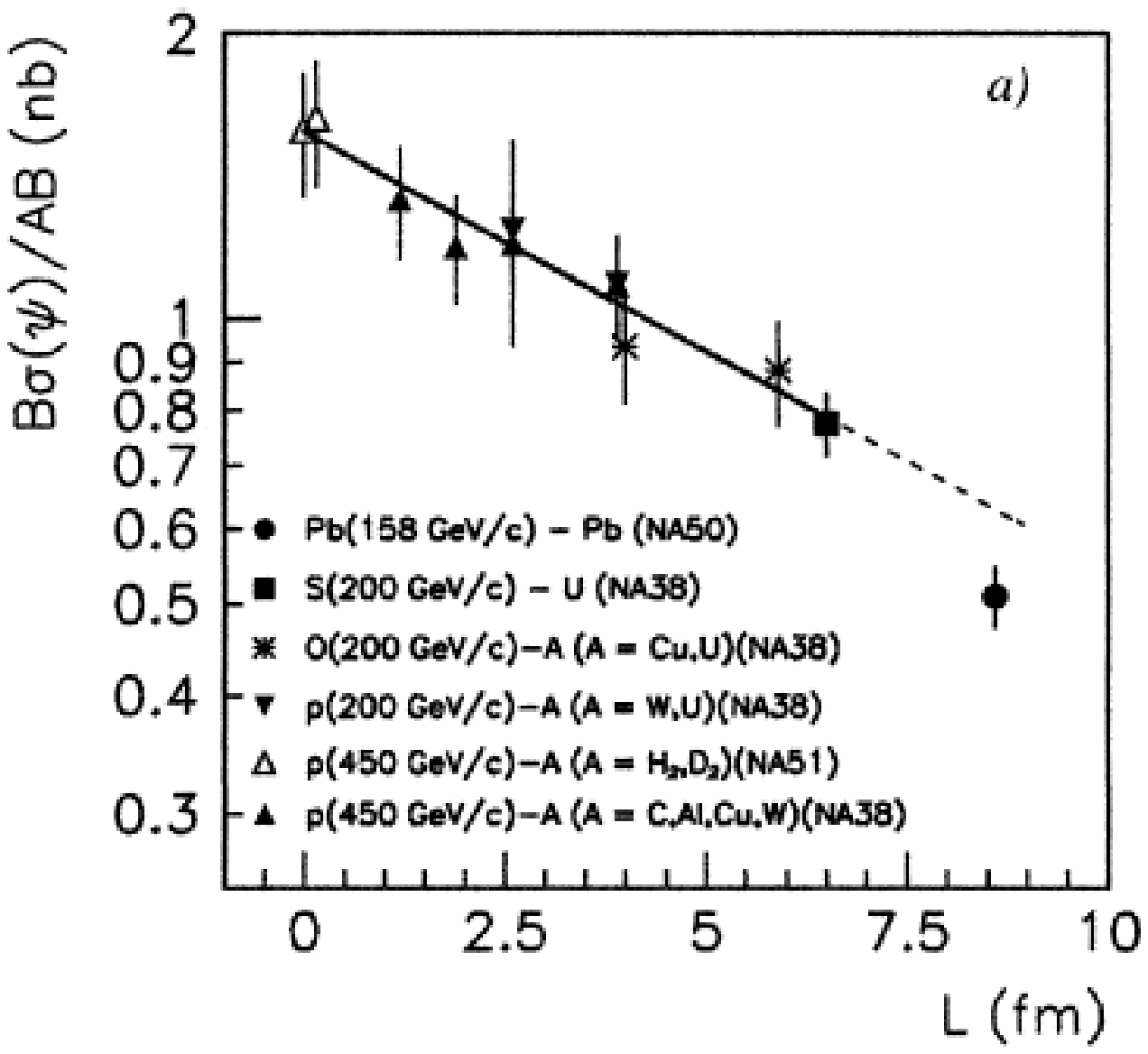}
\end{minipage}
\hspace{0.5cm}
\begin{minipage}{0.5\textwidth}
\includegraphics[width=.80\textwidth]{anomalous1.epsi}
\end{minipage}
\caption{Anomalous $J/\psi$ suppression in Pb(158\,AGeV)-Pb collisions 
at the CERN-SPS measured by the NA50 experiment as a function of the 
Glauber variable $L$ (left panel, from Ref.~\cite{Gonin:1996wn}) and as a
function of the number of charged particles $N_{ch}$ (right panel, from 
Ref.~\cite{Alessandro:2004ap}).
\label{anomalous}
}
\end{figure}

\begin{figure}[!htb]
\includegraphics[width=0.8\textwidth,height=8cm]{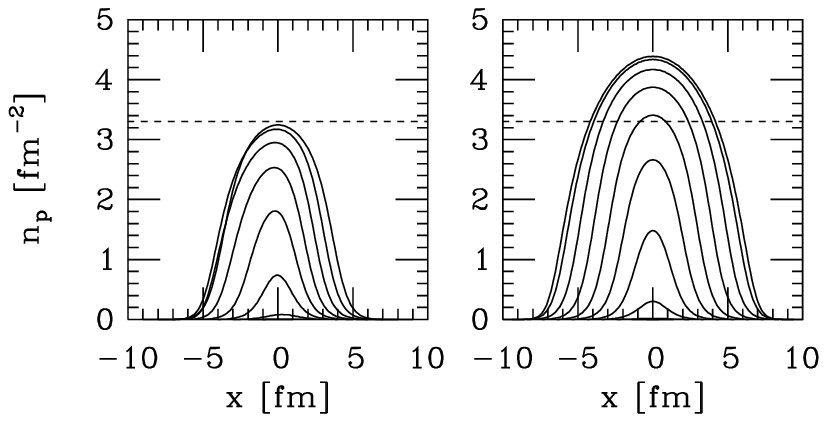}
\caption{The transverse density of participants, $n_p$, in S-U 
(left) and Pb-Pb (right) collisions along the direction of the impact 
parameter vector, $\vec b$, for different values of $b=0,2,4,...$fm. 
The dashed line indicates the adopted threshold value 
$n_p^{\rm crit}=3.3$\,fm$^{-2}$ for the onset of anomalous $J/\psi$ 
suppression. Figure taken from Ref.~\cite{Blaizot:1996nq}.} 
\label{n_part}
\end{figure}

The situation changed when the Pb beam became available at the SPS.
In central Pb(158\,AGeV)-Pb collisions the successor experiment of NA38, 
NA50, found a stronger $J/\psi$ suppression than predicted by the 
extrapolation of the nuclear absorption law (based on the Glauber 
model)~\cite{Gonin:1996wn,Abreu:1997jh}, see Fig.~\ref{anomalous}.
This deviation was dubbed ``anomalous $J/\psi$ suppression", as
opposed to the ``normal suppression" caused by nuclear absorption
(following the simple exponential dependence on the $L$-variable).
A first interpretation of the anomalous suppression as a signal for QGP 
formation was given by Blaizot and Ollitrault~\cite{Blaizot:1996nq} 
within a ``threshold-suppression" scenario. It was assumed that a 
charm-anticharm quark pair cannot hadronize into 
a charmonium state when it is produced in a region where the transverse 
density of participant nucleons exceeds a certain ``critical" value, 
$n_p^{\rm crit}$, see Fig.~\ref{n_part}. This scenario provided a 
consistent explanation of the available data in the sense that the value 
for $n_p^{\rm crit}$ (characterizing the maximal energy density of the
subsequently formed medium) required to describe the suppression in
Pb-Pb was larger than the maximum value $n_p$ reached with light-ion 
projectiles (see also Ref.~\cite{Kharzeev:1996yx}).

This initial interpretation was followed by a wide variety of theoretical
investigations including hadronic comover dissociation, parton-induced 
break-up reactions in the QGP, pre-equilibrium effects, and combinations 
thereof. The NA50 collaboration was led to conclude that the $J/\psi$
suppression pattern provides evidence for the deconfinement of 
quarks and gluons~\cite{Abreu:2000ni}
\footnote{For the final NA50 results concerning $J/\psi$ suppression, see
Ref.~\cite{Alessandro:2004ap}.}.
One of the controversially discussed issues was (and still is) 
whether the data provide evidence for a ``threshold" or ``onset" 
behavior in terms of step-like patterns in the centrality dependence. 
Further clarification was hoped to be gained from studying intermediate
size collision systems, as recently done by the NA60 
collaboration~\cite{Arnaldi:2006ee,Arnaldi:2007aa}. In In(158\,AGeV)-In 
collisions, the ``onset" of anomalous $J/\psi$ suppression does not seem 
to follow 
a scaling with the Glauber variable $L$, but rather with the number of 
participants, $N_{part}$, in the collision. Note that $N_{part}$, 
contrary to $L$, is a quantity closely related to the density of 
secondary particles, potentially forming a QGP, see Fig.~\ref{fig_In-In}.
\begin{figure}[!t] 
\begin{center}
\begin{minipage}[t]{16.5 cm}
\centering
\includegraphics[width=0.85\textwidth]{intro_fig5.epsi}
\end{minipage}
\caption{``Anomalous" $J/\psi$ production (i.e., relative to expectations 
from normal nuclear absorption) in In(158\,AGeV)-In collisions
measured by the NA60 experiment compared to 
other projectile-target combinations at the CERN-SPS.
The ``threshold" for anomalous suppression appears to scale with 
the number of participants (right panel) rather than with the Glauber 
variable $L$ (left panel); figures taken from Ref.~\cite{Arnaldi:2006ee}.}
\label{fig_In-In}
\end{center} 
\end{figure} 

\begin{figure}[!t]
\begin{minipage}{0.5\textwidth}
\includegraphics[width=.95\textwidth,height=0.3\textheight]{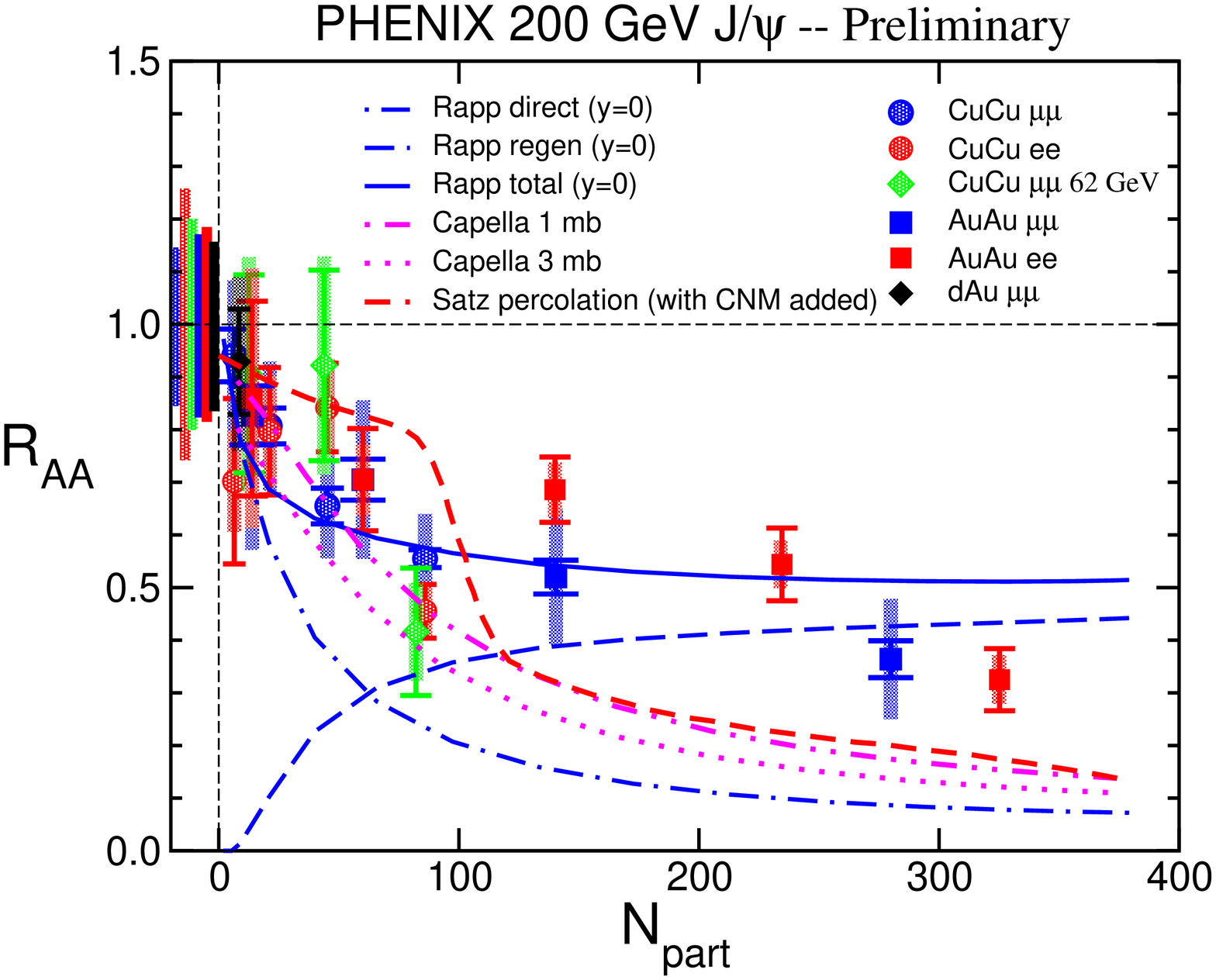}
\end{minipage}
\begin{minipage}{0.5\textwidth}
\includegraphics[width=.95\textwidth]{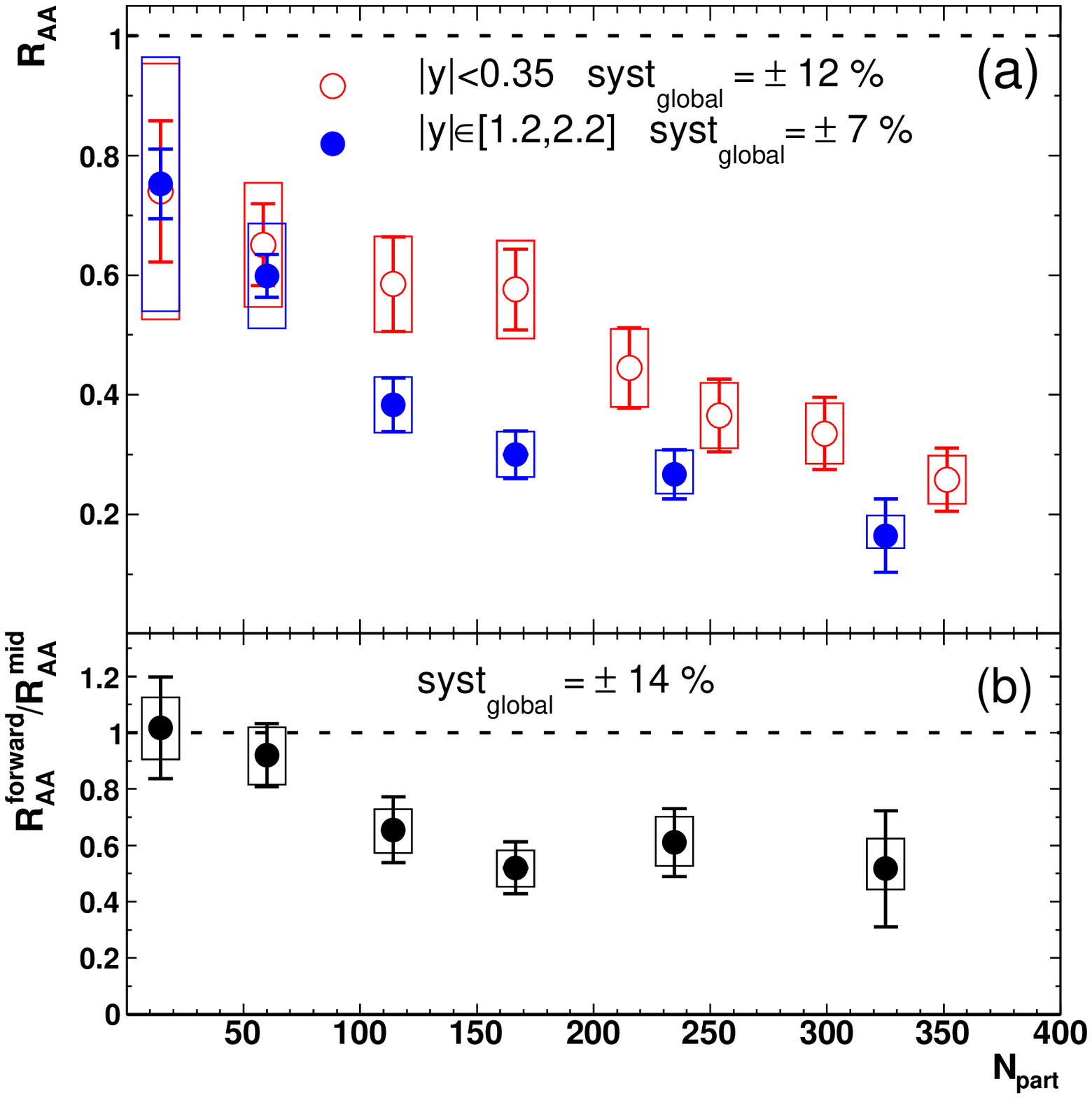}
\end{minipage}
\caption{$J/\psi$ production, relative to the binary-collision scaled
yield in $p$-$p$ collisions, measured in the PHENIX experiment 
in $\sqrt{s_{NN}}=200$\,GeV Au-Au collisions at RHIC.
Left panel: comparison with kinetic 
models~\cite{Grandchamp:2003uw,Thews:2005fs} of charmonium production
at central rapidity, illustrating the importance of secondary 
$J/\psi$ production via charm-anticharm recombination (figure from 
Ref.~\cite{Leitch:2006ff}). 
Right panel: increase of $J/\psi$ suppression at forward relative
to central rapidity, suggestive for regeneration 
mechanisms (from Ref.~\cite{Adare:2006ns}).}
\label{fig_phenix}
\end{figure}

In the year 2000 the experimental program at RHIC commenced. First
$J/\psi$ data for Au-Au collisions became available in 
2003~\cite{Adler:2003rc}, while recent ones may be found
in Ref.~\cite{Adare:2006ns}. 
Figure~\ref{fig_phenix} shows the experimental data for $J/\psi$ production
in terms of the so-called nuclear modification factor,
$R_{AA}$, as a function of centrality. It is defined as the yield observed
in heavy-ion reactions relative to the one in $p$-$p$ collisions scaled 
by the number, $N_{\rm coll}$, of binary $N$-$N$ collisions,
\begin{equation}
R_{AA}(N_{part})
=\frac{N_{AA}^{J/\psi}}{\langle N_{coll}\rangle N_{pp}^{J/\psi}} \ .
\end{equation}
In the absence of medium effects one expects $R_{AA}$=1.
An anomalous suppression of charmonium production was confirmed
in semi-/central Au-Au collisions, with a magnitude  
similar to the one observed at the SPS. This seems rather surprising 
in view of the factor of 10 higher collision energy at RHIC, inducing 
higher (initial) energy densities.
However, it had been predicted~\cite{BraunMunzinger:2000px,Thews:2000rj,
Gorenstein:2000ck,Grandchamp:2001pf} that at RHIC and LHC the copious 
production of open charm leads to an additional source
of $J/\psi$ production, through the recombination of charm quarks
(or charmed mesons) in the hot and dense medium. Calculations including
both suppression and regeneration mechanisms~\cite{Grandchamp:2001pf} 
had anticipated that a stronger ``anomalous" suppression is largely 
compensated by secondary production via charm-anticharm
coalescence. This is illustrated in the left panel of Fig.~\ref{fig_phenix}
where a comparison with reaction kinetic models for $J/\psi$
production~\cite{Grandchamp:2003uw,Thews:2005fs} shows that a na\"ive
extrapolation of ``anomalous" $J/\psi$ suppression in a QGP environment 
from SPS to RHIC fails.
Additional support for a secondary component of the $J/\psi$ yield at RHIC
follows from the observation that $J/\psi$ suppression is less pronounced
at central rapidity relative to forward rapidity, since at central rapidity 
the regeneration effect should be larger due to a larger charm content
in the medium. On the contrary, dissociation mechanisms should increase the
suppression due to larger energy densities at midrapidity.

The above discussion has given a first indication of the complexity in
describing and understanding the various facets figuring into quarkonium
observables in heavy-ion collisions,
let alone their interpretation in terms of properties of the
Quark-Gluon Plasma. It is thus mandatory to identify key concepts
and combine them into a comprehensive approach that allows for a systematic 
interpretation of experimental data under a broad range of conditions
(encompassing, e.g., all available collision energies,
from FAIR via SPS and RHIC to LHC). 
This will be the main objective of this article, by means of a critical 
review of existing approaches and their comparison to data. 
An essential ingredient to such approaches are controlled reference points,
e.g., equilibrium limits for in-medium properties, or $p$-$A$ collisions
to separate ``normal" nuclear effects from hot/dense matter effects.
Our article is roughly organized according to the following decomposition
of the problem: \\
(1) {\bf Equilibrium properties of quarkonia} (Sec.~\ref{sec_equil}): 
these provide the main link
between properties of the QGP and the medium created
in heavy-ion collisions. The basic quantity to be determined is the 
thermodynamic quarkonium spectral function: its pole mass
(binding energy) largely determines the equilibrium abundance (in
connections with the properties of open charm/bottom), while in-/elastic
reaction rates determine thermal and chemical relaxation times
and are encoded in the spectral widths. \\
(2) {\bf Quarkonium transport} (Sec.~\ref{ssec_transport}), 
which is required to evaluate the coupling of the quarkonia
to the medium by computing the evolution of their phase space
distribution, starting from realistic initial conditions.
In principle, this evolution progresses toward equilibration,
e.g., as a function of transverse momentum. The method of choice
is a transport treatment incorporating regeneration reactions 
to enable a relaxation toward thermal and chemical
equilibrium. Since regeneration processes involve
open-charm/-bottom spectra, a reliable assessment of the latter 
becomes mandatory. \\
(3) {\bf Initial spectra and pre-equilibrium interactions} 
(Sec.~\ref{ssec_nuc-abs}): these determine the
initial condition of the transport equation in item (i); they include
modifications relative to spectra in $p$-$p$ collisions ($p_T$ broadening
via Cronin effect, shadowing), as well as absorption and rescattering
on primordial nucleons and secondary particles in the pre-equilibrium
evolution of the medium. \\
(4) {\bf Observables} (Sec.~\ref{ssec_obs}): these need to be computed by 
combining all of the above. An additional input is a realistic modeling 
of the bulk medium evolution, ideally in terms of a locally thermalized 
medium (if applicable), e.g., within a hydrodynamic evolution, 
or in terms of transport models in regimes of incomplete thermalization. \\ 
In Sec.~\ref{sec_exp} we give an outlook on the capabilities of, and 
physics questions to be addressed with, future experiments at LHC, RHIC-II 
and FAIR, while Sec.~\ref{sec_sum} contains a brief summary.

%% file: diss-hg.tex
The theoretical description of the $J/\psi$ in medium is not complete
without an understanding of its properties in hadronic matter. Since
the latter is characterized by confined color charges, color screening
of a tightly bound, spatially compact $c\bar c$ state is believed to be 
negligible below $T_c$ (on the order of 10~MeV or so)~\cite{Morita:2007hv}.
Thus, most of the calculations of $J/\psi$ properties in hadronic matter 
have focused on the role of dissociation reactions of the type $J/\psi + M 
\to D +\bar{D}$ where $M$ denotes a meson from the heat bath (e.g., $\pi$, 
$\rho$) and $D$ generically denotes a $D$-meson (e.g., $D$(1870) or 
$D^*$(2010)); for reaction with baryons ($B$) from the medium one has 
$J/\psi + B \to \Lambda_c,\Sigma_c +\bar{D}$. On the phenomenological side, 
a quantitative assessment of the inelastic reaction rates in the hadronic
phase of a heavy-ion reaction is indispensable for disentangling the
``anomalous suppression" in the QGP. Here, we focus on the most recent
developments over the last 5 years or so (for earlier reviews, see, e.g., 
Refs.~\cite{Blaschke:2002ww,Barnes:2003dg,Rapp:2003vj,Ivanov:2003ge,Bourque:2004av}).

\begin{figure}[!t]
\parbox{9cm}{
\includegraphics[width=8cm,height=4.5cm]{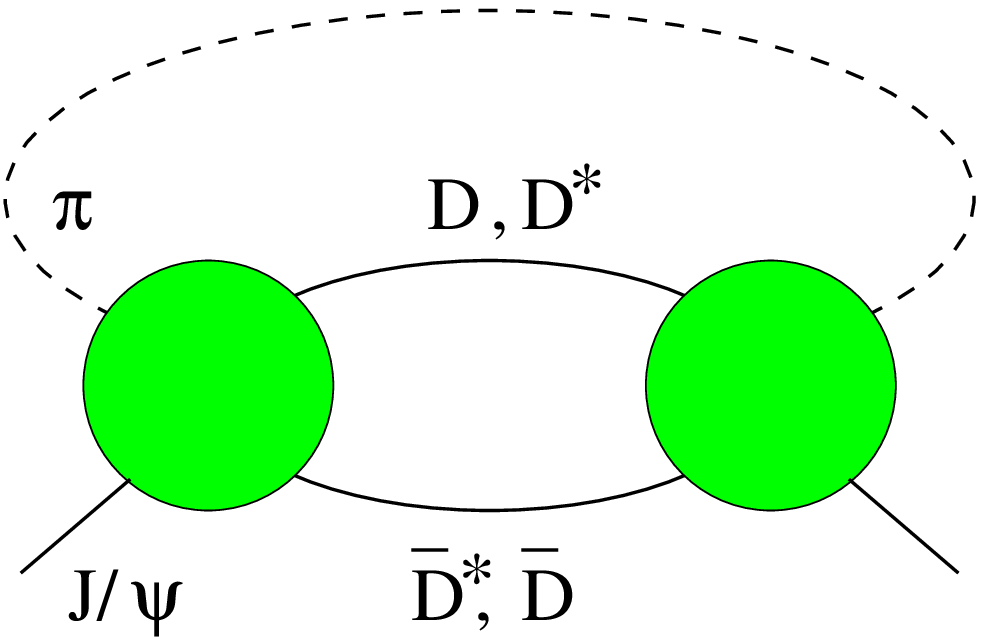}
}\hfill
\parbox{9cm}{
\includegraphics[width=8cm,height=4.5cm]{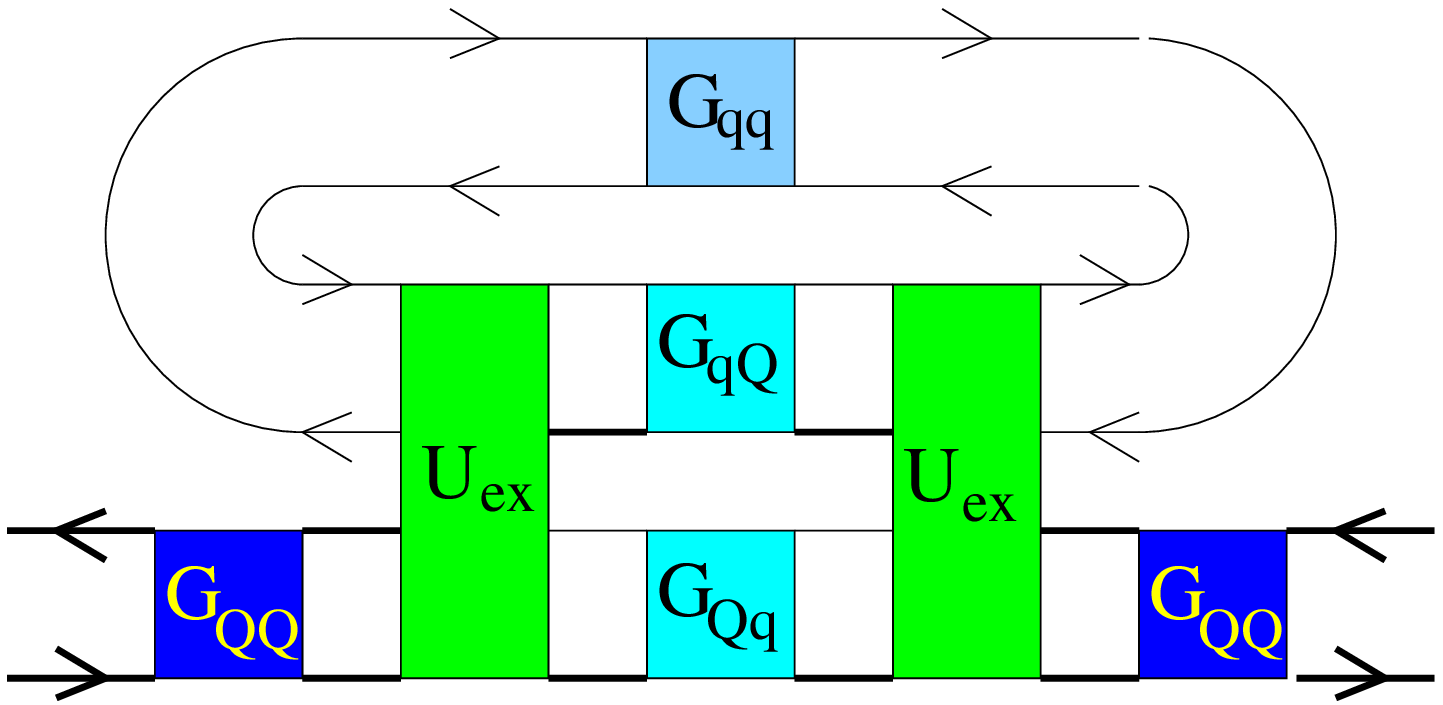}
}
\caption{Diagrammatic representation of the $J/\psi$ self-energy due
to interactions with pions from the hadronic heat bath, with effective 
vertices in a hadronic basis (left panel) and resolved into a
quark basis with microscopic pion ($G_{qq}$), $D$-meson
($G_{Qq}$) and $J/\psi$ ($G_{QQ}$) propagators as well quark-exchange
interactions ($U_{\rm ex}$ (right panel)~\cite{Blaschke:2004dv}. 
The $J/\psi$ dissociation width follows from the imaginary part
of the self-energy, $\Gamma_\Psi = -{\rm Im}\,\Sigma_\Psi /m_\Psi$; the
underlying process may be visualized by cutting the left-hand diagram 
through the middle (note that this cannot be done at the quark level
since confinement forbids external quark lines in the hadronic phase).   
Diagrams of this type naturally appear in the cluster expansion for 
two-particle properties, see Fig.~\ref{fig:cluster-cluster} in
Sec.~\ref{ssec_plasma}.}
\label{fig_jpsidiss-dia}
\end{figure}

The ``natural" framework to evaluate hadronic $J/\psi$ dissociation
are effective hadronic models to compute the pertinent inelastic cross
section for the reactions of the type listed above. However, even after
the exploitation of standard symmetry in hadronic 
physics (such as vector-current conservation, vector-meson 
dominance~\cite{Matinyan:1998cb} and 
chiral~\cite{Navarra:2001pz,Bourque:2008es} or broken 
flavor-$SU$(4)~\cite{Matinyan:1998cb,Haglin:1999xs} symmetry), a 
quantitative control over the dissociation rates could not be achieved 
due to uncertainties in hadronic formfactors.  The latter are required 
to account for the finite size of the vertices in the effective theory. 
When varying the pertinent cutoff parameters over a typical hadronic 
scale, say,  $\Lambda$=1-2\,GeV, the $J/\psi$ dissociation cross sections 
can vary by about 2 orders of magnitude (from sub-mb to tens of mb). 
Alternatively, parton-based calculations have been performed which, in 
principle, could provide a more microscopic description of the effective 
vertices.  Combining the LO gluo-dissociation cross section (as discussed 
in the previous section for deconfined gluons in the 
QGP)~\cite{Bhanot:1979vb} with the gluon distributions within a hadron, 
the resulting hadron-$J/\psi$ break-up cross section was found to be 
small (sub-mb)~\cite{Kharzeev:1994pz},
primarily due to the soft gluon distribution function within the hadrons 
(also, the thermal motion of hadrons below $T_c$ is suppressed compared
to deconfined gluons above $T_c$; medium effects in the bound state have 
been studied in Ref.~\cite{Arleo:2004ge}, and NLO calculations have been 
carried out for the $\Upsilon$~\cite{Song:2005yd} but not for the $J/\psi$). 
However, the hadronic dissociation reactions of charmonia are presumably
dominated by other processes, specifically quark exchanges as 
illustrated in Fig.~\ref{fig_jpsidiss-dia}, which have been addressed
within effective quark 
models~\cite{Martins:1994hd,Wong:1999zb,Blaschke:2000zm,Barnes:2003dg,Ivanov:2003ge,Bourque:2008es}. 
A realistic description of such processes requires non-perturbative
input in terms of both the effective quark propagators (especially
for their momentum-dependent mass functions in the chirally broken and 
confined hadronic phase) and the interaction vertices (which are usually
resummed in ladder approximation). In principle, these quantities can 
be constrained by vacuum phenomenology (hadron spectra, decay constants 
and quark/gluon condensates), but their 
implementation in a hadronic medium poses formidable challenges, 
including a fully relativistic treatment, cancellations due to 
constraints from chiral symmetry, etc.
Note that the quarkonium propagator, $G_{QQ}$, figuring into 
the right-hand-side of Fig.~\ref{fig_jpsidiss-dia}, is closely related
to the potential models discussed in Sec.~\ref{ssec_pot}. For example, 
the $D\bar D$ state appropriate for hadronic matter could be 
implemented into the charmonium $T$-matrix by coupling an additional 
$D\bar D$ channel, $G_{D\bar D}$, into Eq.~(\ref{LS}) using a suitably 
constructed transition potential, 
$U_{\rm ex}\equiv V_{c\bar c \to D\bar D}$.

\begin{figure}[!t]
\includegraphics[width=9.3cm]{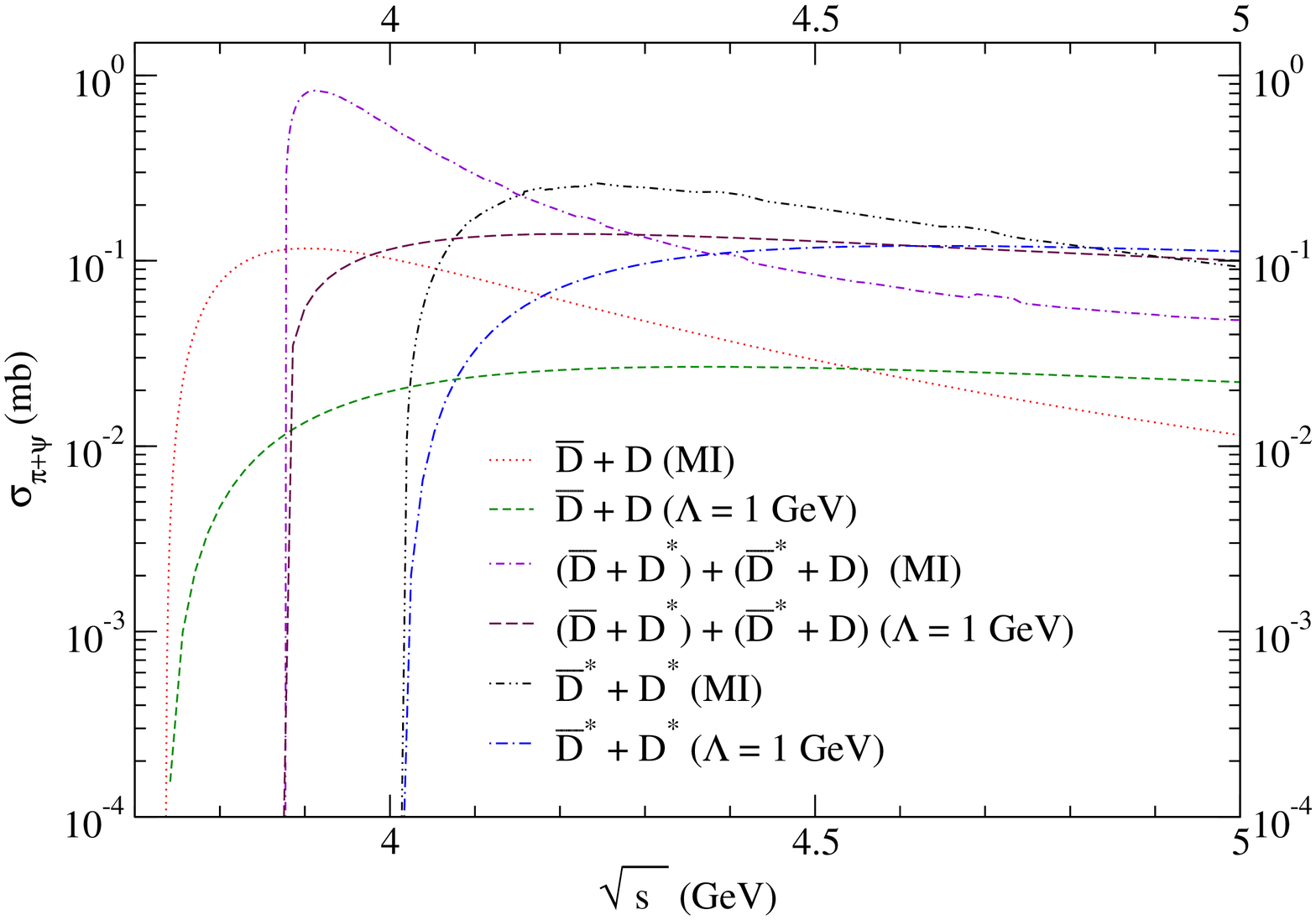}
\includegraphics[width=9.3cm]{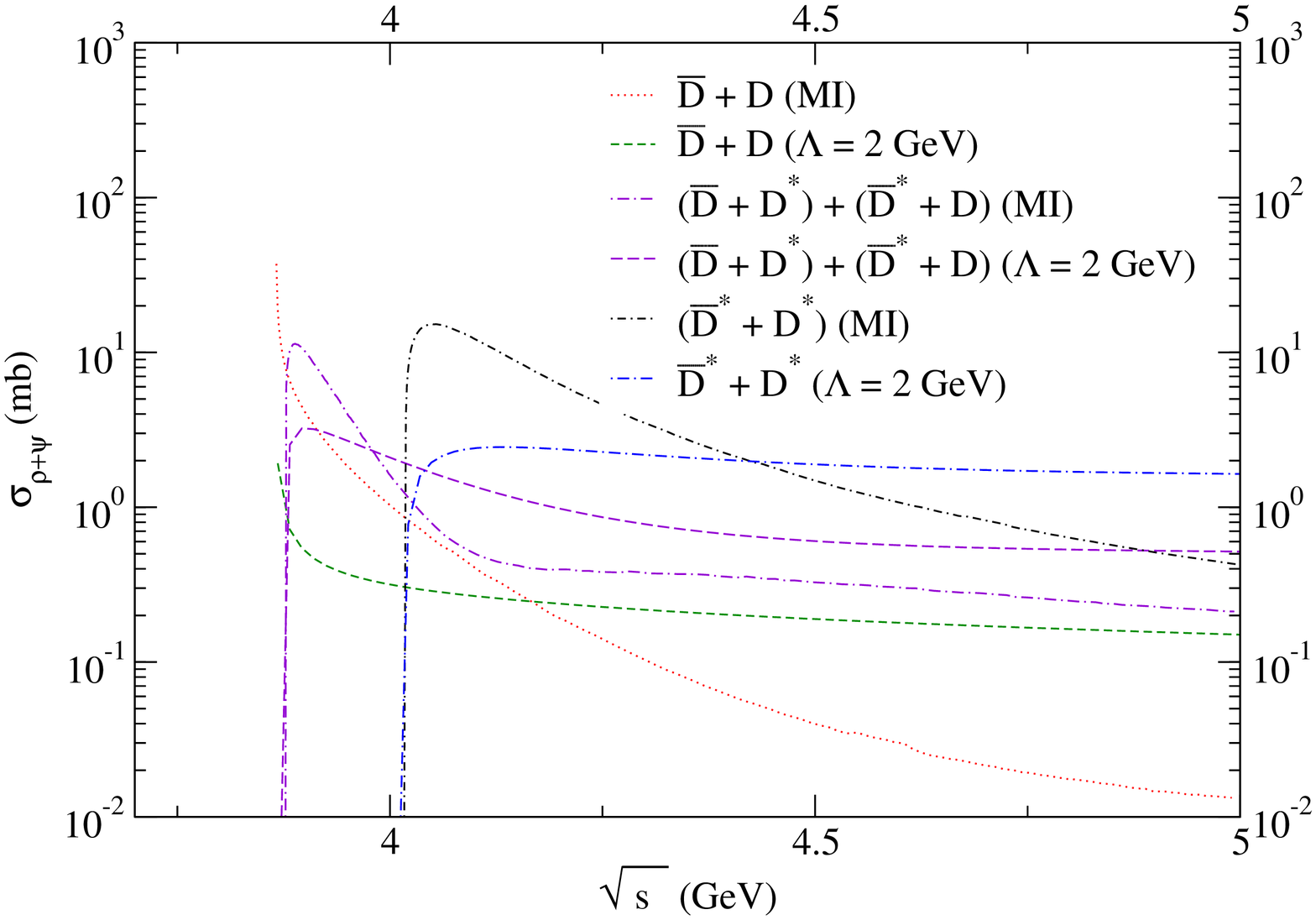}
\caption{Dissociation cross sections of the $J/\psi$ induced by pions
(left panel) and $\rho$-mesons (right panel) as a function of center of
mass energy of the reaction. In each panel, results
for a chiral (Nambu-Jona-Lasinio) quark model (labeled 
``MI")~\cite{Bourque:2008es} are compared to those for a chiral effective 
hadronic model~\cite{Bourque:2008ta} in which the vertex formfactor 
cutoffs (quoted in parenthesis) have been adjusted to approximately match 
the strength in the quark model calculations. The low-energy thresholds
in the left panel correspond to (from left to right) $2m_D\simeq3.75$\,GeV, 
$m_D+m_{D^*}\simeq3.88$\,GeV and $2m_{D^*}\simeq4.02$\,GeV, and
in the right panel to (from left to right) 
$m_{J/\psi}+m_{\rho}\simeq3.87$\,GeV and $2m_{D^*}\simeq4.02$\,GeV.
Figures are taken from Ref.~\cite{Bourque:2008es}.}
\label{fig_jpsi-hg-xsec}
\end{figure}
A recent example for the calibration of hadronic vertices evaluated
in chiral effective theory~\cite{Bourque:2008ta} using a chiral quark 
model~\cite{Bourque:2008es} is shown in Fig.~\ref{fig_jpsi-hg-xsec} for 
the processes $\pi+J/\psi \to D+\bar D$ (left panel) and 
$\rho+J/\psi \to D+\bar D$ (right panel; see also 
Ref.~\cite{Blaschke:2008mu} for
a similar analysis). These plots illustrate several features:
(a) The energy dependence of the cross sections for hadronic 
and quark models does not match well; most notably, the quark-model 
results exhibit threshold peaks which do not appear in the hadronic 
calculations (whether this is a realistic feature remains to be seen);
 nevertheless, with formfactor cutoff values in a 
typical hadronic range, $\Lambda_{\pi J/\psi}$=1\,GeV and 
$\Lambda_{\rho J/\psi}$=1\,GeV, the overall strengths are comparable;
(b) the cross sections for $\rho$-induced dissociation are significantly
larger than the pion-induced ones, and both agree within a factor of
$\sim$2 with their counterparts in quark-model 
calculations~\cite{Blaschke:2002ww,Barnes:2003dg,Ivanov:2003ge}.
This is quite encouraging in view of the much larger uncertainties
alluded to above; similar results are obtained in 
Ref.~\cite{Blaschke:2008mu}. A recent comparison for nucleon-induced 
dissociation can be found in Refs.~\cite{Hilbert:2007hc,Oh:2007ej} (note 
that the low-energy $\Psi$-$N$ break-up cross section, which is relevant 
for $J/\psi$ absorption in an equilibrated hadron gas, is different from 
(and probably much smaller than) the high-energy absorption cross section 
figuring into ``nuclear absorption" in heavy-ion 
collisions discussed in Sec.~\ref{ssec_nuc-abs}).

\begin{figure}[!t]
\parbox{9cm}{
\includegraphics[width=8cm,height=6cm]{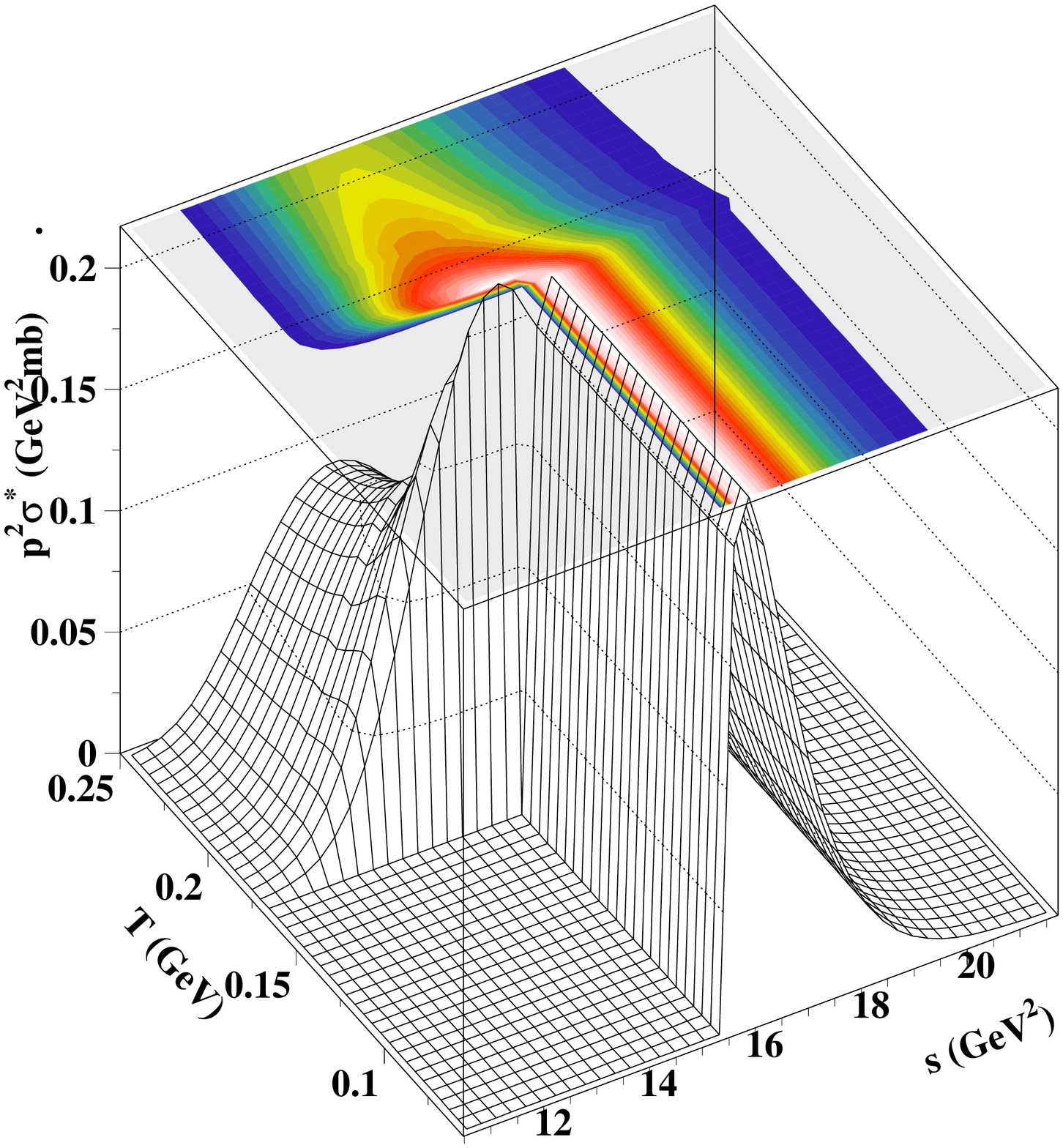}
}\hfill
\parbox{9cm}{
\includegraphics[width=8cm,height=6cm,angle=0]{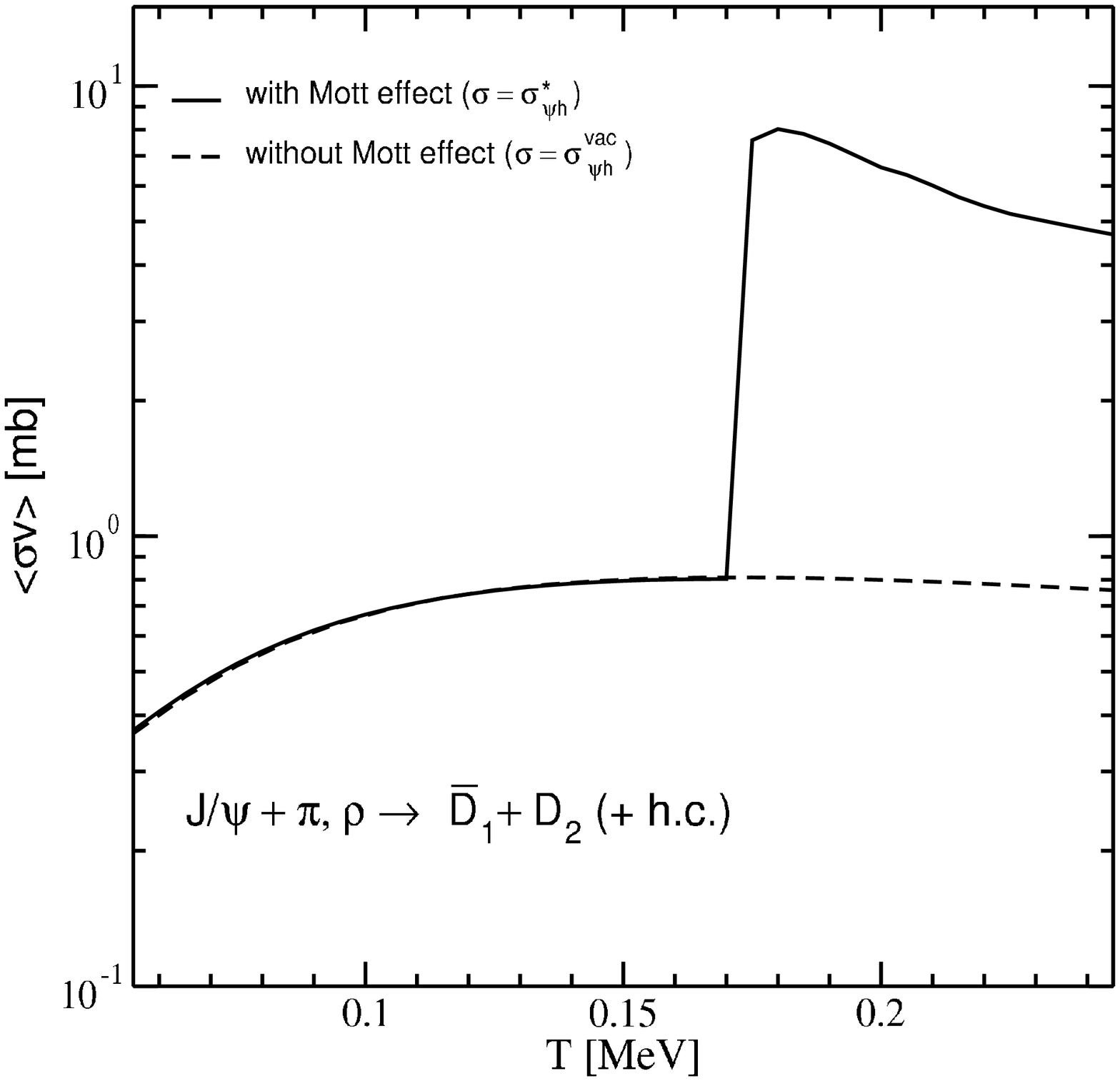}
}
\caption{Left panel: energy ($s$) and temperature ($T$) 
dependence of the effective cross section ($\sigma^*$) for $J/\psi$ breakup 
by $\rho$-meson impact. We display $p^2\sigma^*(s;T)$ ($p$: initial-state
hadron momentum) for better visibility 
of the effective lowering of the breakup threshold when $T$ exceeds the 
$D$-meson Mott temperature $T^{\rm Mott}\simeq 172$\,MeV;
right panel: $T$ dependence of the thermally averaged $J/\psi$ 
breakup cross section in a $\pi$-$\rho$ gas; 
the calculation with vacuum $D$-mesons (dashed line) is compared to one 
with in-medium broadened $D$-meson spectral functions 
(due to the Mott effect at the chiral phase transition) causing
a step-like enhancement (solid line) caused by the reduced 
breakup threshold; from Ref.~\cite{Blaschke:2002ww}.
\label{fig_jpsi-diss-med}
}
\end{figure}
The energy-dependent cross sections for quarkonium break-up by
hadron impact enable to calculate the temperature- (and density-) 
dependent dissociation rates in hadronic matter.
If one assumes the short-distance vertex functions to be unaffected by 
the surrounding medium, the issue remains whether mass and width of the 
final-state open-charm hadrons change with temperature and density. These
medium effects, in particular, imply modifications of the thresholds for 
the breakup processes~\cite{Grandchamp:2003uw,Sibirtsev:1999jr,Tsushima:2000cp,Burau:2000pn,Friman:2002fs,Blaschke:2003ji,Fuchs:2004fh}.
The theoretical basis for the discussion of quark-exchange effects in the 
self-energy of heavy quarkonia in strongly correlated quark matter is given 
by systematic cluster expansion techniques developed in the context of 
plasma physics (more details are contained in the next subsection in
connection with Figs.~\ref{fig:cluster-ex}, \ref{fig:cluster-ex2} and 
\ref{fig:cluster-cluster}). Fig.~\ref{fig_jpsi-diss-med} illustrates 
the effect of the spectral broadening
of $D$-mesons at the chiral phase transition due to the opening of the 
decay channel into their quark constituents (Mott effect) for temperatures
exceeding the $D$-meson Mott temperature, $T^{\rm Mott}\approx 172$\,MeV.
The temperature-dependent lowering of the $J/\psi$ breakup threshold in 
the thermally averaged $J/\psi$ cross section in a $\pi$-$\rho$ meson gas 
induces a step-like enhancement~\cite{Blaschke:2002ww}.
This effect has been discussed as a possible mechanism underlying the
threshold-like anomalous suppression pattern of $J/\psi$'s observed by
the NA50 experiment~\cite{Burau:2000pn,Blaschke:2000er}.

We finally make a rough estimate of the magnitude of the 
dissociation widths of the $J/\psi$ in hadronic matter, based on the
most recent evaluations discussed above, and compare it to the values
found in the QGP as discussed in the previous Section. 
In analogy to Eq.~(\ref{rate}), the $J/\psi$ dissociation rate in the
hadron gas can be written as a convolution of the $h$-$J/\psi$ break-up
cross section with the thermal distributions of all hadrons $h$ in the 
heat bath, 
\begin{equation}
\Gamma_\Psi^{\rm HG} = \sum\limits_h \int \frac{d^3k}{(2\pi)^3} 
\ f^{h}(\omega_k,T)  \ v_{\rm rel} \ \sigma^{\rm diss}_{h\Psi}(s) \ . 
\label{rate-hg}
\end{equation}
Let us approximate this expression by 
$\Gamma_\Psi= \langle \sigma^{\rm diss}\,n_h^{\rm tot}\, v_{\rm rel}\rangle$
with a total density $n_h^{\rm tot}=5\,n_0 =0.8\,{\rm fm}^{-3}$ for a
hadron-resonance gas at $T$=175~MeV. From Figs.~\ref{fig_jpsi-hg-xsec}
and \ref{fig_jpsi-diss-med} we estimate 
$\langle \sigma^{\rm diss}\,v_{\rm rel}\rangle \simeq 0.5$\,mb as an 
estimate for the average over all hadrons (note that mass thresholds, 
the relative velocity $v_{\rm rel}<1$ and an average over the thermal 
motion significantly reduce the average value for 
$\sigma^{\rm diss}\,v_{\rm rel}$, compared to the relatively narrow maxima 
in the cross section plots), yielding $\Gamma_\Psi^{\rm HG}\simeq 8$\,MeV. 
This is significantly smaller than the estimates for the QGP-induced 
suppression shown in the right panel of Fig.~\ref{fig_diss-rates}, even 
close to $T_c$. We can therefore expect that the anomalous suppression of
the $J/\psi$ in heavy-ion collisions is very small (and even less for the
$\Upsilon$). However, the situation may be quite different for excited
charmonia, $\psi'$ and $\chi_c$, which contribute ``indirectly" to
$J/\psi$ production via late decays after freezeout (``feed-down"). 
Unfortunately, much less is known about hadronic dissociation of 
$\psi'$ and $\chi_c$ states.

%% file: plasma.tex
In developing a theoretical approach to heavy quarkonia as messengers of the
deconfinement and/or hadronization transition of a quark-gluon plasma as formed
in a heavy-ion collision, one should aim at a unifying description
where hadrons appear as bound states (clusters) of quarks and gluons.
The situation is reminiscent of the problem of two-particle states in QED
plasmas where a well-developed theory in the framework of the Green function
technique exists.
These methods have been widely elaborated for the case of the hydrogen plasma,
where the electrons and protons as the elementary constituents can form
hydrogen atoms as bound states of the attractive Coulomb interaction.
The problem is tractable analytically for the isolated two-particle system,
with a discrete energy spectrum of bound states and a continuous spectrum of
scattering states. More complex bound states, such as molecular hydrogen 
can also be formed.

In a many-particle system, the problem of bound state formation needs to
account for medium effects. They give contributions to a plasma Hamiltonian,
\begin{equation}
\label{plasmaH}
H^{\rm pl} = H^{\rm Hartree}+ H^{\rm Fock}+ H^{\rm Pauli}+ H^{\rm MW}
+ H^{\rm Debye}+ H^{\rm pp}+ H^{\rm vdW}+ \dots ,
\end{equation}
where the first three terms - the Hartree and Fock energies of one-particle
states and the Pauli blocking for the two-particle states -  are of
first order in the interaction and represent the mean-field approximation.
The following two terms of the plasma Hamiltonian are the Montroll-Ward (MW)
term accounting for the dynamical screening of the interaction in the 
self-energy, and the dynamical screening (Debye) of the interaction between 
the bound particles.  These contributions are related to the polarization 
function and are of particular interest for plasmas due to the long-range 
character of the Coulomb interaction.
In a consistent description, both terms should be treated simultaneously.
The last two contributions to the plasma Hamiltonian are of second order 
in the fugacity (particle density): the polarization potential (pp), 
describing the interaction of a bound state with free charge carriers, and 
the van der Waals (vdW) interaction, accounting for the impact of correlations 
(including bound states) in the medium on the two-particle system under 
consideration~\cite{Redmer:1997,Ebeling:1986}.

Approximations to medium effects in the self-energy and the effective
interaction kernel have to be made in a consistent way, resulting in
predictions for the modification of one- and two-particle states.
On this basis, the kinetics of bound-state formation and breakup processes
can be described, establishing the ionization equilibrium under given
thermodynamic conditions~\cite{Schlanges:1988}, see, e.g., Eq.~(\ref{rate-eq}).
Coulomb systems similar to the hydrogen plasma are electron-hole plasmas in
semiconductors~\cite{Zimmermann:1978}, where excitons and bi-excitons play the
role of the atoms and molecules.
Other systems which have been widely studied are expanded fluids
like alkali plasmas or noble gas plasmas~\cite{Redmer:1997}.
Applications of the plasma physics concepts for cluster formation and 
Mott effect to the rather short-ranged strong interactions have been 
given, e.g., in Refs.~\cite{Ropke:1982,Ropke:1983} for nuclear matter and
in Refs.~\cite{Horowitz:1985tx,Ropke:1986qs} for quark matter.

In this Section, we would like to discuss basic insights from the
investigations of bound-state formation in electromagnetic plasmas, as 
far as they might concern our discussion of heavy-quarkonia formation in 
hot and dense matter. Before going more into the details, let us give a 
brief overview. The bound-state energy remains rather inert to changes of 
the medium since the self-energy and interaction effects partially 
compensate each other to lowest order in density. The small size of the 
bound states plays an important role in this respect.
The compensation is not operative for continuum states which are 
influenced by self-energy effects only, usually resulting in a lowering 
of the in-medium ionization threshold. This often leads to a strong 
enhancement of the transition rates from bound to free states, inducing 
a sequential ``melting'' of different bound-state excitation levels into 
the continuum of scattering states for certain critical plasma parameters
(Mott effect~\cite{Mott:1968}), until eventually the ground state becomes 
unbound.

The theory of plasma correlations has also been developed for strongly
non-ideal systems, where the formation of clusters in the medium needs 
to be taken into account. This situation is reminiscent of a hadronizing 
quark-gluon plasma; we will therefore refer to cluster expansion techniques 
as the theoretical basis.

\begin{figure}[!ht]
\parbox{8.5cm}{
\includegraphics[width=8.5cm,angle=0]{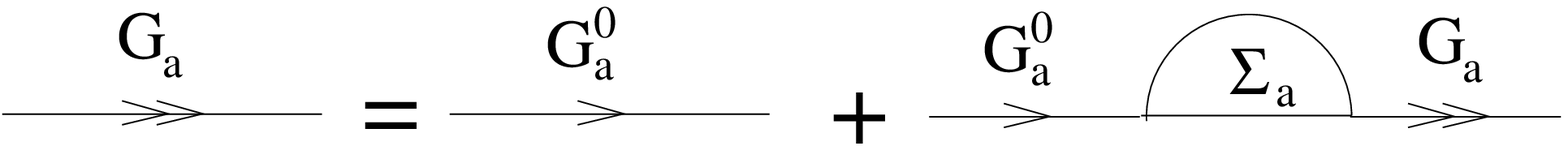}
}\hfill
\parbox{8.5cm}{
\includegraphics[width=8.5cm,angle=0]{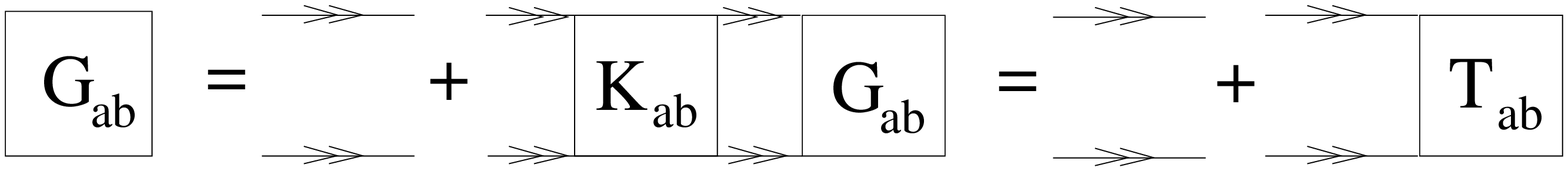}
}
\caption{The coupled one- and two-particle problem in the medium. Dyson equation
(left) and Bethe-Salpeter equation (right) need to be solved in consistent
(conserving) approximations for self-energy ($\Sigma$) and interaction kernel
($K$).
\label{fig:BSE-Dyson} }
\end{figure}

\paragraph{Bethe-Salpeter equation and plasma Hamiltonian.}
A systematic approach to the description of bound states in plasmas starts 
from the Bethe-Salpeter equation (BSE) for the thermodynamic (Matsubara-)
two-particle Green function of particles $a$ and $b$ (in analogy to 
Eq.~(\ref{GTG}) in Sec.~\ref{ssec_pot}),
\begin{equation}
\label{BSE}
G_{ab} = G_{ab}^0 + G_{ab}^0~K_{ab}~G_{ab} 
= G_{ab}^0 + G_{ab}^0~T_{ab}~G_{ab}^0 \ ,
\end{equation}
which is equivalent to the use of the two-particle $T$-matrix $T_{ab}$.
The 2-body equation has to be solved in conjunction with the Dyson equation 
for the full one-particle Green function,
\begin{equation}
\label{Dyson}
G_a = G_a^{0} + G_a^{0} \Sigma_a G_a~,
\end{equation}
defined by the dynamical self-energy $\Sigma_a(p,\omega)$ and the
free one-particle Green function
$G_a^{0}(p,\omega)=[\omega - \varepsilon_a(p)]^{-1}$
for a particle of species $a$ with the dispersion relation 
$\varepsilon_a(p)=\sqrt{p^2+m^2_a}\approx m_a+p^2/(2m_a)$,
see Fig.~\ref{fig:BSE-Dyson}.

The BSE contains the information about the spectrum of two-particle bound as
well as scattering states in the plasma.
A proper formulation of the plasma effects on the two-particle spectrum
is essential to understand why bound and scattering states are influenced in a
different way by the surrounding medium, leading to the Mott-effect for
bound states. 
We here give the essence of a detailed discussion as presented in
Ref.~\cite{Ebeling:1986}.

The homogeneous BSE associated with Eq.~(\ref{BSE}) can be cast into a form 
of an effective Schr\"odinger equation for the wave function 
$\psi_{ab}(p_1,p_2,z)$ of two-particle states in the 
medium~\cite{Zimmermann:1978},
\begin{eqnarray}
\label{wave-eq}
\sum_q \left\{\left[\varepsilon_a(p_1)+ \varepsilon_b(p_2) - z \right] 
\delta_{q,0}
-V_{ab}(q)\right\} \psi_{ab}(p_1+q,p_2-q,z)
=\sum_q H_{ab}^{\rm pl}(p_1,p_2,q,z) \psi_{ab}(p_1+q,p_2-q,z)  ,
\end{eqnarray}
where $a,b$ denote a pair of particles with 3-momenta $p_1$ and $p_2$ which
transfer a 3-momentum $2q$ in their free-space interaction $V_{ab}(q)$, and 
$z$ is a complex two-particle energy variable. The in-medium effects 
described by Eq.~(\ref{wave-eq}) have been singled out in the definition 
of a plasma Hamiltonian, containing all modifications beyond the 
two-body problem in free space~\cite{Ebeling:1986,Zimmermann:1978},
\begin{eqnarray}
\label{plasma-h}
H_{ab}^{\rm pl}(p_1,p_2,q,z)&=&
\underbrace{V_{ab}(q)\left[N_{ab}(p_1,p_2)-1\right]}_{\rm (i)~Pauli~blocking}~
-\underbrace{\sum_{q'} V_{ab}(q')\left[N_{ab}(p_1+q',p_2-q')-1\right]
}_{\rm (ii)~Exchange~self-energy}\delta_{q,0}
\nonumber\\
&+&\underbrace{\Delta V_{ab}(p_1,p_2,q,z)N_{ab}(p_1,p_2)}_{\rm (iii)~ 
Dynamically~screened~potential}
-\underbrace{\sum_{q'} 
\Delta V_{ab}(p_1,p_2,q',z)N_{ab}(p_1+q',p_2-q')
}_{\rm (iv)~Dynamical~self-energy}\delta_{q,0} \ .
\end{eqnarray}
Here, $\Delta V_{ab}(p_1,p_2,q,z)=K_{ab}(p_1,p_2,q,z)-V_{ab}(q)$
stands for the in-medium modification of the bare interaction potential 
to a dynamically screened interaction kernel $K_{ab}(p_1,p_2,q,z)$.
The effects of phase space occupation
are encoded in the function $N_{ab}(p_1, p_2)$, which for the case of an 
uncorrelated fermionic medium takes the form of the Pauli blocking factor
$N_{ab}(p_1, p_2)=1-f_a(p_1)-f_b(p_2)$, where 
$f_a(p)=\{\exp[(\varepsilon_a(p)-\mu_a)/T]+1\}^{-1}$ is the Fermi 
distribution and $\mu_a$ the chemical potential of species $a$; 
cf.~the factors  $\hat f^{Q\bar{Q}}$ in Eq.~(\ref{LS}).
Eq.~(\ref{wave-eq}) is a generalization of the two-particle Schr\"odinger 
equation, where the left-hand side describes the isolated two-particle problem 
while many-body effects due to the surrounding medium are given on 
the right-hand side.  
The in-medium effects denoted in the plasma Hamiltonian (\ref{plasmaH}) can be 
obtained from the one derived in the Bethe-Salpeter approach (\ref{plasma-h}) 
upon proper choice of the interaction kernel $K_{ab}$ so that 
Eq.~(\ref{plasmaH}) appears as a special case of Eq.~(\ref{plasma-h}).

The influence of the plasma Hamiltonian on the spectrum of bound and scattering
states can be qualitatively discussed in perturbation theory.
Since bound states are localized in coordinate space, their momentum-space wave
functions extend over a finite range $\Lambda$ and we may assume them to be 
$q$-independent, 
$\psi_{ab}(p_1+q,p_2-q,z=E_{nl})\approx \psi_{ab}(p_1,p_2,z=E_{nl})$, 
for small momentum transfer, $q < \Lambda$, and to vanish otherwise 
($E_{nl}$ energy of the bound state with quantum numbers $n$ and $l$). 
Assuming further a flat momentum dependence of the Pauli blocking factors  
$N_{ab}(p_1+q, p_2-q) \approx N_{ab}(p_1, p_2)$ for small $q$ where the 
interaction is strong, we obtain a cancellation of the Pauli blocking term (i) 
by the exchange self-energy (ii), and of the dynamically screened potential 
(iii) by the dynamical self-energy (iv). Therefore, the bound-state energy 
remains largely unmodified by medium effects. 
For scattering states which are extended in coordinate space and can be 
represented by a $\delta$-function in momentum space, the above 
cancellations do 
not apply and a shift of the two-particle continuum threshold results.
For this mechanism to work it is important that approximation schemes for the 
self-energy and the interaction kernel are consistent as, e.g., in the 
conserving scheme of $\Phi$-derivable theories~\cite{Baym:1962sx}.

Summarizing the discussion of the plasma Hamiltonian, we state that bound-state 
energies remain unshifted to lowest order in the charge carrier density while the 
threshold for the continuum of scattering states is lowered. This formally 
implies a reduction of the binding energy if the latter is identified with the
difference between the 2-particle threshold and the bound-state energy, 
$\varepsilon_B = 2m_a + m_b - E_{nl}$.
The intersection points of bound state energies and continuum threshold define 
the Mott densities (and temperatures) for bound-state dissociation.

When applying this approach to heavy quarkonia in a medium where heavy 
quarks (either free or bound in heavy hadrons) are rare, then $N_{ab}=1$ 
so that both, (i) and (ii), can be safely neglected.
The effects (iii) and (iv) originate from the dynamical coupling of the
two-particle state to collective excitations (plasmons) in the medium.
In the screened-potential approximation, the interaction kernel
is represented by
\begin{eqnarray}
V^S_{ab}(p_1p_2,q,\omega) &=&
V^S_{ab}(q,\omega)\delta_{P,p_1+p_2}\delta_{2q,p_1-p_2} 
\nonumber\\
V^S_{ab}(q,\omega) &=& V_{ab}(q) + V_{ab}(q)\Pi_{ab}(q,\omega)V^S_{ab}(q,\omega)
=V_{ab}(q)[1-\Pi_{ab}(q,\omega)V_{ab}(q)]^{-1} \ ,
\label{V_S}
\end{eqnarray}
with the total momentum $P$,  momentum transfer $2q$ and energy transfer 
$\omega$ in the two-particle system.  The most frequently used approximation 
for the polarization function, $\Pi_{ab}(q,\omega)$, or for the equivalent 
dielectric function, $\varepsilon_{ab}(q,\omega)=1-\Pi_{ab}(q,\omega)V_{ab}(q)$, 
is the random phase approximation (RPA).
In the next two paragraphs we discuss the static, long wavelength limit of
the RPA and its generalization in a clustered medium.

\paragraph{Example 1: Statically screened Coulomb potential.}
The systematic account of the modification of the interaction potential
between charged particles $a$ and $b$ through polarization of the medium is taken
into account in the dynamical polarization function, $\Pi_{ab}(q,\omega)$, which in
RPA reads~\cite{Ebeling:1986}
\begin{equation}
\label{Pi-RPA}
\Pi_{ab}^{\rm RPA}(q,\omega)=2 \delta_{ab} \int \frac{d^3p}{(2\pi)^3}
\frac{f_a(E_{p}^a)-f_a(E_{p-q}^a)}{E_{p}^a-E_{p-q}^a-\omega} \ .
\end{equation}
For the Coulomb interaction, corresponding to the exchange of a massless
vector boson, the potential is obtained from the longitudinal propagator
in Coulomb gauge, $V_{ab}(q)=e_a e_b/q^2$. 
For a recent discussion in the context of heavy-quarkonium correlators and 
potentials see, e.g., Refs.~\cite{Laine:2006ns,Beraudo:2007ky,Brambilla:2008cx}.
Due to the large masses of the constituents in the heavy-quarkonium system, 
one may use a Born-Oppenheimer expansion to replace the dynamically 
screened interaction by its static ($\omega=0$) and long-wavelength
($q\to 0$) limit.
For nondegenerate systems the distribution functions are Boltzmann 
distributions and their difference can be expanded as
\begin{equation}
f_a(E_{p}^a)-f_a(E_{p-q}^a)=
{\rm e}^{-E_p^a/T}\left(1- {\rm e}^{-(E_{p-q}^a-E_p^a)/T}\right)
\approx - f_a(E_{p}^a) (E_{p}^a-E_{p-q}^a)/T \ ,
\end{equation}
so that the energy denominator gets compensated and the polarization function
becomes 
\begin{equation}
\Pi_{ab}^{\rm RPA}(q,0)=-2 \frac{\delta_{ab}}{T} 
\int \frac{d^3p}{(2\pi)^3}f_a(E_{p}^a)=-\delta_{ab} \frac{n_a(T)}{T} \ .
\end{equation}
The corresponding dielectric function $\varepsilon^{\rm RPA}_{ab}(q,\omega)$
takes the form
\begin{equation}
\lim_{q\to 0}\varepsilon^{\rm RPA}(q,0)=1+\frac{\mu_D^2}{q^2} \ ,
~~~\mu_D^2 = \frac{1}{T}\sum_a e_a^2 n_a(T) \ .
\end{equation}
The screened Coulomb potential in this approximation is therefore 
$V_{ab}^S(q)=V_{ab}(q)/\varepsilon^{\rm RPA}(q,0)=e_a e_b/(q^2+\mu_D^2)$.
In this ``classical'' example  of the statically screened Coulomb 
interaction, the contribution to the plasma Hamiltonian is real and
in coordinate representation given by
\begin{equation}
\label{V-eff}
\Delta V_{ab}(r) = -\frac{\alpha}{r}({\rm e}^{-\mu_D r}-1)
\approx \alpha \mu_D - \frac{\alpha}{2}\mu_D^2 r \ ,
\end{equation}
where $\alpha=e^2/(4\pi)$ is the fine structure constant.
For the change in the Hartree self-energy of one-particle states
due to Debye screening we can perform an estimate in momentum space,
\begin{equation}
\label{Delta-SE}
\Sigma_{a} = \frac{4 \pi \alpha}{(2s_a+1)} \int\frac{d^3q}{(2\pi)^3}
\left[ \frac{1}{q^2+\mu_D^2} - \frac{1}{q^2}\right]f_a(E_q^a)
\approx -\frac{\alpha\mu_D^2}{\pi}\int_0^\infty \frac{dq}{q^2+\mu_D^2}
=-\frac{\alpha\mu_D}{2} \ ,
\end{equation}
which compensates to lowest order in the density the shift of the 
bound-state
energy levels due to the screening of the interaction (\ref{V-eff}),
\begin{equation}
\label{Delta-cont}
\Delta\epsilon_{B,ab} \approx \alpha \mu_D
={\mathcal O}(\sqrt{n a_{\rm B,0}^3}) \ ,
\end{equation}
($a_{\rm B,0}$: Bohr radius in vacuum)
so that~\cite{Ebeling:1986,Ebeling:1989}
\begin{equation}
\label{Delta-En}
\Delta E_{nl} \approx -\frac{\alpha}{2}\mu_D^2 \langle r \rangle_{nl}
= {\mathcal O}({n a_{\rm B,0}^3}) \ .
\end{equation}
The Debye mass, $\mu_D$, is equivalent to the inverse of the Debye radius
$r_{\rm D}$ characterizing the effective range of the interaction, and
depends on the square root of the density $n(T)$ of charge carriers.
It is this different response of bound states and scattering continuum to
an increase of density and temperature in the medium which leads to the
Mott effect for electrons in an insulator~\cite{Mott:1968,Redmer:1997}:
bound states of the Debye potential can only exist when the Debye radius is
larger than
$r_{\rm D,Mott}=0.84~a_{\rm B,0}$~\cite{Rogers:1970}.
This entails that above a certain density even the ground-state electrons
become unbound and form a conduction band, resulting in an insulator-metal
transition (also called Mott-transition).
Further details concerning this example can be found in 
Ref.~\cite{Kraeft:1990}.

In complete analogy to the electronic Mott effect it is expected that
in hadronic matter at high density the hadrons - as bound states of 
quarks - undergo a Mott transition which results in a phase transition 
from the color-insulating phase (hadronic matter) to a color-conducting 
or even color-superconducting phase (deconfined quark matter). This 
applies to light hadrons as well as to heavy quarkonia, whereby due to 
the different scales of Bohr radii the Mott dissociation of heavy
quarkonia is expected to occur at higher densities than for light hadrons.
In some approaches quark self-energy effects are neglected and 
one is only left with the medium effect due to a 
statically screened potential. Consequently, in such a picture the
continuum edge of the scattering states remains unshifted and, due to the lack
of a compensating effect, the effective interaction entails an appreciable 
medium dependence of the bound-state energies (masses). 
For the electron-hole plasma in highly excited semiconductors 
it could be shown experimentally, however, that the
compensation picture is correct and the bound state energies remain almost
unshifted~\cite{Fehrenbach:1981}. In the context of heavy quarkonia in
the QGP this could, at least qualitatively, provide a natural explanation
of the approximate constancy of the Euclidean correlator ratios found
in thermal lattice QCD. 

One may of course absorb the self-energy effects into a redefinition of the
effective interaction, by adding a homogeneous mean-field contribution. This 
is equivalent to the use of the Ecker-Weitzel potential~\cite{Ecker:1956},
\begin{equation}
\label{E-W}
V_{\rm Ecker-Weitzel}(r) = 
-\frac{\alpha}{r}{\rm e}^{-\mu_D r} - \alpha \mu_D \ .
\end{equation}
Recent investigations of the screening problem
in the context of Debye-H\"uckel theory~\cite{Dixit:1989vq}
and $Q\bar Q$ correlators~\cite{Brambilla:2008cx,Beraudo:2007ky}
have obtained this continuum shift ($-\alpha \mu_D$) as a homogeneous
background field contribution.
According to the above lesson from plasma physics, however, this contribution
should be attributed to the self-energy of the constituents rather than
to the interaction kernel, since it determines the shift of the continuum 
edge (cf.~Eq.~(\ref{mcstar})).

\paragraph{Example 2: Heavy quarkonia in a relativistic quark plasma.}

We consider the problem of a heavy quark-antiquark pair interacting via (an
ansatz for) the statically screened Cornell potential (\ref{Vqq-med}) in a 
relativistic quark plasma. The thermodynamics of this medium is described in 
good agreement with lattice QCD data by a Nambu-Jona-Lasinio model coupled 
to a Polyakov-loop potential (PNJL model~\cite{Ratti:2005jh} which allows us 
to estimate the Debye mass, $\mu_D(T)$, by evaluating the RPA polarization 
function, Eq. (\ref{Pi-RPA}), for $N_c$ colors and $N_f$ flavors of massless 
quarks with $E_p^a=|p|$ as
\begin{equation}
\Pi_{ab}^{\rm RPA}(q\to 0,0)
=2 \delta_{ab}\int \frac{d^3p}{(2\pi)^3}\frac{d f_a(E_p^a)}{d E_p^a}
=-2\delta_{ab} \int_0^\infty \frac{dp~p}{\pi^2}f_\Phi(p)
=-\frac{\delta_{ab}}{6 \pi^2} I(\Phi) T^2 \ . 
\end{equation}
Here, $I(\Phi)=(12/\pi^2)\int_0^\infty dx x f_\Phi(x)$ and 
$f_\Phi(x)=[\Phi(1+2e^{-x})e^{-x}+e^{-3x}]/[1+3\Phi(1+e^{-x})e^{-x}+e^{-3x}]$
is the generalized quark distribution function~\cite{Hansen:2006ee}
which in the case of a deconfined medium ($\Phi=1$) is just the Fermi 
function, yielding $I(1)=1$. 
In the deconfinement transition region, where $0<\Phi<1$,  quark 
excitations are strongly suppressed, e.g., $I(0)=1/9$. 
Taking as the bare potential a color singlet one-gluon exchange
$V(q)=-4\pi \alpha_s/q^2$, $\alpha_s=g^2/(4\pi)$, the Fourier transform of the
Debye potential, $V^S(r)=-\alpha \exp(-\mu_D(T)r)/r$, results as a statically 
screened potential with the Debye mass $\mu_D(T)=4\pi \alpha_s I(\Phi) T^2 $.
The Hartree self-energy for heavy quarks in a PNJL  quark plasma is then given
by Eq.~(\ref{Delta-SE}).

In the spirit of Eq.~(\ref{Delta-SE}), one can evaluate the Hartree 
shift due to the screened confinement part, 
$V_{\rm conf}^S(r)=(\sigma/\mu_D)(1-\exp(-\mu_D r))$, of the potential 
(\ref{Vqq-med}) and show that it vanishes, $\Sigma_{\rm conf}=0$.

\begin{figure}[!t]
\parbox{8.5cm}{
\includegraphics[width=8.5cm,angle=0]{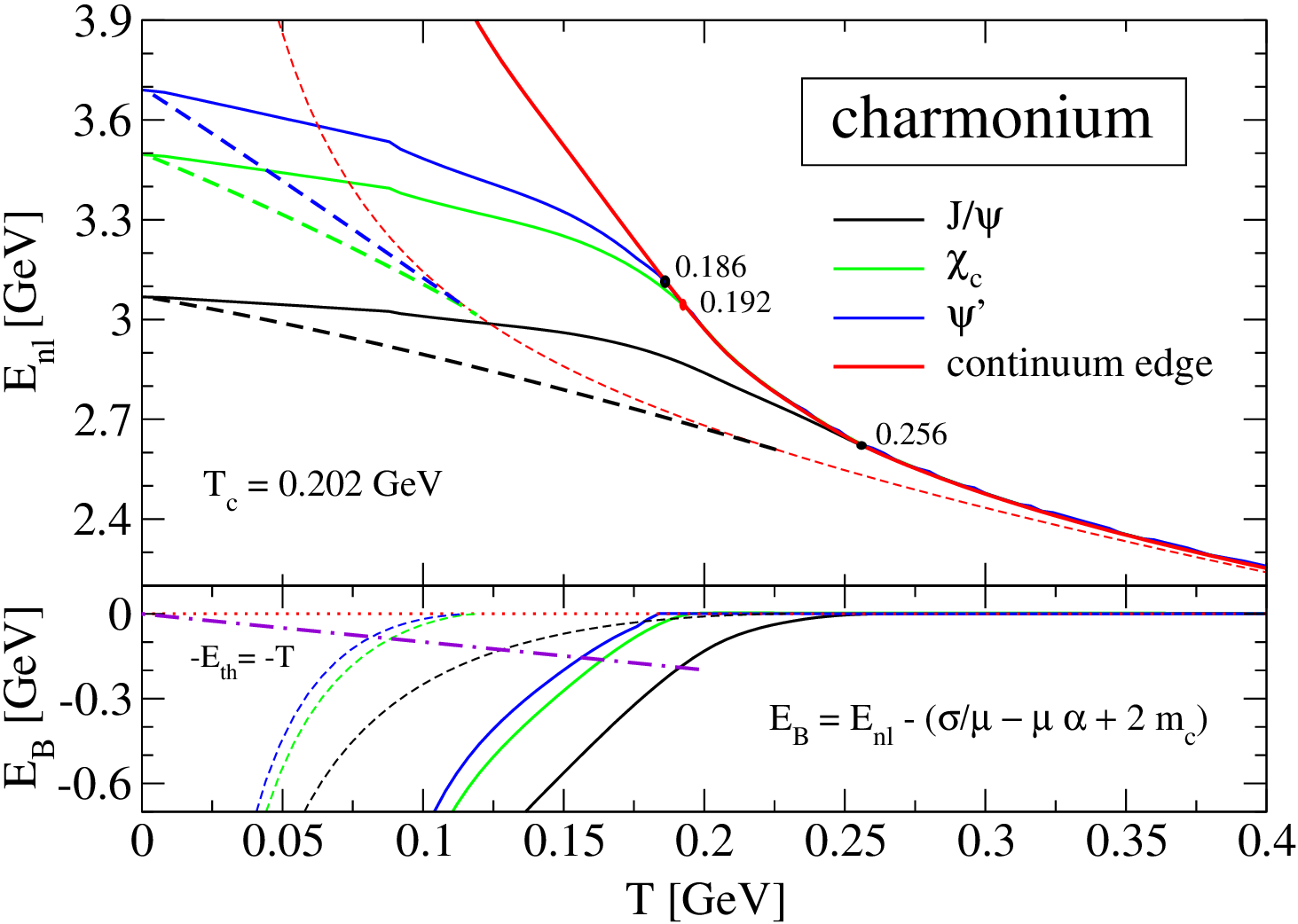}
}
\hfill
\parbox{8.5cm}{
\includegraphics[width=8.5cm,angle=0]{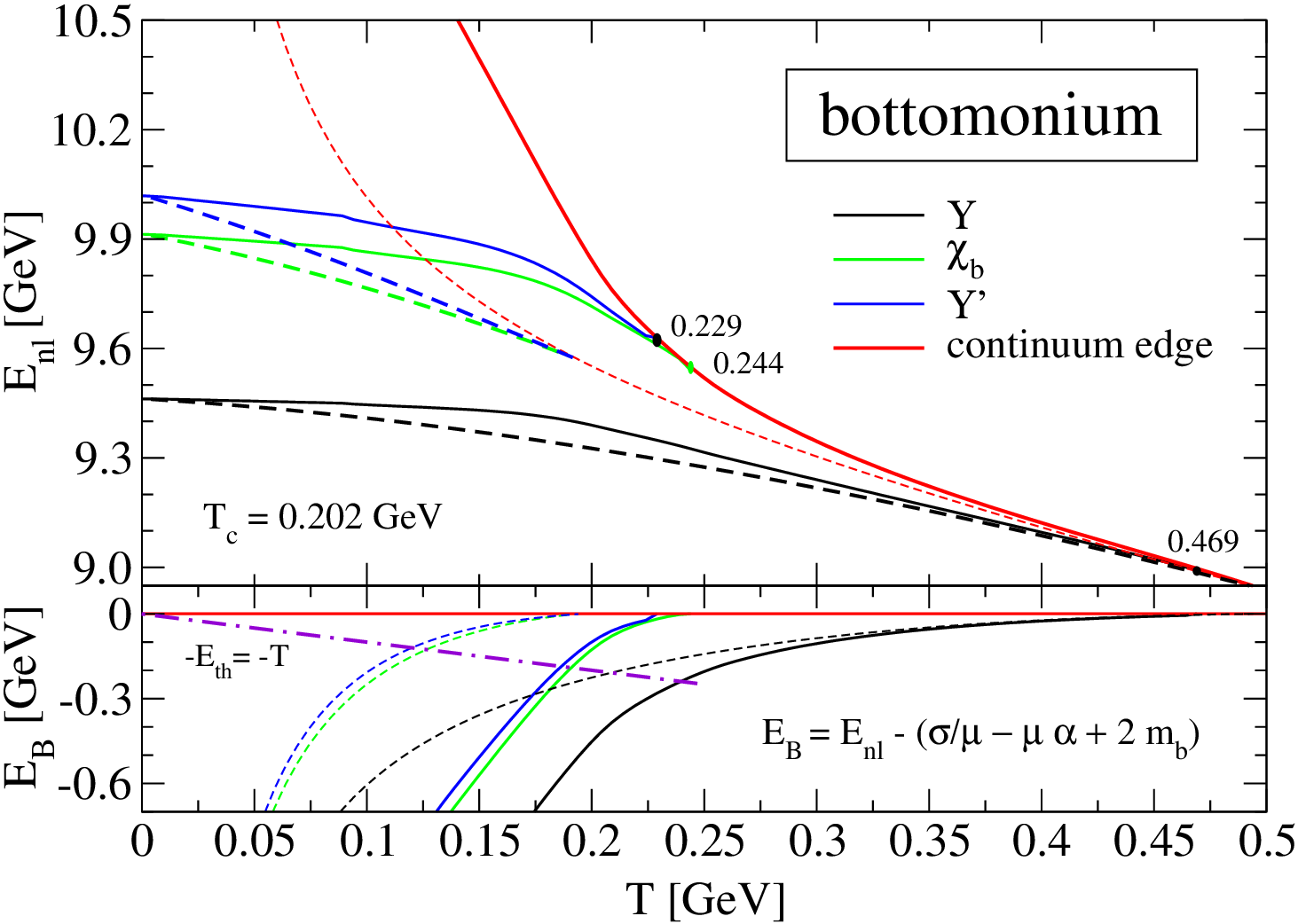}
}
\caption{Two-particle energies of charmonia (left panel) and bottomonia 
(right panel) in a statically screened potential, from 
Ref.~\cite{Jankowski:2009kr}. The full results of the PNJL model (solid
lines) are compared to results without the coupling to the Polyakov loop
potential (dashed lines). 
\label{fig:quarkonia-pnjl}
}
\end{figure}

In Fig.~\ref{fig:quarkonia-pnjl} we show the temperature dependence of the 
two-particle energies for heavy quarkonia as a solution of the effective 
Schr\"odinger equation 
\begin{equation}
\label{SEq}
H^{\rm pl}(r;T) \phi_{nl}(r;T)=E_{nl}(T) \phi_{nl}(r;T)
\end{equation}
for the plasma Hamiltonian
\begin{equation}
\label{H_QQ_pl}
H^{\rm pl}(r;T)=2m_Q-\alpha \mu_D(T)-\frac{\vec{\nabla}^2}{m_Q}
+V_{Q\bar{Q}}(r;T) \ ,
\end{equation}
as evaluated recently in Ref.~\cite{Jankowski:2009kr}.

We can draw three conclusions from this example: 
(1) Due to the suppression of light-quark excitations by the confining effects 
of the Polyakov loop the lowering of the continuum edge and the merging of 
two-particle bound states with it (Mott effect) occur at higher temperatures 
when compared to the simple NJL model case; 
(2) The inclusion of the Hartree selfenergy for heavy quarks in the plasma 
Hamiltonian, Eq.~(\ref{H_QQ_pl}), leads to a compensation and even 
overcompensation of the upwards shift of quarkonia masses obtained from a 
solution of the Schr\"odinger equation with the naive Hamiltonian, 
Eq.~ (\ref{H_QQ}); see, e.g., upper left panel of Fig.~\ref{fig_G-tmat} 
or upper right panel Fig.~\ref{fig_MP08} for the effect of neglecting 
the downwards shift of the continuum due to the heavy quark self-energy 
on the temperature dependence of the quarkonium spectrum. The compensation
of a decreasing $m_c^*$ and decreasing $\varepsilon_B$ indeed play an
important role in understanding the rather constant euclidean correlator
ratios from lattice QCD.
(3) From Fig.~\ref{fig:quarkonia-pnjl} one can also read off the in-medium 
lowering of the dissociation threshold $k_0^{\rm diss}=\varepsilon_B$, 
which is the energy difference between the considered bound-state level and 
the continuum edge. This quantity is indifferent to the inclusion of a rigid
self-energy shift.
The lowering of $k_0^{\rm diss}$ with increasing density and/or temperature
leads to a strong increase in the quarkonium breakup cross section, 
Eq.~(\ref{sigma-diss}), by thermal impact,
and to bound-state dissociation even before the binding energies vanish 
at the critical Mott densities/temperatures for the corresponding states.
To estimate this effect  we show in the lower panels of the 
graphs in Fig.~\ref{fig:quarkonia-pnjl} the thermal energy $E_{\rm th}=T$ 
together with the in-medium binding energy $E_B\equiv -\varepsilon_B$.
The temperatures for the onset of bound-state dissociation by
thermal collisions with medium particles are obtained from the crossing points 
of these curves which are considerably lower than the naive Mott temperatures 
corresponding to vanishing binding energies. 
\\

\begin{figure}[!t]
\begin{center}
\parbox{14cm}{
\includegraphics[width=14cm,angle=0]{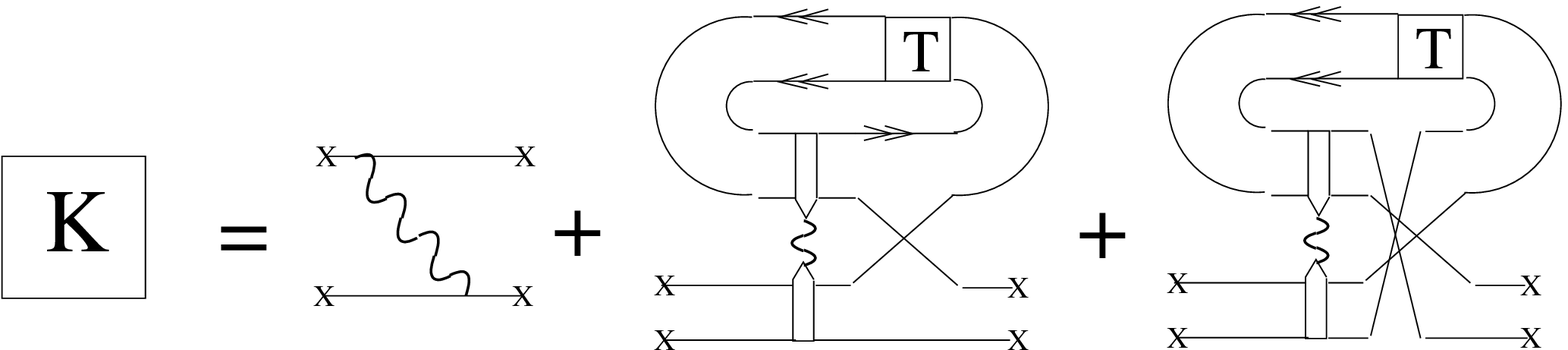}
\\[5mm]}
\parbox{10cm}{
\includegraphics[width=10cm,angle=0]{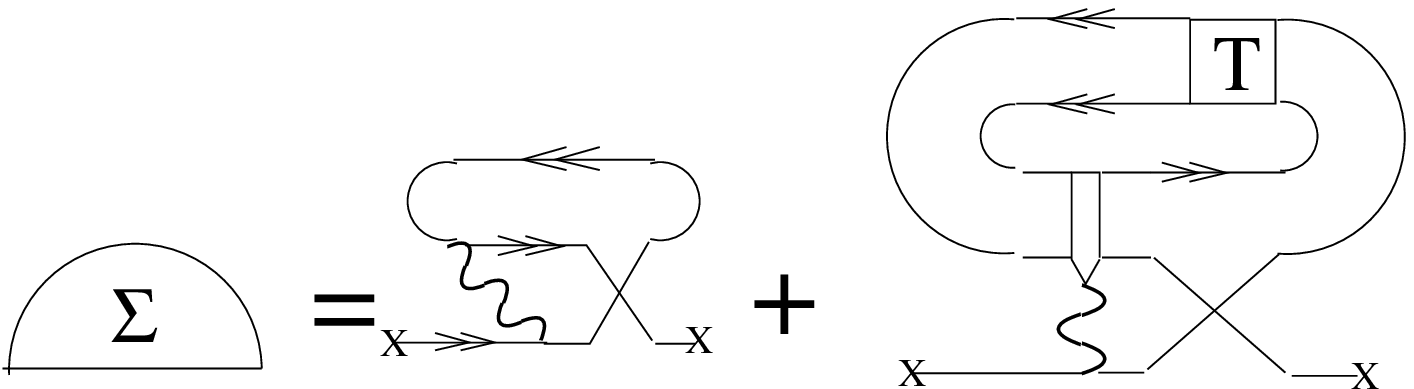}
}\hfill
\parbox{5cm}{
\includegraphics[width=5cm,angle=0]{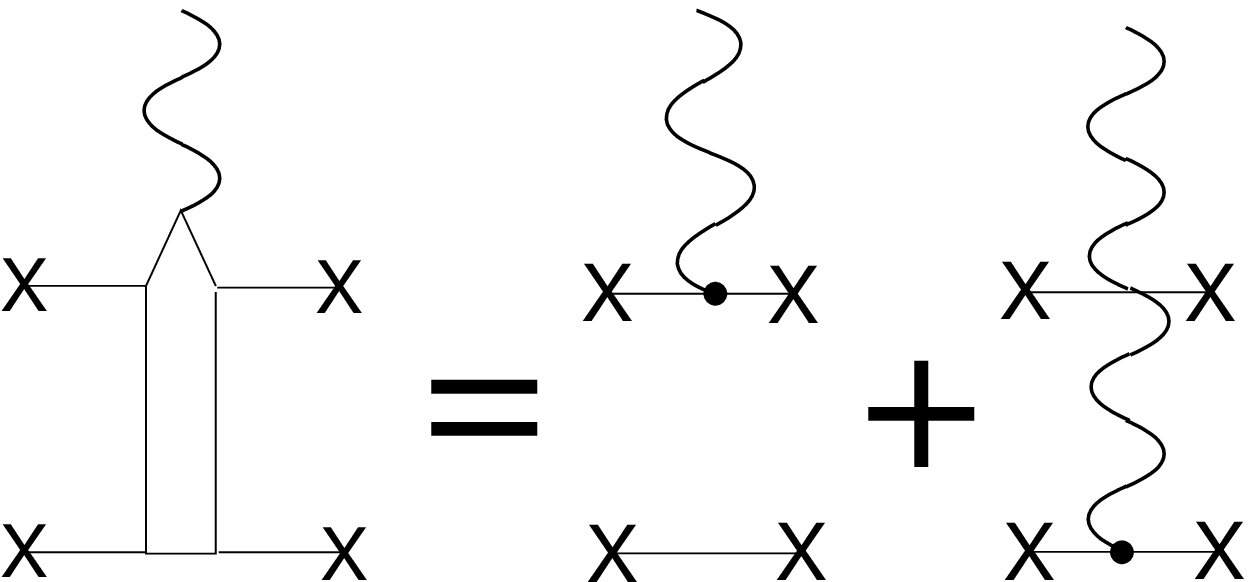}
}
\end{center}
\caption{Cluster expansion for the interaction kernel for the 
two-particle problem 
in a strongly correlated medium (upper equation) and the corresponding
self-energy (lower left equation) with a dipole ansatz for the vertex 
(lower right equation).
\label{fig:cluster-ex} }
\end{figure}
\begin{figure}[!t]
\begin{center}
\parbox{14cm}{
\includegraphics[width=14cm,angle=0]{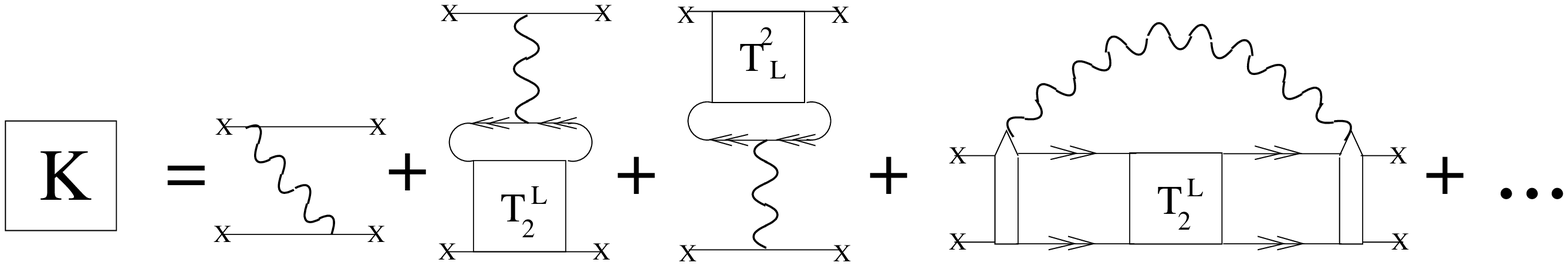}
\\[5mm]}
\parbox{14cm}{
\includegraphics[width=12cm,angle=0]{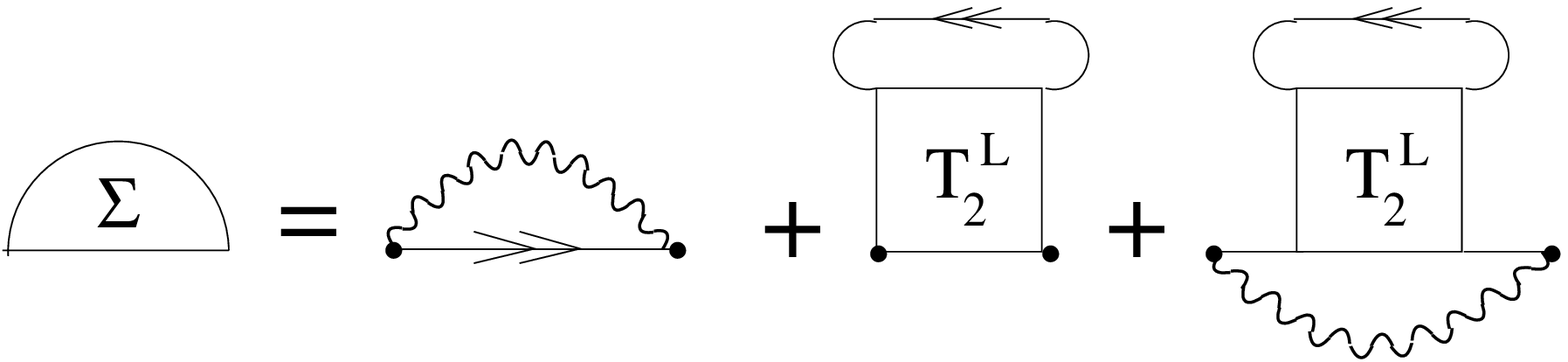}
}
\end{center}
\caption{Alternative way of drawing the diagrams for the cluster expansion 
of the interaction kernel and the corresponding self-energy of 
Fig.~\ref{fig:cluster-ex} in a form familiar in plasma and nuclear physics; 
$T_2^L$ denotes the 2-body $T$-matrix computed in ladder approximation.} 
\label{fig:cluster-ex2} 
\end{figure}
\begin{figure}[!t]
\vspace{0.5cm}
\parbox{6cm}{
\includegraphics[width=6cm,angle=0]{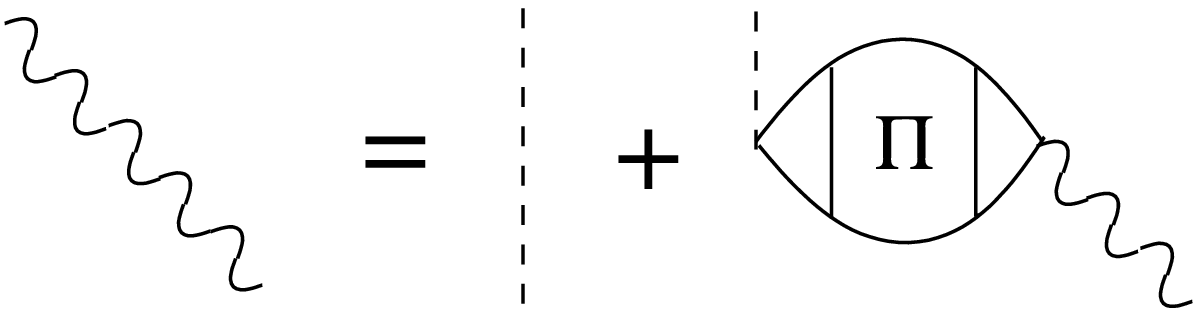}
}\hfill
\parbox{9cm}{
\includegraphics[width=9cm,angle=0]{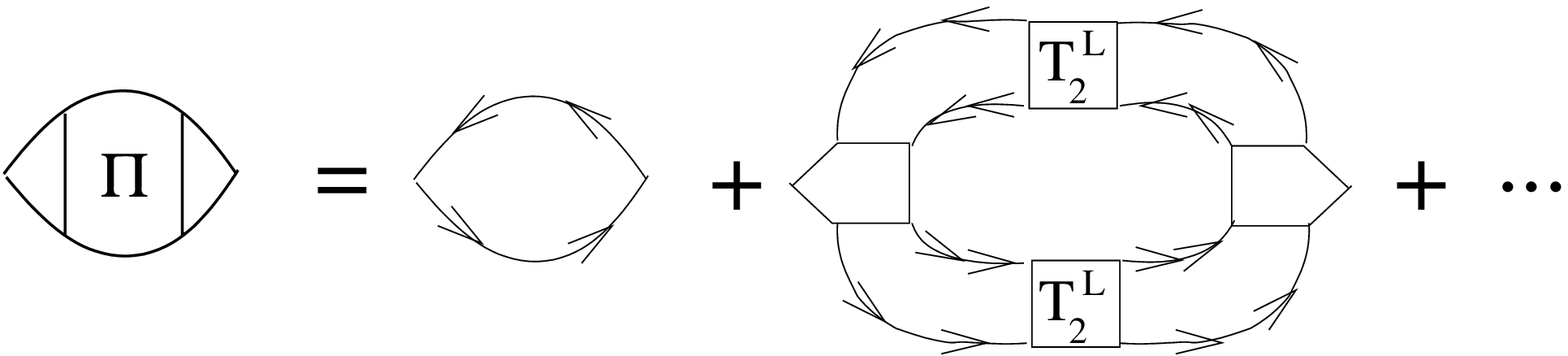}
}
\caption{Left panel: The dynamically screened interaction potential, 
$V_{ab}^S(\omega, q)$ (wavy line), determined by the bare potential 
(dashed line) and the polarization function, $\Pi_{ab}(\omega, q)$.
Right panel: Cluster expansion for the generalized RPA, where in addition
to free particles (RPA) also two-particle states (cluster-RPA) contribute 
to the polarizability of the medium~\cite{Ropke:1979}.}
\label{fig:cluster-screen}
\end{figure}

For the development of a comprehensive approach to heavy quarkonia in
hadronizing hot/dense QCD matter another insight from plasma physics
may be of relevance and will be discussed next: the effect of strong 
correlations (bound states) in the bulk of the medium.
To this end, the bound states will be treated like a new species occurring in 
the system. Accordingly, additional diagrams have to be taken into account
which arise from a cluster expansion of the interaction kernel $K_{ab}$ and 
the corresponding self-energy $\Sigma_a$, see 
Figs.~\ref{fig:cluster-ex}-\ref{fig:cluster-screen}. 
In the plasma Hamiltonian, $H^{\rm pl}$, this leads 
to a generalization of the self-energy contributions 
(cluster Hartree-Fock approximation), the distribution functions 
in the Pauli-blocking factors and the dynamical screening (cluster-RPA). 
The van-der-Waals interaction in Eq.~(\ref{plasmaH}) appears naturally as a 
contribution to the cluster expansion, describing polarization effects due to 
bound states in the medium.

\paragraph{Two-particle states in the medium: Cluster expansion.}
In the vicinity of the plasma phase transition, correlations play an important
role, and their proper accounting requires rather sophisticated theoretical
methods such as cluster expansion techniques.
For the problem of charmonium in dense hadronic matter at the deconfinement
transition, i.e., in the strong-coupling case, we suggest a systematic Born
series expansion of collisions with free and bound states in the surrounding
matter so that all terms linear in the density of free particles and bound
states are taken into account.

We here describe the cluster expansion in terms of its diagrammatic
expressions for the interaction kernel and the corresponding self-energy.
The $1^{st}$ Born approximation diagrams of this expansion are given
in Fig.~\ref{fig:cluster-ex}, see also the monograph~\cite{Ebeling:1986}.
The wavy lines denote the dynamically screened interaction $V_{ab}^S$,
which in a strongly correlated plasma receives contributions from the 
polarization of the medium beyond the RPA, denoted as generalized (cluster-)
RPA in Fig.~\ref{fig:cluster-screen}~\cite{Ropke:1979}.
Bound and scattering states are described consistently in the two-particle 
$T$-matrices. For a generalization to higher $n$-particle correlations, see
Refs.~\cite{Ropke:1983,Ropke:1984,Ebeling:1986}.
The diagrams containing $T$-matrices do not contribute to the charmonium 
spectrum as long as the densities of the charmed quarks and of charmed hadrons 
in the medium are negligible. 
This is the situation expected for FAIR, SPS and RHIC energies, but 
for the expected copious charm production at LHC these terms may become
significant.

\begin{figure}[!t]
\vspace{0.5cm}
\parbox{7cm}{
\includegraphics[width=7cm,height=4cm,angle=0]{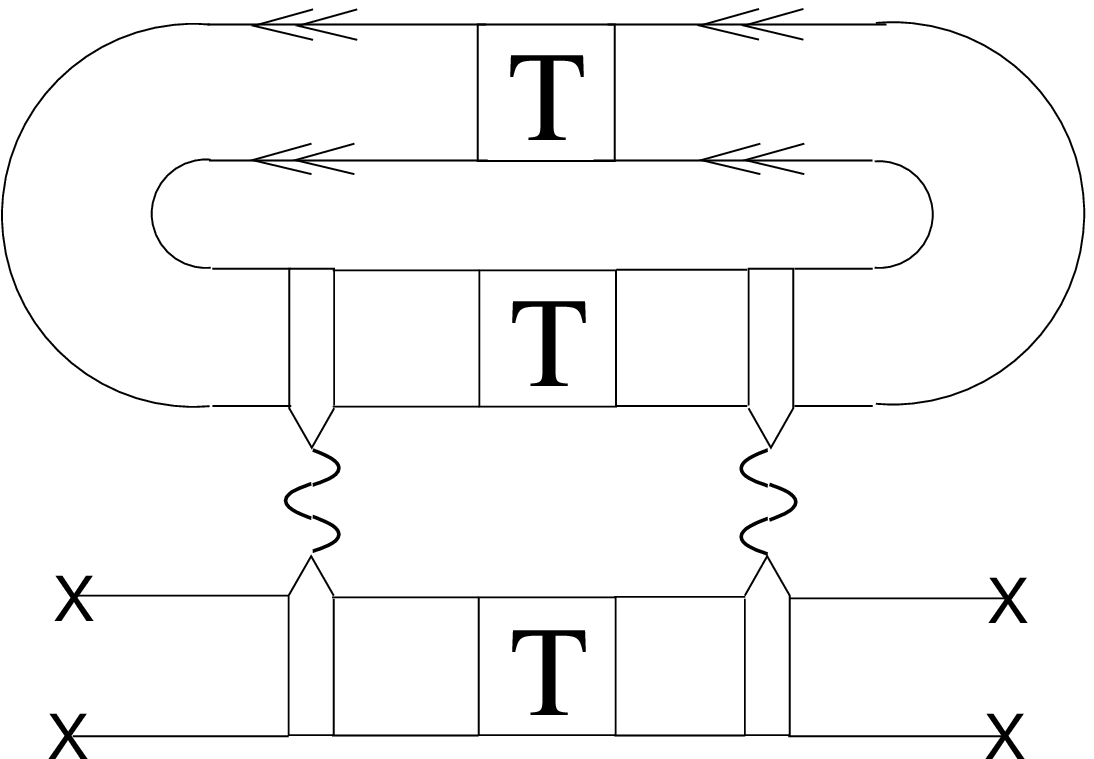}
}\hfill
\parbox{7cm}{
\includegraphics[width=7cm,height=4cm,angle=0]{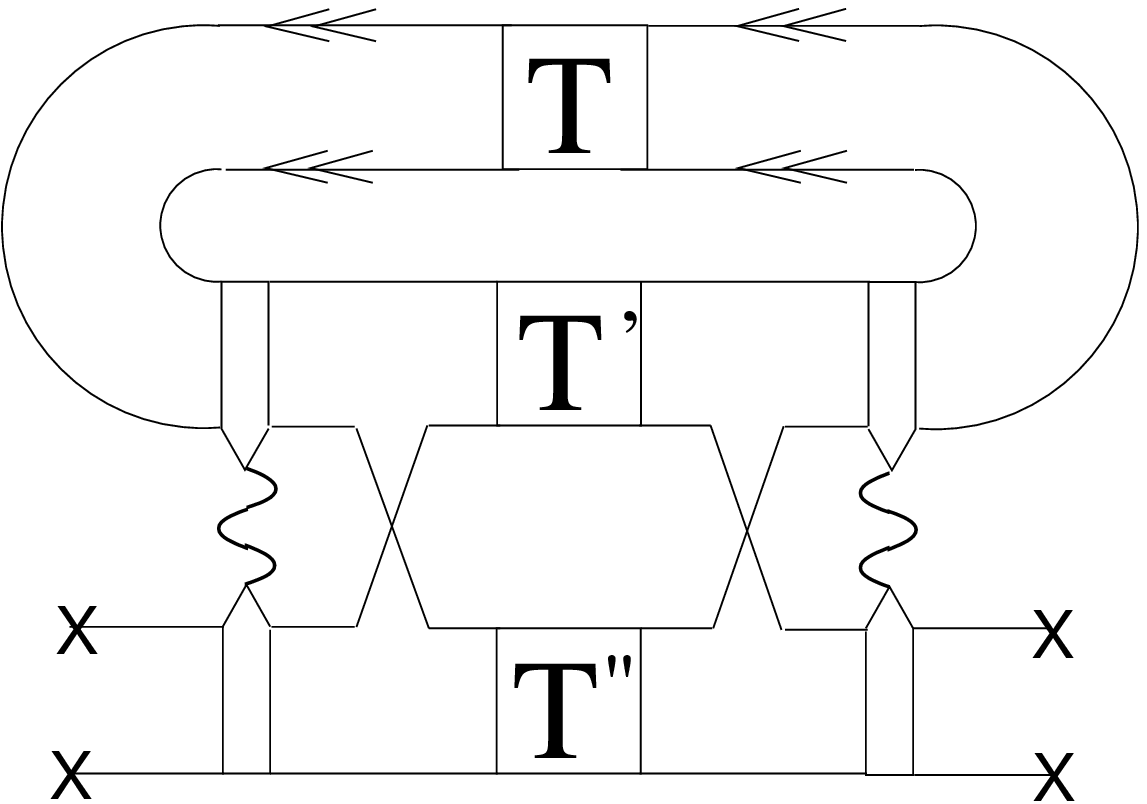}
}
\caption{Contributions to the dynamical self-energy of a two-particle
system in a correlated medium at  $2^{nd}$ Born order.
Left panel: impact by two-particle states without constituent exchange
(van-der-Waals or dipole-dipole interaction).
Right panel: constituent-rearrangement collisions (string-flip process),
from Ref.~\cite{Blaschke:2006ct}, see also Ref.~\cite{Satz:2006uh}.
\label{fig:cluster-cluster}
}
\end{figure}

At the $2^{nd}$ Born order, we distinguish two classes of collisions
with light clusters (hadrons) that can give rise to spectral broadening 
of the charmonia.
The first class concerns hadron impact without quark rearrangement inducing 
transitions to excited states, shown in the left panel of  
Fig.~\ref{fig:cluster-cluster}.
These processes have been considered for charmonium-hadron interactions 
within the operator product expansion techniques following Peskin and 
Bhanot~\cite{Bhanot:1979vb,Peskin:1979va}, 
see Refs.~\cite{Kharzeev:1994pz,Arleo:2001mp}.
The result is a deformation of the charmonium spectrum under conservation 
of the spectral weight integrated over all charmonia states.
In the second class are quark rearrangement (string-flip) processes,
as indicated in the right panel of Fig. \ref{fig:cluster-cluster}.
They induce transitions to open-charm hadrons which are also responsible 
for charmonium dissociation in hadronic matter, cf. Sec.~\ref{ssec_diss-hg}.
%

\paragraph{Example 3: String-flip model of charmonium dissociation.}
As a third example for utilizing insights from plasma physics we 
discuss charmonium dissociation within the string-flip model of quark 
matter~\cite{Horowitz:1985tx,Ropke:1986qs,Miyazawa:1979vx,Blaschke:1984yj}.
In this model string-type color interactions between quarks are saturated
within the sphere of nearest neighbors so that in a dense system of 
overlapping quark-antiquark pairs frequent string-flip processes occur
to drive the system into its minimal energy configuration at any time. 
The microscopic evaluation of those processes requires diagrams as the one
shown in the right panel of Fig.~\ref{fig:cluster-cluster} involving at least 
four quarks. Evaluating the imaginary part of this diagram, in 
connection with the optical theorem ($\sigma_{\rm diss} \propto {\rm Im}~T$), gives
access to the heavy-quarkonium dissociation rate in hadronic matter,
cf.~Eq.~(\ref{rate-hg}) in the previous section \ref{ssec_diss-hg}.
When  considering a heavy quark-antiquark pair in dense matter with 
negligible heavy-flavor fraction, the Pauli blocking and exchange 
self-energy contributions are negligible, but the strong correlations 
with light quarks of complementary color within the nearest neighbor sphere 
will result in a mean-field self-energy shift (Hartree shift $\Delta^H$) for 
all quarks~\cite{Ropke:1988bx} which determines the shift of the continuum 
edge. 
Because of the compensation in the Bethe-Salpeter
kernel between the effects of screening of the interaction and self-energy 
shifts calculated with it (see discussion above), it is suggested that 
to lowest order the bound-state energies remain unshifted when increasing the 
temperature and/or density of the medium.
In contrast to the first example of Debye screening of long-range Coulombic 
interactions, the screening mechanism in the string flip model is color 
saturation within nearest neighbors, applicable for strong, short-range 
interactions as appropriate for the case of the sQGP at RHIC or dense systems
at CBM-FAIR.
The resulting two-particle energy spectrum for charmonium and bottomonium 
states is discussed in Ref.~\cite{Ropke:1988zz}, where the 
static screening picture is compared to the string-flip picture.
Within the latter the  in-medium lowering of the dissociation threshold 
$k_0^{\rm diss}$ is directly given by the behavior of the continuum edge, 
i.e. by the Hartree shift, as known for rate coefficients in strongly coupled 
plasmas \cite{Schlanges:1988}.

%% file: nuc-abs.tex
\begin{figure}[!tb]
\begin{center}
\begin{tabular}{ccc}
\includegraphics[width=0.3\textwidth,height=0.3\textwidth]{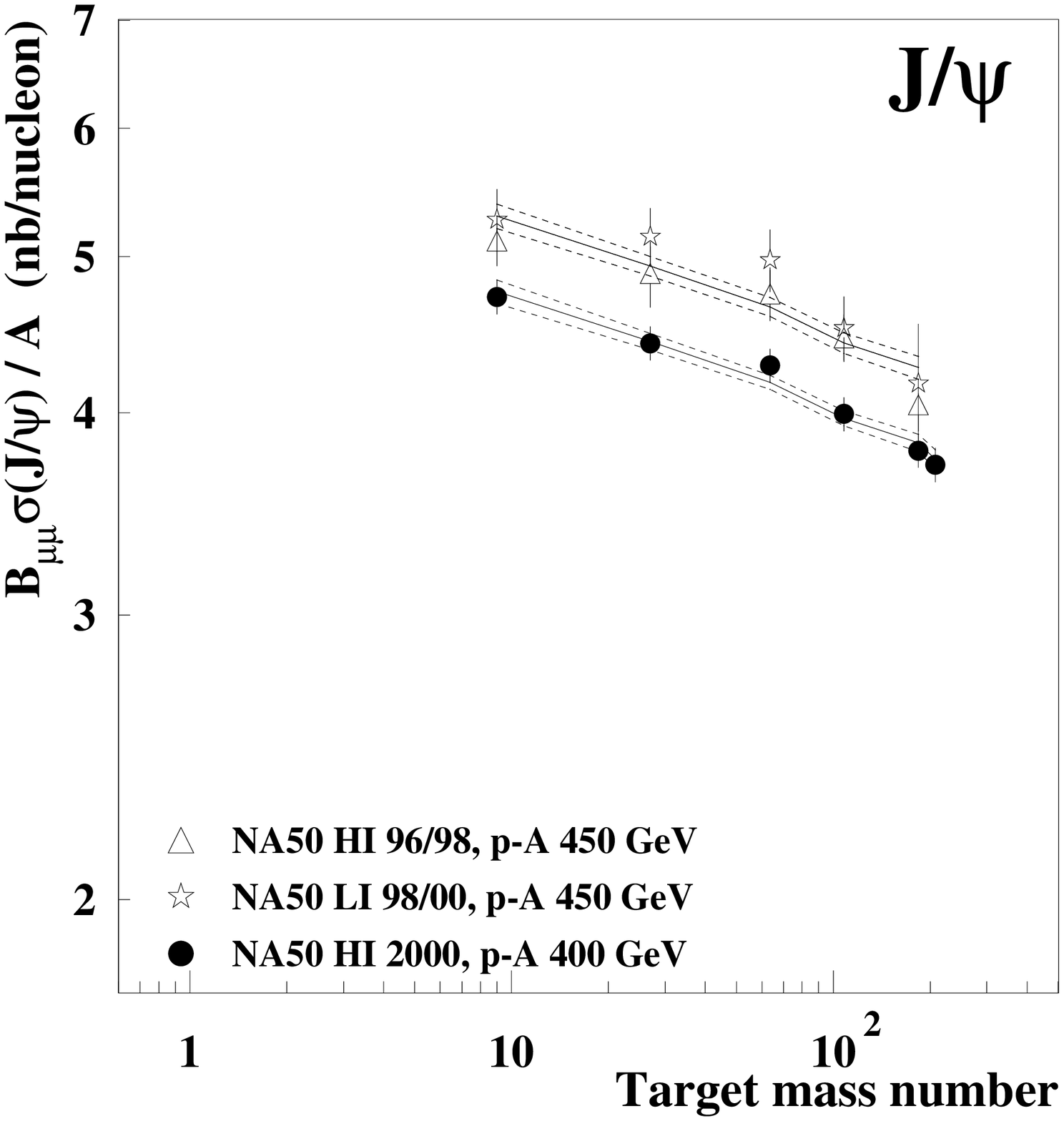}&
\includegraphics[width=0.33\textwidth,height=0.32\textwidth]{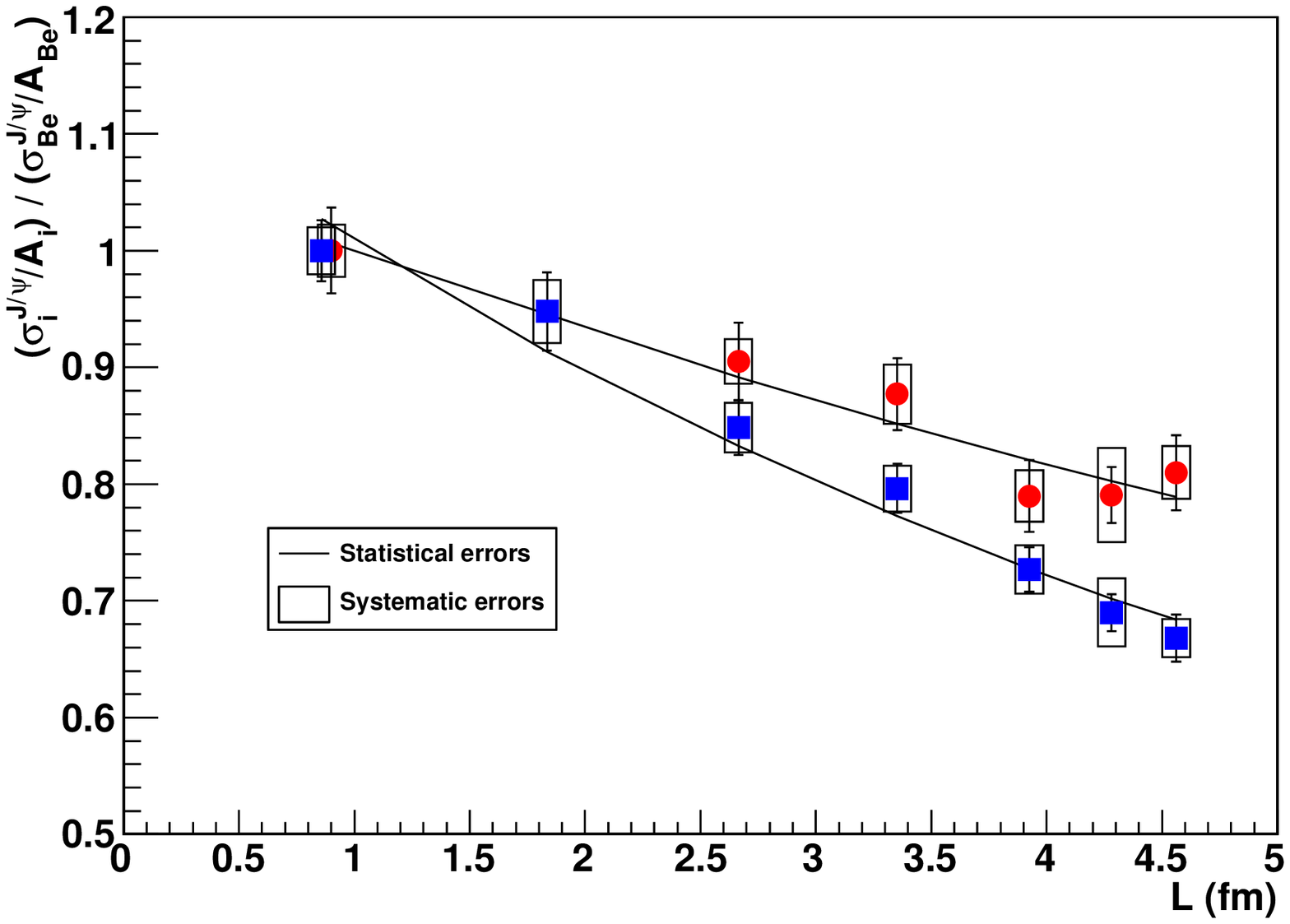}&
\includegraphics[width=0.3\textwidth,height=0.3\textwidth]{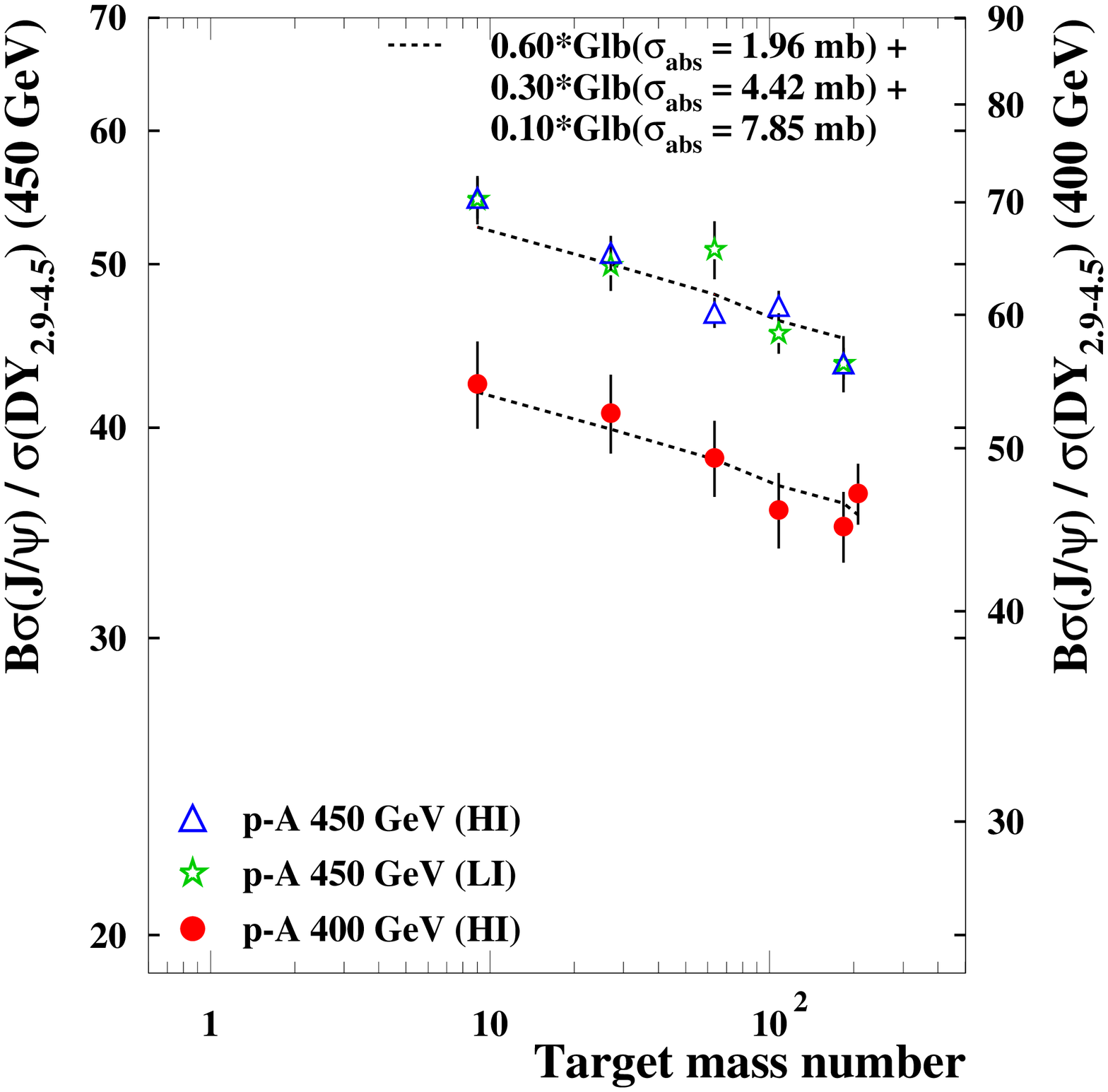}
\end{tabular}
\end{center}
\caption{Left panel: $J/\psi$ cross section in $p$-$A$ collisions from three 
different NA50 data sets \cite{Alessandro:2003pc} at 400\,GeV and 450\,GeV
together with a Glauber fit of an effective nuclear-absorption cross section, 
$\sigma_{\rm abs}=4.2 \pm 0.5$\,mb.
Middle panel: Cross sections for $J/\psi$ production in $p$-$A$ collisions 
normalized to $p$-$Be$ for 158\,GeV (squares) and 400\,GeV (circles) from the 
NA60 experiment~\cite{Arnaldi:2010ky}. The lines are Glauber model fits with
$\sigma_{\rm abs}(158~{\rm GeV})=7.6\pm0.7({\rm stat})\pm0.6({\rm syst})$\,mb 
and 
$\sigma_{\rm abs}(400~{\rm GeV})=4.3\pm0.8({\rm stat})\pm0.6({\rm syst})$\,mb.
Right panel: $A$-dependence of the  $J/\psi$ over Drell-Yan (DY) cross section 
ratio, compared to a Glauber model calculation taking into account feed-down 
from higher $c\bar c$ states~\cite{Lourenco:2006sr}.}
\label{fig_sig-abs-Glauber}
\end{figure}

A commonly employed strategy to isolate cold-nuclear-matter (CNM) 
effects on the production of heavy quarkonia in heavy-ion collisions, 
encoded in the suppression factor $S_{\rm nuc}$ in Eq.~(\ref{supp-fac}), 
is to study its $A$-dependence in $p$-$A$ collisions relative to 
$p$-$p$ collisions. The pertinent effects on quarkonium production
may be classified into three different stages: 
(1) {\em initial state}, prior to $Q\bar Q$ production by parton fusion, 
due to the nuclear modification of parton distribution functions (PDFs)
and/or parton energy loss; 
(2) {\em formation stage} of the quarkonium states from the 
initial  $Q\bar Q$ pair; and 
(3) {\em final-state} inelastic interactions leading to 
quarkonia absorption in the target matter.
It is at present an open problem how to disentangle these three CNM effects 
from the experimental results for the $A$-dependence measured in $p$-$A$ 
collisions. 
To obtain a simplified baseline for predictions and applications in 
heavy-ion collisions, it is assumed that absorption is the dominant 
process in all three stages and may be analyzed using Glauber theory in 
terms of a survival probability for a $J/\psi$ produced in a $p$-$A$ 
collision~\cite{Gerschel:1993uh,Kharzeev:1996yx},
\begin{eqnarray}
S_{\rm nuc}^{pA}&=&\frac{\sigma_{pA\to \psi}}{A\sigma_{pp\to \psi}}
= \int d^2 b~\int_{-\infty}^\infty 
dz~n_N(b,z) \exp\left\{ -(A-1)\int_{z}^\infty dz'~n_N(b,z')
\sigma_{\rm abs}(z'-z)\right\} \ ,
\label{sigma-pA}
\end{eqnarray}
where $b$ is the impact parameter and $n_N(b,z)$ is the density profile
of the target nucleus. 
As a result a ``convoluted" $J/\psi$ absorption cross section, 
$\sigma_{\rm abs}$, is extracted from experimental data for the $A$-dependence 
of the survival probability, Eq.~(\ref{sigma-pA}), parameterizing the 
combined result of all CNM effects. 
A Glauber model fit to three different NA50 data sets~\cite{Alessandro:2003pc}
gives an effective nuclear absorption cross section of 
$\sigma_{\rm abs}=4.2 \pm 0.5$\,mb (see left panel of 
Fig.~\ref{fig_sig-abs-Glauber}) for both 400\,GeV and 450\,GeV incident proton 
energies. Recently, the NA60 collaboration has published results for the 
$A$-dependence of $J/\psi$ production in $p$-$A$ collisions at 158\,GeV and 
400\,GeV~\cite{Arnaldi:2010ky} which show a larger effect at lower incident 
energy:   
$\sigma_{\rm abs}(158~{\rm GeV})=7.6\pm0.7({\rm stat})\pm0.6({\rm syst})$\,mb 
and 
$\sigma_{\rm abs}(400~{\rm GeV})=4.3\pm0.8({\rm stat})\pm0.6({\rm syst})$\,mb, 
see middle panel of Fig.~\ref{fig_sig-abs-Glauber}.
The origin of the increase of $\sigma_{\rm abs}$ towards lower collision 
energies is presently under investigation.
One may speculate that an energy dependence of the  $J/\psi$ breakup cross 
section on nucleons could be responsible for such a behavior (note that the 
high-energy absorption cross section relevant in the present context might be 
quite different from the low-energy dissociation cross section discussed in 
Sec.~\ref{ssec_diss-hg} as relevant for the hadronic phase of heavy-ion 
collisions). Another option are formation time effects which we will return to 
below. 

In extracting absorption cross sections from experiment one has to keep in 
mind that only about 60\% of the observed 
$J/\psi$ mesons are directly produced, while about 30\% (10\%) arise  
from $\chi_c $ ($\psi'$) feed-down. 
If one approximates $\sigma_{\psi N}$ geometrically, as $\pi r^2$, with 
$r_{J/\psi}=0.25$\,fm,  $r_{\psi'}=2\times r_{J/\psi}$ and 
$r_{\chi_c}=1.5\times r_{J/\psi}$, and repeats the Glauber-model calculation, 
an equally acceptable description of the $A$-dependence for 400\,GeV and 
450\,GeV is obtained~\cite{Lourenco:2006sr}, see the right panel of 
Fig.~\ref{fig_sig-abs-Glauber}.
This reasoning would point to a smaller absorption cross section for the
$J/\psi$ on nucleons, on the order of 2\,mb.
Formation-time effects have been evaluated in Ref.~\cite{He:1999aj} within 
a quantum mechanical description of the formation of charmonia from an 
initially small-sized ``pre-meson'' state. An energy- and time-dependent cross
section has been derived. At a collision energy of $\sqrt{s}=10$\,GeV, the 
initial pre-meson starts out with $\sigma_{\rm abs}\simeq3$\,mb, evolving into 
$\sigma_{J/\psi~N}=2.8\pm 0.3$\,mb and $\sigma_{\psi'N}=10.5\pm 3.6$\,mb
at asymptotic times ($t\to \infty$). In this framework, which is in 
accordance with geometric scaling, a good description of the ($x_F$-dependence 
of the) E866 data~\cite{Leitch:1999ea} can be achieved, and an apparent 
discrepancy between the $\sigma_{\rm abs}$ values extracted from 
hadro- and photo-production data can be resolved~\cite{Hufner:1997jg}.
A characteristic feature of the quantum mechanical treatment of charmonium 
formation in $p$-$A$ (also applicable to $AA$) 
collisions~\cite{Matsui:1989ig,Cugnon:1993ye,Cugnon:1993yf,Koudela:2003yd} 
is the possibility for an oscillatory behavior of the survival 
probability with time, most pronounced for the $\psi'$. 
This is different from the classical picture of a $c\bar c$ state 
expanding in time~\cite{Blaizot:1989de,Gavin:1990gm,Arleo:1999af,Arleo:2001nr}
and could be tested via detailed kinematic (e.g., $x_F$) dependencies
of production cross sections~\cite{Koudela:2003yd}.
\begin{figure}[!t]
\vspace{0.5cm}
\begin{tabular}{ccc}
\includegraphics[width=0.4\textwidth]{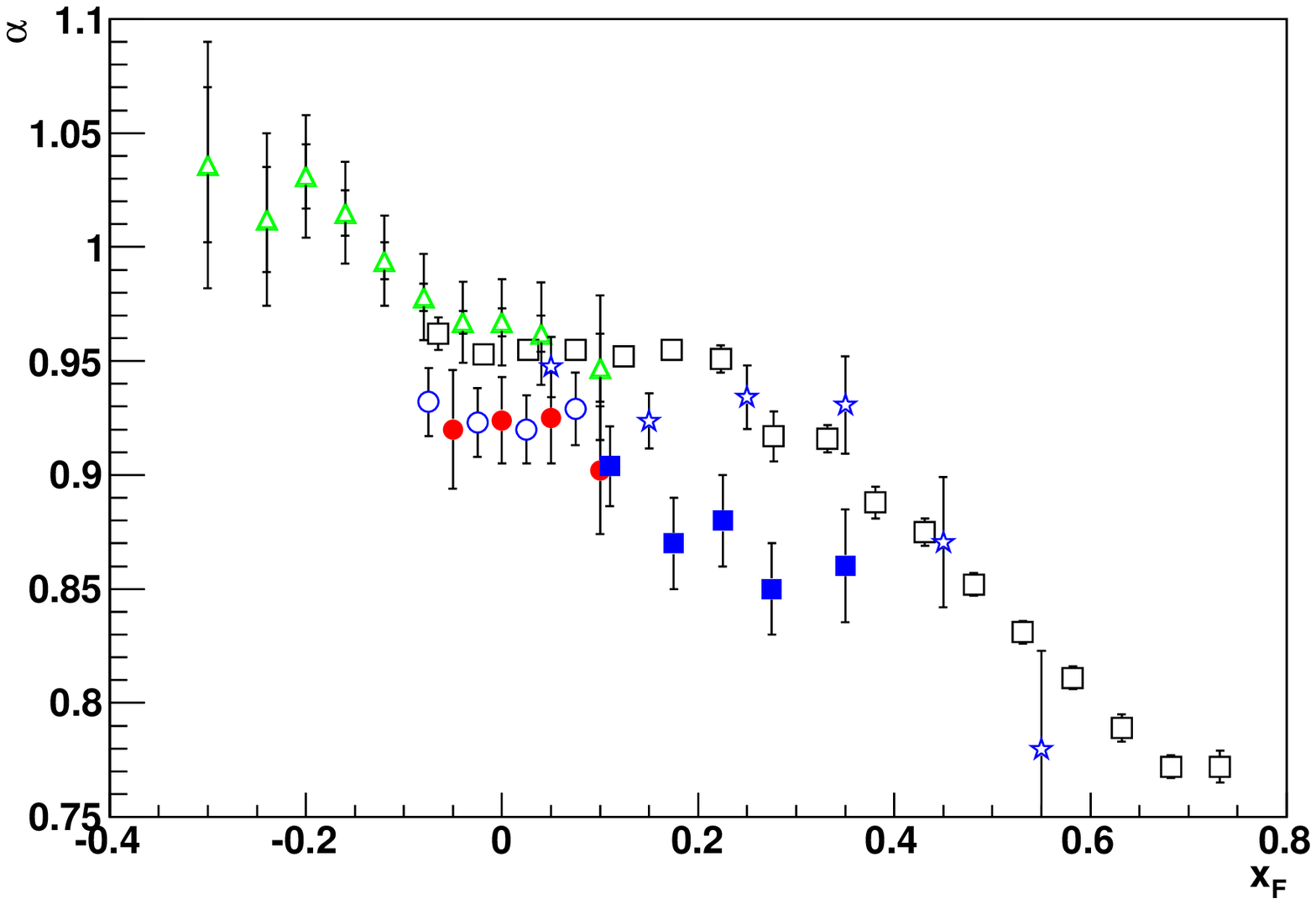}&
\includegraphics[width=0.26\textwidth]{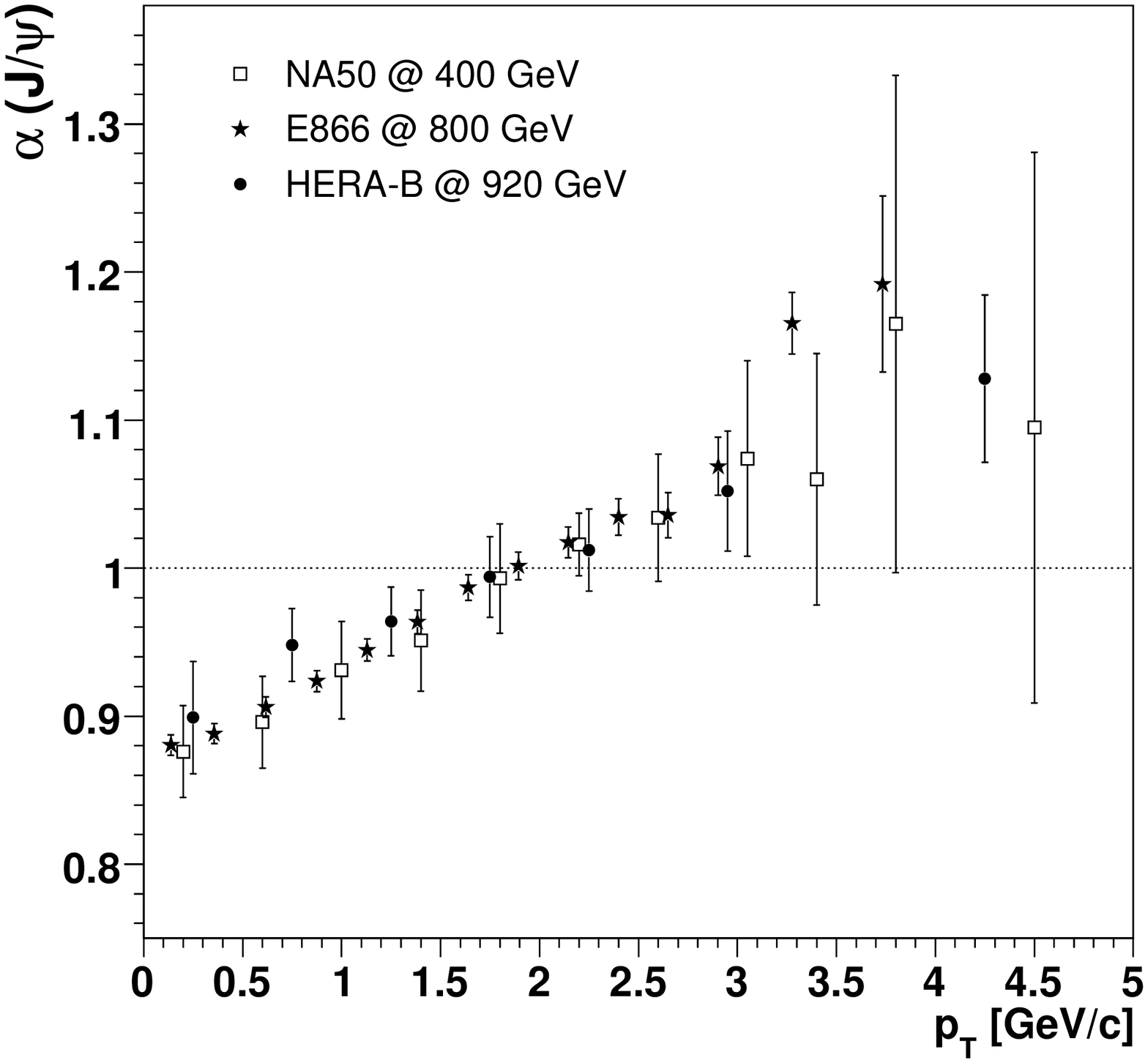}&
\includegraphics[width=0.26\textwidth]{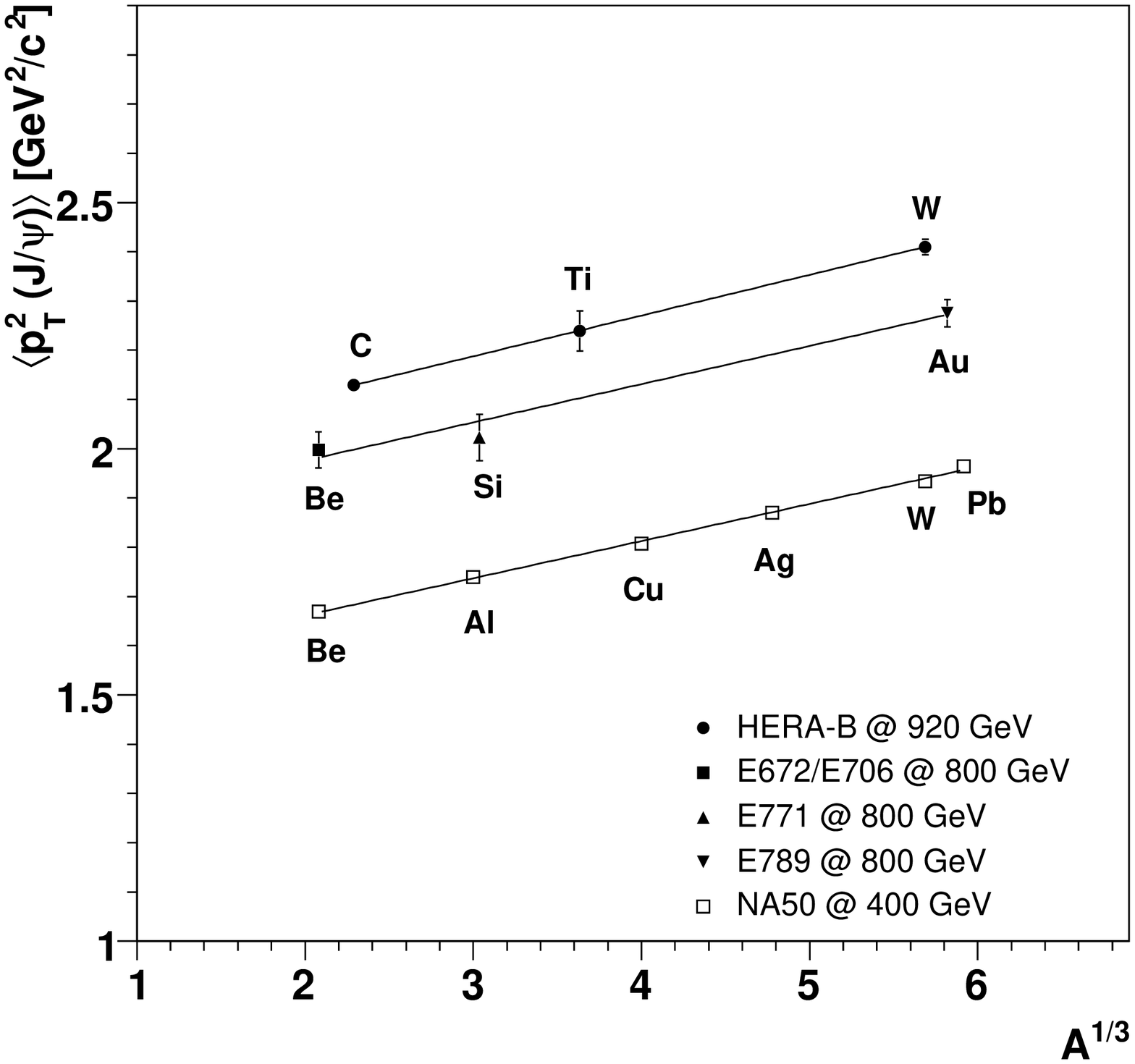}
\end{tabular}
\caption{The nuclear-dependence parameter $\alpha$ as a function of $x_F$ 
  (left panel) and of $p_T$ (middle panel), for $J/\psi$ mesons measured 
  in $p$-$A$ collisions at three different projectile energies (the
  symbols in the left panel correspond to data from HERA-B (open 
  triangles)~\cite{Faccioli:2006ty}, E866 (open squares)~\cite{Leitch:1999ea},
  NA50 at 450\,GeV (open circles)~\cite{Alessandro:2003pc,Ramello:2006db}, 
  NA60 at 400\,GeV and 158\,GeV (closed circles and squares, 
  respectively)~\cite{Arnaldi:2010ky} and NA3 (open stars)~\cite{Badier:1983dg}.
  A measure for $p_T$ broadening (Cronin effect), $\langle p_T^2\rangle$,
  increases with nuclear target size and with collision energy (right); 
  compilation of data in the left panel taken from Ref.~\cite{Arnaldi:2010ky} 
  and in the middle and right panels from Ref.~\cite{Lourenco:2006sr}.}
\label{fig_nuc-abs-alpha}
\end{figure}

The kinematic dependencies of CNM effects on $x_F$ and $p_T$ have been 
measured for charmonium 
production~\cite{Badier:1983dg,Leitch:1999ea,Alessandro:2003pc,Ramello:2006db,Faccioli:2006ty,Arnaldi:2010ky} 
and are often parametrized with the nuclear-dependence parameter, 
$\alpha(x_F,p_T)$, defined via
\begin{equation}
\label{alpha-par}
\sigma_{pA\to \psi}=\sigma_{pp\to \psi} ~A^\alpha \ .
\end{equation}
Results from fixed-target experiments at different center-of-mass (cm) 
energies are compared in Fig.~\ref{fig_nuc-abs-alpha}.
The values of $\alpha$ at $x_F\approx 0$ (relevant for heavy-ion collision 
experiments) tend to increase with collision energy. Since the scaling of 
$\alpha$ with $p_T$ (middle panel of Fig.~\ref{fig_nuc-abs-alpha}) appears to
be independent of energy, the increase with cm energy might originate from 
the increase of the average $p_T$ with collision energy, see right panel of 
Fig.~\ref{fig_nuc-abs-alpha}.

The two parameterizations of CNM effects, Eqs.~(\ref{alpha-par}) and 
(\ref{sigma-pA}), can be related to each other
\cite{Vogt:1999cu,Gerschel:1998zi}. 
Introducing the effective path length $L$ of the $J/\psi$ trajectory inside 
the target nucleus with nuclear matter density $n_0$, Eq.~(\ref{sigma-pA})
can be approximated as
\begin{eqnarray}
S_{\rm nuc}^{pA}&=&\exp(-L n_0 \sigma_{\rm abs})~.
\label{sigma-LA}
\end{eqnarray}
Comparing with Eq.~(\ref{alpha-par}) one finds 
\begin{eqnarray}
\sigma_{\rm abs}=\frac{1-\alpha}{L n_0}\ln A \ .
\label{sigma-alpha}
\end{eqnarray}
If one furthermore assumes a uniform density profile, one has 
$L=3r_0 A^{1/3}/4$, and with $A^{1/3} \approx \ln A$ (valid for large $A>50$), 
the relation (\ref{sigma-alpha}) takes the form  
\begin{equation}
\sigma_{\rm abs}=\frac{16 \pi}{9}r_0^2 (1-\alpha) \ \ , \ \ 
r_0=1.2\,{\rm fm} \ .
\end{equation}
With this ``pocket formula" a range of $\alpha$-values, 
$\alpha=0.95\pm 0.02$, translates into a range of absorption cross 
sections of $\sigma_{\rm abs}=4.0\pm 1.6$\,mb. For a similar example, 
see Ref.~\cite{Gerschel:1998zi}.
\begin{figure}[!t]
\begin{center}
\includegraphics[width=0.4\textwidth,height=0.3\textwidth]{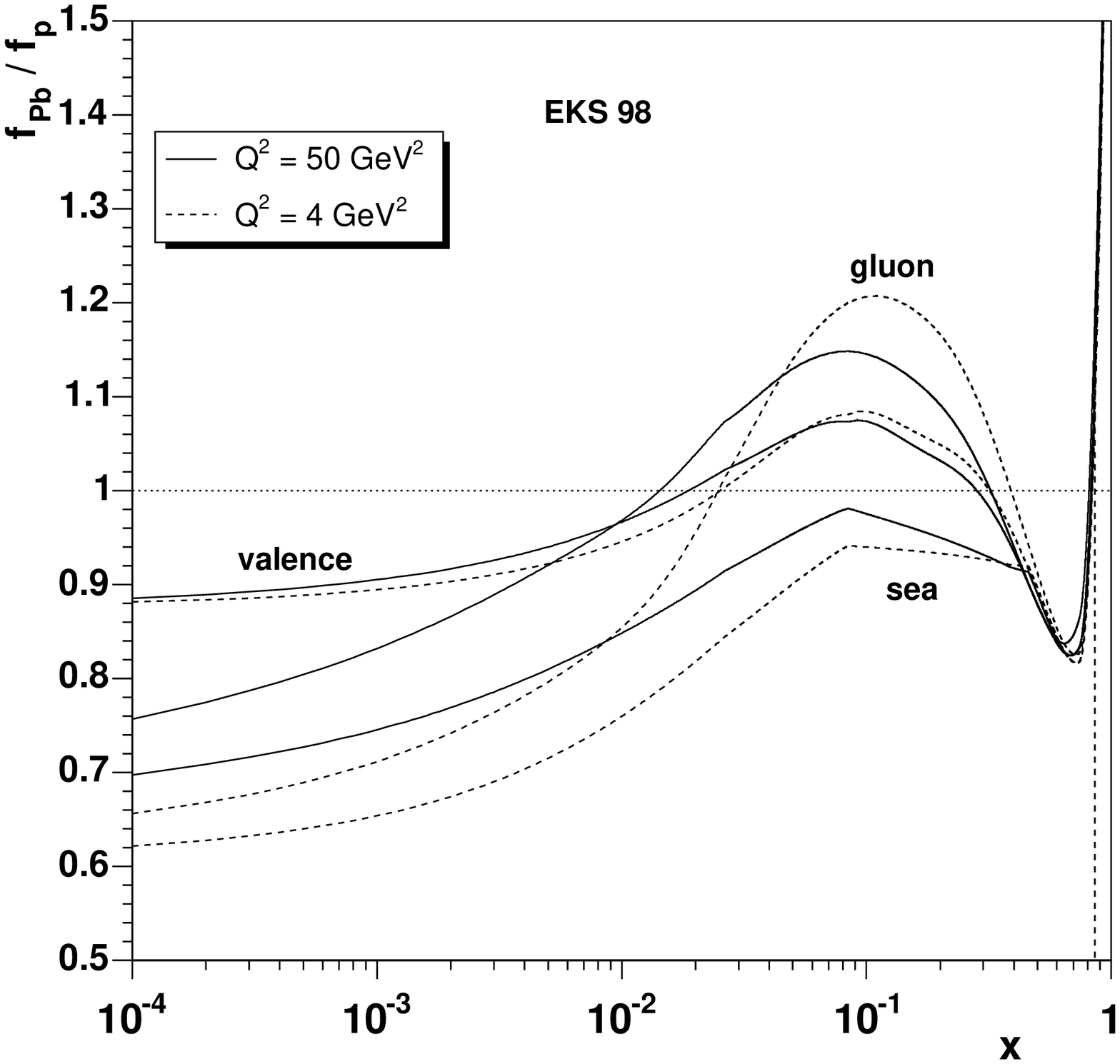}
\includegraphics[width=0.4\textwidth,height=0.3\textwidth]{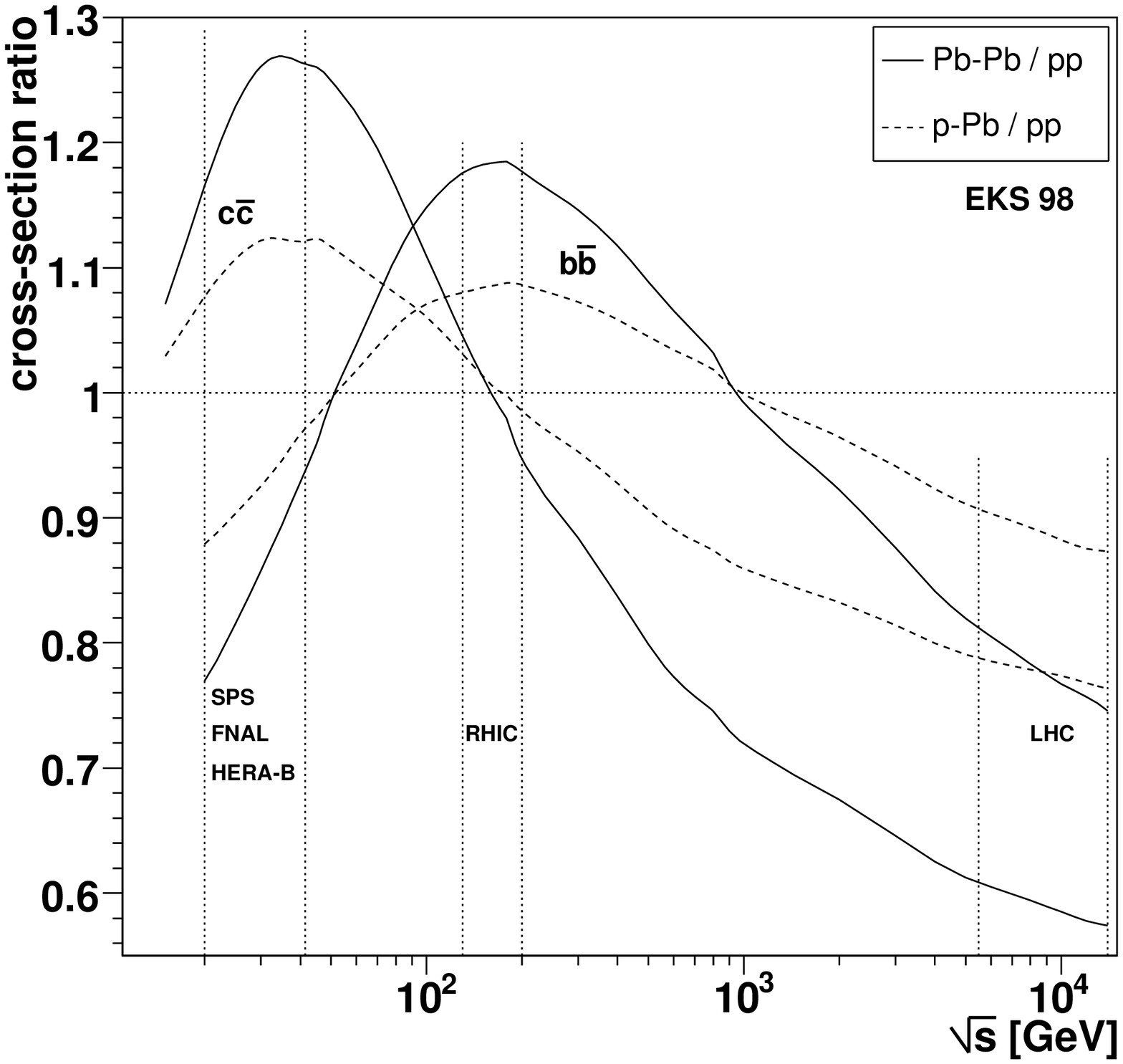}
\end{center}
\caption{Left panel: nuclear modification of gluon distribution functions 
  in a Pb nucleus,
  according to the EKS98~\cite{Eskola:1998df,Eskola:2001gt} weight functions,
  from Ref.~\cite{Lourenco:2006vw}.
  Right panel: changes of the $c\bar c$ and $b\bar b$ cross sections induced by
  the nuclear modification of the PDFs at mid-rapidity, 
  from Ref.~\cite{Lourenco:2006vw}.
\label{nuc-effect-PDF}
}
\end{figure}

Next we turn to CNM effects in nuclear PDFs (nPDFs), 
$f_i^A(x,Q^2)$, characterized by the ratio~\cite{Accardi:2004be} 
\begin{equation}
R_i^A(x,Q^2)=f_i^A(x,Q^2)/f_i^p(x,Q^2) \ . 
\end{equation}
The left panel of Fig.~\ref{nuc-effect-PDF} shows this ratio for a lead 
nucleus within the EKS98 parametrization~\cite{Eskola:1998df,Eskola:2001gt},
for two values of the momentum transfer, $Q^2=4(50)$\,GeV$^2$, 
appropriate for $c\bar c$ ($b\bar b$) production. Relevant values for the 
Bjorken variable, $x=M/\sqrt{s}$, at central rapidity ($y=0$)
and for charm (bottom) production, $M\sim m_Q=1.5(5)$\,GeV, are in the range of
$0.03<x<0.3$ from SPS to HERA-B energies ($200<E_{\rm lab}<920$\,GeV, 
corresponding to $\sqrt{s}=$~20-40\,GeV).
Of particular interest are gluons, which are expected to dominate the 
production cross section for $Q\bar Q$ pairs via the two-gluon fusion 
process (except for close to the production threshold).  
Typical nuclear effects are: (a) shadowing at low $x$, where  
$R^A(x,Q^2)<1$, (b) anti-shadowing at intermediate $x$, where $R^A(x,Q^2)>1$,
(c) EMC effect at $0.25< x<0.8$, where $R^A(x,Q^2)<1$, and (d) Fermi motion at
$x\approx 1$, where $R^A(x,Q^2)>1$. The right panel of 
Fig.~\ref{nuc-effect-PDF} indicates a 10\% $c\bar c$ production cross 
section enhancement at midrapidity for SPS energies,
while there is essentially no effect at RHIC.
\begin{figure}[!t]
\begin{tabular}{ccc}
\includegraphics[width=0.3\textwidth]{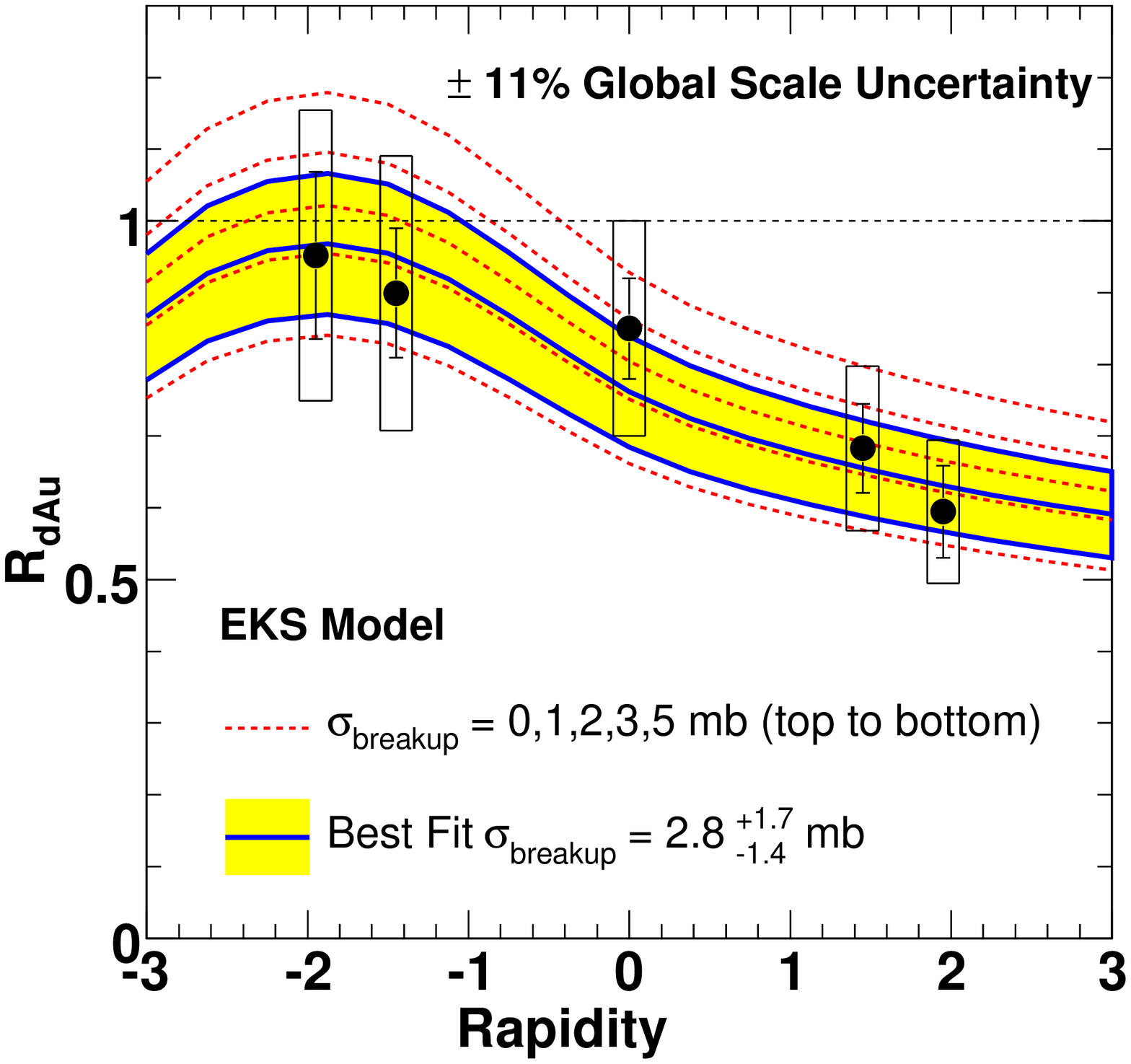}&
\includegraphics[width=0.3\textwidth]{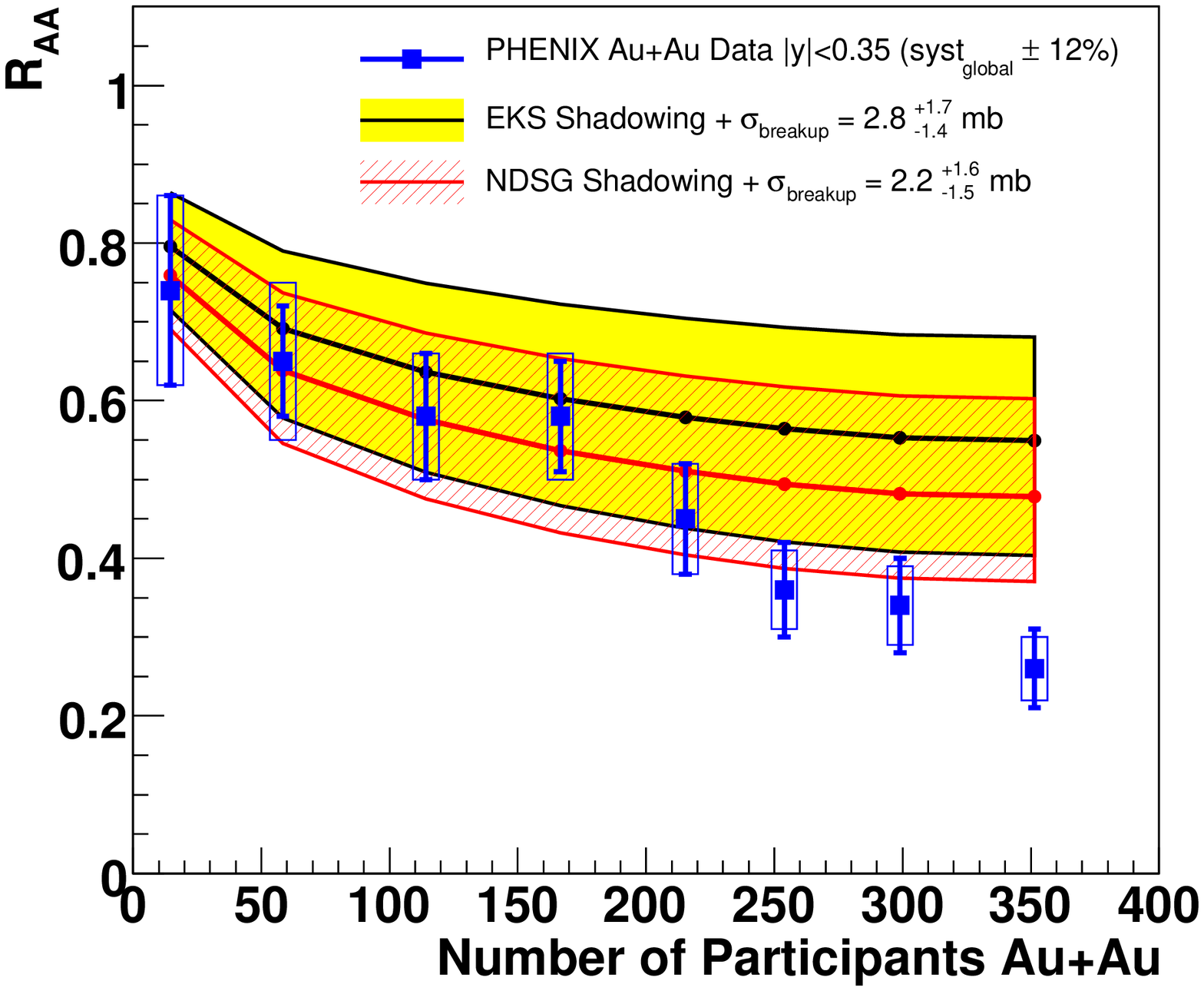}&
\includegraphics[width=0.3\textwidth]{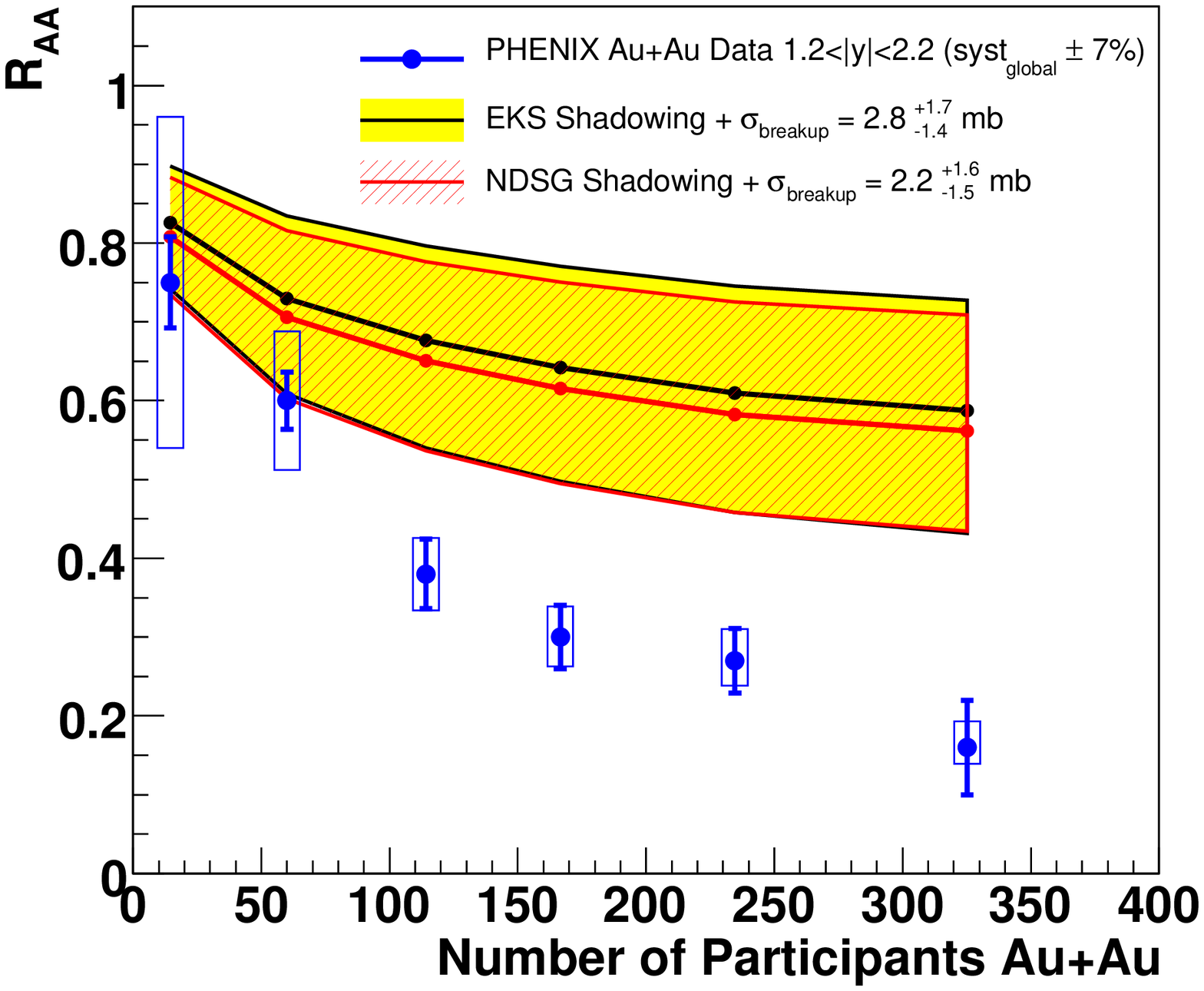}
\end{tabular}
\caption{Left panel: PHENIX data for the nuclear modification factor, 
$R_{\rm dAu}$, as a function of rapidity, compared to theoretical curves
using the EKS shadowing model~\cite{Eskola:1998df,Eskola:2001gt} for different 
values of the nuclear absorption cross section $\sigma_{\rm abs}$
(=$\sigma_{\rm breakup}$). 
The best fit to these data gives $\sigma_{\rm abs} = 2.8 ^{+1.7}_{-1.4}$\,mb.
Middle panel: $R_{AA}$ for Au-Au collisions at mid-rapidity 
compared to projections of CNM effects extrapolated 
from the d-Au data, for EKS98 and nDSg~\cite{deFlorian:2003qf}
shadowing. Right panel:
$R_{AA}$ at forward rapidity; from Ref.~\cite{Adare:2007gn}.
\label{rdau-rauau}
}
\end{figure}
On the other hand, for d-Au collisions at backward rapidities ($y=-2.0$)
at RHIC, a 15\% antishadowing is predicted, compared to 20\% shadowing
at forward rapidities of $y=2.0$ (in the direction of the Au beam).
This trend for the $y$ dependence can be seen in the recent PHENIX 
data at $\sqrt{s_{NN}}=200$\,GeV~\cite{Adare:2007gn}, see 
left panel of Fig.~\ref{rdau-rauau}.
Based on EKS98 PDFs, a best fit to the measured nuclear modification factor, 
$R_{\rm dAu}$, results in a ``residual" nuclear absorption cross section of 
$\sigma_{\rm abs} = 2.8 ^{+1.7}_{-1.4}$\,mb. For the 
nDSg PDF~\cite{deFlorian:2003qf} a smaller cross 
section of $\sigma_{\rm abs} = 2.2 ^{+1.8}_{-1.5}$\,mb has been extracted.
This is an example of how, in principle, a ``de-convolution'' of 
initial-state effects due to the nuclear dependence of PDFs (in particular 
for gluons) may be performed. 
A recent systematic study of the ``residual" $J/\psi$-nucleon absorption 
cross section was performed in Ref.~\cite{Arleo:2006qk}. After factorization 
of the charmonium production process, a global fit involving data from SPS 
to RHIC energies resulted in $\sigma_{\rm abs} = 3.4 {\pm 0.2}$\,mb 
when neglecting uncertainties in the gluon distribution functions. 
It was furthermore concluded that present uncertainties in the gluon (anti-) 
shadowing do not allow for a precise determination of $\sigma_{\rm abs}$. 
On the one hand, the new EPS08 nPDFs~\cite{Eskola:2008ca}, which, for the 
first time, included RHIC results for forward-rapidity hadron production in
d-Au~\cite{Arsene:2004ux} into the analysis, show a strong 
enhancement of gluon anti-shadowing (shadowing) in the SPS (LHC) domain.  
On the other hand, the rather mild $x$-dependence of $J/\psi$ production 
in $p$-$A$ collisions at $\sqrt{s_{NN}}=38.8$\,GeV in E866~\cite{Leitch:1999ea}
disfavors the nDSg, EKS98 and EPS08 parameterizations~\cite{Arleo:2008zc}. 
This might suggest that the suppression observed in the BRAHMS 
data~\cite{Arsene:2004ux} may not be entirely attributed to nPDFs alone, but also 
to other mechanisms such as initial-state parton energy loss~\cite{Vitev:2003xu}.

Another aspect which complicates the ``de-convolution'' of CNM effects is
the microscopic nature of the $J/\psi$ production process. Recent reanalyses 
of the two-gluon fusion process have been performed in 
Refs.~\cite{Haberzettl:2007kj,Artoisenet:2009mk,Brodsky:2009cf,Lansberg:2010vq} 
(see also Ref.~\cite{Khoze:2004eu}). 
These  works suggest that production processes involving a hard gluon recoiling
off the $J/\psi$ can account for the low- and intermediate-transverse
momentum spectra from the Fermilab Tevatron (CDF) to RHIC (PHENIX), as opposed 
to the soft-gluon emission mechanism of the color evaporation model (CEM) at 
leading order (LO) or to the color-octet model (COM) of NRQCD at LO. 
In this context, one distinguishes the kinematical effect of an ``extrinsic" 
source of $p_T$ due to a recoiling gluon from a ``primary" source of $p_T$ due
to an ``intrinsic" $p_T$ inherited from the initial gluon pair (as in the COM 
and CEM). In the "extrinsic" case, a shift of the rapidity dependence of the 
$J/\psi$ $R_{\rm dAu}$ occurs, not inconsistent with PHENIX data~\cite{Adare:2007gn}.
In connection with the EKS98 nPDFs this requires a larger absorption cross section, 
$\sigma_{\rm abs}=4.2$\,mb, to be compatible with the data~\cite{Ferreiro:2008wc}.

It is important to assess the limits of a classical description 
of CNM effects, i.e., whether a quantum mechanical approach to the
propagation of a charm quark pair (pre-meson) in the nuclear environment 
is warranted. The pre-meson evolution is governed by two time scales: 
the proper time for the formation of a spectrum of charmonium eigenstates
typically estimated as 
$\tau_{J/\psi}\simeq (m_{\psi'}-m_{J/\psi})^{-1}\simeq 0.35$\,fm/$c$,
and the so-called coherence time, $\tau_c=2E_{J/\psi}/m_{J/\psi}^2$,
representing the lifetime of a $c\bar{c}$ fluctuation in the lab frame. 
When the formation length becomes larger than the size of the Lorentz
contracted nucleus, the phenomenon of color transparency 
occurs~\cite{Brodsky:1988xz,Kopeliovich:1991pu}, also interpreted 
geometrically by a small-size pre-meson state~\cite{Blaschke:1992pw}.
When the coherence length becomes comparable to this size (e.g., by 
Lorentz time dilation), the amplitudes for processes on different 
nucleons interfere destructively, producing additional suppression. 
A review of the quantum mechanical treatment including applications to 
heavy-flavor production in a nuclear environment can be found in
Ref.~\cite{Kopeliovich:2003cn}.
 
An example for the extrapolation of the CNM effects from d-Au to Au-Au
collisions is shown in Fig.~\ref{rdau-rauau} for mid- (middle panel) 
and forward rapidity (right panel) in comparison to PHENIX data
\cite{Adare:2007gn}. While at forward rapidity the onset of effects beyond
CNM suppression appears at $N_{\rm part}\approx 70$ (which roughly coincides
with the onset of anomalous suppression in the NA50 experiment), there is 
almost no deviation from the CNM baseline at midrapidity, even for central 
collisions. This counter-intuitive result points to the existence of an 
additional $J/\psi$ production process which is predominantly operative 
in the central-rapidity region. 
In the previous Section, charm recombination in the medium has been 
discussed as a gain process in the kinetic equation for charmonium production. 
Its role in the interpretation of present and future experiments will be 
discussed quantitatively in the following Section.

%% file: exp.tex
This chapter is devoted to the presentation of future opportunities of 
quarkonia measurements in heavy-ion collisions.
Detectors are currently being designed, built, installed, commissioned
or upgraded around three accelerators which will deliver heavy-ion beams 
in three different energy domains:
\begin{itemize}
\item{RHIC-II (starting time scheduled for 2013) 
is an upgrade program of the existing RHIC (Relativistic 
Heavy Ion Collider) at Brookhaven National Laboratory, USA, which delivers
beams up to $\sqrt{s_{NN}} =$~200\,GeV for Au. The upgrade of the machine 
consists of an increase of the present luminosity by an order of magnitude as 
well as a low-energy program ($\sqrt{s_{NN}} =$~5-15\,GeV).
Whereas the low-energy physics program will mainly focus on evidence
for the critical point, the high-luminosity high-energy physics program
aims at completing existing measurements with a strong emphasis on the
heavy-flavor sector.
To achieve this goal, an upgrade program of two detectors (PHENIX and STAR)
is underway;}
\item{The LHC (Large Hadron Collider) at CERN will provide, from 2009 on, 
the largest energy
ever delivered by an accelerator with proton beams up to 
$\sqrt{s} = 14$\,TeV and Pb beams up to $\sqrt{s_{NN}} = 5.5$\,TeV.
Three of the four LHC experiments (ALICE, ATLAS and CMS) will take heavy-ion
data with an important part of their physics program devoted to quarkonia 
measurements.
Whereas ATLAS and CMS are designed for proton-proton physics, ALICE is 
dedicated to the study of heavy-ion collisions;}
\item{The SIS-300, located at the FAIR (Facility for Antiproton and 
Ion Research) at GSI, Darmstadt, will deliver, from 2015 on, heavy ion 
beams in fixed target mode with beam energy from
10\,AGeV to 45\,AGeV ($\sqrt{s_{NN}} =$~4.5-9.3\,GeV).
The Compressed Baryonic Matter (CBM) detector will allow to explore, 
with high-statistics measurements 
of rare signals like quarkonia, the moderate temperature and large
baryon chemical potential region of the QCD phase diagram.}
\end{itemize}
In two of these energy ranges (LHC and SIS-300)
heavy quark resonances will be measured for the first time.
This ensures a rich and exciting future physics program.
In the following, we present the motivations for studying quarkonium production
in these three energy ranges as well as a short overview of the detector 
capabilities needed to achieve this goal.
All detectors have in common the goal of performing quarkonium measurements 
as a function of centrality and reaction plane, 
in a broad acceptance, with large statistics and low background, to ensure
a mass resolution good enough to separate all vector resonances and 
to identify feed-down from excited states.
For more details we refer 
to~\cite{Frawley:2008kk}, to
\cite{Carminati:2004fp,Alessandro:2006yt,D'Enterria:2007xr,ATLAS-HI} 
and to \cite{CBM-HI} for the RHIC-II, the LHC and the SIS-300 
experimental programs, respectively. 

\subsection{Quarkonia in heavy-ion collisions at the RHIC-II}
The RHIC has started delivering heavy-ion beams in 2000.
Since then, the four heavy-ion experiments (BRAHMS, PHENIX, PHOBOS and STAR)
have collected a large set of data for various systems at energies
$\sqrt{s_{NN}} = 20 - 200~{\rm GeV}$.
The status of information gathered up to 2004 has been summarized both
from the theoretical side~\cite{Wang2004dn} and from the experimental 
side~\cite{Adams2005dq}.
Hard probe precision measurements have become accessible at RHIC with the 
high statistics runs performed since 2004.
The measurements, done by the PHENIX and STAR collaborations,
have revealed some of the most exciting aspects of the hot and dense matter
produced at RHIC.
A particularly striking and unexpected observation in the quarkonium sector
is the $J/\psi$ suppression whose magnitude in mid-central and central 
heavy-ion collisions is similar to that 
measured at SPS energy for the Pb-Pb system (recall Figs.~\ref{fig_raa-na50}
and \ref{fig_raa-phenix}).
As a consequence, several models which offer a successful quantitative 
description of $J/\psi$ suppression at SPS energy (assuming that the 
suppression is driven by the system temperature or its 
energy density) over-predict the suppression observed at RHIC.
Furthermore, the observed $J/\psi$ suppression is larger in the forward region
than in the central region (Fig.~\ref{fig_y-rhic}). 
This observation is again striking as most of the suppression models predict an
opposite effect, i.e., the higher the energy density the larger the suppression.
The interpretation of these observations has been discussed in 
Secs.~\ref{ssec_nuc-abs} and~\ref{ssec_obs}.

Further detailed measurements are mandatory in order to disentangle
the different theoretical interpretations and to achieve a comprehensive 
understanding of quarkonia production in heavy-ion collisions at RHIC.
Indeed, direct evidence for resonance dissociation by color screening
requires the centrality dependence of $J/\psi$, $\chi_c$ and $\psi^\prime$
to be measured simultaneously.
Regeneration models can be constrained from the inspection of $J/\psi$ 
rapidity distributions which are expected to 
narrow~\cite{Andronic:2006ky,Thews:2005vj} (recall Fig.~\ref{fig_y-rhic}),
as well as from a precise measurement of the (open) charm production
cross-section.
Trends in the excitation function of $J/\psi$ suppression could also
reveal the relative importance of the regeneration process as it should 
increase with the multiplicity of $c\bar{c}$ pairs~\cite{Grandchamp:2001pf}.
More stringent tests include measurements of $J/\psi$ elliptic flow 
since, as open charm shows non-zero flow\footnote{In contrast to 
light-flavor hadrons, the heavy-flavor flow is, so far, not measured 
experimentally through identified hadrons, but in an inclusive way via 
the flow of non-photonic electrons.  The latter is obtained from full 
distributions of electrons after subtraction of Dalitz-decay electrons 
from light hadrons and photon-conversion electrons.},
$J/\psi$, if produced by regeneration, should inherit this flow (recall
Fig.~\ref{fig_v2-rhic}).
In addition, both $J/\psi$ flow and suppression should exhibit a specific 
$p_T$ dependence marked by QGP effects at relatively low $p_T$
whereas the high $p_T$ region should be mostly populated by 
$J/\psi$ which could escape the medium unaffected.
In this respect it is also important to be able to identify
secondary $J/\psi$ from
bottom decay whose yield might be sizable at high $p_T$.
Furthermore, detailed measurements in $p$(-like)-$A$ collisions are mandatory
in order to disentangle cold and hot nuclear effects.
We finally note the possibility to identify the presence of the deconfined 
medium from the measurement of $J/\psi$ polarization~\cite{Ioffe:2003rd}.
In addition to the above observables based on charmonia, the measurement of 
bottomonium states should provide additional insights.
The temperature of the medium at RHIC is expected to be large enough to 
induce the break-up of $\Upsilon(2S)$ and $\Upsilon(3S)$ whereas the 
$\Upsilon(1S)$ is supposed to melt at higher temperatures (it 
might still be affected by inelastic reactions with partons in the QGP, as 
discussed in Secs.~\ref{ssec_diss-qgp} and~\ref{ssec_bottom}).

The previously exposed physics program can partially be covered in the
next years with the existing RHIC machine and detectors.
However, it is clear that the full physics program
requires high statistics and excellent data quality.
This quality will be achieved by means of an upgrade of the 
machine luminosity and of the detectors.

\subsubsection{Experimental conditions}
With the luminosity upgrade of the machine, the integrated luminosity
per week is expected to be 2500\,$\mu$b$^{-1}$ and 33\,pb$^{-1}$ for 
Au-Au and $p$-$p$ collisions at $\sqrt{s_{NN}} = 200$\,GeV, respectively.
A typical run at RHIC consists of 24 weeks of data taking. This is 
significantly longer than a LHC run in the heavy-ion mode (see below).
As a consequence, although the heavy flavor production cross sections are 
much bigger at the LHC, the lower cross sections at RHIC-II are 
compensated by the integrated luminosity so that the heavy-flavor yields 
for one year of running are expected to be similar at RHIC-II and at 
the LHC.

\begin{figure}[!th] 
\begin{center} 
\begin{minipage}[t]{16.5 cm} 
\centering
\epsfig{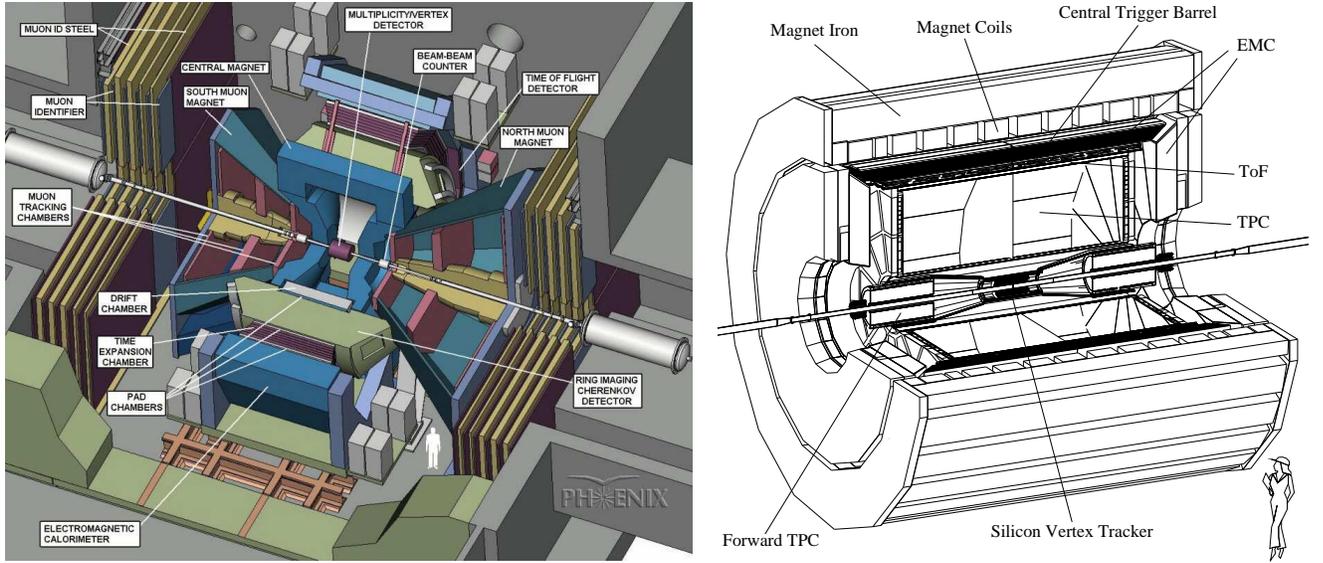} 
\end{minipage} 
\caption{Longitudinal view of the PHENIX (left) and STAR (right) detectors.}
\label{phenix_star_display}
\end{center} 
\end{figure} 

\subsubsection{The PHENIX experiment}
The PHENIX experiment is made of four spectrometers and a set of specialized
detectors to determine the collision centrality and to provide triggers
(Fig.~\ref{phenix_star_display} left).
In the central region, two arms allow to detect electrons, photons and 
charged hadrons. 
They consist of a complex arrangement of drift chambers, pad chambers,
Time Expansion Chambers (TEC) operated as transition detectors, 
time-of-flight (ToF) detectors, RICH detectors,
aerogel detectors and electromagnetic calorimeters (EMCs).
A magnet supplying a field parallel to the beam is placed around the 
interaction vertex. In the forward and backward regions, two 
spectrometers allow muon measurements
in the rapidity range $-2.25<y<-1.15$ and $1.15<y<2.44$.
They consist of a muon tracker (three stations of multi-plane drift 
chambers) placed inside a radial magnetic field, followed by a 
muon identifier (alternating layers of steel absorbers and streamer tubes for
tracking) both with full azimuthal coverage.

The upgrade program of PHENIX~\cite{Frawley:2008kk} consists of a barrel 
and two endcaps Silicon Vertex Detectors (SVDs), two Nose Cone 
Calorimeters (NCCs) and an upgrade of the muon trigger.
The SVD will provide inner tracking with full azimuthal 
coverage for $|\eta|<2.5$.
This will improve the dilepton mass resolution and reduce the background 
both in the electron and in the muon channels.
It will further allow to measure displaced vertices which is mandatory
for identification of secondary $J/\psi$'s from $B$-meson decay.
The NCCs contain electromagnetic and hadronic 
sections covering the acceptance $0.9<|\eta|<3.5$.
These are key detectors for measuring $\chi_c\rightarrow J/\psi + \gamma$ 
decays.
The muon trigger upgrade consists of adding three layers of RPC detectors 
in each muon arm together with associated front-end electronics and 
trigger logic.
This will improve the level-1 trigger selection for both single and 
di-muons.

Typical expected yields with the upgraded PHENIX detector for a 12-week 
physics run with Au beams at RHIC-II are $45000$ ($395000$) 
$J/\psi$ and 400 (1040) $\Upsilon$ in the electron (muon) channel.

\subsubsection{The STAR experiment}
The different sub-systems of the STAR experiment 
(Fig.~\ref{phenix_star_display} right) are placed in a solenoidal magnet 
operating at $0.5~{\rm T}$.
The main detectors are 
(i) a large azimuthally symmetric Time Projection Chamber (TPC)
 providing charged particle tracking within $|\eta| < 1.8$, 
(ii) the Silicon Vertex Tracker (SVT) and Silicon Strip Detector (SSD) 
for charge particle tracking close to the interaction point 
and vertexing, 
(iii) the Forward TPCs (FTPCs) covering $2.8<|\eta|<3.8$
to track particles at forward and backward rapidities,
(iv) the ToF system providing additional particle identification
in the acceptance of the central detectors and 
(v) the EMC (both in the central region and at
forward angles) which ensures neutral particle measurements.
Additional specialized detectors complement the setup for centrality 
measurements and trigger.

The upgrade program of STAR~\cite{Frawley:2008kk} includes a 
ToF system covering the full acceptance of the central 
barrel, new front-end electronics for the existing TPC, 
an upgrade of the data acquisition system and 
a new inner tracking system based on two layers of silicon pixel sensors
and three layers of silicon strip detectors.
The ToF will extend the momentum range for hadron
identification and, in conjunction with the electromagnetic calorimeter,
provide a level-2 trigger for $J/\psi\rightarrow e^+e^-$ measurements.
The new inner tracking system will make displaced vertex measurements 
accessible. 
As for the PHENIX experiment, this will be used to tag $J/\psi$ from 
$B$ hadron decays, to reconstruct charm hadron in their hadronic decay 
channels and will allow open heavy flavor measurements from single electrons.
Finally, the DAQ upgrade will allow to collect rare signals with
large statistics.

Simulations indicate that up to $220\cdot10^{3}$ $J/\psi$ and $11200$ 
$\Upsilon$ could be reconstructed in STAR during a 12 week physics run at
RHIC-II in Au-Au collisions.

\subsection{Quarkonia in heavy ion collisions at the LHC}
With a nucleus-nucleus center-of-mass energy nearly 30 times larger than 
the one reached at RHIC, the LHC will provide the biggest step in energy in 
the history of heavy-ion collisions and open a new era for studying the 
properties of strongly interacting matter under extreme thermodynamical 
conditions.
This new energy regime will lead to a much higher energy density, to faster 
equilibration and to a longer lifetime of the deconfined system,
resulting in an enhanced role of the QGP over final-state hadronic 
interactions~\cite{Schukraft2001vg}.
The high temperature and close to vanishing baryon chemical potential of 
the system will render it close to the conditions of the primordial universe.
In addition, heavy-ion collisions at the LHC access unprecedented
small Bjorken-$x$ values where low-momentum gluons are expected to be close
to saturation and lead to a significant shadowing effect.
As a consequence high-density parton distributions could be responsible 
for a large fraction of particle production~\cite{Kharzeev:2004if}.
Another exciting aspect of this new energy regime is the massive production 
rate of hard processes.
They will provide an ideal tool for a detailed characterization 
of the deconfined medium.
The heavy-flavor sector at LHC energy is also subject to significant 
differences with respect to SPS and RHIC energies.
First, the large production rate offers the possibility to use new 
and a large variety of observables\footnote{From RHIC to LHC, it is expected 
that the number of $c\bar{c}$ ($b\bar{b}$) pairs produced in central 
nucleus-nucleus collisions will increase by a factor 
10 (100)~\cite{Alessandro:2006yt}: up to 115 $c\bar{c}$ (5 $b\bar{b}$)
pairs are expected to be produced in 5\,\% central Pb-Pb collisions
at $5.5~{\rm TeV}$.
This estimate is based on next-to-leading order pQCD calculations and 
includes shadowing; see Refs.~\cite{Alessandro:2006yt,Bedjidian:2003gd} 
for more details.}.
The magnitude of most of the in-medium effects is therefore anticipated 
to be dramatically enhanced.  Some of these aspects are discussed in
the following.

\paragraph{New observables.}
The $\Upsilon(1S)$ state is expected to dissolve significantly above the 
critical temperature, at  
$\sim$3-4~$T_c$~\cite{Digal:2001ue,Cabrera:2006wh,Mocsy:2007jz}, 
which presumably comes into reach only with the LHC.
The spectroscopy of the $\Upsilon$ family at LHC should then reveal unique 
characteristics of the QGP.
In particular, the $p_T$ dependence of the $\Upsilon(2S)/\Upsilon(1S)$
ratio presents significant sensitivity to the dissociation 
temperatures~\cite{Gunion:1996qc} (see 
Refs.~\cite{Grandchamp:2005yw,Abreu:2007kv} for updates).
Measuring the $\Upsilon(2S)$ is also particularly interesting in order to 
unravel $J/\psi$ suppression versus regeneration:
the $\Upsilon(2S)$ and $J/\psi$ dissociation temperatures are predicted
to be similar~\cite{Digal:2001ue} whereas, in contrast to charmonia, 
bottomonia are expected to be little affected by regeneration 
processes~\cite{Grandchamp:2005yw}.
Electroweak ${\rm Z}^0$ and ${\rm W}^\pm$ bosons will be available with 
large statistics and can serve as reference processes for quarkonium
suppression studies.

\paragraph{Large resonance dissociation rate.}
In addition to comoving hadrons and color screening, quarkonia can be 
destroyed by parton ionization~\cite{Xu:1995eb}.
This mechanism, induced by thermal partons in the QGP,
starts being effective for temperatures above the critical temperature but 
below the temperature of resonance dissociation by color screening
(cf.~Sec.~\ref{ssec_diss-qgp}). Recall, however, the interplay between 
the two mechanisms on the final quarkonium yields: 
a large dissociation by color screening implies a low binding
which facilitates a large break-up rate by parton ionization.
Recent estimates~\cite{Bedjidian:2003gd} (see Ref.~\cite{Blaschke:2004dv}
for an update) of the quarkonium dissociation cross sections show that 
none of directly produced the $J/\psi$ survives the deconfined phase 
at LHC, and that about 20\% of the $\Upsilon$ are destroyed, possibly 
more if color-screening is strong, 
cf.~Fig.~\ref{fig_ups-lhc}~\cite{Grandchamp:2005yw}. 

\paragraph{Large secondary charmonium production.}
Secondary charmonia yields can arise due to  
statistical hadronization~\cite{Braun-Munzinger:2000px}
and/or kinetic recombination~\cite{Thews:2000rj}.
These processes result in an increase of $J/\psi$ yield 
with collision centrality roughly proportional to $N_{c\bar{c}}^2$
(grand-canonical limit). Again, this effect is expected to entail 
dramatic consequences at the LHC.
Due to the large number of $c\bar{c}$ pairs produced in heavy-ion 
collisions at LHC, models predict a qualitatively different centrality 
dependence of the $J/\psi$ yield~\cite{Bedjidian:2003gd,Andronic:2007bi},
recall right panel of Fig.~\ref{fig_excit}.  

\paragraph{Large charmonium rate from $B$ hadron decay.}
Another source of non-direct charmonia arises from 
the decay of $B$ mesons.  The ratio 
$N(B\rightarrow J/\psi)/N({\rm direct}~J/\psi)$ can be 
determined as follows.
The number of directly produced $J/\psi$ in central (5\%) Pb-Pb collisions
at $5.5~{\rm TeV}$ is 0.5~\cite{Bedjidian:2003gd}\footnote{Including 
shadowing and feed-down from higher states.}.
The corresponding number of $b\bar{b}$ pairs (with shadowing) amounts to 
4.6~\cite{Bedjidian:2003gd}.
The $B\rightarrow J/\psi X$ branching ratio is 
$1.16\pm0.10\%$~\cite{Yao:2006px}.
Therefore 
$N(B\rightarrow J/\psi)/N({\rm direct}~J/\psi) = 20\%$
in $4\pi$\,\footnote{This ratio is subject to large uncertainties since
(i) the predictions from the color-evaporation model (CEM) on $J/\psi$ 
cross sections differ by up to a factor of 2 at LHC 
energies~\cite{Bedjidian:2003gd}, and
(ii) the total $b\bar b$ production 
cross section in $p$-$p$ collisions at LHC is predicted within a factor of
2-3 uncertainty resulting from the choice of the quark mass, the 
renormalization and factorization scales and the parton distribution 
function.  In addition, resonance suppression and/or regeneration, 
heavy-quark energy loss and 
other effects not taken into account here could play a significant role.}.
The secondary $J/\psi$ from $B$-meson decay must be subtracted from the 
primary $J/\psi$ yield prior to $J/\psi$ suppression studies.
They can further be used in order to measure the $B$-meson production 
cross section in $p$-$p$ collisions~\cite{Acosta:2004yw}, 
to estimate shadowing in $p$-$A$ collisions and to probe
the medium-induced $b$-quark energy loss in $A$-$A$ collisions.
Indeed, it has been shown~\cite{Lokhtin:2001nh} that the $p_T$ and 
$\eta$ distributions of those $J/\psi$ exhibit pronounced sensitivity to
$b$-quark energy loss.


\subsubsection{Experimental conditions}
The LHC will be operated seven months per year in the $p$-$p$ mode
and one month per year in the heavy-ion mode.
The corresponding estimated effective running time is $10^7$\,s 
and $10^6$\,s for $p$-$p$ collisions and $A$-$A$ collisions, respectively.
The expected luminosity for Pb-Pb collisions is about
$5 \cdot 10^{26}~{\rm cm}^{-2} {\rm s}^{-1}$ which results in a minimum-bias
interaction rate of $4~{\rm kHz}$. 
As described in Ref.~\cite{Carminati:2004fp}, the heavy- (and light-) ion 
runs include, over the first five years of operation, one Pb-Pb run at low 
luminosity, two Pb-Pb runs at high luminosity, one $p$-$A$ run and one light
ion-ion run.
In the following years different options are considered depending on the first
results.


\subsubsection{The ALICE experiment}
ALICE (A Large Ion Collider Experiment) is the only LHC experiment
dedicated to the study of nucleus-nucleus  
collisions~\cite{Carminati:2004fp,Alessandro:2006yt}.
The ALICE physics program also includes the study of $p$-$p$ collisions 
which will provide the reference for heavy-ion data.
The ALICE experiment is designed to perform high-precision 
measurements of numerous observables based on hadrons, leptons and 
photons, in a broad acceptance.

\begin{figure}[!t] 
\begin{center} 
\begin{minipage}[t]{16.5 cm} 
\centering
\epsfig{file=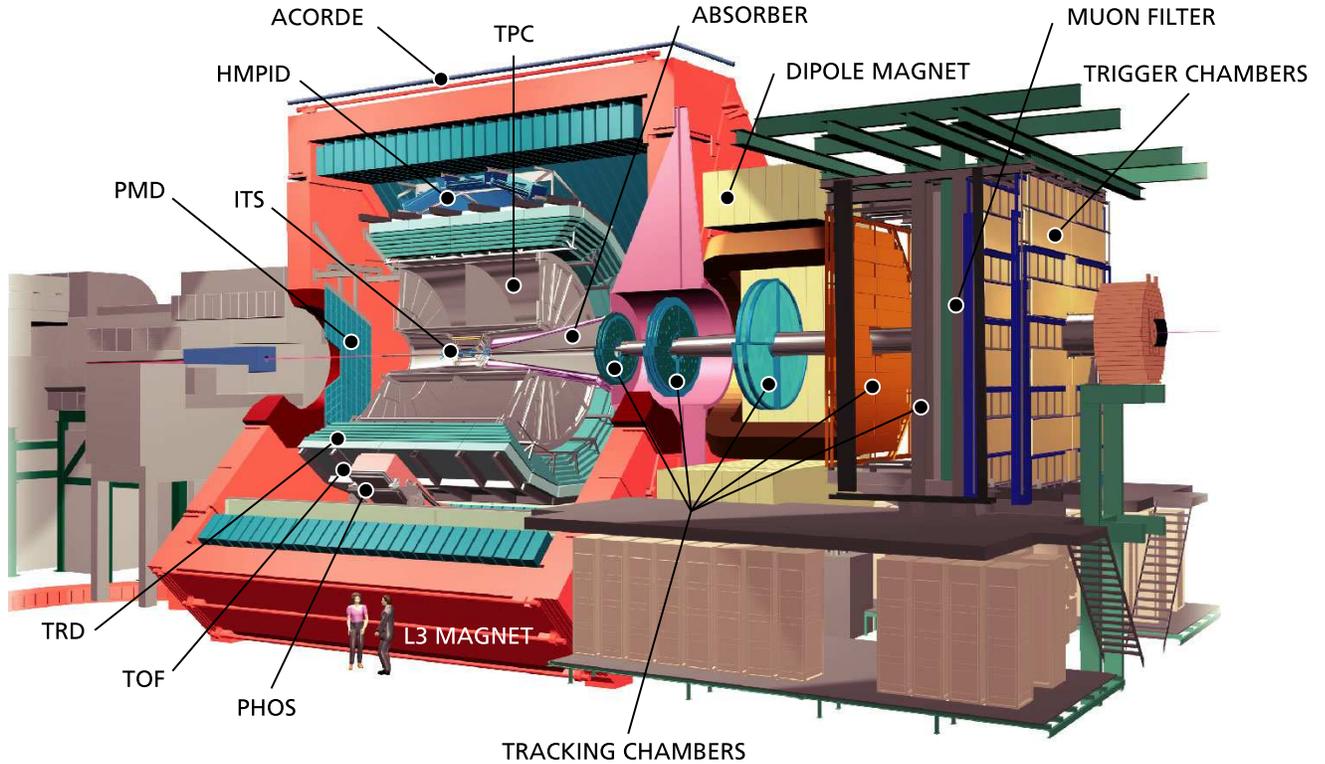,width=1.05\textwidth} 
\end{minipage} 
\caption{Longitudinal view of the ALICE detector.}
\label{alice_display}
\end{center} 
\end{figure}

The detector (Fig.~\ref{alice_display}) consists of a central part, 
a forward muon
spectrometer and forward/backward small acceptance detectors.
The central part of ALICE consists of four layers of detectors 
placed in the solenoidal field ($B \leq 0.5~{\rm T}$) 
provided by the L3 magnet previously used at the Large Electron Positron 
collider (LEP). From inside out, these detectors are 
(i) the Inner Tracker 
System (ITS) consisting of six layers of silicon detectors, 
(ii) the TPC, 
(iii) the Transition Radiation Detector (TRD) and 
(iv) the ToF system based on multi-gap resistive plate chambers.
They provide charged particle reconstruction and identification in the 
pseudo-rapidity range $|\eta|<0.9$, with full azimuthal coverage and a 
broad $p_T$ acceptance.
The ALICE central barrel will later be equipped with a large acceptance 
($|\eta|<1.4$, $\Delta\Phi = 110^\circ$) 
EMC (not shown in Fig.~\ref{alice_display}).
These large area devices are complemented by two smaller acceptance detectors: 
the High Momentum Particle IDentification (HMPID) and the PHOton Spectrometer
(PHOS).
In the forward/backward region, additional detectors (T0, V0 and FMD,
not shown in Fig.~\ref{alice_display}) enable a fast characterization and 
selection of the events, as well as charged-particle measurements in the 
pseudo-rapidity range $-3.4 <\eta <5.1$.
At large rapidities, photon multiplicity and spectator nucleons in heavy-ion
collisions will be measured by the Photon Multiplicity Detector (PMD) and the 
Zero-Degree Calorimeters (ZDC; not shown in Fig.~\ref{alice_display}), 
respectively.
A forward muon spectrometer covering the pseudo-rapidity range 
$-4<\eta <-2.5$ complements the central part. 
It consists of a front absorber, a dipole magnet, ten high-granularity
tracking chambers, a muon filter and four large area trigger chambers.

\begin{figure}[!th] 
\begin{center} 
\begin{minipage}[t]{16.5 cm} 
\centering
\epsfig{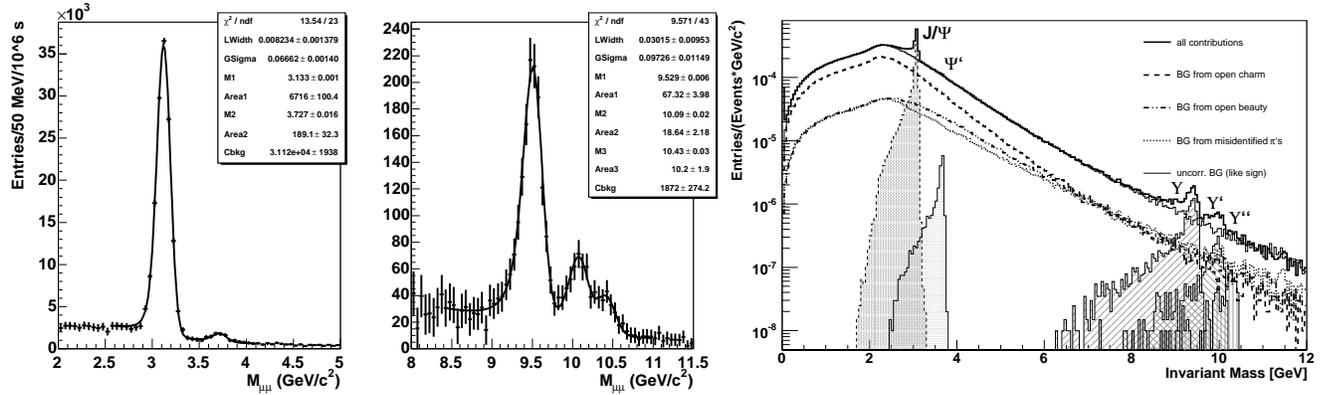} 
\end{minipage} 
\caption{Invariant-mass spectra of low-mass dimuons (left), high-mass dimuons 
	(middle), and dielectrons (right), expected to be measured with the 
	ALICE detector within one month of Pb beam running 
        (extracted from Ref.~\cite{Alessandro:2006yt}).
	In the left and middle plots, the non-correlated background is 
	subtracted from the total spectrum assuming a 
	perfect subtraction, i.e., the statistical error of the ``full" 
        spectrum is assigned to the remaining spectrum of the sum of the 
        correlated sources.}
\label{alice_imass}
\end{center} 
\end{figure}

The performance of the ALICE detector in the quarkonium sector is
summarized in Fig.~\ref{alice_imass}.
Quarkonium states will be measured in both electron 
($|\eta | < 0.9$) and muon channels ($-4<\eta <-2.5$).
The acceptance will allow reconstruction of differential distributions 
down to very low transverse momentum in most cases. The resolution 
of the apparatus, better than $100~{\rm MeV/}c^2$ for invariant
masses around $10~{\rm GeV/}c^2$, allows for a separation of all 
quarkonium vector resonance states.
The expected number of reconstructed $J/\psi$ and $\Upsilon$ during one 
month of data taking is $677\cdot 10^3$ and $9600$, respectively, in 
the muon channel for minimum-bias Pb-Pb collisions.
The corresponding numbers in the electron channel for central (10\%) 
collisions are $121\cdot 10^3$ and 1800.

\subsubsection{The CMS experiment}
\begin{figure}[!t] 
\begin{center} 
\begin{minipage}[t]{16.5 cm} 
\centering
\epsfig{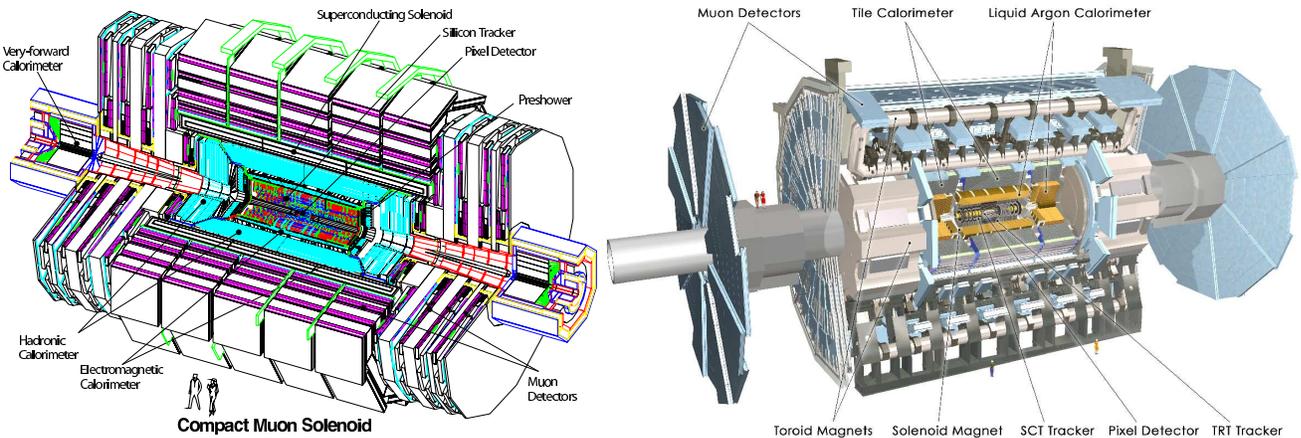} 
\end{minipage} 
\caption{Longitudinal view of the CMS (left) and of the ATLAS (right) 
	detectors.}
\label{cms_atlas_display} 
\end{center} 
\end{figure}

CMS (Compact Muon Solenoid) is a general-purpose detector
(Fig.~\ref{cms_atlas_display} left) designed to measure muons, 
electrons, photons and jets.
Although the detector is optimized for $p$-$p$ collisions, a strong 
heavy-ion program has been developed~\cite{D'Enterria:2007xr}.
CMS is composed, from the interaction point to the outer side, of a tracking
system, an EMC, a hadronic calorimeter and muon
chambers arranged in a central barrel and two endcaps.
The central element of CMS is a 13\,m long, 3\,m diameter
magnet which delivers a $B = 4~{\rm T}$ solenoidal field surrounding the 
tracking and calorimetric systems.
The tracker is based on several layers of silicon pixel and strip counters 
and covers the pseudo-rapidity region $|\eta|<2.5$.
The EMC, made of lead-tungstenate crystals, covers
the pseudo-rapidity region $|\eta|<1.5$ in the central barrel.
This coverage is extended to $|\eta|<3$ with the endcaps.
The hadronic calorimeter is made of copper plates and plastic scintillator
sandwiches.
Its acceptance is $|\eta|<2$ in the central barrel and reaches $|\eta|<5.3$
with the endcaps.
Two additional very forward calorimeters ensure coverage in the pseudo-rapidity
range $3 < |\eta| <  5$.
The muon system is located outside the central magnet.
It consists of four layers of detectors (three for tracking and one for 
trigger) covering the pseudo-rapidity range $|\eta|<2.4$ ($|\eta|<1.5$ 
in the barrel).
Very forward calorimeters, including two ZDCs
($3 < |\eta| <  5.2$) and a quartz fiber calorimeter 
($ 5.3 < |\eta| < 6.7$) allow measurements of the collision 
centrality and electromagnetic energy 
(these detectors are not shown in Fig.~\ref{cms_atlas_display}).

\begin{figure}[!t] 
\begin{center} 
\begin{minipage}[t]{16.5 cm} 
\centering
\epsfig{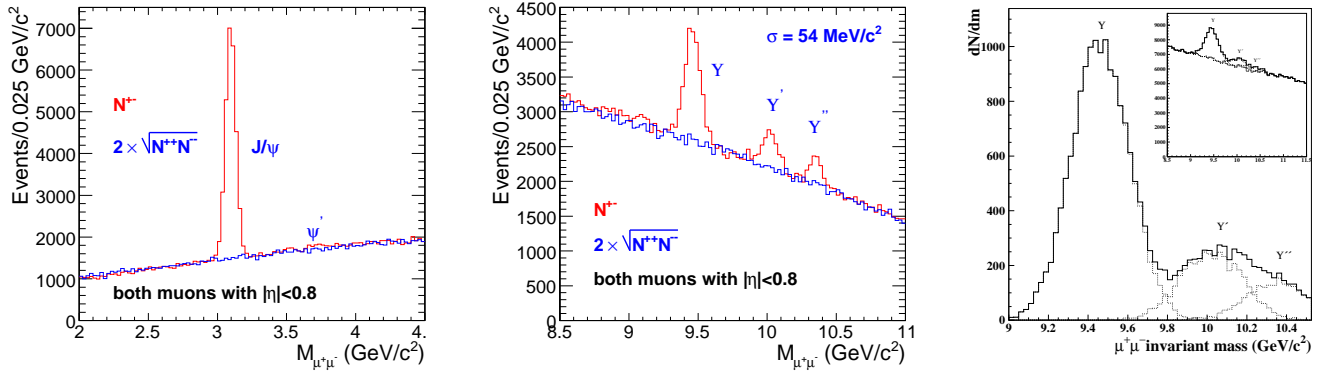} 
\end{minipage} 
\caption{Invariant-mass spectra of opposite-sign and like-sign muon pairs in 
	the $J/\psi$ (left) and $\Upsilon$ (middle) mass regions expected to 
	be measured with the central barrel of the CMS detector within one 
	month of Pb-Pb collisions~\cite{D'Enterria:2007xr}.
	Right: corresponding invariant-mass spectrum of high-mass 
	opposite-sign muon pairs reconstructed with the ATLAS 
	detector~\cite{ATLAS-HI}.}
\label{cms_atlas_imass}
\end{center} 
\end{figure}

Figure~\ref{cms_atlas_imass} shows the dimuon invariant-mass spectra 
expected to be measured for charmonia (left) and bottomonia (middle).
The muon acceptance ranges in $\eta$ from $-0.8$ to $+0.8$ in the 
central barrel and from $\pm 0.8$ to $\pm 2.4$ in the endcaps.
The excellent mass resolution of the apparatus will allow a very clean
separation of quarkonia from the $J/\psi$ to the $\Upsilon(3S)$.
In the acceptance of the central barrel, $J/\psi$ measurements will be 
limited to $p_T > 4$\,GeV/$c$ due to a high $p_T$ threshold induced
by the calorimeters on single muons.
This can be extended to lower $p_T$ when detecting muons in the endcaps
thanks to the extra longitudinal Lorentz boost.
Due to its higher mass, bottomonium will be measured down to zero $p_T$
in the whole $\eta$ acceptance.
Simulation results indicate that the typical number of reconstructed 
$J/\psi$ and $\Upsilon$ will be $184\cdot 10^3$ and $37.7\cdot 10^3$ 
in minimum-bias Pb-Pb collisions for one month of data taking.

\subsubsection{The ATLAS experiment}
Like CMS, ATLAS (A Toroidal LHC ApparatuS) is designed for $p$-$p$ physics. 
The detector capabilities for heavy ion physics have been recently 
investigated~\cite{ATLAS-HI}. 
The design of the detector (Fig.~\ref{cms_atlas_display} right) is similar 
to that of CMS with a tracking system, an EMC, 
a hadronic calorimeter and muon chambers 
placed in a barrel and two endcaps.
The tracking system of ATLAS is composed of silicon pixel detectors,
the SemiConductor Tracker (SCT) made of silicon strip detectors, and the 
Transition Radiation Tracker (TRT).
It is placed inside a $B = 2$\,T solenoidal magnet and covers the 
pseudo-rapidity range $|\eta|<2.5$.
The electromagnetic calorimeter is a liquid-argon device covering 
$|\eta|<4.9$.
It is surrounded by the hadronic calorimeter which consists 
of lead scintillators in the barrel ($|\eta|<1.7$) and liquid argon 
in the endcaps ($1.5 < |\eta| < 3.2$).
Two additional (electromagnetic and hadronic) calorimeters cover the very 
forward region ($3.2 < |\eta| < 4.9$).
The muon spectrometer consists of a toroidal magnet providing a 
$B = 4$\,T field and several muon chambers with different 
technologies in the barrel and in the endcaps.
The acceptance of the muon spectrometer covers $|\eta|<1.0$ in the barrel and
extends to $\eta = \pm 2.5$ in the endcaps.

As an illustration of the ATLAS performance for measuring quarkonium signals, 
the right panel of Fig.~\ref{cms_atlas_imass} shows the invariant-mass 
spectrum of high-mass muon pairs.
In the full $\eta$ acceptance, the mass resolution is 
145\,MeV/$c^2$ at around 10\,GeV/$c^2$.
The expected statistics for $\Upsilon$ for one month of Pb-Pb minimum-bias 
collisions is 10-15\,$\cdot 10^3$ (with a muon-$p_T$ threshold of 
3\,GeV/$c$).
The number of reconstructed $J/\psi$ ranges from 8000 to 
$216\cdot 10^3$ depending on the trigger threshold.

\subsection{Quarkonia in heavy-ion collisions at the SIS-300}
While the future heavy-ion experiments at RHIC-II and at the LHC will focus 
on the study of the QCD phase diagram at large temperatures and small 
chemical potential, the SIS-300 accelerator at FAIR aims at exploring 
the region of moderate temperatures and large baryon chemical potentials.
In this region of the phase diagram, lQCD calculations predict a critical 
endpoint whose location is not precisely known.
Beyond this critical endpoint, for higher baryon chemical potentials and 
lower temperatures, one expects 
a first order phase transition from hadronic to partonic matter.
This regime has been very little explored experimentally so far.
The corresponding beam energy range (from $\sim$10-45\,AGeV) has only 
been partially covered in the past by pioneering experiments at the AGS 
in Brookhaven and, more recently, by experiments at CERN-SPS.
A non-monotonous behavior was observed in the excitation function of several
observables, such as the $\langle K^+\rangle/\langle \pi^+\rangle$ ratio, at a
beam energy of around 30\,AGeV~\cite{:2007fe}.
An interpretation of this intriguing finding as a signal for the onset of
deconfinement is still under debate.
A large low-mass dielectron enhancement was observed at 
40\,AGeV by the CERES experiment~\cite{Adamova:2002kf}, 
corroborating the trend in the 
$\langle K^+\rangle/\langle \pi^+\rangle$ excitation function.
In fact, experimental limitations, both from the accelerators (beam 
intensities) and from the detectors (capabilities to measure rare signals)
prevented comprehensive measurements to be performed at 
such low beam energies.  
Consequently,
charm hadrons (and charmonia)
have never been measured in this energy range.
Such measurements could be especially interesting as this beam-energy 
range is close to the production threshold for $c\bar c$ pairs.
Therefore, the characteristics of the produced charm hadrons are expected
to be particularly sensitive to the properties of the medium in the early 
stage of the collision.
We further note that this region of the phase diagram is well suited
for investigating properties of hadrons in a dense baryonic medium.
This is based on the increasing gap between the putative phase  
boundary and the chemical freeze-out line when decreasing the beam energy
from RHIC and top-SPS to FAIR.
It furthermore suggests that at FAIR energies, contrary to RHIC and 
top-SPS, the produced system spends a relatively long time in a dense 
baryonic phase\footnote{Note, however, that the uncertainty from
lQCD calculations on the phase boundary is still large.}.

\subsubsection{Experimental conditions}
The SIS-300 synchrotron at the FAIR facility will deliver heavy-ion beams 
from 10-35\,AGeV for uranium and up to 45\,AGeV for lighter ions with 
${Z/A} = 0.5$.
Proton beams will be available up to 90\,GeV.
Thanks to the unprecedented high beam intensities of $2\cdot 10^9/{\rm s}$, 
rare probes such as $J/\psi$, whose typical production rate is $2\cdot 10^{-5}$
per central Au-Au collision at 25\,AGeV, will become available.
Detection of such a small signal in such a high-intensity
environment, and among about 1000 charged tracks per event, is highly 
challenging.
It requires excellent detector performance in terms of radiation hardness,
read-out speed, online-event selection, particle identification and 
data processing.
The CBM detector is currently being designed to meet these requirements.

\begin{figure}[!t] 
\begin{center} 
\centering
\epsfig{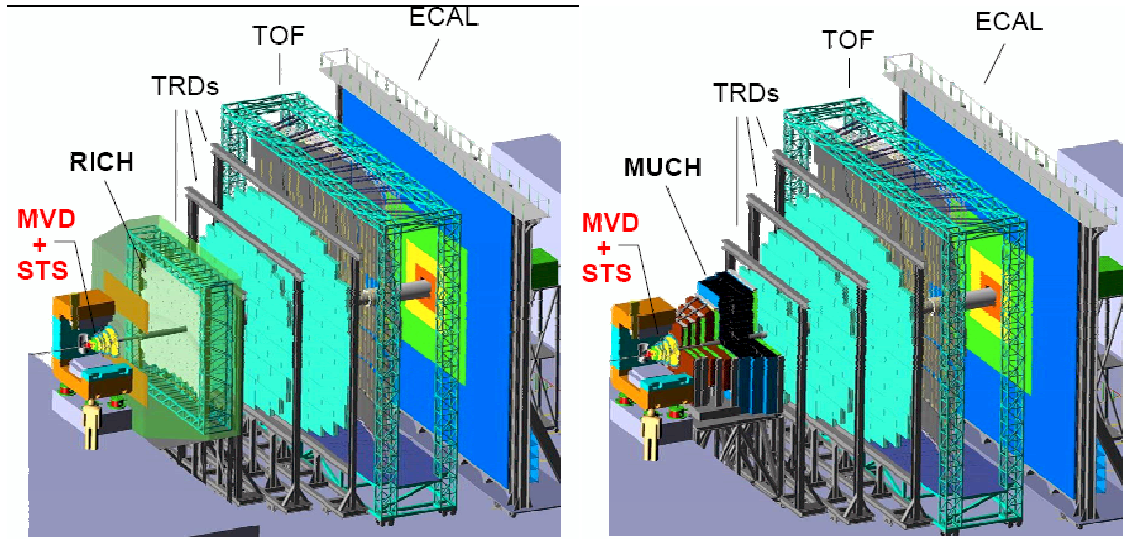} 
\caption{Longitudinal view of the CBM detector. 
	Left: configuration for 
	electron and hadron measurements. 
	Right: configuration for muon measurements.
	\label{cbm_display}} 
\end{center} 
\end{figure} 

\subsubsection{The CBM experiment}
The CBM experiment is a next generation fixed-target universal detector 
capable of measuring hadronic, leptonic and photonic probes in a large 
geometrical acceptance with good vertexing.
Its modular configuration (Fig.~\ref{cbm_display}) will allow quarkonia 
measurements both in the electron and muon channels.
Hadron-track reconstruction and momentum measurements will be performed 
by the Silicon Tracking System (STS) which consists of several stations of 
radiation hard silicon strips located inside the aperture of a 
1\,Tm bending dipole magnet.
Displaced vertices will be measured with high precision by means of a Micro 
Vertex Detector (MVD) based on monolithic active pixel sensors operated in 
vacuum close to the target.
Electrons will be identified by a combination of RICH and TRD detectors 
placed downstream.
For muon measurements, the RICH detector will be replaced by an active 
absorber system (MUCH) made of sandwiches of detection planes and iron 
layers.  The setup is completed by a RPC-based ToF system for 
charged-hadron measurements and by an EMC for neutral particle measurements.

\begin{figure}[h] 
\begin{minipage}[t]{16.5 cm} 
\centering
\epsfig{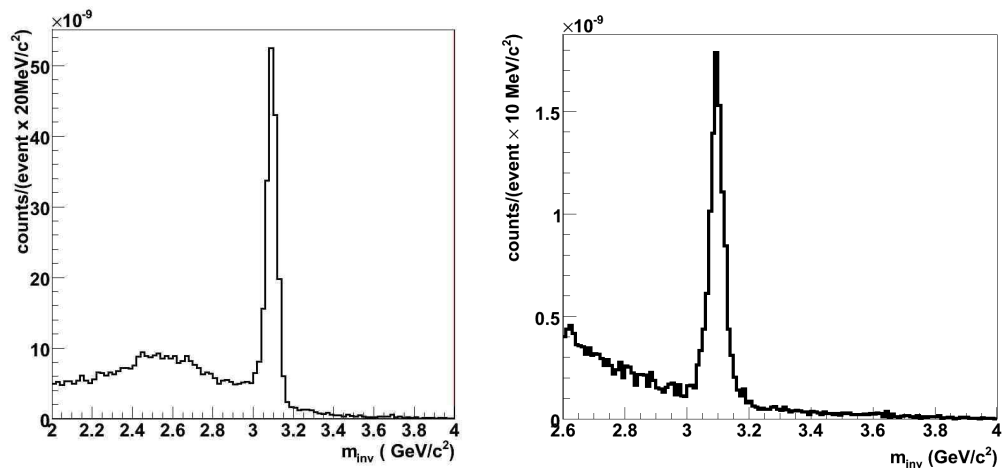} 
\end{minipage} 
\caption{Invariant-mass spectrum showing $J/\psi$ reconstruction in 
	central Au-Au collisions at 25\,AGeV with the 
	CBM detector in the electron (left) and muon (right) 
	channels~\cite{Claudia}.}
\label{cbm_imass}
\end{figure}

Figure~\ref{cbm_imass} shows the $J/\psi$ signal extracted from 
$4\cdot 10^{10}$ central Au-Au collisions at 25\,AGeV.
The corresponding expected data taking time is 11 hours in the muon 
channel and 55 hours in the electron channel.
The simulation assumes a $\pi^\pm$ suppression of $10^4$ from the 
RICH, TRD and ToF systems in the electron channel and a very good suppression
of muons from $\pi^\pm,K^\pm$ decay in the muon channel thanks to kink 
detection with the STS.
The obtained reconstruction performance is comparable in both channels.

%% file: sum.tex
The description of quarkonium production in ultrarelativistic heavy-ion 
collisions remains a challenging task, but new insights keep emerging 
on the three major frontiers, i.e., theoretical, phenomenological
and experimental. In this review we have mostly focused on recent
developments and have attempted to combine the current knowledge toward 
developing a consistent framework which can be checked against lattice
QCD (lQCD) ``data" and realistically applied to experimental data. In 
each of the following paragraphs, we briefly summarize the main points
for the three frontiers and subsequently identify directions of future work.  

First principle lQCD computations of quarkonium correlation
functions at finite temperature, now also available including dynamical 
light quarks, have consolidated the finding that ground-state charmonia 
(bottomonia) are little affected up to temperatures of $\sim$2(3)\,$T_c$. 
This translates into the presence of bound (or resonance) states in the 
pertinent spectral functions when extracted from the correlators 
using probabilistic methods. On the other hand, lQCD computations of
the free energy of a heavy-quark (HQ) pair in a Quark-Gluon Plasma (QGP) 
clearly exhibit the effect of color screening, increasing with
temperature. When these free energies are injected as a potential 
into a Schr\"odinger equation, the $J/\psi$ melts at significantly
smaller temperatures than suggested by the correlators, slightly above 
$T_c$. Higher melting temperatures are obtained from potential models 
when employing the lQCD-based {\em internal} energies, which apparently 
agrees better with the spectral function results. However, in both cases, 
i.e. using free or internal energies, potential models are able to 
reproduce the approximate constancy of the correlation functions. 
This has recast some doubt on whether ground-state charmonia really
survive well above $T_c$. One of the main reasons for this redundancy 
is that a reduced binding can be compensated by a reduced open-charm 
threshold, corresponding to a reduced thermal charm-quark mass - an
effect which is well established for bound states in electromagnetic 
plasmas. Furthermore, nonperturbative rescattering effects typically 
lead to a large enhancement in the continuum part of the quarkonium 
spectral functions, especially close to the quark-antiquark threshold. 
Thus a reliable determination of the in-medium charm-quark mass, as well 
as of finite-width effects (for both charmonia and open-charm states), 
is needed to improve the evaluation of quarkonium spectral functions. 
In addition, it will be important to develop better criteria for
the applicability of potential models at finite temperature, e.g., by
setting up effective field theories if suitable scale separations
can be applied. 

In-medium binding energies turn out to play an important role in 
quarkonium dissociation reactions. For strong binding, leading-order 
gluo-dissociation is the most important process. For weakly bound
quarkonia, next-to-leading order processes (``quasifree" destruction) 
become dominant, especially close to the ``melting" temperatures; both
thermal quarks and gluons contribute to these reactions. The pertinent 
inelastic widths of charmonia can reach up to several hundred MeV for
temperatures as low as 1.5\,$T_c$ (depending on the value of
$\alpha_s$), translating into lifetimes (well) below 1\,fm/$c$. 
For phenomenology in heavy-ion collisions this implies that quarkonium
dissociation will be operative well before the pertinent bound states
dissolve due to color screening. It also reiterates that a consistent 
treatment of screening and finite-width effects is mandatory. 
Furthermore, the spectral properties of open heavy-flavor states are expected 
to change appreciably in the medium, with direct consequences for quarkonia:
while effective HQ (or heavy-meson) masses determine the onset of the open
heavy-flavor continuum, finite widths can open additional quarkonium
decay channels even below threshold. Taking guidance from well-developed 
concepts in electromagnetic plasma physics, 
finite temperature $T$-matrix approaches seem to be a suitable tool
to incorporate all of these aspects. 

Equilibrium properties of quarkonia in medium (masses and widths) are 
key ingredients to transport theoretical evaluations of quarkonium 
observables in heavy-ion collisions, especially if carried out in a 
thermally evolving bulk medium (which appears to be a good approximation 
at SPS and RHIC). A quantitative understanding of cold-nuclear-matter 
effects on initial quarkonium spectra is mandatory to extract 
``anomalous" suppression in the subsequently formed hot and dense 
medium. Cold-nuclear-matter effects include modifications of the nuclear 
parton distribution functions (``shadowing"), initial-state energy loss, 
Cronin effect and nuclear absorption. Experimental evidence is mounting 
that the relative importance of these effects changes significantly with 
collision energy and rapidity, and thus empirical constraints from 
$p$-$A$ (or $d$-$A$) data are indispensable. In addition, secondary 
charmonium production (``regeneration") via $c$-$\bar c$ coalescence (in 
the QGP or at hadronization) is expected to become increasingly relevant 
with increasing collision energy. This implies a coupling of charmonium 
yields and spectra to the abundance and spectra of open charm, which 
themselves are subject to medium modifications. This further adds to the 
complexity of the calculations reinforcing the need of comprehensive 
analyses of open- and hidden-flavor observables in collision centrality, 
energy, system size and 3-momentum. The current state of phenomenology 
suggests that (a) the observed $J/\psi$ suppression at the SPS mostly 
occurs in a hot/dense medium at energy densities above the critical one 
found in lQCD, with initial temperatures of $T_0$$\simeq$220-250~MeV;
(b) at RHIC energies, a significant regeneration component develops, 
accounting for roughly 50\% of the measured $J/\psi$ yield in central 
Au-Au; this assertion is corroborated by the observed narrowing in the 
rapidity distributions and an approximately constant average transverse 
momentum with centrality, with regeneration contributions 
prevalent at low $p_T$. If the regeneration yield becomes dominant at the 
LHC, a qualitatively new (increasing) centrality dependence of the nuclear 
modification factor may emerge (depending on the open-charm cross section). 
A more direct probe of color screening is provided by the $\Upsilon$: if 
it retains its vacuum binding energy up to $\sim$2\,$T_c$, it will be rather 
inert under RHIC conditions; however, an in-medium reduction in its binding 
increases the dissociation rates noticeably which might suppress $\Upsilon$ 
production as much as $J/\psi$ production. Such a signal benefits from 
the absence of bottomonium regeneration at RHIC. A similar feature could 
be even more pronounced at the LHC, especially if regeneration of the 
$J/\psi$ is large, while it is still expected to be small for $\Upsilon$. 
Transverse-momentum spectra will further illuminate the different 
mechanisms, and/or reveal new ones. E.g., first RHIC data for $J/\psi$'s 
at high $p_T\ge5$\,GeV indicate a nuclear modification factor close 
to one, very different from light-hadron spectra. Elliptic flow could be
another good discriminator between direct and secondary production, but 
only if the regeneration component in semi-central collision persists to 
sufficiently large $p_T$, above at least 2\,GeV.
Finally, excited $\Psi$ and $\Upsilon$ states are of great interest to
test suppression patterns as regeneration contributions are usually
suppressed due to their larger mass.

The quantitative realization of the above observables requires an 
advanced experimentation which will become available at RHIC-II and the 
LHC. While the former appears to operate at an energy suited to study 
the properties of a strongly coupled, liquid-like QGP, the latter may, 
for the first time, reach into a weakly coupled gas-like regime. At 
lower energies, the CBM experiment at FAIR will explore strongly
interacting matter at maximum net baryon density; little (nothing) is 
known theoretically (experimentally) about charmonium properties under 
these conditions.  
In view of these developments, and with an adequate theoretical support, 
quarkonia promise to remain a rich, and eventually quantitative, probe 
of high-energy density matter in heavy-ion collisions.

\vspace{0.5cm}

\noindent
{\bf Acknowledgment}\\
We are indebted to our colleagues A.~Beraudo for a careful reading of
sections 2 and 3, to M.~Laine and G.~R\"opke for valuable comments on 
section 2, to J.-P.~Lansberg for useful comments on section 3.2, and to 
A.~Andronic, R.~Averbeck and P.~Senger for useful comments on section 4. 
The work of RR has been supported by the U.S. National Science Foundation
(NSF) with a CAREER award under grant no.~PHY-0449489, by NSF grant 
no.~PHY-0969394, and by a Bessel Research Award from the A.-v.-Humboldt 
foundation.
The work of DB has been supported in part by the Polish
Ministry for Science and Higher Education under grant No.
N N 202 0953 33 and by RFBR grant No. 08-02-01003-a.

%% file: references.tex

%% file: PPNP-june2010.bbl
\begin{thebibliography}{99}
\setlength{\itemsep}{-0.0cm}
\bibitem{Matsui:1986dk}
  T.~Matsui and H.~Satz,
 Phys.\ Lett.\ B \textbf{178} (1986) 416.
\bibitem{Gerschel:1998zi}
  C.~Gerschel and J.~H\"ufner,
  Ann.\ Rev.\ Nucl.\ Part.\ Sci.\  {\bf 49} (1999) 255.
\bibitem{Vogt:1999cu}
  R.~Vogt,
  Phys.\ Rept.\  {\bf 310} (1999) 197.
\bibitem{Satz:2005hx}
  H.~Satz,
  J.\ Phys.\ G {\bf 32} (2006) R25.
\bibitem{Kharzeev:2007ej}
  D.~E.~Kharzeev,
  J.\ Phys.\ G {\bf 34} (2007) S445.
\bibitem{Kluberg:2005yh}
  L.~Kluberg,
  Eur.\ Phys.\ J.\  C {\bf 43} (2005) 145.
\bibitem{Amsler:2008zzb}
  C.~Amsler {\it et al.}  [Particle Data Group],
  Phys.\ Lett.\  B {\bf 667} (2008) 1.
\bibitem{Eichten:1979ms}
  E.~Eichten, K.~Gottfried, T.~Kinoshita, K.~D.~Lane and T.~M.~Yan,
  Phys.\ Rev.\  D {\bf 21} (1980) 203.
\bibitem{Brambilla:2004wf}
  N.~Brambilla {\it et al.},
CERN Yellow Rep. 2005-005, [{\sf arXiv:hep-ph/0412158}].
\bibitem{Bali:2000gf}
  G.~S.~Bali,
  Phys.\ Rept.\  {\bf 343} (2001) 1.
\bibitem{Kaczmarek:2005ui}
  O.~Kaczmarek and F.~Zantow,
  Phys.\ Rev.\  D {\bf 71} (2005) 114510.
\bibitem{Adler:2004ta}
  S.~S.~Adler {\it et al.}  [PHENIX Collaboration],
  Phys.\ Rev.\ Lett.\  {\bf 94} (2005) 082301. 
\bibitem{Shahoyan:2007zz}
  R.~Shahoyan  [NA60 Collaboration],
  J.\ Phys.\ G {\bf 34} (2007) S1029.
\bibitem{Levai:1994dx}
P.~Levai, B.~M{\"u}ller and X.-N. Wang, {Phys. Rev. C} 
\textbf{51} (1995) 3326.
\bibitem{Shuryak:1978ij}
  E.~V.~Shuryak,
  Phys.\ Lett.\  B {\bf 78} (1978) 150
  [Sov.\ J.\ Nucl.\ Phys.\  {\bf 28} (1978) 408 / YAFIA 28 (1978) 796]. 
\bibitem{Cleymans:1984zs}
  J.~Cleymans and C.~Vanderzande,
  Phys.\ Lett.\  B {\bf 147} (1984) 186.
\bibitem{Mott:1968}
N.~F.~Mott,
Rev. Mod. Phys. {\bf 40}, 677 (1968).
\bibitem{Redmer:1997}
R.~Redmer, Phys. Rep. {\bf 282} (1997) 35.
\bibitem{Baglin:1990iv}
  C.~Baglin {\it et al.}  [NA38 Collaboration],
  Phys.\ Lett.\  B {\bf 220} (1989) 471.
\bibitem{Gerschel:1988wn}
  C.~Gerschel and J.~H\"ufner,
  Phys.\ Lett.\  B {\bf 207} (1988) 253.
\bibitem{Gavin:1988hs}
  S.~Gavin, M.~Gyulassy and A.~Jackson,
  Phys.\ Lett.\  B {\bf 207} (1988) 257.
\bibitem{Vogt:1988fj}
  R.~Vogt, M.~Prakash, P.~Koch and T.~H.~Hansson,
  Phys.\ Lett.\  B {\bf 207} (1988) 263.
\bibitem{Badier:1983dg}
  J.~Badier {\it et al.}  [NA3 Collaboration],
  Z.\ Phys.\  C {\bf 20} (1983) 101.
\bibitem{Abreu:1998ee}
  M.~C.~Abreu {\it et al.}  [NA38 Collaboration],
  Phys.\ Lett.\  B {\bf 444} (1998) 516.
\bibitem{Alde:1990wa}
  D.~M.~Alde {\it et al.},
  Phys.\ Rev.\ Lett.\  {\bf 66} (1991) 133.
\bibitem{Baglin:1991vb}
  C.~Baglin {\it et al.},
  Phys.\ Lett.\  B {\bf 270} (1991) 105.
\bibitem{Gerschel:1993uh}
  C.~Gerschel and J.~H\"ufner,
  Z.\ Phys.\  C {\bf 56} (1992) 171.
\bibitem{Gonin:1996wn}
  M.~Gonin {\it et al.}  [NA50 Collaboration],
  Nucl.\ Phys.\  A {\bf 610} (1996) 404c.
\bibitem{Abreu:1997jh}
  M.~C.~Abreu {\it et al.}  [NA50 Collaboration],
  Phys.\ Lett.\  B {\bf 410} (1997) 337.
\bibitem{Blaizot:1996nq}
  J.~P.~Blaizot and J.~Y.~Ollitrault,
  Phys.\ Rev.\ Lett.\  {\bf 77} (1996) 1703.
\bibitem{Kharzeev:1996yx}
  D.~Kharzeev, C.~Lourenco, M.~Nardi and H.~Satz,
  Z.\ Phys.\  C {\bf 74} (1997) 307.
\bibitem{Alessandro:2004ap}
  B.~Alessandro {\it et al.}  [NA50 Collaboration],
  Eur.\ Phys.\ J.\  C {\bf 39} (2005) 335.
\bibitem{Arnaldi:2006ee}
  R.~Arnaldi {\it et al.}  [NA60 Collaboration],
  Nucl.\ Phys.\  A {\bf 774} (2006) 711.

\bibitem{Arnaldi:2007aa}
  R.~Arnaldi {\it et al.}  [NA60 Collaboration],
  Nucl.\ Phys.\  A {\bf 783} (2007) 261.
\bibitem{Abreu:2000ni}
  M.C.~Abreu {\it et al.}  [NA50 Collaboration],
  Phys.\ Lett.\  B {\bf 477} (2000) 28.
\bibitem{Adler:2003rc}
  S.S.~Adler {\it et al.}  [PHENIX Collaboration],
  Phys.\ Rev.\  C {\bf 69} (2004) 014901. 
\bibitem{Adare:2006ns}
A.~Adare {\it et al.}  [PHENIX Collaboration],
  Phys.\ Rev.\ Lett.\  {\bf 98} (2007) 232301.
\bibitem{Grandchamp:2003uw}
  L.~Grandchamp, R.~Rapp and G.~E.~Brown,
  Phys.\ Rev.\ Lett.\  {\bf 92} (2004) 212301.
\bibitem{Thews:2005fs}
  R.~L.~Thews,
  Eur.\ Phys.\ J.\  C {\bf 43} (2005) 97.
\bibitem{Leitch:2006ff}
  M.~J.~Leitch,
  AIP Conf.\ Proc.\  {\bf 892} (2007) 404.
\bibitem{BraunMunzinger:2000px}
  P.~Braun-Munzinger and J.~Stachel,
  Phys.\ Lett.\  B {\bf 490} (2000) 196.
\bibitem{Thews:2000rj}
  R.~L.~Thews, M.~Schroedter and J.~Rafelski,
  Phys.\ Rev.\  C {\bf 63} (2001) 054905.
\bibitem{Gorenstein:2000ck}
  M.~I.~Gorenstein, A.~P.~Kostyuk, H.~St\"ocker and W.~Greiner,
  Phys.\ Lett.\  B {\bf 509} (2001) 277.
\bibitem{Grandchamp:2001pf}
  L.~Grandchamp and R.~Rapp,
  Phys.\ Lett.\  B {\bf 523} (2001) 60.
\bibitem{DeTar:1987xb}
  C.~E.~DeTar and J.~B.~Kogut,
  Phys.\ Rev.\  D {\bf 36} (1987) 2828.
\bibitem{Born:1991zz}
 K.~D.~Born, S.~Gupta, A.~Irb\"ack, F.~Karsch, E.~Laermann,
B.~Petersson and H.~Satz [MT(c) Collaboration],
  Phys.\ Rev.\ Lett.\  {\bf 67} (1991) 302.
\bibitem{Asakawa:2000tr}
  M.~Asakawa, T.~Hatsuda and Y.~Nakahara,
  Prog.\ Part.\ Nucl.\ Phys.\  {\bf 46} (2001) 459.
\bibitem{Asakawa:2003re}
  M.~Asakawa and T.~Hatsuda,
  Phys.\ Rev.\ Lett.\  {\bf 92} (2004) 012001.
\bibitem{Datta:2003ww}
  S.~Datta, F.~Karsch, P.~Petreczky and I.~Wetzorke,
  Phys.\ Rev.\  D {\bf 69} (2004) 094507.
\bibitem{Morrin:2005zq}
 R.~Morrin, A.~P.~O Cais, M.~B.~Oktay, M.~J.~Peardon,
J.~I.~Skullerud, G.~Aarts and C.~R.~Allton,
  PoS {\bf LAT2005} (2006) 176.
\bibitem{Jakovac:2006sf}
  A.~Jakovac, P.~Petreczky, K.~Petrov and A.~Velytsky,
  Phys.\ Rev.\  D {\bf 75} (2007) 014506.
\bibitem{Aarts:2007pk}
  G.~Aarts, C.~Allton, M.~B.~Oktay, M.~Peardon and J.~I.~Skullerud,
  Phys. Rev.  D {\bf 76} (2007) 094513;\\
  M.~B.~Oktay, M.~J.~Peardon, J.~I.~Skullerud, G.~Aarts and C.~R.~Allton,
  PoS LAT2007 (2007) 227.
\bibitem{Petrov:2005ej}
  K.~Petrov, A.~Jakovac, P.~Petreczky and A.~Velytsky,
  PoS {\bf LAT2005} (2006) 153.
\bibitem{McLerran:1980pk}
  L.~D.~McLerran and B.~Svetitsky,
  Phys.\ Lett.\  B {\bf 98} (1981) 195.
\bibitem{Philipsen:2008qx}
  O.~Philipsen,
  Nucl.\ Phys.\  A {\bf 820} (2009) 33C.
\bibitem{Kaczmarek:2002mc}
  O.~Kaczmarek, F.~Karsch, P.~Petreczky and F.~Zantow,
  Phys.\ Lett.\  B {\bf 543} (2002) 41.
\bibitem{Petreczky:2004pz}
  P.~Petreczky and K.~Petrov,
  Phys.\ Rev.\  D {\bf 70} (2004) 054503.
\bibitem{Escobedo:2008sy}
  M.~A.~Escobedo and J.~Soto,
  arXiv:0804.0691 [hep-ph].
\bibitem{Brambilla:2008cx}
  N.~Brambilla, J.~Ghiglieri, A.~Vairo and P.~Petreczky,
  Phys.\ Rev.\  D {\bf 78} (2008) 014017.
\bibitem{Karsch:1987pv}
  F.~Karsch, M.T.~Mehr and H.~Satz,
  Z.\ Phys.\  C {\bf 37} (1988) 617.
\bibitem{Kaczmarek:2004gv}
  O.~Kaczmarek, F.~Karsch, F.~Zantow and P.~Petreczky,
  Phys.\ Rev.\  D {\bf 70} (2004) 074505
  [Erratum-ibid.\  D {\bf 72} (2005) 059903].
\bibitem{Digal:2001ue}
  S.~Digal, P.~Petreczky and H.~Satz,
  Phys.\ Rev.\  D {\bf 64} (2001) 094015.
\bibitem{Shuryak:2004tx}
  E.V.~Shuryak and I.~Zahed,
  Phys.\ Rev.\  D {\bf 70} (2004) 054507.
\bibitem{Wong:2004zr}
  C.Y.~Wong,
  Phys.\ Rev.\  C {\bf 72} (2005) 034906.
\bibitem{Alberico:2005xw}
  W.M.~Alberico, A.~Beraudo, A.~De Pace and A.~Molinari,
  Phys.\ Rev.\  D {\bf 72} (2005) 114011.
\bibitem{Mannarelli:2005pz}
  M.~Mannarelli and R.~Rapp,
{Phys.\ Rev.\ C} \textbf{72} (2005) 064905.
\bibitem{Satz:2008zc}
  H.~Satz,
  J.\ Phys.\ G {\bf 36} (2009) 064011. 
\bibitem{Rapp:2002pn}
  R.~Rapp,
 Eur. Phys. J A {\bf 18} (2003) 459.
\bibitem{Mocsy:2005qw}
  A.~Mocsy and P.~Petreczky,
  Phys.\ Rev.\  D {\bf 73} (2006) 074007.
\bibitem{Alberico:2006vw}
  W.M.~Alberico, A.~Beraudo, A.~De Pace and A.~Molinari,
  Phys.\ Rev.\  D {\bf 75} (2007) 074009.
\bibitem{Cabrera:2006wh}
  D.~Cabrera and R.~Rapp,
  Phys.\ Rev.\  D {\bf 76} (2007) 114506.
\bibitem{Ebeling:1986}
  W.~Ebeling, W.-D.~Kraeft, D.~Kremp, G.~R\"opke,
  {\it Quantum Statistics of Charged Many-Particle Systems},
  Plenum, New York (1986).
\bibitem{Machleidt:1989tm}
  R.~Machleidt,
  Adv.\ Nucl.\ Phys.\  {\bf 19} (1989) 189.
\bibitem{Kaczmarek:2005gi}
  O.~Kaczmarek and F.~Zantow,
  arXiv:hep-lat/0506019.
\bibitem{Rapp:2008fv}
  R.~Rapp, D.~Cabrera, V.~Greco, M.~Mannarelli and H.~van Hees,
  arXiv:0806.3341 [hep-ph].
\bibitem{Umeda:2007hy}
  T.~Umeda,
  Phys.\ Rev.\  D {\bf 75} (2007) 094502.
\bibitem{Petreczky:2005nh}
  P.~Petreczky and D.~Teaney,
  Phys.\ Rev.\  D {\bf 73} (2006) 014508.
\bibitem{Mocsy:2007jz}
  A.~Mocsy and P.~Petreczky,
  Phys.\ Rev.\ Lett.\  {\bf 99} (2007) 211602;
  Phys.\ Rev.\  D {\bf 77} (2008) 014501.
\bibitem{Alberico:2007rg}
  W.M.~Alberico, A.~Beraudo, A.~De Pace and A.~Molinari,
  Phys.\ Rev.\  D {\bf 77} (2008) 017502.
\bibitem{Datta:2006ua}
  S.~Datta, A.~Jakovac, F.~Karsch and P.~Petreczky,
  AIP Conf.\ Proc.\  {\bf 842} (2006) 35.
\bibitem{Laine:2006ns}
  M.~Laine, O.~Philipsen, P.~Romatschke and M.~Tassler,
  JHEP {\bf 0703} (2007) 054.
\bibitem{Beraudo:2007ky}
  A.~Beraudo, J.P.~Blaizot and C.~Ratti,
  Nucl.\ Phys.\  A {\bf 806} (2008) 312.
\bibitem{Laine:2007qy}
  M.~Laine, O.~Philipsen and M.~Tassler,
  JHEP {\bf 0709} (2007) 066.
\bibitem{Laine:2007gj}
  M.~Laine,
  JHEP {\bf 0705} (2007) 028.
\bibitem{Burnier:2007qm}
  Y.~Burnier, M.~Laine and M.~Vepsalainen,
  JHEP {\bf 0801} (2008) 043.
\bibitem{Petreczky:2008px}
  P.~Petreczky,
   Eur.\ Phys.\ J.\  C {\bf 62} (2009) 85.
\bibitem{Wong:2006bx}
  C.~Y.~Wong and H.~W.~Crater,
  Phys.\ Rev.\  D {\bf 75} (2007) 034505.
\bibitem{Bhanot:1979vb}
  G.~Bhanot and M.~E.~Peskin,
  Nucl.\ Phys.\  B {\bf 156} (1979) 391.
\bibitem{Peskin:1979va}
  M.~E.~Peskin,
  Nucl.\ Phys.\  B {\bf 156} (1979) 365.
\bibitem{Arleo:2004ge}
  F.~Arleo, J.~Cugnon and Y.~Kalinovsky,
  Phys.\ Lett.\  B {\bf 614} (2005) 44.
\bibitem{Blaschke:2004dv}
  D.~Blaschke, Y.~Kalinovsky and V.~Yudichev,
  Lect.\ Notes Phys.\  {\bf 647} (2004) 366.
\bibitem{Grandchamp:2002wp}
  L.~Grandchamp and R.~Rapp,
  Nucl.\ Phys.\  A {\bf 709} (2002) 415.
\bibitem{Ropke:1988zz}
  G.~R\"opke, D.~Blaschke and H.~Schulz,
  Phys.\ Rev.\  D {\bf 38} (1988) 3589.
\bibitem{Kharzeev:1995ju}
  D.~Kharzeev, L.~D.~McLerran and H.~Satz,
  Phys.\ Lett.\  B {\bf 356} (1995) 349.
\bibitem{Park:2007zza}
  Y.~Park, K.I.~Kim, T.~Song, S.H.~Lee and C.Y.~Wong,
  Phys.\ Rev.\  C {\bf 76} (2007) 044907.
\bibitem{Song:2007gm}
  T.~Song, Y.~Park, S.H.~Lee and C.Y.~Wong,
  Phys.\ Lett.\  B {\bf 659} (2008) 621.
\bibitem{Zhao:2007hh}
  X.~Zhao and R.~Rapp,
  Phys.\ Lett.\  B {\bf 664} (2008) 253.
\bibitem{Combridge:1978kx}
  B.~L.~Combridge,
  Nucl.\ Phys.\  B {\bf 151} (1979) 429.
\bibitem{Grandchamp:2005yw}
  L.~Grandchamp, S.~Lumpkins, D.~Sun, H.~van Hees and R.~Rapp,
  Phys.\ Rev.\  C {\bf 73} (2006) 064906.
\bibitem{Dumitru:2007hy}
  A.~Dumitru, Y.~Guo and M.~Strickland,
  Phys.\ Lett.\  B {\bf 662}, 37 (2008)
\bibitem{Burnier:2009yu}
  Y.~Burnier, M.~Laine and M.~Vepsalainen,
  Phys.\ Lett.\  B {\bf 678}, 86 (2009)
\bibitem{Philipsen:2009wg}
  O.~Philipsen and M.~Tassler,
  arXiv:0908.1746 [hep-ph].
\bibitem{Riek:2010fk}
  F.~Riek and R.~Rapp,
  arXiv:1005.0769 [hep-ph].
\bibitem{Morita:2007hv}
  K.~Morita and S.~H.~Lee,
  Phys.\ Rev.\  C {\bf 77}, 064904 (2008)
  [arXiv:0711.3998 [hep-ph]].
\bibitem{Blaschke:2002ww}
  D.~Blaschke, G.~Burau, Yu.~Kalinovsky and T.~Barnes,
  Eur.\ Phys.\ J.\  A {\bf 18} (2003) 547.
\bibitem{Barnes:2003dg}
  T.~Barnes, E.S.~Swanson, C.Y.~Wong and X.M.~Xu,
  Phys.\ Rev.\  C {\bf 68} (2003) 014903.
\bibitem{Ivanov:2003ge}
  M.~A.~Ivanov, J.~G.~K\"orner and P.~Santorelli,
  Phys.\ Rev.\  D {\bf 70} (2004) 014005.
\bibitem{Bourque:2004av}
  A.~Bourque, C.~Gale and K.L.~Haglin,
  Phys.\ Rev.\  C {\bf 70} (2004) 055203.
\bibitem{Bourque:2008ta}
  A.~Bourque and C.~Gale,
  Phys.\ Rev.\  C {\bf 78} (2008) 035206.
  [arXiv:0802.2738 [hep-ph]].
\bibitem{Oh:2007ej}
  Y.~Oh, W.~Liu and C.M.~Ko,
  Phys.\ Rev.\  C {\bf 75} (2007) 064903.
  [arXiv:nucl-th/0702077].
\bibitem{Navarra:2001pz}
  F.S.~Navarra, M.~Nielsen and M.~R.~Robilotta,
  Phys.\ Rev.\  C {\bf 64} (2001) 021901.
  [arXiv:nucl-th/0103051].
\bibitem{Bourque:2008es}
  A.~Bourque and C.~Gale,
  Phys.\ Rev.\  C {\bf 80} (2009) 015204.
\bibitem{Matinyan:1998cb}
  S.~G.~Matinyan and B.~M\"uller,
  Phys.\ Rev.\  C {\bf 58} (1998) 2994.
\bibitem{Haglin:1999xs}
  K.L.~Haglin,
  Phys.\ Rev.\  C {\bf 61} (2000) 031902.
  [arXiv:nucl-th/9907034].
\bibitem{Song:2005yd}
  T.~Song and S.~H.~Lee,
  Phys.\ Rev.\  D {\bf 72} (2005) 034002.
\bibitem{Wong:1999zb}
  C.Y.~Wong, E.S.~Swanson and T.~Barnes,
  Phys.\ Rev.\  C {\bf 62} (2000) 045201.
\bibitem{Martins:1994hd}
  K.~Martins, D.~Blaschke and E.~Quack,
{  Phys.\ Rev.\ C} \textbf{51} (1995) 2723.
\bibitem{Hilbert:2007hc}
  J.P.~Hilbert, N.~Black, T.~Barnes and E.S.~Swanson,
  Phys.\ Rev.\  C {\bf 75} (2007) 064907

\bibitem{Prorok:2008zm}
  D.~Prorok, L.~Turko and D.~Blaschke,
  AIP Conf.\ Proc.\  {\bf 1038} (2008) 73.
\bibitem{Blaschke:2000zm}
 D.~B.~Blaschke, G.~R.~G.~Burau, M.~A.~Ivanov, Yu.~L.~Kalinovsky and 
 P.~C.~Tandy,
  arXiv:hep-ph/0002047.
\bibitem{Blaschke:2008mu}
  D.B.~Blaschke, H.~Grigorian and Yu.L.~Kalinovsky,
  arXiv:0808.1705 [hep-ph].
\bibitem{Sibirtsev:2000aw}
  A.~Sibirtsev, K.~Tsushima and A.W.~Thomas,
  Phys.\ Rev.\  C {\bf 63} (2001) 044906.
\bibitem{Sibirtsev:1999jr}
  A.~Sibirtsev, K.~Tsushima, K.~Saito and A.~W.~Thomas,
  Phys.\ Lett.\  B {\bf 484} (2000) 23.
\bibitem{Tsushima:2000cp}
  K.~Tsushima, A.~Sibirtsev, K.~Saito, A.~W.~Thomas and D.~H.~Lu,
  Nucl.\ Phys.\  A {\bf 680} (2001) 280.
\bibitem{Burau:2000pn}
  G.R.G.~Burau, D.B.~Blaschke and Y.L.~Kalinovsky,
  Phys.\ Lett.\ B {\bf 506} (2001) 297.
\bibitem{Friman:2002fs}
  B.~Friman, S.~H.~Lee and T.~Song,
  Phys.\ Lett.\  B {\bf 548} (2002) 153.
\bibitem{Blaschke:2003ji}
D.~Blaschke, G.~Burau, Yu.~Kalinovsky, V.~Yudichev,
Prog.\ Theor.\ Phys.\ Suppl. {\bf 149} (2003) 182.
\bibitem{Fuchs:2004fh}
  C.~Fuchs, B.V.~Martemyanov, A.~Faessler and M.I.~Krivoruchenko,
  Phys.\ Rev.\  C {\bf 73} (2006) 035204.
\bibitem{Blaschke:2000er}
  D.~Blaschke, G.~Burau and Yu.L.~Kalinovsky,
  arXiv:nucl-th/0006071.
\bibitem{Schlanges:1988}
 M.~Schlanges, Th.~Bornath and D.~Kremp,
 Phys. Rev. A {\bf 38} (1988) 2174.
\bibitem{Zimmermann:1978}
  R.~Zimmermann, K.~Kilimann, W.~D.~Kraeft, D.~Kremp and G.~R\"opke,
  Phys. Stat. Sol. (b) {\bf 90} (1978) 175.
\bibitem{Ropke:1982}
  G.~R\"opke {\it et al.}, 
Nucl. Phys. A {\bf 379} (1982) 536.
\bibitem{Ropke:1983}
  G.~R\"opke {\it et al.}, 
Nucl. Phys. A {\bf 399} (1983) 587.
\bibitem{Horowitz:1985tx}
  C.~J.~Horowitz, E.~J.~Moniz and J.~W.~Negele,
  Phys.\ Rev.\  D {\bf 31} (1985) 1689.
\bibitem{Ropke:1986qs}
  G.~R\"opke, D.~Blaschke and H.~Schulz,
  Phys.\ Rev.\  D {\bf 34} (1986) 3499.
\bibitem{Baym:1962sx}
  G.~Baym,
  Phys.\ Rev.\  {\bf 127} (1962) 1391.
\bibitem{Ebeling:1989}
  W.~Ebeling and K.~Kilimann,
  Z. Naturforsch. {\bf 44a} (1989) 519.
\bibitem{Rogers:1970}
  F.~J.~Rogers, H.~C.~Garboske~Jr., and D.~J.~Harwood,
  Phys. Rev. A {\bf 1} (1970) 1577.
\bibitem{Kraeft:1990}
  W.~D.~Kraeft, D.~Kremp, K.~Kilimann, and H.~E.~DeWitt, 
  Phys. Rev. A {\bf 42} (1990) 2340.
\bibitem{Fehrenbach:1981}
G.~W.~Fehrenbach, W.~Sch\"afer, J.~Treusch, and R.~G.~Ulbrich,
Phys. Rev. Lett. {\bf 49} (1981) 1281.
\bibitem{Ecker:1956}
G.~Ecker and W.~Weitzel, Ann. Phys. (Lpz.) {\bf 17} (1956) 126.
\bibitem{Dixit:1989vq}
  V.~V.~Dixit,
  Mod.\ Phys.\ Lett.\  A {\bf 5} (1990) 227.

\bibitem{Ratti:2005jh}
 C.~Ratti, M.~A.~Thaler and W.~Weise,
  Phys.\ Rev.\  D {\bf 73} 014019 (2006).
\bibitem{Hansen:2006ee}
  H.~Hansen {\it et al.}, 
  Phys.\ Rev.\  D {\bf 75} 065004 (2007).
\bibitem{Jankowski:2009kr}
  J.~Jankowski, D.~Blaschke and H.~Grigorian,
  arXiv:0911.1534 [hep-ph].
\bibitem{Ropke:1979}
  G.~R\"opke and R.~Der,
  Phys. stat. sol. (b) {\bf 92} (1979) 501.
\bibitem{Ropke:1984}
  G.~R\"opke {\it et al.}, 
Nucl. Phys. A {\bf 424} (1984) 594.
\bibitem{Kharzeev:1994pz}
  D.~Kharzeev and H.~Satz,
  Phys.\ Lett.\  B {\bf 334} (1994) 155.
\bibitem{Arleo:2001mp}
  F.~Arleo, P.B.~Gossiaux, T.~Gousset and J.~Aichelin,
  Phys.\ Rev.\  D {\bf 65} (2002) 014005.
\bibitem{Blaschke:2006ct}
  D.~Blaschke and V.~L.~Yudichev,
  AIP Conf.\ Proc.\  {\bf 842} (2006) 38.
\bibitem{Satz:2006uh}
  H.~Satz,
  arXiv:hep-ph/0602245.
\bibitem{Miyazawa:1979vx}
  H.~Miyazawa,
  Phys.\ Rev.\  D {\bf 20} (1979) 2953.
\bibitem{Blaschke:1984yj}
  D.~Blaschke, F.~Reinholz, G.~R\"opke and D.~Kremp,
  Phys.\ Lett.\  B {\bf 151} (1985) 439.
\bibitem{Ropke:1988bx}
  G.~R\"opke, D.~Blaschke and H.~Schulz,
  Phys.\ Lett.\  B {\bf 202} (1988) 479.
\bibitem{Capella:2000zp}
  A.~Capella, E.~G.~Ferreiro and A.~B.~Kaidalov,
  Phys.\ Rev.\ Lett.\  {\bf 85} (2000) 2080.
\bibitem{Cassing:1996zb}
  W.~Cassing and C.~M.~Ko,
  Phys.\ Lett.\  B {\bf 396} (1997) 39.
\bibitem{Spieles:1999pm}
  C.~Spieles {\it et al.},
  Phys.\ Lett.\  B {\bf 458} (1999) 137.
\bibitem{Chaudhuri:2003nj}
  A.K.~Chaudhuri,
  Phys.\ Rev.\  C {\bf 68} (2003) 037901.
\bibitem{Ko:1998fs}
  C.~M.~Ko, B.~Zhang, X.~N.~Wang and X.~F.~Zhang,
  Phys.\ Lett.\  B {\bf 444} (1998) 237.
\bibitem{BraunMunzinger:2000dv}
  P.~Braun-Munzinger and K.~Redlich,
  Eur.\ Phys.\ J.\  C {\bf 16} (2000) 519.
\bibitem{Gazdzicki:1999rk}
  M.~Gazdzicki and M.~I.~Gorenstein,
  Phys.\ Rev.\ Lett.\  {\bf 83} (1999) 4009.
\bibitem{Andronic:2006ky}
  A.~Andronic, P.~Braun-Munzinger, K.~Redlich and J.~Stachel,
  Nucl.\ Phys.\  A {\bf 789} (2007) 334.
\bibitem{vanHees:2007me}
  H.~van Hees, M.~Mannarelli, V.~Greco and R.~Rapp,
  Phys.\ Rev.\ Lett.\  {\bf 100} (2008) 192301.
\bibitem{Greco:2003vf}
  V.~Greco, C.~M.~Ko and R.~Rapp,
  Phys.\ Lett.\  B {\bf 595} (2004) 202.
\bibitem{Yan:2006ve}
  L.~Yan, P.~Zhuang and N.~Xu,
  Phys.\ Rev.\ Lett.\  {\bf 97} (2006) 232301.
\bibitem{Rafelski:1996hf}
  J.~Rafelski, J.~Letessier and A.~Tounsi,
  Acta Phys.\ Polon.\  B {\bf 27} (1996) 1037.
\bibitem{Bugaev:2001sj}
  K.~A.~Bugaev, M.~Gazdzicki and M.~I.~Gorenstein,
  Phys.\ Lett.\  B {\bf 523} (2001) 255.
\bibitem{Thews:2005vj}
  R.~L.~Thews and M.~L.~Mangano,
  Phys.\ Rev.\  C {\bf 73} (2006) 014904.
\bibitem{Adare:2006hc}
  A.~Adare {\it et al.}  [PHENIX Collaboration],
  Phys.\ Rev.\ Lett.\  {\bf 97} (2006) 252002.
\bibitem{Zhong:2007iq}
  C.~Zhong  [STAR Collaboration],
  J.\ Phys.\ G {\bf 34} (2007) S741.
\bibitem{Frixione:1997ma}
  S.~Frixione, M.~L.~Mangano, P.~Nason and G.~Ridolfi,
  Adv.\ Ser.\ Direct.\ High Energy Phys.\  {\bf 15} (1998) 609.
\bibitem{Lourenco:2006vw}
  C.~Lourenco and H.~K.~W\"ohri,
  Phys.\ Rept.\  {\bf 433} (2006) 127.
\bibitem{Rapp:2003vj}
  R.~Rapp and L.~Grandchamp,
  J.\ Phys.\ G {\bf 30} (2004) S305.
\bibitem{Andronic:2007zu}
  A.~Andronic, P.~Braun-Munzinger, K.~Redlich and J.~Stachel,
  Phys.\ Lett.\  B {\bf 659} (2008) 149.
\bibitem{Greco:2007sz}
  V.~Greco, H.~van Hees and R.~Rapp,
  arXiv:0709.4452 [hep-ph].
\bibitem{Gossiaux:2004qw}
  P.~B.~Gossiaux, V.~Guiho and J.~Aichelin,
  J.\ Phys.\ G {\bf 31} (2005) S1079.
\bibitem{Karsch:1987uk}
F.~Karsch and R.~Petronzio,
Phys.\ Lett.\  B {\bf 193} (1987) 105.
\bibitem{Blaizot:1987ha}
  J.~P.~Blaizot and J.~Y.~Ollitrault,
  Phys.\ Lett.\  B {\bf 199} (1987) 499.
\bibitem{Chu:1987sy}
  M.~C.~Chu and T.~Matsui,
  Phys.\ Rev.\  D {\bf 37} (1988) 1851.
\bibitem{Karsch:1990wi}
  F.~Karsch and H.~Satz,
  Z.\ Phys.\  C {\bf 51} (1991) 209.
\bibitem{Cugnon:1995fu}
  J.~Cugnon and P.~B.~Gossiaux,
  Phys.\ Lett.\  B {\bf 359} (1995) 375.
\bibitem{Zhang:2002ug}
  B.~Zhang, C.~M.~Ko, B.~A.~Li, Z.~W.~Lin and S.~Pal,
  Phys.\ Rev.\  C {\bf 65} (2002) 054909.
\bibitem{Linnyk:2006ti}
  O.~Linnyk, E.~L.~Bratkovskaya, W.~Cassing and H.~St\"ocker,
  Nucl.\ Phys.\  A {\bf 786} (2007) 183.
\bibitem{Linnyk:2007zx}
  O.~Linnyk, E.~L.~Bratkovskaya, W.~Cassing and H.~St\"ocker,
  Phys.\ Rev.\  C {\bf 76} (2007) 041901.
\bibitem{Bratkovskaya:2003ux}
  E.~L.~Bratkovskaya, W.~Cassing and H.~St\"ocker,
  Phys.\ Rev.\  C {\bf 67} (2003) 054905.
\bibitem{Capella:2007jv}
  A.~Capella, L.~Bravina, E.~G.~Ferreiro, A.~B.~Kaidalov, K.~Tywoniuk
  and E.~Zabrodin,
 Eur.\ Phys.\ J.\  C {\bf 58} (2008) 437.
\bibitem{Alessandro:2003pc}
  B.~Alessandro {\it et al.}  [NA50 Collaboration],
  Eur.\ Phys.\ J.\  C {\bf 33} (2004) 31.
\bibitem{Arnaldi:2010ky}
  R.~Arnaldi {\it et al.}  [NA60 Collaboration],
  arXiv:1004.5523 [nucl-ex].
\bibitem{Lourenco:2006sr}
  C.~Lourenco,
  Nucl.\ Phys.\  A {\bf 783} (2007) 451.
\bibitem{He:1999aj}
  Y.~B.~He, J.~H\"ufner and B.~Z.~Kopeliovich,
  Phys.\ Lett.\  B {\bf 477} (2000) 93.
\bibitem{Hufner:1997jg}
  J.~H\"ufner and B.~Z.~Kopeliovich,
  Phys.\ Lett.\  B {\bf 426} (1998) 154.
\bibitem{Matsui:1989ig}
  T.~Matsui,
  Annals Phys.\  {\bf 196} (1989) 182.
\bibitem{Cugnon:1993ye}
  J.~Cugnon and P.~B.~Gossiaux,
  Z.\ Phys.\  C {\bf 58} (1993) 77.
\bibitem{Cugnon:1993yf}
  J.~Cugnon and P.~B.~Gossiaux,
  Z.\ Phys.\  C {\bf 58} (1993) 95.
\bibitem{Koudela:2003yd}
  D.~Koudela and C.~Volpe,
  Phys.\ Rev.\  C {\bf 69} (2004) 054904.
\bibitem{Blaizot:1989de}
  J.~P.~Blaizot and J.~Y.~Ollitrault,
  Phys.\ Lett.\  B {\bf 217} (1989) 386.
\bibitem{Gavin:1990gm}
  S.~Gavin, R.~Vogt,
  Nucl.\ Phys.\  B {\bf 345} (1990) 104.
\bibitem{Arleo:1999af}
  F.~Arleo, P.~B.~Gossiaux, T.~Gousset and J.~Aichelin,
  Phys.\ Rev.\  C {\bf 61} (2000) 054906.
\bibitem{Arleo:2001nr}
  F.~Arleo, P.~B.~Gossiaux and J.~Aichelin,
  Phys.\ Rev.\  C {\bf 65} (2002) 054911.
\bibitem{Leitch:1999ea}
  M.~J.~Leitch {\it et al.}  [FNAL E866/NuSea collaboration],
  Phys.\ Rev.\ Lett.\  {\bf 84} (2000) 3256.
\bibitem{Faccioli:2006ty}
  P.~Faccioli  [HERA-B Collaboration],
  AIP Conf.\ Proc.\  {\bf 814} (2006) 545.
\bibitem{Ramello:2006db}
  L.~Ramello  [NA50 Collaboration],
  Nucl.\ Phys.\  A {\bf 774} (2006) 59.
\bibitem{Eskola:1998df}
  K.~J.~Eskola, V.~J.~Kolhinen and C.~A.~Salgado,
  Eur.\ Phys.\ J.\  C {\bf 9} (1999) 61.
\bibitem{Eskola:2001gt}
  K.~J.~Eskola, V.~J.~Kolhinen and R.~Vogt,
  Nucl.\ Phys.\  A {\bf 696} (2001) 729.
\bibitem{Accardi:2004be}
  A.~Accardi {\it et al.},
  arXiv:hep-ph/0308248.
\bibitem{deFlorian:2003qf}
  D.~de Florian and R.~Sassot,
  Phys.\ Rev.\  D {\bf 69} (2004) 074028.
\bibitem{Adare:2007gn}
  A.~Adare {\it et al.}  [PHENIX Collaboration],
  Phys.\ Rev.\  C {\bf 77} (2008) 024912.
\bibitem{Arleo:2006qk}
  F.~Arleo and V.~N.~Tram,
  Eur.\ Phys.\ J.\  C {\bf 55} (2008) 449.
\bibitem{Eskola:2008ca}
  K.~J.~Eskola, H.~Paukkunen and C.~A.~Salgado,
  JHEP {\bf 0807} (2008) 102.
\bibitem{Arsene:2004ux}
  I.~Arsene {\it et al.}  [BRAHMS Collaboration],
  Phys.\ Rev.\ Lett.\  {\bf 93} (2004) 242303.
\bibitem{Arleo:2008zc}
  F.~Arleo,
  Phys.\ Lett.\  B {\bf 666} (2008) 31.
\bibitem{Vitev:2003xu}
  I.~Vitev,
  Phys.\ Lett.\  B {\bf 562} (2003) 36.
\bibitem{Haberzettl:2007kj}
  H.~Haberzettl and J.~P.~Lansberg,
  Phys.\ Rev.\ Lett.\  {\bf 100} (2008) 032006.
\bibitem{Artoisenet:2009mk}
  P.~Artoisenet and E.~Braaten,
  Phys.\ Rev.\  D {\bf 80}, 034018 (2009).
\bibitem{Brodsky:2009cf}
  S.~J.~Brodsky and J.~P.~Lansberg,
  Phys.\ Rev.\  D {\bf 81}, 051502 (2010).
\bibitem{Lansberg:2010vq}
  J.~P.~Lansberg,
  arXiv:1003.4319 [hep-ph].
\bibitem{Khoze:2004eu}
  V.~A.~Khoze, A.~D.~Martin, M.~G.~Ryskin and W.~J.~Stirling,
  Eur.\ Phys.\ J.\  C {\bf 39}, 163 (2005).
\bibitem{Ferreiro:2008wc}
  E.~G.~Ferreiro, F.~Fleuret, J.~P.~Lansberg and A.~Rakotozafindrabe,
  Phys.\ Lett.\  B {\bf 680} (2009) 50.
\bibitem{Brodsky:1988xz}
  S.~J.~Brodsky and A.~H.~Mueller,
  Phys.\ Lett.\  B {\bf 206} (1988) 685.
\bibitem{Kopeliovich:1991pu}
  B.~Z.~Kopeliovich and B.~G.~Zakharov,
  Phys.\ Rev.\  D {\bf 44} (1991) 3466.
\bibitem{Blaschke:1992pw}
  D.~Blaschke and J.~H\"ufner,
  Phys.\ Lett.\  B {\bf 281} (1992) 364.
\bibitem{Kopeliovich:2003cn}
  B.Z.~Kopeliovich and J.~Raufeisen,
  Lect.\ Notes Phys.\  {\bf 647} (2004) 305.
\bibitem{Abt:2002vq}
  I.~Abt {\it et al.}  [HERA-B Collaboration],
  Phys.\ Lett.\  B {\bf 561} (2003) 61.
\bibitem{Abt:2006va}
  I.~Abt {\it et al.}  [HERA-B Collaboration],
  Eur.\ Phys.\ J.\  C {\bf 49} (2007) 545.
\bibitem{Leitch:2008vw}
  M.J.~Leitch  [PHENIX Collaboration],
  arXiv:0806.1244 [nucl-ex].
\bibitem{Acosta:2004yw}
  D.~E.~Acosta {\it et al.}  [CDF Collaboration],
  Phys.\ Rev.\  D {\bf 71} (2005) 032001.
\bibitem{Zhu:2004nw}
  X.~l.~Zhu, P.~f.~Zhuang and N.~Xu,
  Phys.\ Lett.\  B {\bf 607} (2005) 107.
\bibitem{Karsch:2007dt}
  F.~Karsch,
  PoS {\bf LAT2007} (2007) 015.
\bibitem{Spieles:1999kp}
 C.~Spieles, R.~Vogt, L.~Gerland, S.~A.~Bass, M.~Bleicher,
 H.~St\"ocker and W.~Greiner,
  Phys.\ Rev.\  C {\bf 60} (1999) 054901.
\bibitem{Arnaldi:2007zz}
  R.~Arnaldi {\it et al.}  [NA60 Collaboration],
  Phys.\ Rev.\ Lett.\  {\bf 99} (2007) 132302.
\bibitem{Digal:2003sg}
  S.~Digal, S.~Fortunato and H.~Satz,
  Eur.\ Phys.\ J.\  C {\bf 32} (2004) 547.
\bibitem{Rapp:2005rr}
  R.~Rapp,
  Eur.\ Phys.\ J.\  C {\bf 43} (2005) 91.
\bibitem{Capella:2005cn}
  A.~Capella and E.~G.~Ferreiro,
  Eur.\ Phys.\ J.\  C {\bf 42} (2005) 419.
\bibitem{Arnaldi:2006it}
  R.~Arnaldi {\it et al.}  [NA60 Collaboration],
  J.\ Phys.\ G {\bf 32} (2006) S51.
\bibitem{Lourenco:2005zz}
  C.~Lourenco {\it et al.}  [NA60 Collaboration],
  PoS {\bf HEP2005} (2006) 133.
\bibitem{PereiraDaCosta:2005xz}
  H.~Pereira Da Costa  [PHENIX Collaboration],
  Nucl.\ Phys.\  A {\bf 774} (2006) 747.
\bibitem{Gunji:2007uy}
  T.~Gunji, H.~Hamagaki, T.~Hatsuda and T.~Hirano,
  Phys.\ Rev.\  C {\bf 76} (2007) 051901.
\bibitem{Zhao:2010}
  X.~Zhao and R.~Rapp, in preparation (2010). 
\bibitem{Young:2008he}
  C.~Young and E.~Shuryak,
  Phys.\ Rev.\  C {\bf 79} (2009) 034907.
\bibitem{Zhuang:2003fu}
  P.~F.~Zhuang and X.~L.~Zhu,
  Phys.\ Rev.\  C {\bf 67} (2003) 067901.
\bibitem{Blaizot:1988ec}
  J.~P.~Blaizot, J.~Y.~Ollitrault,
  Phys.\ Rev.\  D {\bf 39} (1989) 232.
\bibitem{Gerland:1998bz}
  L.~Gerland, L.~Frankfurt, M.~Strikman, H.~St\"ocker and W.~Greiner,
  Phys.\ Rev.\ Lett.\  {\bf 81}, 762 (1998)
\bibitem{Blaizot:1988hh}
  J.~P.~Blaizot and J.~Y.~Ollitrault,
  Phys.\ Lett.\  B {\bf 217} (1989) 392.
\bibitem{Hufner:1988wz}
  J.~H\"ufner, Y.~Kurihara and H.~J.~Pirner,
  Phys.\ Lett.\  B {\bf 215} (1988) 218.
  [Acta Phys.\ Slov.\  {\bf 39} (1989) 281].
\bibitem{Gavin:1988tw}
  S.~Gavin and M.~Gyulassy,
  Phys.\ Lett.\  B {\bf 214} (1988) 241.
\bibitem{Abreu:2000xe}
  M.~C.~Abreu {\it et al.}  [NA50 Collaboration],
  Phys.\ Lett.\  B {\bf 499} (2001) 85.
\bibitem{Gorenstein:2001ti}
  M.~I.~Gorenstein, K.~A.~Bugaev and M.~Gazdzicki,
  Phys.\ Rev.\ Lett.\  {\bf 88} (2002) 132301.
\bibitem{Pal:2000zm}
  D.~Pal, B.~K.~Patra and D.~K.~Srivastava,
  Eur.\ Phys.\ J.\  C {\bf 17} (2000) 179.
\bibitem{Tang:2008uy}
  Z.~Tang [STAR Collaboration], Proc. of 24. Winter Workshop on Nuclear
  Dynamics, South Padre Island (TX, USA), April 05-12, 2008;
  arXiv:0804.4846 [nucl-ex].
\bibitem{Zhao:2008vu}
  X.~Zhao and R.~Rapp, Proc. of 24. Winter Workshop on Nuclear
  Dynamics, South Padre Island (TX, USA), April 05-12, 2008; 
  arXiv:0806.1239 [nucl-th].
\bibitem{Wang:2002ck}
  X.~N.~Wang and F.~Yuan,
  Phys.\ Lett.\  B {\bf 540} (2002) 62.
\bibitem{Lin:2003jy}
  Z.~W.~Lin and D.~Molnar,
  Phys.\ Rev.\  C {\bf 68} (2003) 044901.
\bibitem{vanHees:2005wb}
  H.~van Hees, V.~Greco and R.~Rapp,
  Phys.\ Rev.\  C {\bf 73} (2006) 034913.
\bibitem{Adare:2006nq}
  A.~Adare {\it et al.}  [PHENIX Collaboration],
  Phys.\ Rev.\ Lett.\  {\bf 98} (2007) 172301.
\bibitem{Ravagli:2007xx}
  L.~Ravagli and R.~Rapp,
  Phys.\ Lett.\  B {\bf 655} (2007) 126;
  L.~Ravagli, H.~van Hees and R.~Rapp,
  Phys.\ Rev.\  C {\bf 79} (2009) 064902.
\bibitem{Zhao:2008pp}
  X.~Zhao and R.~Rapp,
  Eur. Phys. J. C {\bf 62} (2009) 109.
\bibitem{Frawley:2009}
A.~Frawley, talk at int. workshop on ``Heavy Quarkonia Production
in Heavy-Ion Collisions, ECT* (Trento, Italy), May 25-29, 2009;
and in preparation.
\bibitem{Cacciari:2007xx}
M.~Cacciari, P.~Nason and R.~Vogt,
Phys. Rev. Lett. {\bf 95} (2005) 122001.
\bibitem{Adare:2006kf}
A.~Adare {\it et al.}  [PHENIX Collaboration],
  Phys.\ Rev.\ Lett.\  {\bf 98} (2007) 232002.
\bibitem{Karsch:2005nk}
  F.~Karsch, D.~Kharzeev and H.~Satz,
  Phys.\ Lett.\  B {\bf 637} (2006) 75.
\bibitem{Abreu:1998vw}
  M.C.~Abreu {\it et al.}  [NA50 Collaboration],
  Nucl.\ Phys.\  A {\bf 638} (1998) 261.
\bibitem{Sorge:1997bg}
H.~Sorge, E.~V.~Shuryak and I.~Zahed,
Phys.\ Rev.\ Lett.\  {\bf 79} (1997) 2775.
\bibitem{Linnyk:2008uf}
  O.~Linnyk, E.~L.~Bratkovskaya and W.~Cassing,
  Nucl.\ Phys.\  A {\bf 807} (2008) 79.
\bibitem{Morino:2008nc}
  Y.~Morino  [PHENIX Collaboration],
  J.\ Phys.\ G {\bf 35} (2008) 104116.
\bibitem{Cosentino:2008qn}
  M.~R.~Cosentino  [STAR Collaboration],
  arXiv:0806.0353 [nucl-ex].
\bibitem{Gunion:1996qc}
J.~F.~Gunion and R.~Vogt,
  Nucl.\ Phys.\  B {\bf 492} (1997) 301.
\bibitem{Affolder:1999wm}
  A.~A.~Affolder {\it et al.}  [CDF Collaboration],
  Phys.\ Rev.\ Lett.\  {\bf 84} (2000 2094.
\bibitem{Frawley:2008kk}
  A.~D.~Frawley, T.~Ullrich and R.~Vogt,
  Phys.\ Rept.\  {\bf 462} (2008) 125.
\bibitem{Carminati:2004fp}
  F.~Carminati {\it et al.}  [ALICE Collaboration],
  J.\ Phys.\ G {\bf 30} (2004) 1517.
\bibitem{Alessandro:2006yt}
  B.~Alessandro {\it et al.}  [ALICE Collaboration],
  J.\ Phys.\ G {\bf 32} (2006) 1295.
\bibitem{D'Enterria:2007xr}
  D.~G.~d'Enterria {\it et al.},
  J.\ Phys.\ G {\bf 34} (2007) 2307.
\bibitem{ATLAS-HI}
 ATLAS Collaboration, CERN/LHCC/{\bf 2004--009}.
\bibitem{CBM-HI}
CBM Collaboration, Technical Status Report (2005), 
{\tt http://www.gsi.de/fair/experiments/CBM/}
\bibitem{Wang2004dn}
        T.D.~Lee, Nucl. Phys. A 750 (2005) 1;
        T.~Ludlam, Nucl. Phys. A 750 (2005) 9;
        M.~Gyulassy and L.~McLerran, Nucl. Phys. A 750 (2005) 30;
        E.~Shuryak, Nucl. Phys. A 750 (2005) 64;
        B.~M\"uller, Nucl. Phys. A 750 (2005) 84;
        X.-N.~Wang, Nucl. Phys. A 750 (2005) 98;
        H.~St\"ocker, Nucl. Phys. A 750 (2005) 121;
        J.-P.~Blaizot and F.~Gelis, Nucl. Phys. A 750 (2005) 148.
\bibitem{Adams2005dq}
        I.~Arsene et al. (BRAHMS Collaboration), Nucl. Phys. A 757 (2005) 1;
        R.R.~Betts et al. (PHOBOS Collaboration), Nucl. Phys. A 757 (2005) 28;
        J.~Adams et al. (STAR Collaboration), Nucl. Phys. A 757 (2005) 102;
        K.~Adcox et al. (PHENIX Collaboration), Nucl. Phys. A 757 (2005) 184.
\bibitem{Ioffe:2003rd}
  B.~L.~Ioffe and D.~E.~Kharzeev,
  Phys.\ Rev.\  C {\bf 68} (2003) 061902.
\bibitem{Schukraft2001vg}
  J.~Schukraft,
  Nucl.\ Phys.\ A {\bf 698} (2002) 287.
\bibitem{Kharzeev:2004if}
  D.~Kharzeev, E.~Levin and M.~Nardi,
  Nucl.\ Phys.\  A {\bf 747} (2005) 609
\bibitem{Bedjidian:2003gd}
M.~Bedjidian {\it et al.},
arXiv:hep-ph/0311048.
\bibitem{Abreu:2007kv}
  N.~Armesto {\it et al.},
  J.\ Phys.\ G {\bf 35} (2008) 054001.
\bibitem{Xu:1995eb}
X.M.~Xu, D.~Kharzeev, H.~Satz and X.N.~Wang,
Phys.\ Rev.\ C {\bf 53} (1996) 3051.
\bibitem{Braun-Munzinger:2000px}
  P.~Braun-Munzinger and J.~Stachel,
  Phys.\ Lett.\ B {\bf 490} (2000) 196.
\bibitem{Andronic:2007bi}
  A.~Andronic, P.~Braun-Munzinger, K.~Redlich and J.~Stachel,
  Phys.\ Lett.\  B {\bf 652} (2007) 259.
\bibitem{Yao:2006px}
  W.~M.~Yao {\it et al.}  [Particle Data Group],
  J.\ Phys.\ G {\bf 33} (2006) 1.
\bibitem{Lokhtin:2001nh}
I.~P.~Lokhtin and A.~M.~Snigirev,
Eur.\ Phys.\ J.\ C {\bf 21} (2001) 155.
\bibitem{:2007fe}
  C.~Alt {\it et al.}  [NA49 Collaboration],
  Phys.\ Rev.\  C {\bf 77} (2008) 024903.
\bibitem{Adamova:2002kf}
  D.~Adamova {\it et al.}  [CERES/NA45 Collaboration],
  Phys.\ Rev.\ Lett.\  {\bf 91} (2003) 042301.
\bibitem{Claudia}
        C.~H\"ohne for the CBM Collaboration,
        Int. Journ. of Mod. Phys. E {\bf 16} (2008) 2419.

\end{thebibliography}
